\documentclass[10pt]{article}

\usepackage{graphicx}

\usepackage[all,cmtip]{xy}

\usepackage{setspace}
\usepackage{bbm,youngtab}
\usepackage{amscd,mathrsfs}
\usepackage{amssymb,amsmath,dsfont,amsfonts,amsthm}
\usepackage{euscript,enumerate}
\usepackage{stmaryrd}%for double brackets
\usepackage[cbgreek]{textgreek} %For non-italic greek letters
\usepackage{booktabs}
\usepackage{multirow}
\usepackage{hhline} %For double horizontal lines in tables
\usepackage{color}

\usepackage[bb=stix]{mathalpha}

\DeclareMathAlphabet{\mathpzc}{OT1}{pzc}{m}{it}

\allowdisplaybreaks %For breaking long equations and putting them on different pages

\usepackage{authblk}
\usepackage{natbib}
\usepackage{color, soul}
\usepackage{enumerate}
\usepackage{empheq}
\usepackage{rotating}
\usepackage{ftnxtra}
\usepackage{fnpos}
\usepackage[titletoc,toc,title]{appendix}
\usepackage{euscript}
\usepackage{graphicx}
\usepackage{epsfig}
\usepackage{epstopdf}
\DeclareGraphicsExtensions{.pdf,.png,.jpg,.eps}
\usepackage{pstool}
\usepackage{upgreek}
\usepackage{mathrsfs}
\usepackage{tikz-cd}

\usepackage{psfrag}

\numberwithin{equation}{section}

\theoremstyle{plain}	
\newtheorem{thm}{Theorem}[section]

\newtheorem*{prop*}{Proposition}
\theoremstyle{definition}	

\newtheorem{remark}[thm]{Remark}
\newtheorem{example}[thm]{Example}

\usepackage{caption}
\usepackage{subcaption}

\usepackage{hyperref}
\hypersetup{colorlinks=true, linkcolor=blue}
\hypersetup{colorlinks=true,citecolor=blue}

\DeclareMathAlphabet{\mathpzc}{OT1}{pzc}{m}{it}

\usepackage{amsmath, amsthm, amssymb}

\usepackage{cleveref}

\DeclarePairedDelimiter\abs{\lvert}{\rvert}
\makeatletter
\newsavebox{\@brx}
\newcommand{\llangle}[1][]{\savebox{\@brx}{\(\m@th{#1\langle}\)}%
  \mathopen{\copy\@brx\mkern2mu\kern-0.9\wd\@brx\usebox{\@brx}}}
\newcommand{\rrangle}[1][]{\savebox{\@brx}{\(\m@th{#1\rangle}\)}%
  \mathclose{\copy\@brx\mkern2mu\kern-0.9\wd\@brx\usebox{\@brx}}}%
\let\oldabs\abs
\def\abs{\@ifstar{\oldabs}{\oldabs*}}
\makeatother

\usepackage{tabularx,booktabs,caption,ragged2e}
\newcolumntype{L}{>{\RaggedRight\arraybackslash}X}
\usepackage{accents}
\newcommand{\Fe}{\accentset{e}{\mathbf{F}}}

\newcommand{\Ft}{\accentset{t}{\mathbf{F}}}
\newcommand{\Ce}{\accentset{e}{\mathbf{C}}}
\newcommand{\Go}{\mathring{\mathbf{G}}}

\newcommand{\GL}{\accentset{\mathcal{L}}{\mathbf{G}}}
\newcommand{\GS}{\accentset{\mathcal{B}}{\mathbf{G}}}

\usepackage{mathtools}

    {\end{bmatrix}}%

\usepackage[all,cmtip]{xy}
\usepackage{bm}

\usepackage{enumitem}

\begin{document}

\title{\textbf{Nonlinear mechanics of phase-change-induced accretion
}}

\author[1]{Satya Prakash Pradhan}
\author[1,2]{Arash Yavari\thanks{Corresponding author, e-mail: arash.yavari@ce.gatech.edu}}
\affil[1]{\small \textit{School of Civil and Environmental Engineering, Georgia Institute of Technology, Atlanta, GA 30332, USA}}
\affil[2]{\small \textit{The George W. Woodruff School of Mechanical Engineering, Georgia Institute of Technology, Atlanta, GA 30332, USA}}

\maketitle

%-----------------------------
%-----------------------------
\begin{abstract}
In this paper, we formulate a continuum theory of solidification within the context of finite-strain coupled thermoelasticity. We aim to fill a gap in the existing literature, as the existing studies on solidification typically decouple the thermal problem (the classical Stefan's problem) from the elasticity problem, and often limit themselves to linear elasticity with small strains. Treating solidification as an accretion problem, with the growth velocity correlated with the jump in the heat flux across the boundary, it presents an initial boundary-value problem (IBVP) over a domain whose boundary location is a priori unknown. This IBVP is solved numerically for the specific example of radially inward solidification in a spherical container. Several parametric studies are conducted to compare the numerical results with the rigid cases in the literature and gain insights into the role of elastic deformations in solidification.
\end{abstract}

\begin{description}
\item[Keywords:] Accretion, ablation, surface growth, nonlinear elasticity, thermoelasticity, phase change, solidification, Stefan's problem.
\end{description}

\tableofcontents

%-----------------------------
%-----------------------------
\section{Introduction}  \label{Sec:Introduction}

%-----------------------------
Various types of phase changes are observed in our surroundings, ranging from the freezing of seas \citep{stefan1891theorie} and the polymerization of proteins within living cells \citep{fedosejevs2022sharp,jiang2015phase} to the ongoing solidification process in the Earth's core \citep{buffett1992analytical,buffett1996thermal,buffett1993forced,labrosse1997cooling,labrosse2007crystallizing}. 
In engineering, phase transitions are highly relevant in various contexts, including concrete solidification \citep{bavzant1997microprestress}, the  shape memory effect observed in polymers and alloys \citep{zarek20163d, elahinia2016fabrication}, cryopreservation \citep{mazur1970cryobiology,coussy2005poromechanics}, as well as the applications of phase change materials in thermal energy storage and photonics \citep{pielichowska2014phase, wuttig2017phase}. Several theoretical studies comprehensively categorize all such phase transition phenomena that are observed in Nature \citep{landau1936theory,jaeger1998ehrenfest,binder1987theory,stanley1971phase}. Without delving into excessive detail, we specify that  in this work our focus is on the liquid-to-solid phase transition, which is classified as a first-order phase transition. These transitions are characterized by a finite discontinuity in the first derivative of the free energy with respect to a specific thermodynamic variable. In the case of solidification, this discontinuity manifests as a change in density, which can be heuristically related to the derivative of free energy with respect to pressure. Such transitions involve the release of latent heat while the temperature remains constant. This latent heat release causes a jump in the heat flux across the moving boundary, which is typically known as Stefan's condition. 

The term \emph{Stefan's problem} broadly refers to the family of mathematical models describing physical processes involving heat transfer, diffusion, and latent heat, which feature a moving boundary with an \emph{a priori} unknown location. The earliest known work in this field was a study conducted by \citet{lame1831memoire} on the cooling of a half-space filled with a homogeneous liquid at its solidification temperature. They demonstrated that the thickness of the solidified crust is proportional to the square root of time. However, it was when \citet{stefan1891theorie} published his work on the formation of ice in polar seas that this type of problem caught the attention of many researchers, and the field was named after him.
The history of what is now known as the Stefan's problem has been meticulously compiled in several texts \citep{rubinshteuin1971stefan,rubinstein1979stefan,danilyuk1985stefan,vuik1993some,visintin2008introduction,gupta2017classical}, all of which provide extensive and comprehensive bibliographies on the subject matter. Therefore, we do not attempt to provide a historical survey here. 

Over the past century, research on Stefan-type problems has predominantly fallen into the following categories: mathematical modeling of natural and engineering processes involving moving interfaces \citep{horvay1962freezing,chambre1956dynamics,crank1972moving}, investigations into the existence and uniqueness of solutions \citep{rubinstein1947solution,evans1951note,douglas1957uniqueness,MR0125341}, development of efficient numerical techniques for solutions \citep{lotkin1960calculation,melamed1958reduction,budak1965difference,fasano1979free} of problems with an unknown moving boundary, and generalizations such as extensions to higher dimensions.

\paragraph{Motivation of this study.} Solidification plays a vital role in several manufacturing processes that constitute the backbone of modern-day industries, such as traditional casting \citep{kou2015criterion}, injection molding \citep{isayev1984residual, yang2009injection}, selective laser sintering \citep{mercelis2006residual}, vat photopolymerization \citep{deore2021direct, bachmann2021cavity}, and ice-templating \citep{shao2020freeze}. 
However, within the setting of fully nonlinear and coupled thermoelasticity, there is a scarcity of studies addressing the mathematical modeling of deformations and stresses during the solidification process. Such modeling is of important for the design and analysis of manufacturing processes involving solidification, where molten materials cool to ambient temperatures. The substantial temperature drop in this process can result in severe part distortion and the development of high residual stresses. 
It is equally important to obtain the continuous evolution of thermal stresses and deformations throughout the manufacturing process to assess the potential occurrence of mechanical instabilities and failures \citep{debroy2018additive}. Residual stresses play a vital role, as they dictate how manufactured components respond to external stimuli, including service loads \citep{withers2001residualb}. Excessive residual and thermal stresses can give rise to issues such as layer delamination during deposition and the formation of cracks as the part cools down \citep{debroy2018additive}. Moreover, thermal contraction can distort parts made through these processes, affecting their geometric tolerance \citep{klingbeil2002residual}.
While many methods exist for measuring thermal stresses during fabrication or residual stresses post-fabrication, they typically measure the values at specific locations due to the cost and time constraints \citep{withers2001residuala}. Thus, understanding the continuous evolution of thermal stresses and residual stress distribution, whether through numerical or analytical tools \citep{mukherjee2017improved}, is critical for designing manufacturing processes to mitigate geometric inaccuracies, instabilities, and failures \citep{mukherjee2017mitigation}.

The aim of the present work is to analyze stress and deformation during solidification and their residual effects in a nonlinear thermoelastic framework. As new layers are deposited onto the surface of a solidifying body, it gives rise to an accretion problem. Accretion refers to the growth of a deformable body through the addition of material points on its boundary.
Drawing inspiration from \citet{Eckart1948} and \citet{Kondo1949}, a natural approach to modeling accreting bodies is to treat them as time-dependent Riemannian manifolds. The Riemannian metric for the new material points depends on the state of deformation at that point during the accretion process. If the source of anelasticity in the problem is time-independent; the metric at each point remains constant after attachment. However, in the case of thermoelastic accretion, this metric is temperature-dependent and therefore evolves with time at each material point.
The geometric theory of accretion was initially formulated by \citet{Sozio2017} for surface growth in cylindrical and spherical bodies. Several theoretical results related to accretion boundary-value problems were discussed in \citep{Sozio2019}. This theory was later extended by \citet{pradhan2023accretion} to include ablation, which refers to the removal of material points from the boundary.
Accretion of circular cylindrical bars under finite extension and torsion has been explored in studies by \citet{Yavari2023Accretion} and \citet{Yavari2022Torsion}. Further, \citet{Sozio2020} formulated a thermoelastic accretion boundary-value problem using the geometric theory of thermoelasticity proposed by \citet{ozakin2010geometric} and \citet{Sadik2017Thermoelasticity}. In their work, \citet{Sozio2020} modeled the effects of heat conduction and thermal expansion in an infinite cylinder and a $2$D block undergoing accretion through the addition of hot molten layers. However, the effect of phase transition was not taken into consideration, and the accretion surface velocities were assumed to be externally controlled. In this paper, we model accretion induced by solidification as a Stefan's problem, where the accretion velocity is a priori unknown. We take into account the effects of latent heat released during solidification, and the accretion velocity is related to the heat flux through Stefan's condition.

\paragraph{Existing literature.}
One of the earliest studies of solidification that focused on mechanical stresses was conducted by \citet{rongved1954residual}, who examined the residual stresses generated during the quenching of glass spheres.
He modeled the viscoelastic behavior of glass similar to that of a Maxwell material with temperature-dependent viscosity and provided an explicit solution for transient thermal stresses in a compressible sphere.
\citet{weiner1963elasto} studied the one-dimensional growth of an elastic/perfectly-plastic slab that started solidifying as the surface temperature of a molten liquid pool at one end was dropped below the melting point. The liquid melt was assumed to be at a fixed temperature initially. The time evolution and spatial variation of temperature in both phases were considered. They utilized Neumann's solution \cite[p.~283]{Carslaw1959} for the temperature field and the location of the moving boundary in one-dimensional phase change problems. The slab was assumed to have vanishing stress at the moving interface and was constrained against bending. Their problem was inspired by the early stages of solidification during the metal casting process where temperatures are close to the melting point. Their findings revealed that plastic flow can initiate right from the beginning on both the casting and solidification surfaces. Moreover, they observed that the stresses at the casting surface were compressive.  
%-----------------------------

\citet{chambre1956dynamics} conducted one of the earliest studies on the dynamics of liquid-to-solid phase change, considering the density changes induced during solidification. He considered the convective motion in the fluid near the interface, arising from the large density jumps across it, and modeled it using the incompressible Navier-Stokes equations at constant pressure. Further, he assumed the solid to be rigid and have infinite thermal conductivity so that it remained at the constant solidification temperature throughout the process. However, in the present work, we neglect inertial effects in both the solid and the liquid phases, while still considering a moderate density change across the solidification interface.
%-----------------------------

\citet{tien1969thermal} studied the thermal stresses developed during the solidification of an elastic beam with a temperature-dependent Young's modulus. \citet{richmond1971theory} considered a nonlinear viscoelastic model with a temperature-dependent Young's modulus and viscosity to study the early stages of solidification inside a rectangular mold with a uniform non-steady surface temperature and pressure. In particular, they computed the stresses and deformations in the solidifying skin for slow cooling processes and calculated the time required for the formation of air gap between the mold and the skin.
%-----------------------------
\citet{o1983boundary} used a boundary integral element method to study moving boundaries in phase change heat transfer problems. The analysis was limited to problems with a very low Stefan's number, meaning that the heat capacity effects were negligible compared to the latent heat effects. In such cases, the temperature profiles within the individual phases remain relatively constant over time. They investigated the radial freezing around a pipe with a thin initial frozen layer surrounded by the unfrozen liquid initially at the freezing temperature. The temperature history of the surface of the pipe was considered to be known and was assumed to decrease with time. They examined the evolution of the freezing front radius until it became considerably large compared to the pipe radius. They compared their numerical solution with the semi-analytical solution for phase change around an annulus with an infinitesimally small radius. Although the semi-analytical solution considered the transient heat equation in both phases while the numerical solution considered the steady state heat equation only in the frozen state, there was still good agreement between the two. They also studied the radial ablation of a pre-existing frozen layer around the same pipe, melting due to a specified impinging surface flux. Furthermore, they studied radially-asymmetric freezing around a cold pipe passing eccentrically through a drum containing fluid and compared their numerical solution with experimental results. However, they did not consider stresses due to solidification and heat transfer.

%-----------------------------
\citet{heinlein1986boundary} investigated solidification stresses generated during $1$D solidification of aluminum bars using the boundary element method. An aluminum  bar is assumed to be solidifying as it is chilled at one end where the temperature is given as a function of time. The other end of the bar is the moving solidification front, which is exposed to the hydrostatic pressure exerted by the liquid aluminum. They solved the $1$D transient heat equation by modifying the boundary integrals prescribed in \citep{o1983boundary}. For the elastic analysis, they worked in the setting of small strains and linear elasticity theory. They assumed an additive decomposition of the total strain into elastic, thermal and other non-elastic strains.
%-----------------------------
\citet{zabaras1987analysis} analyzed the motion of the phase-change interface in $2$D problems. They considered the inward solidification of a liquid melt initially at its melting temperature inside a square cavity whose surface is suddenly cooled down and maintained at a colder temperature.  They used the boundary element method to solve the transient heat equation. \citet{zabaras1990front} studied the evolution of deformations and thermal stresses induced during radially inward solidification of a hypoelastic-viscoplastic circular cylinder. They used finite elements that continuously move and deform to analyze boundary-value problems with an evolving domain. They assumed an additive decomposition of the strain rate into elastic and non-elastic components with a hydrostatic state of stress at the solidification interface. \citet{zabaras1991calculation} examined the residual stresses generated during axially-symmetric solidification of cylinders for different cooling conditions using the same FEM formulation.

%-----------------------------
Inspired by applications in cryobiology, \citet{rubinsky1980thermal} and \citet{rabin1998thermal} examined the stresses generated during inward freezing of a sphere. The stresses induced due to the freezing of water in the biological material can be a source of damage in the organ. \citet{rubinsky1980thermal} considered a homogeneous spherical organ, initially near its freezing temperature, which is frozen by the application of a constant cooling rate on its outer surface. They modeled ice as a perfectly elastic medium and computed the temperature and stress distributions.
\citet{rabin1998thermal} considered an inviscid liquid initially at its solidification temperature occupying a spherical domain whose outer surface is subsequently cooled and forcibly maintained at a fixed temperature. They regarded the frozen portion as an elastic/perfectly-plastic material, and conducted parametric studies to examine the mechanical stresses within the solid and the hydrostatic pressure within the fluid as the freezing front advances. They showed that in materials with physical properties resembling water, the stresses arising from thermal expansion in the solid state were notably lower in comparison to the stresses resulting from volumetric expansion during phase transition. 
They demonstrated that following the completion of the freezing process, a substantial portion of the frozen region is occupied by a plastic zone. They concluded that the potential for tissue destruction was inevitable, regardless of the speed at which the freezing process was conducted, as long as there was a substantial expansion associated with phase transition.

%-----------------------------
\citet{chan2006solidification} conducted experiments to study the solidification of $n-$hexadecane inside a sphere by keeping the surface temperature constant. They observed that the solidification front starts to propagate inward in a spherically-symmetric fashion. Later, the phase-change interface loses its spherical symmetry and develops some irregularity/eccentricity as the shrinkage in the solidified material causes the formation of voids. The rate of solidification is very high initially and reduces subsequently. However, they did not consider stresses generated during the process.
%-----------------------------
Numerous studies have explored the inward solidification of a spherical liquid domain initially at its freezing temperature \citep{pedroso1973inward,riley1974inward,stewartson1976stefan,soward1980unified}. However, their main focus was to improve the approximation of the temperature profile as the phase change interface neared the center of the sphere. Another example of such a study is the asymptotic analysis conducted by \citet{mccue2003extinction} who investigated the $2$D inward solidification of a melt within a rectangular domain at its fusion temperature. For a large Stefan's number, they computed the time required for complete solidification and observed that the phase change interface forms an exact ellipse as it approaches the center. In none of these studies, mechanical stresses were taken into account in their analyses.
%-----------------------------
\citet{pedroso1973state} studied the stresses generated during the inward solidification of spheres. The state of stress at the freezing front was assumed to be hydrostatic, determined by the corresponding pressure in the fluid, and the stress inside the solid was modeled using linear isotropic thermoelasticity equations. They showed that the solid is residually stressed after the inner liquid pressure and outer tractions were removed. They also investigated the effects of different liquid compressibility, freezing temperature, and liquid pressure.
%-----------------------------

\citet{abeyaratne1993continuum} investigated solid-solid phase transitions in a one-dimensional domain by deriving a kinetic relation for the motion of the phase-change interface that allowed them to be influenced by local stress states. This kinetic equation related the interface velocity to the thermodynamic driving force using the second law of thermodynamics. They also analyzed the onset of thermally or mechanically induced phase transitions in thermoelastic solids via a nucleation criterion. 
\citet{Tomassetti2016} studied accretion-ablation induced by diffusion. They considered a thick permeable spherical shell that has grown on the surface of a rigid spherical substrate. The spherical shell is surrounded by a fluid medium with free particles that diffuse into the permeable shell to reach the surface of the spherical substrate where they polymerize and attach to the spherical shell. As this accretion occurs at the fixed inner boundary some of the particles on the outer boundary are ablated out into the fluid medium. Accretion and ablation are governed by the following factors: strain energy of the solid shell, external mechanical power, and the difference in chemical potential of the particles when they are free as compared to when they are attached to the solid shell. The driving force---a measure of deviation from thermodynamic equilibrium---is assumed to have a linear relationship with the flux of particles at the accretion-ablation boundaries. The accretion-ablation rates are thermodynamically determined in terms of the chemical potential of the particles and strain energy density of the shell. They studied accretion-ablation in a treadmilling regime and simply considered the steady state solution of the diffusion equation. A more general analysis would take into account the time-evolution of particle flux and stresses in the solid due to transient diffusion.
%-----------------------------
\citet{abeyaratne2022surface} studied the stability of a pre-stressed elastic $2$D-half space accreting due to steady-state diffusion of free particles from the other half space. They reported that such surface growth of a half space with surface tension is not always stable if the accretion interface is traction-free. \citet{abeyaratne2022stability} examined the stability of a similar prestressed elastic half space accreting by diffusion, while the other half space containing the free particles is assumed to be compliant and provide some resistance to growth.
%-----------------------------

\citet{fekry2023thermal} examined the evolution of stresses in a thermoviscoelastic cylinder manufactured via the process of selective laser melting. He modeled the process of additive manufacturing as the accretion of discrete layers on the cylindrical boundary. 
Furthermore, \citet{lychev2023evaluation1,lychev2023evaluation2} studied the evolution of temperature and stress, as well as residual stresses and distortions in a thermoelastic cylindrical bar manufactured by lateral sintering. In the context of small deformations and temperature gradients, they formulated discrete accretion as a recursive problem in terms of strain and stress increments.
However, the effects of latent heat during solidification was not considered in these works.

%-----------------------------
%\todo{These are some of the the works related to the interface mechanics that discuss Stefan's condition:\\
%\citet{fernandez1979generalized} used the surface thermodynamics theory developed by \citet{murdoch1976thermodynamical} to derive jump conditions across a melting surface, specifically Stefan's condition for the heat flux. They postulated the interface to posses surface tension and some stored energy that could induce heat transfer within it, all while disregarding bulk velocity and jump in the density change across the interface. \\
%\cite{visintin2009phase}\\
%\cite{visintin1985stefan}\\
%\cite{baldoni1997conditions}: They arrive at the Clapeyron's equation using the Hadamard compatibility condition for the Gibbs potential across the solid-liquid interface.  \\
%\cite{gill1981rapid}: \\
%}

\citet{rejovitzky2015theory} formulated a continuum theory to study the stresses generated during the deposition of solid electrolyte interphase layers, which play a significant role in the degradation of Li-ion batteries. 
Based on the experimental results of \citet{smith2011interpreting}, they assumed the thickness of the accumulated layer to be proportional to the square root of time, thus avoiding the complexity of obtaining it through the use of the diffusion equation and reaction kinetics. 
They modeled the electrolyte as a linear elastic material with a small Young's modulus and vanishing Poisson's ratio. The state of the attaching layers at the time of deposition was considered as the stress-free reference configuration for the points within it. 
The deformation gradient of the attaching layer, relative to this configuration, was then decomposed multiplicatively into growth and elastic parts, with the growth part corresponding to the inelastic deformation induced within each layer during attachment. 
They demonstrated the capability of their formulation by simulating the evolution of Cauchy stress in the deposited layers during charging and discharging cycles on deformable spheroidal anodes.
%-----------------------------

Polymerization is an example where phase transformation occurs as the result of an exothermic reaction converting a partially cured gel to fully cured polymer. \citet{kumar2021analytical} provided analytical estimates for the velocity of the reaction front propagating steadily in a $1$D adiabatic domain. \citet{kumar2022surface} studied the evolution of mechanical stresses and large deformations that are induced due to phase transformation by polymerization. Their thermo-chemo-mechanical model involves a coupled system of the following equations: the balance of linear momentum, the transient heat equation and the reaction kinetics, where the unknowns are the deformation field, the temperature field, and the degree of cure. The reaction kinetics are assumed to be unaffected by the mechanical deformations while the thermo-mechanical coupling is evident in their formulation. They considered the example of a $2$D adiabatic domain and observed that the reaction interface travels at an almost constant speed.
%-----------------------------
\citet{li2024mechanical} examined the propagation of reaction fronts in the process of polymerization, where a minimal energy input transforms monomers at a soft gel-like state to a stiffer solid polymer. In a slender one-dimensional body under axial load, they studied the influence of mechanical properties on the propagation of the reaction front, considering the effects of thermal expansion and density changes resulting from the reaction. Using both experimental and theoretical analyses, they demonstrated that the propagation of the reaction front can be quenched by the application of mechanical loads, establishing a clear thermo-mechanical coupling. In particular, they observed that below a critical applied load, the reaction front moves at an almost constant speed, but slows down abruptly above this critical load.

%-----------------------------
\paragraph{Problem overview.}In this paper, we consider the solid and liquid phases as homogeneous, isotropic, compressible, hyperelastic materials that are rigid heat conductors. We neglect the inertial effects in both phases. Additionally, we do not account for the influence of pressure on the phase change temperature. To be specific, our study focuses on the inward solidification of a liquid inclusion initially at its solidification temperature, trapped within a deformable solid body that is being externally cooled, with both phases composed of the same material. 
For such problems, we calculate the evolution of deformation, stresses, and temperature field inside the solid, as well as the location of the phase change front as it progresses inward. 
We consider the solidification process until the radius of the inclusion reaches a certain small value. This is because surface stresses are known to dominate when the inclusion size decreases beyond a certain limit \citep{Bico2018}. Since surface stresses have not been considered in this work, the numerical solutions that are too close to the center would be physically irrelevant. Furthermore, in materials where the liquid phase is denser than the solid phase near the melting point, the pressure in the liquid inclusion induced by compressive stresses significantly increases as the phase change front approaches the center of the sphere. Therefore, the accretion process is terminated with a certain time margin prior to achieving full solidification. Finally, the resulting body is detached from the rigid container, drained of any remaining liquid, and then cooled to an ambient temperature. The residual stresses and distortions are subsequently computed for this configuration.

%-----------------------------
\paragraph{Outline.}
This paper is organized as follows. First, the general theory of thermoelastic accretion is described in \S2. The balance laws---including the conservation of mass, linear and angular momenta, the heat equation, and Stefan's condition---are discussed in \S3. In \S4, radially inward solidification in a cold rigid container is modeled as a thermoelastic accretion problem with an unknown accretion velocity, and the numerical results for the corresponding non-dimensionalized moving boundary value problem are discussed in detail. Finally, conclusions are given in \S5.

%-----------------------------
\section{Thermoelastic accretion induced by phase change} \label{Sec:thermoelasticaccretion}
%-----------------------------

This section provides a concise overview of nonlinear thermoelasticity, the mechanics of accretion and the application of Stefan's condition in solidification problems. 
For a thorough analysis of geometric thermoelasticity, see \citep{ozakin2010geometric,Sadik2017Thermoelasticity}. 
In-depth insights into accretion mechanics are available in \citep{Sozio2019}. For a comprehensive understanding of the Stefan's problem the reader is referred to the texts by \citet{rubinshteuin1971stefan} and \citet{gupta2017classical}.

Consider the phase transition of a finite quantity of liquid undergoing cooling and solidification, either within a rigid container or as an inclusion within a deformable solid. 
As the liquid solidifies and attaches to the surface of the container or the deformable body, the body grows via accretion. 
In other words, the solidifying body undergoes accretion and the adjacent fluid undergoes ablation, while the material points in the solid-liquid system as a whole remain conserved. 

Let $\mathcal{S}$ denote the three-dimensional ambient Euclidean space, with $\mathbf{g}$ representing its standard flat metric. 
Both the solid and liquid phases assume their respective deformed configurations endowed with this ambient Euclidean metric. 
Those parts of the solid-liquid (pair) composite body which remain unaffected by phase transformation are equipped with a temperature-dependent metric, which is flat at the initial temperature. 
The individual reference configurations of the solid and liquid phases evolve as  material points are transferred from one phase to the other. 
The material metric for an accreting layer is a priori an unknown field and is determined by its temperature and state of deformation at the time of attachment.

%-----------------------------
%-----------------------------
\subsection{The solidifying body}

\begin{figure}[t!]
\centering
\vskip 0.0in
\includegraphics[width=.64\textwidth]{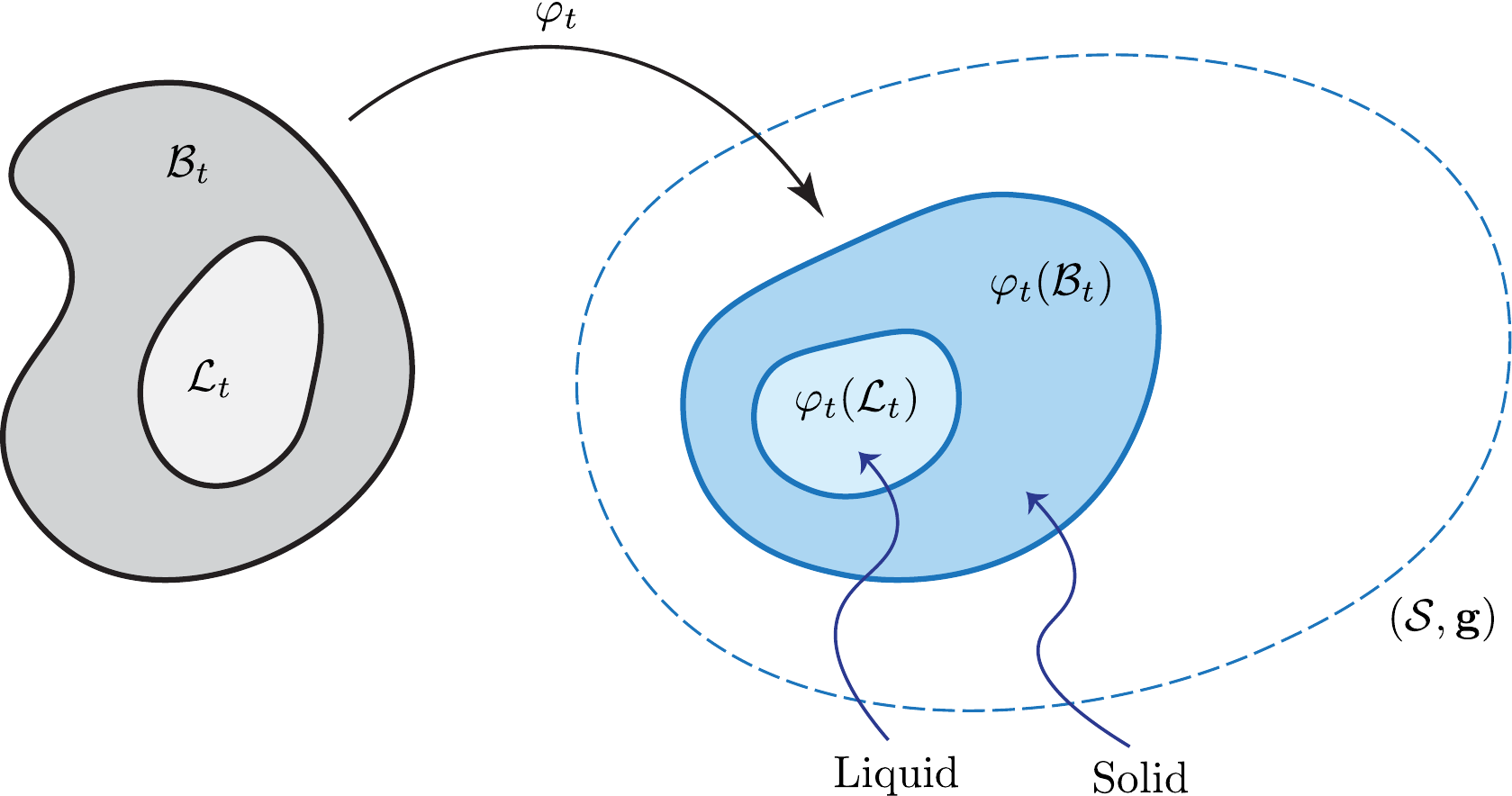}
\vskip 0.1in
\caption{Deformation of an elastic solid with a liquid inclusion.}
\label{fig:twophase}
\end{figure}

Consider a solid body $\mathcal{B}_0$ with a liquid inclusion $\mathcal{L}_0$, both initially stress-free.\footnote{The effect of gravity is neglected, and hence, there is no pressure caused by the self-weight of the liquid.} 
The initial solid-liquid body $\mathcal{Z}=\mathcal{B}_0 \cup \mathcal{L}_0$ inherits a flat metric from the ambient Euclidean space. Assume that solidification (accretion) begins at $t=0$. Let $\mathcal{M}  \supset \mathcal{Z}$ denote the ambient material space, which is a connected and orientable three-dimensional manifold embeddable in $\mathbb{R}^3$. 
Let the map $\tau: \mathcal{L}_0 \rightarrow [0,\infty)$ assign a time of solidification (attachment) to every fluid point. 
The accreting solid and the ablating fluid are identified with their respective time-dependent material manifolds $\mathcal{B}_t$ and  $\mathcal{L}_t$ (Figure \ref{fig:twophase}). 
They are defined as follows
%-----------------------------
\begin{equation}
 	\mathcal{B}_t
	 = \mathcal{B}_0 \cup \tau^{-1}[0,t] \,, \qquad
  	\mathcal{L}_t
  	= \mathcal{L}_0 \setminus \tau^{-1}[0,t)\,.
\end{equation}
%-----------------------------
Note that $\mathcal{B}_t \cup \mathcal{L}_t=\mathcal{B}_0 \cup \mathcal{L}_0$, although $(\mathcal{B}_t , \mathcal{L}_t) \neq (\mathcal{B}_0, \mathcal{L}_0)$. It is assumed that the differential $\text{d}\tau$ never vanishes. Let $\Omega_t \subset \partial \mathcal{B}_t $ be the accretion surface where the solidifying material is about to attach. The level sets $\Omega_t=\tau^{-1}(t)$ are assumed to be $2-$manifolds, which are diffeomorphic to each other for all $t\geq 0$. This assumption ensures the existence of a material motion.

%-----------------------------
%-----------------------------
\subsection{Kinematics of accretion}

For an accreting body, the deformation map $\varphi_t : \mathcal{B}_t \rightarrow \mathcal{S}$ is assumed to be a $C^1$ homeomorphism for each $t$. 
The deformation gradient $\mathbf{F}_t=T\varphi_t $ is a two-point tensor $\mathbf{F}_t : T_X \mathcal{B}_t \rightarrow T_ {x} \mathcal{C}_t$, where $x=\varphi_t(X)$ and $\mathcal{C}_t=\varphi_t(\mathcal{B}_t)$.\footnote{ Let $\{X^A\}$ and $\{x^a\}$ be local coordinate charts on $\mathcal{B}_t$ and $\mathcal{C}_t\subset \mathcal{S}$, respectively. The deformation gradient is represented as
%-----------------------------
\begin{equation} \label{Deformation-Gradient}
    \mathbf{F}
    =\frac{\partial \varphi^a}{\partial X^A}\,\frac{\partial}{\partial x^a}\otimes dX^A\,.
\end{equation}
%-----------------------------
We use the flexible notations $\varphi_t(X)=\varphi(X,t)$ and  $\mathbf{F}_t(X)=\mathbf{F}(X,t)$. Note that  $\big\{\frac{\partial}{\partial x^a}\big\}$ and $\{dX^A\}$ form the bases for $T_x\mathcal{C}_t$ and $T^*_X\mathcal{B}_t$, respectively.} 
The material and spatial velocity fields are defined as $\mathbf{V}(X,t)=\frac{\partial}{\partial t} \varphi(X,t)$ and $\mathbf{v}_t =\mathbf{V}_t \circ \varphi_t^{-1}$, respectively. 
Similarly, the material and spatial acceleration fields are defined as $\mathbf{A}(X,t)=\frac{\partial}{\partial t} \mathbf{V}(X,t)$ and $\mathbf{A}_t =\mathbf{a}_t \circ \varphi_t^{-1}$, respectively.\footnote{ In components, $A^a=\frac{\partial V^a}{\partial t}+\gamma^a{}_{bc}V^bV^c$ and $a^a=\frac{\partial v^a}{\partial t}+\frac{\partial v^a}{\partial x^b}v^b+\gamma^a{}_{bc}v^bv^c$. Here, $\gamma^a{}_{bc}$ denote the Christoffel symbols for the Levi-Civita connection $\nabla^\mathbf{g}$, i.e., ${\nabla^\mathbf{g}}_{\frac{\partial}{\partial x^b}}\,\frac{\partial}{\partial x^c}=\gamma^a{}_{bc}\,\frac{\partial}{\partial x^a}$. Similarly, the Christoffel symbols of $\nabla^\mathbf{G}$ are denoted as $\Gamma^A{}_{BC}$, i.e., ${\nabla^\mathbf{G}}_{\frac{\partial}{\partial X^B}}\,\frac{\partial}{\partial X^C}=\Gamma^A{}_{BC}\,\frac{\partial}{\partial X^A}$.}

The map $\bar{\varphi} (X) = \varphi(X,\tau(X))$ records the placement of $X$ at its time of attachment. 
In general, $\bar{\varphi}: \mathcal{L}_0 \to \mathcal{S}$ is not injective. 
Moreover, the \emph{frozen deformation gradient} $\bar{\mathbf{F}}(X)=\mathbf{F}_{\tau(X)}(X)$, which captures the deformation gradient at the time of attachment, is not the tangent of an embedding, in general. 
Even when $\bar{\varphi}$ is an embedding, $T\bar{\varphi}$ is not equal to $\bar{\mathbf{F}}$. 
In fact $T\bar{\varphi}=\bar{\mathbf{F}}+\bar{\mathbf{V}}\otimes d\tau$. 
While $\bar{\mathbf{F}}$ is compatible within each individual layer $\Omega_t$, it is incompatible, in general. 
The incompatibility of the $\bar{\mathbf{F}}$ is the fundamental reason behind the existence of local anelastic distortions in accreting bodies, and hence the presence of residual stresses.  

Let $\omega_t = \varphi_t(\Omega_t)$ denote the accretion surface in the deformed configuration. 
The growth (accretion) velocity $\mathbf{u}_t$ is a vector field that describes the velocity at which new material is being added onto $\omega_t$, i.e., $-\mathbf{u}_t$ is the velocity of accreting particles relative to $\omega_t$ just before attachment. 
The material growth velocity, denoted as $\mathbf{U}_t$, is a vector field that characterizes the time evolution of the layers $\Omega_t$ within the material ambient space. 
The vector field $\mathbf{U}_t$ is not uniquely determined and can be selected from an equivalence class of material growth velocities that correspond to isometric material manifolds. 
Let $\mathbf{w}_t$ and $\mathbf{W}_t$ be the spatial and referential depictions of the total velocity of the accretion surface $\omega_t$, i.e., $\mathbf{W}_t \circ \varphi_t =\mathbf{w}_t $. 
It can be shown that $\mathbf{W}_t=\bar{\mathbf{F}}\mathbf{U}_t + \mathbf{V}_t$, where the term $\bar{\mathbf{F}}\mathbf{U}_t $ accounts for the influence of accretion. 

The accretion-induced anelasticity is modeled by the accretion tensor $\mathbf{Q}$, which is a time-independent two-point tensor, defined as
%----------------
\begin{equation} \label{eq:Qdef}
	\mathbf Q (X) 
	= \bar{\mathbf{F}} (X) 
	+ \left[ \mathbf u_{\tau(X)}(\bar\varphi(X)) 
	- \bar{\mathbf{F}}(X) \mathbf U_{\tau(X)}(X) \right] 
	\otimes \mathrm d \tau(X) \,,\qquad  X\in \mathcal{L}_0\,.
\end{equation}
%----------------
Since $\langle d\tau,\mathbf{U} \rangle=1$, it follows that $\mathbf Q \mathbf U = \mathbf u$. 
Although the accretion tensor $\mathbf Q$ is compatible within each individual layer, it is not the tangent map of any embedding. 
For more details, see \citep{Sozio2019}.
%-----------------------------
\begin{remark}
%---------------------
Let us consider a foliation chart $\{\Xi^1, \Xi^2, \tau\}$ induced by the time of attachment map $\tau$ in the ambient material manifold $\mathcal{M}$ and a local chart $\{x^1,x^2,x^3\}$ in the ambient Euclidean manifold $\mathcal{S}$. The accretion tensor has the following representation with respect to the frames $\left\{\frac{\partial}{\partial \Xi^1},\frac{\partial}{\partial \Xi^2},\frac{\partial}{\partial \tau} \right\}$ and $\left\{\frac{\partial}{\partial x^1},\frac{\partial}{\partial x^2},\frac{\partial}{\partial x^3}\right\}$ \citep{Sozio2019}
%-----------------------------
\begin{equation}
	\left[ {Q^i}_J  \right]
	=
	\begin{bmatrix}
    \frac{\partial \bar{\varphi}^1 }{\partial \Xi^1 } & \frac{\partial \bar{\varphi}^1}{\partial \Xi^2 }  & u^1_{\tau}\circ \bar{\varphi}  \\[5 pt]
    \frac{\partial \bar{\varphi}^2}{\partial \Xi^1 } &  \frac{\partial \bar{\varphi}^2}{\partial \Xi^2 }  &  u^2_{\tau}\circ \bar{\varphi}  \\[5 pt]
    \frac{\partial \bar{\varphi}^3}{\partial \Xi^1 } &  \frac{\partial \bar{\varphi}^3}{\partial \Xi^2 }  &  u^3_{\tau}\circ \bar{\varphi}    \end{bmatrix} \,.
\end{equation}
%-----------------------------
%------------
\end{remark}
%-----------------------------
%---------------------
%---------------------

%-----------------------------
%-----------------------------
\subsection{Material metric for thermoelastic accretion} 

Consider a time-dependent material manifold $(\mathcal{B}_t,\mathbf{G})$, where the metric $\mathbf{G}$ measures distances corresponding to the relaxed state, taking into account the thermal history of the body. In geometric thermoelasticity, the metric $\mathbf{G}$ is a function of temperature $T(X,t)$, and is given by \citep{Sadik2017Thermoelasticity,Sozio2020}
%-----------------------------
\begin{equation} \label{tempmetrics}
\mathbf{G}(X,T)=e^{\bm{\omega}^\star(X,T)} \mathbf{G}_0(X)e^{\bm{\omega}(X,T)} \,, \end{equation}
%-----------------------------
where $\bm{\omega}$ is a $1 \choose 1$-tensor characterizing thermal expansion properties in the solid and $\mathbf{G}_0$ is a temperature independent metric.\footnote{The adjoint of deformation gradient $\mathbf{F}^{\star}(X,t): T^*_x\mathcal{C}_t\to T^*_X\mathcal{B}_t$ is defined such that 
%---------------------
\begin{equation} 
	\langle \boldsymbol{\alpha},\mathbf{F}\mathbf{W}\rangle
	=\langle \mathbf{F}^{\star}\boldsymbol{\alpha},\mathbf{W}\rangle\,,\quad 
	\forall\; \mathbf{W}\in T_X\mathcal{B}_t\,,~\boldsymbol{\alpha}\in T^*_x \mathcal{C}_t
	\,,
\end{equation}
%---------------------
where $\langle\,.\,,.\,\rangle$ is the natural paring of $1$-forms in $T^*_x \mathcal{C}_t$ with vectors in $T_x \mathcal{C}_t$, i.e., $\langle \boldsymbol{\alpha},\mathbf{w}\rangle=\alpha_a\,w^a$. $\mathbf{F}^{\star}$ is a $1 \choose 1$-tensor with the following coordinate representation 
%---------------------
\begin{equation} 
	\mathbf{F}^{\star}=\frac{\partial\varphi^a}{\partial X^A} \,dX^A 
	\otimes \frac{\partial}{\partial x^a}\,.
\end{equation}
%---------------------
}
The volumetric coefficient of thermal expansion $\beta(X,T)$ is given by
%-----------------------------
\begin{equation}
 \beta(X,T)
 = \frac{\partial}{\partial T} \,\text{tr}\,\bm{\omega}(X,T) \,.
\end{equation}
%-----------------------------
For a time-independent reference temperature field $T_0(X)$, it is assumed that $\bm{\omega}\big(X,T_0(X)\big)=\mathbf{0}$, and hence $\mathbf{G}\big(X,T_0(X)\big)=\mathbf{G}_0(X)$. In the thermally accreted part of the body, $T_0(X)$ is assumed to be the temperature of the attached material at its time of attachment. However, in the initial body $\mathcal{B}_0$, $T_0(X)$ represents the initial temperature. The material metric $\mathbf{G}\big(X,T_0(X)\big)$ for the accreted portion  is calculated by pulling back the Euclidean ambient metric $\mathbf{g}$ via the accretion tensor $\mathbf{Q}$: 
%-----------------------------
\begin{equation}
	\mathbf{G}\big(X,T_0(X)\big)
	=\mathbf{Q}^\star(X) \mathbf{g}\big(\bar{\varphi}(X)\big)\mathbf{Q}(X)\,.
\end{equation}
%-----------------------------
The temperature-dependent material metric is therefore given by 
%-----------------------------
\begin{equation}
	\mathbf{G}(X,T)
	=e^{\bm{\omega}^\star(X,T)} \mathbf{Q}^\star(X) 
	\mathbf{g}\big(\bar{\varphi}(X)\big)\mathbf{Q}(X)e^{\bm{\omega}(X,T)} \,.
\end{equation}
%-----------------------------
Let $\text{d}v$ and $\text{d}V$ denote the spatial and material volume elements, respectively. They are related via the Jacobian as $\text{d}v=J\,\text{d}V$, where
%-----------------------------
\begin{equation}
	J 
	= \sqrt{\frac{\det \mathbf{g}}{\det \mathbf{G}} } \det \mathbf{F}
	= \sqrt{\frac{\det \mathbf{g}}{\det \mathbf{g}\circ \bar{\varphi}} } \,
	\frac{\det \mathbf{F}}{\det \mathbf{Q}} e^{-\text{tr}\,\bm{\omega}} \,.
\end{equation}
%-----------------------------
\begin{remark}
%-----------------------------
For a thermally isotropic and homogeneous body, \eqref{tempmetrics} is simplified as
%-----------------------------
\begin{equation}
	\mathbf{G}(X,T)
	=e^{2 \omega (T)} \mathbf{G}_0(X) \,,
\end{equation}
%----------------------------- 
where the scalar $ \omega (T)$ is related to the coefficient of thermal expansion $\alpha(T)$ as 
%-----------------------------
\begin{equation}
	\omega\big(T(X)\big)
	=\int_{T_0(X)}^{T(X)} \alpha (\eta)\, \text{d}\eta\,.
\end{equation}
%----------------------------- 
The volumetric coefficient of thermal expansion in dimension three is $\beta=3 \alpha$.
%-----------------------------
\end{remark}
%-----------------------------
%-----------------------------
\begin{remark}
 Let $\mathbf{n}_{\tau}$ and $\mathbf{N}_{\tau}$ denote the unit normals to $\omega_{\tau}$ and $\Omega_{\tau}$, with respect to the metrics $\mathbf{g}$ and $\mathbf{G}$, respectively. The growth (accretion) velocities in the deformed and material configurations can be decomposed as follows
%-----------------------------
\begin{equation}
	\mathbf{u}= \mathbf{u}^\parallel+ u^n \, \mathbf{n} \,,\qquad
	\mathbf{U}= \mathbf{U}^\parallel + U^N  \,\mathbf{N}\,, 
\end{equation}
%-----------------------------
where $\llangle \mathbf{u}^\parallel,\mathbf{n} \rrangle_{\mathbf{g}}=\llangle \mathbf{U}^\parallel,\mathbf{N} \rrangle_{\mathbf{G}}=0$. Moreover, $\mathbf{Q}\mathbf{U}^\parallel=\mathbf{u}^\parallel$, $\mathbf{Q}\mathbf{N}=\mathbf{n}$ and $U^N=u^n\circ\varphi >0$ \citep{Sozio2019}.
%------------
\end{remark}
%-----------------------------

%-----------------------------
%-----------------------------
\section{Balance laws}
\label{Sec:balancelaws}
%-----------------------------
%-----------------------------
\subsection{Conservation of mass}

Let the material and spatial mass densities be denoted by $\rho (X,t)$ and $\varrho(x,t)$, respectively. Let $\rho_0(X)$ represent the material mass density corresponding to the flat metric $\mathbf{G}_0$. The mass of a sub-body $\mathcal{U}\subset \mathcal{P}_t$ is calculated as
%-----------------------------
\begin{equation}  \label{eq:MassCons}
	\int_{\varphi_t(\mathcal{U})} \varrho\, \text{d}v
  = \int_{\mathcal{U}} \rho\, \text{d}V
  =\int_{\mathcal{U}} \rho_0 \text{d}V_0
	 \,,
\end{equation}
%-----------------------------
where $\text{d}V_0$ is the volume element corresponding to the stress-free material metric, and is related to $\text{d}V$ as $\text{d}V=\sqrt{\frac{\det \mathbf{G}}{\det \mathbf{G}_0} } \, \text{d}V_0 $. The mass densities are related as $J \varrho	 = \rho$ and $\rho_0 \sqrt{\det \mathbf{G}_0} = \rho \sqrt{\det \mathbf{G}}$, i.e.,
%-----------------------------
\begin{equation}
	 J(X,t) \,\varrho\big(\varphi(X,t),t\big) 
	 = \rho(X,t)  
	 =e^{-\operatorname{tr}\bm{\omega}\left(X,T(X,t)\right)}\rho_0 (X) \,.
\end{equation}
%-----------------------------
The material mass continuity equation is written as 
%-----------------------------
\begin{equation} \label{eq:MatCont}
	 \dot{\rho} +   \frac{1}{2}\rho \operatorname{tr}_ \mathbf{G} \dot{\mathbf{G}}
	 =0 \,,
\end{equation}
%-----------------------------
while the spatial mass continuity equation reads
%-----------------------------
\begin{equation}
	\dot{\varrho} +  \varrho \operatorname{div}_\mathbf{g} \mathbf{v} 
	=\varrho_{,t} +  \operatorname{div}_\mathbf{g}(\varrho \mathbf{v}) 
	=0 \,.
\end{equation}
%-----------------------------
Here, $(\,)\dot{}=\frac{\partial}{\partial t}\big|_{X}(\,)$ represents the material time derivative, while $(\,)_{,t}$ represents the partial derivative $\frac{\partial}{\partial t}\big|_x(\,)$.
%-----------------------------

%-----------------------------
\subsection{Stress and strain tensors}

%-----------------------------
The right and left Cauchy-Green strains are defined as $\mathbf{C}=\mathbf{F}^{\mathsf{T}}\mathbf{F}$ and $\mathbf{b}=\mathbf{F}\mathbf{F}^{\mathsf{T}}$, respectively.\footnote{The transpose of the deformation gradient $\mathbf{F}^{\mathsf{T}}(X,t): T_x\mathcal{C}_t\to T_X\mathcal{B}$ is defined such that
%---------------------
\begin{equation} 
	\llangle \mathbf{F}\mathbf{W},\mathbf{w} \rrangle_{\mathbf{g}}
	=\llangle \mathbf{W},\mathbf{F}^{\mathsf{T}}\mathbf{w} \rrangle_{\mathbf{G}}\,,\quad 
	\forall\;\mathbf{W}\in T_X\mathcal{B}_t\,,~\mathbf{w}\in T_x\mathcal{C}_t
	\,.
\end{equation}
%---------------------
In components, $\big(F^{\mathsf{T}}\big)^A{}_a=G^{AB}F^b{}_B\,g_{ba}$. Thus, $\mathbf{F}^{\star}$ and $\mathbf{F}^{\mathsf{T}}$ are related as $\mathbf{F}^{\mathsf{T}}=\mathbf{G}^{\sharp}\mathbf{F}^{\star}\mathbf{g}$. } In components
%-----------------------------
\begin{equation}
	{C^A}_B
	=G^{AM}{F^a}_M\,g_{ab}{F^b}_B \,,\qquad
	{b^a}_b
	={F^a}_A G^{AB} {F^c}_B g_{cb} \,.
\end{equation}
%-----------------------------
Further, their inverses are denoted by $\mathbf{B}= \mathbf{C}^{-1}$ and $\mathbf{c}= \mathbf{b}^{-1}$, respectively.
Note that $\mathbf{C}^\flat$ is the pull-back of the spatial metric $\mathbf{g}$ via the deformation map $\varphi$ (i.e. $\mathbf{C}^\flat=\varphi^*\mathbf{g}$) and $\mathbf{c}^\flat$ is the push-forward of the material metric $\mathbf{G}$ via $\varphi$ (i.e. $\mathbf{c}^\flat=\varphi_*\mathbf{G}$). Moreover,  
$\mathbf{b}^\sharp=\varphi_*(\mathbf{G}^\sharp)$ and $\mathbf{B}^\sharp=\varphi^*(\mathbf{g}^\sharp)$.\footnote{Here, the musical symbols ${}^\flat$ and ${}^\sharp$ denote the flat and sharp operators that lower and raise tensor indices, respectively.} In components
%-----------------------------
\begin{equation}
	C_{AB}
	= g_{ab}{F^a}_A{F^b}_B \,,\quad
	c_{ab}
	= G_{AB}{F^{-A}}_a{F^{-B}}_b\,,\quad
	B^{AB}
	=g^{ab}{F^{-A}}_a{F^{-B}}_b \,,\quad
	b^{ab}
	= G^{AB}{F^a}_A{F^b}_B \,.
\end{equation}
%-----------------------------
The principal invariants of the right Cauchy-Green strain read
%-----------------------------
\begin{equation}
	I_1
	={ C^{A}}_A \,,\qquad
	I_2
	= \frac{1}{2}(I_1^2- { C^{A}}_B{ C^{B}}_A)\,,\qquad
	I_3
	=\det \mathbf{C} \,.
\end{equation}
%-----------------------------
Note that $\sqrt{I_3}=J=\sqrt{\frac{\det \mathbf{g}}{\det \mathbf{G}}}\det \mathbf{F}$.
The constitutive model for hyperelastic materials is given by an energy density function $W=\tilde{W}(X,T,\mathbf{F},\mathbf{g},\mathbf{G})=\hat{W}(X,T,\mathbf{C}^{\flat},\mathbf{G})$, per unit undeformed volume. The Cauchy stress tensor $\boldsymbol{\sigma}$, the first Piola-Kirchhoff stress tensor $\mathbf{P}$, and the second Piola-Kirchhoff stress tensor $\mathbf{S}$ are related to the energy function as 
%-----------------------------
\begin{equation} 
    \boldsymbol{\sigma}	
	= \frac{2}{J}\frac{\partial \tilde{W}}{\partial \mathbf{g}}\,,\qquad
	\mathbf{P}
	= \mathbf{g}^\sharp \frac{\partial \tilde{W}}{\partial \mathbf{F}}\,,\qquad	
	\mathbf{S}
	= 2\frac{\partial \hat{W}}{\partial \mathbf{C}^\flat} \,.
\end{equation}
%-----------------------------
Note that $\mathbf{P} = J \boldsymbol{\sigma} \mathbf{F}^{-\star}$ and $\mathbf{S} = \mathbf{F}^{-1} \mathbf{P}$.
%-----------------------------
\begin{remark}
%-----------------------------
The energy function for an isotropic solid is a function the principal invariants of $\mathbf{C}$, i.e., $W=\check{W}(X,T,I_1,I_2,I_3)$, and the Cauchy stress is represented as \citep{DoyleEricksen1956,SimoMarsden1983}
%---------------------------------
\begin{equation}
	\bm{\sigma} =
	 \frac{2}{\sqrt{I_3}}
	  \left[ \left(I_2\,\frac{\partial \check{W}}{\partial I_2}
	  +I_3\,\frac{\partial \check{W}}{\partial I_3}\right)
	\mathbf{g}^{\sharp} 
	+\frac{\partial \check{W}}{\partial I_1}\,\mathbf{b}^{\sharp}
	-I_3\,\frac{\partial \check{W}}{\partial I_2}\,\mathbf{c}^{\sharp} \right]\,.
\end{equation}
%---------------------------------
\end{remark}
%-----------------------------
\begin{remark}
%-----------------------------
The energy function for hyperelastic fluids has the functional form $W=\breve{W}(X,T,J)$, where $\breve{W}$ is a smooth, strictly convex function of $J$ that diverges as $J$ approaches $0$ \citep{podio1985cavitation}. Thus, the Cauchy, the first and the second Piola-Kirchhoff stress tensors are written as
%-----------------------------
\begin{equation}
\bm{\sigma}= \frac{\partial\breve{W}}{\partial J} \,\mathbf{g}^\sharp  
\,,\qquad 
\mathbf{P}=J \frac{\partial\breve{W}}{\partial J} \,\mathbf{g}^\sharp \mathbf{F}^{-\star}
\,,\qquad 
\mathbf{S}=J \frac{\partial\breve{W}}{\partial J} \,\mathbf{F}^{-1}\mathbf{g}^\sharp \mathbf{F}^{-\star}\,.
\end{equation}
%-----------------------------
Note that one must have $\frac{\partial\breve{W}}{\partial J}<0$, as hydrostatic stresses are compressive in fluids.
\end{remark}
%-----------------------------
The energy function for homogeneous materials is independent of $X$, i.e., $W=\hat{W}(T,\mathbf{C}^{\flat},\mathbf{G})$ for hyperelastic solids and $W=\breve{W}(T,J)$ for hyperelastic fluids.

%-----------------------------
\subsection{Balance of linear and angular momenta}

%-----------------------------
The localized forms of the balance of linear momentum in terms of the Cauchy and the first Piola-Kirchhoff stress read\footnote{In coordinates
%-----------------------------
\begin{equation} \label{eq:Divsigma}
\begin{aligned}
	\left(\mathrm{div}_\mathbf{g}\boldsymbol \sigma\right)^a
	&={\sigma^{ab}}_{|b}
	=\frac{\partial \sigma^{ab}}{\partial x^b}
	+\sigma^{ac}{\gamma^b}_{cb} 
	+\sigma^{cb}{\gamma^a}_{cb} \,,\\
	\left(\operatorname{Div}\mathbf{P}\right)^a
	&={P^{aA}}_{|A}
	=\frac{\partial P^{aA}}{\partial X^A}
	+P^{aB}{\Gamma^A}_{BA} 
	+{F^b}_A P^{cA}\gamma^a{}_{cb}  \,,
\end{aligned}
\end{equation}
%-----------------------------
where $\gamma^a{}_{bc}=\frac{1}{2}g^{ak}\left(g_{kb,c}+g_{kc,b}-g_{bc,k}\right)$ and $\Gamma^A{}_{BC}=\frac{1}{2}G^{AM}\left(G_{MB,C}+G_{MC,B}-G_{BC,M}\right)$. Note that $\operatorname{Div}$ depends on both the metrics $\mathbf{g}$ and $\mathbf{G}$.}
%-----------------------------
\begin{equation} \label{eq:MatLinMomCons}
	\operatorname{div}_{\mathbf{g}} \bm{\sigma} 
	+ \varrho\, \mathbf{b}
	=\varrho\, \mathbf{a} \,,\qquad
	\operatorname{Div} \mathbf{P} 
	+ \rho \,\mathbf{B}
	=\rho\, \mathbf{A}  \,,
\end{equation}
%-----------------------------
where $\mathbf{b}$ is the spatial body force (per unit mass), while $\mathbf{B}$ is body force referred in material coordinates, i.e., $\mathbf{B}(X,t)=\mathbf{b}(\varphi_t(X),t)$. In components
%-----------------------------
\begin{equation} 
	{\sigma^{ij}}_{|j} + \varrho\, b^i
	=\varrho\, a^i \,,\qquad
	{P^{iI}}_{|I} + \rho\, B^i
	=\rho\, A^i  \,.
\end{equation}
%-----------------------------
The balance of angular momentum in local form reads
%-----------------------------
\begin{equation} 
	\sigma^{ij}=\sigma^{ji}\,,\qquad
	P^{iJ}{F^j}_{J} =P^{jJ}{F^i}_{J} \,.
\end{equation}
%-----------------------------
Note that for slow accretion, the inertial effects can be disregarded.

Let $\mathbf{t}$ be the spatial traction field and $\mathbf{T}$ denote the material traction field. Consider a material surface element $\text{d}A$  with unit normal $\mathbf{N}$, which gets mapped to the element $\text{d}a$ with unit normal $\mathbf{n}$ in the deformed configuration. The traction is related to the Cauchy stress as $\mathbf{t}=\boldsymbol{\sigma}\mathbf{n}^\flat=\llangle \boldsymbol{\sigma},\mathbf{n}\rrangle_{\mathbf{g}}$, and to the first Piola-Kirchhoff stress as $\mathbf{T}=\mathbf{P}\mathbf{N}^\flat=\llangle \mathbf{P},\mathbf{N}\rrangle_{\mathbf{G}}$.\footnote{Note that $n_b=g_{bc} n^c$ and $ N_B= G_{BC} N^C$. Thus, $t^a=\sigma^{ab}\,n_b=\sigma^{ab}g_{bc}\,n^c$ and $T^a=P^{aA}\,N_A=P^{aA}G_{AB}\,N^B$.} Note that $\mathbf{t}\,\text{d}a=\bm{\sigma}\mathbf{n}^\flat\,\text{d}a=J\bm{\sigma}\mathbf{F}^{-\star}\mathbf{N}^\flat\,\text{d}A=\mathbf{P}\mathbf{N}^\flat\,\text{d}A=\mathbf{T}\,\text{d}A$.\footnote{The Nanson's formula $\mathbf{n}^{\flat}da=J\mathbf{F}^{-\star}\mathbf{N}^{\flat}dA$ has been used here. In components, the $1$-forms $n_b \text{d}a$ and $N_B \text{d}A$ are related as $n_b \text{d}a=J {F^{-B}}_b N_B \text{d}A$. } Therefore, the force on the surface element in consideration is $\mathbf{t}\,\text{d}a=\mathbf{T}\,\text{d}A$.

%-----------------------------
\subsection{The heat equation}

Let $\mathcal{T}(x,t)$ be the temperature field defined with respect to the current configuration and let $T(X,t)$ be the temperature field defined with respect to the reference configuration. Since $x= \varphi(X,t)$, it follows that $T(X,t)=\mathcal{T}(\varphi(X,t),t)$, i.e., $T_t=\mathcal{T}_t\circ\varphi_t$, or equivalently, $T_t=\varphi_t^*\mathcal{T}_t$.
Recall that $\text{d}\mathcal{T}=\frac{\partial \mathcal{T}}{\partial x^a}\text{d}x^a$, and $\text{d}T=\frac{\partial T}{\partial X^A}\text{d}X^A$. 
Note that $\text{d}\mathcal{T}$ is a $1$-form in the Euclidean ambient manifold with components $(\text{d}\mathcal{T})_a=\frac{\partial \mathcal{T}}{\partial x^a}$, while $\text{d}T$ is a $1$-form in the material manifold with components $(\text{d}T)_A=\frac{\partial T}{\partial X^A}$ related to $\text{d}\mathcal{T}$  via pull back, i.e., $\text{d}T= \varphi_t^* \text{d}\mathcal{T}$. Thus
%-----------------------------
\begin{equation} \label{eq:Oneforms}
	\frac{\partial T}{\partial X^A}
	= \frac{\partial \mathcal{T}}{\partial x^a} \frac{\partial x^a}{\partial X^A}
	= {F^a}_A  (\text{d}\mathcal{T})_a 
	=(\varphi_t^* \text{d}\mathcal{T})_A
	 \,.
\end{equation}
%-----------------------------
Let $\mathbf{h}(x,t)$ denote the heat flux in the current configuration. Note that $\llangle \mathbf{h},\mathbf{n} \rrangle_{\mathbf{g}} \text{d}a$ is interpreted as the flux through the surface element $\text{d}a$ with unit normal $\mathbf{n}$. The material heat flux vector $\mathbf{H}$ is defined via the Piola transform as 
%-----------------------------
\begin{equation}
	\mathbf{H}(X,t)=J(X,t)\, \mathbf{F}^{-1}(X,t)\, \mathbf{h}(\varphi(X,t),t) \,.
\end{equation}
%----------------------------- 
In components,
%-----------------------------
\begin{equation} \label{eq:Fluxpiola}
	H^A= J {F^{-A}}_a \,h^a \,.
\end{equation}
%----------------------------- 
Let $\mathbf{N}$ denote unit normal to the material surface element $\text{d}A$,  which gets mapped to the deformed element $\text{d}a$. Using \eqref{eq:Fluxpiola} and the Nanson's formula, it is implied that $H^B N_B\, \text{d}A=J {F^{-B}}_b \,h^b N_B\, \text{d}A=h^b n_b \,\text{d}a $, i.e., $\llangle \mathbf{H},\mathbf{N} \rrangle_{\mathbf{G}}\,\text{d}A=\llangle \mathbf{h},\mathbf{n} \rrangle_{\mathbf{g}}\,\text{d}a$. Thus, $\llangle \mathbf{H},\mathbf{N} \rrangle_{\mathbf{G}}$ is interpreted as the heat flux per unit undeformed area. 

The generalized Fourier's law of thermal conduction in the deformed configuration reads
%-----------------------------
\begin{equation} \label{eq:FourierCurrent}
	\mathbf{h} 
	= -\mathbf{k}\, \text{d}\mathcal{T} 
	\,,
\end{equation}
%----------------------------- 
where the $ 2 \choose 0$-tensor $\mathbf{k}$ represents the spatial thermal conductivity. In components, $h^a= -k^{ab} \frac{\partial \mathcal{T}}{\partial x^b}$.  The Fourier's law in the reference configuration is written as 
%-----------------------------
\begin{equation} \label{eq:FourierMaterial}
	\mathbf{H} = -\mathbf{K}\, \text{d}T 	\,,\quad \text{or}\qquad
	H^A 	= -K^{AB} \frac{\partial T}{\partial X^B}\,,
\end{equation}
%-----------------------------
where $\mathbf{K}$ denotes the material thermal conductivity. Note that the material and spatial thermal conductivity tensors are related as $\mathbf{K}= J \varphi_t^*\mathbf{k}$.\footnote{
Using \eqref{eq:Fluxpiola}, \eqref{eq:FourierCurrent} and \eqref{eq:Oneforms}, it is implied that 
%-----------------------------
\begin{equation} \label{eq:RefFluxComp}
	H^A
	=- J{F^{-A}}_a\, k^{ab} \frac{\partial \mathcal{T}}{\partial x^b}
	=- J{F^{-A}}_a\, k^{ab} \frac{\partial  T }{\partial X^B} {F^{-B}}_b \,.
\end{equation}
%-----------------------------
Hence, it can be inferred from \eqref{eq:RefFluxComp} and \eqref{eq:FourierMaterial}$_2$ that $K^{AB}=J {F^{-A}}_a \,k^{ab}{F^{-B}}_b$.
} In components, $K^{AB}=J {F^{-A}}_a \,k^{ab}{F^{-B}}_b$.
Furthermore, upon substituting \eqref{eq:FourierMaterial}$_1$ into the reduced form of the Clausius-Duhem inequality $\langle \text{d}T, \mathbf{H} \rangle \leq 0$, it can be deduced that $\mathbf{K}$ is a positive semi-definite tensor.
The spatial heat equation reads 
%-----------------------------
\begin{equation}
    \varrho \, c_E \dot{\mathcal{T}}
	  +\operatorname{div}_\mathbf{g} \mathbf{h} 
	  = \mathcal{T}  \mathbf{m} \!:\!  \mathbf{d} 
	  + r \,,
\end{equation}
%-----------------------------
where $c_E$ is the specific heat capacity at constant strain, $\mathbf{m}$ is the spatial thermal stress coefficient, $\mathbf{d}= \frac{1}{2}\mathfrak{L}_\mathbf{v} \mathbf{g}$ denotes the rate of deformation  tensor and $r$ represents a heat source (per unit deformed volume) term, see Appendix \ref{AppendixA2}. Equivalently, the material heat equation is written as
%-----------------------------
\begin{equation}
     \rho \, c_E \,\dot{T} 
	  +\operatorname{Div}\mathbf{H} 
	  =T\, \mathbf{M} \!:\!  \mathbf{D} 
	  +R\,,
\end{equation}
%-----------------------------
where $\mathbf{M}$ is the material thermal stress coefficient, $\mathbf{D}= \frac{1}{2}\dot{\mathbf{C}}^\flat$ denotes the material rate of deformation tensor and $R$ represents a heat source (per unit undeformed volume) term, see Appendix \ref{AppendixA1}. The material and spatial thermal stress coefficients are related as $\mathbf{M}=J {\varphi_t} ^* \mathbf{m} $  (see Appendix \ref{AppendixA2}). 
Note that the term $\mathcal{T} \mathbf{m} \!:\! \mathbf{d}$ (or equivalently, $T \mathbf{M}\!:\! \mathbf{D}$) can be omitted if there is no thermoelastic coupling in the material under consideration.\footnote{A classic illustration of thermoelastic coupling is the \emph{Gough-Joule effect}, observed in vulcanized rubber, where the temperature of a rubber band changes during adiabatic stretching \citep{gough1805description,joule1859v}.}  
In the absence of heat sources, the spatial heat equation for a rigid heat conductor is written as 
%-----------------------------
\begin{equation}
 	\varrho \,c_E\, \dot{\mathcal{T}} 
 	= \operatorname{div}_\mathbf{g} (\mathbf{k}\, \text{d}\mathcal{T}) \,,
\end{equation}
%----------------------------- 
or, in components, $\varrho \,c_E \dot{\mathcal{T}}= \left(  k^{ab} \frac{\partial \mathcal{T}}{\partial x^b}\right)_{|a}$. The equivalent material heat equation is written as 
%-----------------------------
\begin{equation} \label{eq:MatHeatEqRigidCond}
    \rho \,c_E\, \dot{T}  
 	= \operatorname{Div} \,(\mathbf{K}\, \text{d}T)
 	\,.
\end{equation}
%-----------------------------
In components, $\rho \,c_E\, \dot{T}= \left(  K^{AB} \frac{\partial T}{\partial X^B}\right)_{|A}$. The heat flux in thermally isotropic solids has the following representation
%-----------------------------
\begin{equation}
 	\mathbf{H}
 	= \left(\phi_{-1} \mathbf{B}^\sharp 
 	+ \phi_{0} \mathbf{G}^\sharp
 	+\phi_{1} \mathbf{C}^\sharp \right)\,\text{d}T \,,
\end{equation}
%-----------------------------  
where $\phi_k = \hat{\phi}_k (X, T, \text{d}T,\mathbf{C}^\flat, \mathbf{G} )$, $k=-1,0,1$, are scalar response functions \citep{truesdell2004non}. We consider the model $\mathbf{H}=-K \mathbf{G}^\sharp \text{d}T$ for our numerical examples, where $K=K(T)$ denotes the heat conduction coefficient.\footnote{Note that $\operatorname{grad}_{\mathbf{G}} T= \mathbf{G}^\sharp \text{d}T$.} Further, $D= \frac{K}{c_E \rho}$ is the traditional thermal diffusivity.

%-----------------------------
%-----------------------------
\subsection{Stefan's condition}

Let $\Gamma_t \subset \partial \mathcal{B}_t$ and $\gamma_t=\varphi_t(\Gamma_t)$. Let $\accentset{+}{\mathbf{h}}$ and $\accentset{-}{\mathbf{h}}$ denote the heat flux per unit area on the opposite sides of the interface $\gamma_t$ in the current configuration. In the absence of any phase change or heat source/sink, the jump in the normal heat flux across $\gamma_t$ vanishes, i.e.,
%-----------------------------
\begin{equation}
	\llangle \accentset{+}{\mathbf{h}}-\accentset{-}{\mathbf{h}} ,\mathbf{n}
	\rrangle_{\mathbf{g}}=0 \quad\text{on}~ \gamma_t\,,
\end{equation}
%-----------------------------
where $\mathbf{n}$ is the outward unit normal to $\gamma_t$. 
%---------------------
\begin{figure}[t!]
\centering
\vskip 0.0in
\includegraphics[width=.32\textwidth]{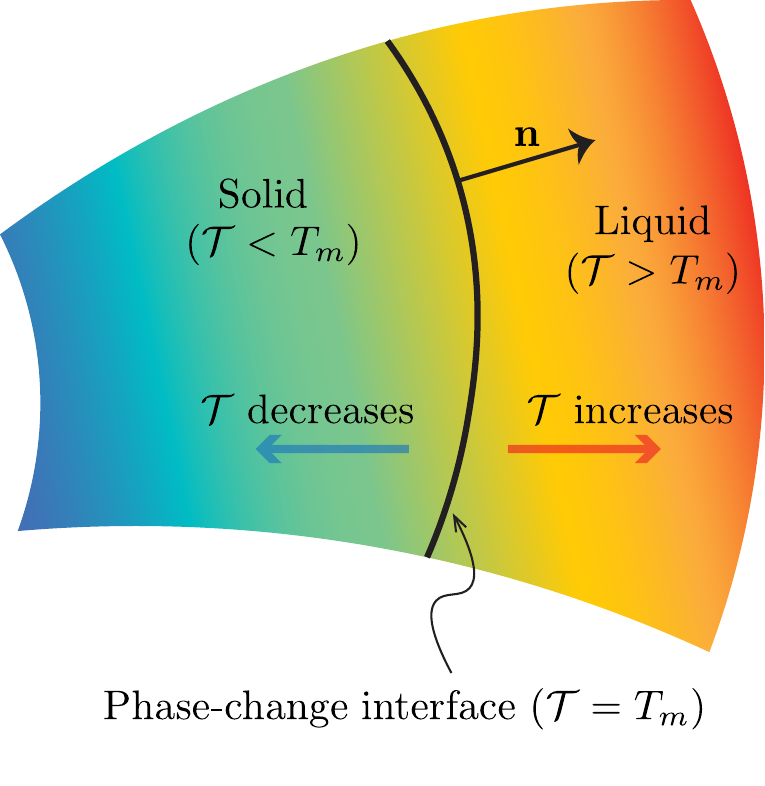}
\vskip 0.1in
\caption{The interface between the (cold) solid and (hot) liquid phases forms an isothermal surface at the solidification temperature $T_m$. The unit normal $\mathbf{n}$ to the phase-change interface is assumed to point from solid to liquid.}
\label{fig:interface1}
\end{figure}
%---------------------

%\todo{Arash: In Figure 2 you are showing the unit normal $\mathbf{n}$. It would be good to have it in Figure 3 as well. That would help the reader.}

%---------------------
\begin{figure}[t!]
\centering
\vskip 0.2in
\includegraphics[width=.64\textwidth]{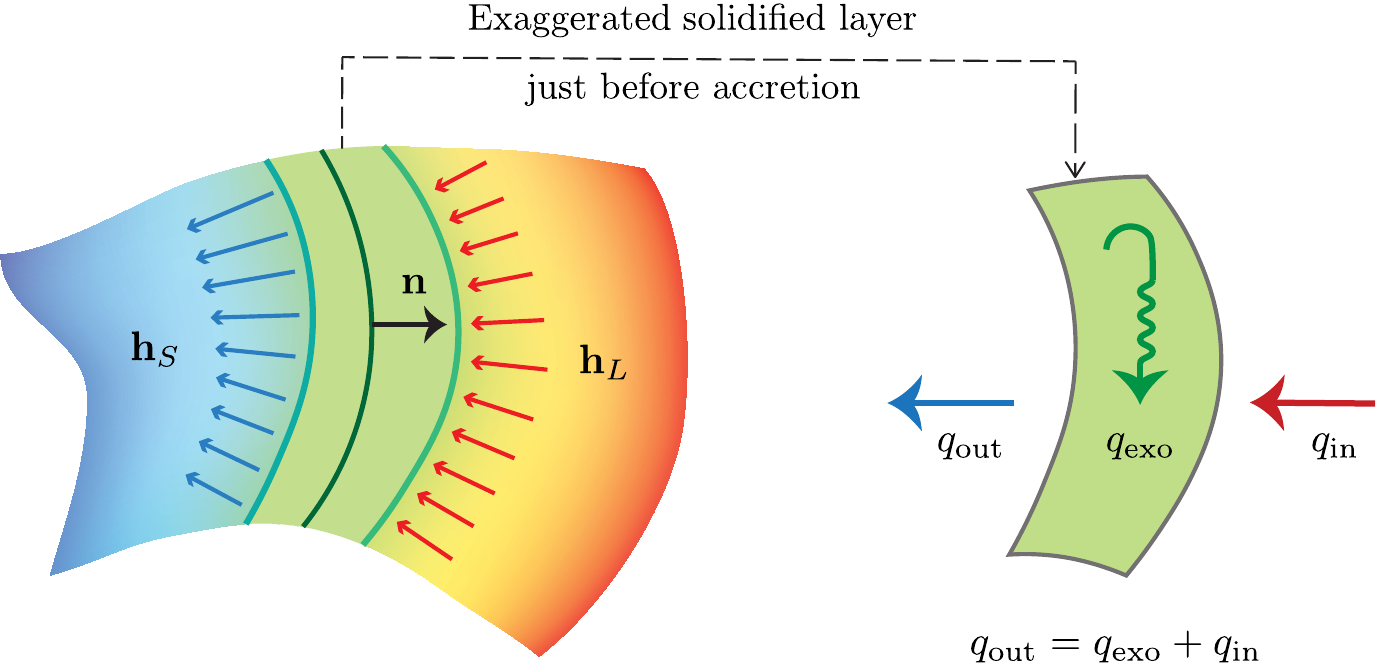}
\vskip 0.1in
\caption{The heat flux $\mathbf{h}_L$ is directed into the interface in the fluid medium due to its higher temperature compared to the melting point. However, in the solid, where the temperature is lower than the melting point, $\mathbf{h}_S$ is directed into the solid. The heat entering the solid comprises of the heat released during solidification and the heat transferred from the liquid medium.}
\label{fig:interface2}
\end{figure}
%---------------------
%---------------------
When $\gamma_t$ is the solidification interface between the (cold) solid and (hot) liquid phases, it forms an isothermal surface at the melting point $T_m$ (Figure \ref{fig:interface1}). Let $\mathbf{n}$ be the unit normal to $\gamma_t$, pointing from solid to liquid, and let $\upsilon$ be an arbitrary subset of $\gamma_t$.
Let $\mathbf{h}_S$ and $\mathbf{h}_L$ be the heat flux in the solid and liquid phases, respectively, in the current configuration. As one moves into the liquid phase from the solidification interface, $\mathcal{T}$ increases, implying that $\mathbf{h}_L$ points towards the solid (Figure \ref{fig:interface2}). Hence, $q_\text{in} = -\int_{\upsilon} \llangle \mathbf{h}_L,\mathbf{n} \rrangle_{\mathbf{g}}\,\text{d}a$ represents the rate of normal heat inflow into the subset $\upsilon$ on the interface from the liquid via conduction. There is a decrease in $\mathcal{T}$ as one moves from the phase change interface into the solid, indicating that $\mathbf{h}_S$ also points into the solid. Thus, the rate of heat flowing out normally from the subset $\upsilon$ on the interface into the solid is $q_\text{out} = -\int_\upsilon \llangle \mathbf{h}_S,\mathbf{n} \rrangle_{\mathbf{g}}\,\text{d}a$. The rate of mass solidified on the subset $\upsilon$ of the interface is represented by the integral $\int_{\upsilon} \rho\,  u^n \text{d}a$  \citep{Sozio2019}. As solidification is exothermic, the rate of heat released in the process is expressed as $q_\text{exo}=l\int_{\upsilon} \rho \, u^n \text{d}a$, where $l$ is the specific latent heat of solidification. The heat flowing into the solid consists of two components: the heat released during solidification and the heat transferred from the surrounding liquid medium \citep{rubinshteuin1971stefan,gupta2017classical}. In terms of heat flow per unit time, $q_\text{out}=q_\text{exo}+q_\text{in}$ (Figure \ref{fig:interface2}), i.e.,
%-----------------------------
\begin{equation}
 	- \int_{\upsilon} \llangle \mathbf{h}_S ,\mathbf{n} \rrangle_{\mathbf{g}}\, \text{d}a 
 = l \int_{\upsilon} \varrho\, J u^n \text{d}a 
 -\int_{\upsilon} \llangle \mathbf{h}_L ,\mathbf{n} \rrangle_{\mathbf{g}}\, \text{d}a\,,
\end{equation}
%-----------------------------
where $\upsilon \subset\gamma_t$ is an arbitrary subset. The localized Stefan's condition therefore reads 
%-----------------------------
\begin{equation}
	\llangle \mathbf{h}_S - \mathbf{h}_L ,\mathbf{n} \rrangle_{\mathbf{g}} 
 	+ l \varrho \, J u^n  =0 \qquad\text{on}~ \gamma_t\,,
\end{equation}
%-----------------------------
where $\varrho \,J u^n = \rho\, U^N$ represents the mass accreted, per unit area, per unit time. The localized Stefan's condition in the reference configuration is expressed as 
%-----------------------------
\begin{equation}
	\llangle \mathbf{H}_S - \mathbf{H}_L ,\mathbf{N} \rrangle_{\mathbf{G}} 
	 + l \rho \, U^N  	=0 \qquad\text{on}~ \Gamma_t\,.
\end{equation}
%-----------------------------
Note that if the liquid is initially at the solidification temperature,  there is no heat flux in the liquid phase, i.e., $\mathbf{h}_L=\mathbf{0}$. In this case, Stefan's  condition is simplified as 
%-----------------------------
\begin{equation}
	-\llangle \mathbf{H}_S,\mathbf{N} \rrangle_{\mathbf{G}} 
	 = l \rho \, U^N  \quad\text{on}~ \Gamma_t \,,\quad \text{or}\quad
	 -\llangle \mathbf{h}_S ,\mathbf{n} \rrangle_{\mathbf{g}} 
 	= l \varrho \, J u^n  \quad\text{on}~ \gamma_t\,.
\end{equation}
%-----------------------------
Thus, the heat entering the solid from the phase change interface is equal to the heat generated in the process of solidification.

%-----------------------------
%-----------------------------------------------------------
\section{Radially inward solidification in a cold rigid container}
\label{Sec:example1}
\begin{figure}[t!]
\centering
\vskip 0.0in
\includegraphics[width=.45\textwidth]{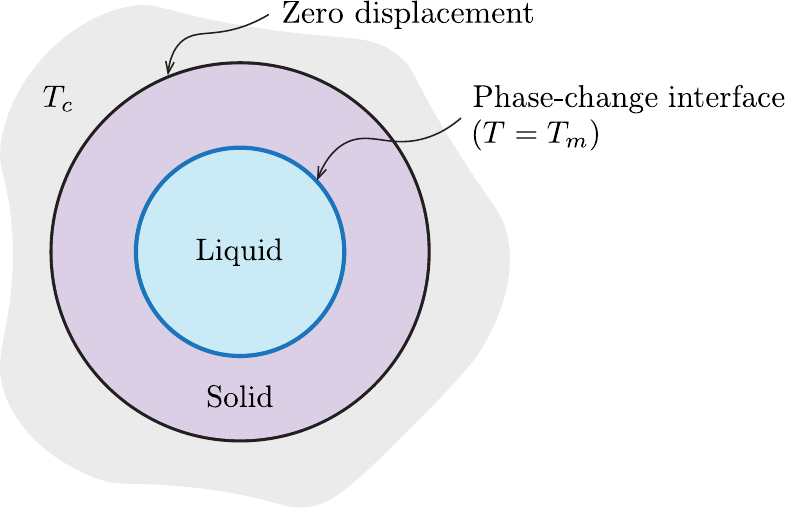}
\vskip 0.1in
\caption{Schematic representation of radially inward solidification in a spherical rigid container with cold walls maintained at the temperature $T_c < T_m$.}
\label{fig:problem1}
\end{figure}

Consider a spherical container of radius $R_0$, filled with a liquid initially at uniform temperature $T_0$. The inner wall of the container is maintained at a constant temperature $T_c$. Let $T_m$ denote the melting point of the material, satisfying the condition $T_c < T_m \leq T_0$. At $t=0$, the outermost layer of liquid begins to cool down and solidifies when the melting point is reached. The container wall acts as a rigid substrate to which the outermost accreted layer firmly attaches, resulting in no displacement of the outer boundary of the accreting body. Layers of liquid solidify and attach to the inner surface of the accreting body---a spherical shell, causing the solidification front to progress inward (Figure \ref{fig:problem1}). The temperature fields within the accreting body and the liquid are both unknowns.\footnote{This problem draws inspiration from the experiments conducted by \citet{chan2006solidification} who investigated the inward solidification of an $n-$hexadecane in a spherical enclosure (capsule) with walls maintained at a constant temperature. They placed this capsule in a cool water tank that was consistently stirred and supplied with cold water from a refrigerated bath. They attached thermocouples to the capsule walls to track its temperature and ensure that it remains constant throughout the process.}

We model both the liquid and solid phases as isotropic compressible hyperelastic materials. To simplify the analysis, an assumption can be made that $T_m = T_0$, allowing for solidification to initiate near the container wall at $t=0$ \citep{stewartson1976stefan,rabin1998thermal}.

%---------------------
%---------------------

%-----------------------------
%-----------------------------
\subsection{Kinematics}

The ambient space has the Euclidean metric 
%-----------------------------
\begin{equation} \label{g-metric}
    \mathbf{g}=\begin{bmatrix}
  1 & 0  & 0  \\
  0 & r^2  & 0  \\
  0 & 0  & r^2 \sin^2 \theta
\end{bmatrix}  \,,
\end{equation}
%-----------------------------
in terms of the spherical coordinates $(r,\theta,\phi)$, where $r \geq 0$, $0\leq \theta \leq \pi$ and $0\leq \phi < 2\pi$. 
%-----------------------------------------------------------

Let $S(t)$ denote the material radius corresponding to the inner surface of the solid at any time $t\leq t_f$, where $t_f$ is the time taken for the completion of freezing.  Note that $S(t)$ is assumed to be a continuous bijective map on $[0,t_f]$ with the initial condition $S(0)=R_0$. The inverse map $\tau=S^{-1}$ assigns the time of attachment to each spherical layer in the material manifold. The accreting solid and the ablating fluid are identified with the following time-dependent material manifolds
%-----------------------------
\begin{equation}
\begin{aligned}
	\mathcal{B}_t
	&=\{ (R,\Theta,\Phi): 
	S(t)\leq R \leq R_0  \,,\; 
	0\leq \Theta \leq \pi  \,,\;
	0\leq \Phi < 2\pi\} \,, \\
		\mathcal{L}_t
	&=\{ (R,\Theta,\Phi): 
	0\leq R \leq S(t)  \,,\; 
	0\leq \Theta \leq \pi  \,,\;
	0\leq \Phi < 2\pi \} \,.
\end{aligned}
\end{equation}
%-----------------------------
Let the temperature field be denoted as $T(R,t)$, and defined piece-wise as follows
%-----------------------------
\begin{equation}
   T(R,t)
   =\begin{cases}
    T^\text{s}(R,t)  \,, & S(t)\leq  R \leq  R_0 \,, \\
     T^\text{f}(R,t)   \,, & 0\leq  R \leq  S(t) \,.
    \end{cases}
\end{equation}
%-----------------------------
Note that $T_t(R)$ is continuous at the solidification interface $R=S(t)$ \citep{caffarelli1983continuity}, i.e., $T^\text{s}(S(t),t)=T^\text{f}(S(t),t)=T_m$. 
The material metric for the liquid phase in its initial state reads
%-----------------------------
\begin{equation} \label{Metrics}
    \GL_0
    =\begin{bmatrix}
  1 & 0  & 0  \\
  0 & R^2  & 0  \\
  0 & 0  & R^2  \sin^2 \Theta
\end{bmatrix} \,, 
\end{equation}
%----------------------------- 
where $(R,\Theta,\Phi)$ are the material spherical coordinates. Thus, the temperature-dependent material metric for the liquid phase is written as
%-----------------------------
\begin{equation} \label{eq:LiquidMatMetric}
  \GL
   =e^{2 \omega^\text{f} (T^\text{f}(R,t))}
  \begin{bmatrix}
  1 & 0  & 0  \\
  0 & R^2  & 0  \\
  0 & 0  & R^2  \sin^2 \Theta
\end{bmatrix}\,,
\end{equation}
%-----------------------------
where the scalar function $\omega^\text{f} (T^\text{f})$ characterizes isotropic and homogeneous thermal expansion in the liquid phase.
We consider radial deformations $\varphi_t(R, \Theta, \Phi)= \left(r(R,t),\theta, \phi\right)$, where $\theta= \Theta$ and $\phi=\Phi$, and
%-----------------------------
\begin{equation}
   r(R,t)
   =\begin{cases}
    r^\text{s}(R,t)  \,, & S(t)\leq  R \leq  R_0 \,,\\
     r^\text{f}(R,t)   \,, & 0\leq  R \leq  S(t) \,,
    \end{cases}
\end{equation}
%-----------------------------
and $r^\text{s}(R_0,t)=R_0$, $r^\text{f}(0,t)=0$.\footnote{%\citet{podio1985cavitation} have investigated cavitation in hyperelastic fluids undergoing similar radial deformations, where the condition $r^\text{f}(0,t)=0$ (in the absence of any cavitation) was referred to as a regular deformation. The cases where $r^\text{f}(0,t)>0$ was referred to as an irregular deformation that created a cavity of radius $r^\text{f}(0,t)$.
\citet{podio1985cavitation} investigated cavitation in hyperelastic fluids undergoing similar radial deformations.
%They termed the deformations satisfying the condition $r^\text{f}(0,t)=0$ to be \emph{regular}, and those with $r^\text{f}(0,t)>0$ as \emph{irregular} deformations that created a cavity (hole) of radius $r^\text{f}(0,t)$.
They termed the deformations satisfying the condition $r^\text{f}(0,t)=0$ \emph{regular}, and those with $r^\text{f}(0,t)>0$, \emph{irregular} deformations corresponding to a cavity (hole) of radius $r^\text{f}(0,t)$.}
Note that $r(R,t)$ is continuous at $R=S(t )$ for all $t \geq 0$.
Let $s(t):= r^\text{s}(S(t),t) = r^\text{f}(S(t),t) $ and $\bar{r}(R):= r^\text{s}(R,\tau(R)) = r^\text{f}(R,\tau(R)) $. Thus, $\bar{r}=s\circ \tau$, or $s= \bar{r} \circ S$.\footnote{Note that
%-----------------------------
\begin{equation} \label{eq:DeformedinterfaceVelocity}
   \dot{s}(t) 
   = {r^\text{s}}_{,R}(S(t),t)\, \dot{S}(t) 
   + {r^\text{s}}_{,t}(S(t),t) 
   = {r^\text{f}}_{,R}(S(t),t)\, \dot{S}(t) 
   + {r^\text{f}}_{,t}(S(t),t)\,,
\end{equation}
%-----------------------------
and thus
%-----------------------------
\begin{equation}
    \left [{r^\text{s}}_{,R}(S(t),t) - {r^\text{f}}_{,R}(S(t),t)\right]\, \dot{S}(t) 
   + {r^\text{s}}_{,t}(S(t),t) - {r^\text{f}}_{,t}(S(t),t)
   = 0 \,.
\end{equation}
%-----------------------------
Hence, the velocity field $r_{,t}(R,t)$ is continuous at $R=S(t)$ if and only if the partial derivative $r_{,R}(R,t)$ is also continuous at $R=S(t)$.} 
The moving phase-change interface in the reference and current configurations are represented as 
%-----------------------------
\begin{equation}
\begin{aligned}
	\Omega_t
	&=\{ \left(S(t),\Theta,\Phi\right): 
	0\leq \Theta \leq \pi  \,,\;
	0\leq \Phi < 2\pi\} \,, \\
	\omega_t
	&=\{ \left(s(t),\theta,\phi\right): 
	0\leq \theta \leq \pi  \,,\;
	0\leq \phi < 2\pi \} \,.
\end{aligned}
\end{equation}
%-----------------------------
The respective deformation gradients in the solid and liquid phases read
%-----------------------------
\begin{equation}
   \mathbf{F}^\text{s}(R,t)=\begin{bmatrix}
  {r^\text{s}}_{,R}(R,t) & 0  & 0  \\
  0 & 1  & 0  \\
  0 & 0  & 1
\end{bmatrix}
\,,\quad\quad
   \mathbf{F}^\text{f}(R,t)=\begin{bmatrix}
  {r^\text{f}}_{,R}(R,t) & 0  & 0  \\
  0 & 1  & 0  \\
  0 & 0  & 1
\end{bmatrix}
\,.
\end{equation}
%-----------------------------
Let $U(t):=\dot{S}(t)<0$ be the material accretion velocity. Let $u(t)$ denote the growth velocity in the current configuration, i.e., $-u(t)$ is the relative velocity of the accreting particles with respect to the interface $\omega_t$. Further define $\bar{u}(R):=u(\tau(R))$ and $\bar{U}(R):=U(\tau(R))$.  
Thus, the accretion tensor $\mathbf{Q}$ has the following representation with respect to the frames $\big\{\frac{\partial}{\partial R}, \frac{\partial}{\partial \Theta},\frac{\partial}{\partial \Phi}\big\}$ and $\big\{\frac{\partial}{\partial r}, \frac{\partial}{\partial \theta},\frac{\partial}{\partial \phi}\big\}$:\footnote{In our example, $\mathbf{u}_{\tau(R)}(\varphi_{\tau(R)}(R,\Theta,\Phi))=\bar{u}(R) \frac{\partial}{\partial r} $ and $\mathbf{U}_{\tau(R)}(R,\Theta,\Phi)=\bar{U}(R) \frac{\partial}{\partial R} $. Recall that the components of the accretion tensor are defined as
%-----------------------------
\begin{equation}
   {Q^i}_I(X)
   ={\bar{F}^i}_I(X) 
   + \Big[u^i\left(\bar{\varphi}(X), \tau(X)\right)-{\bar{F}^i}_J(X) \, U^J\left(X,\tau(X)\right)\Big] (\text{d}\tau)_I(X)\,.
\end{equation}
%-----------------------------
Further, $  \frac{\text{d} \tau(R) }{ \text{d} R} = \frac{1}{\dot{S}\left(\tau(R)\right)} =  \frac{1}{\bar{U}(R)}$. Thus, the nonzero components of $\mathbf{Q}$ are
%-----------------------------
\begin{equation}
\begin{aligned}
   {Q^r}_R  
   = {\bar{F}^r}_R 
   + \left[\bar{u} - {\bar{F}^r}_R \bar{U} \right] \frac{\text{d} \tau }{ \text{d} R} 
   =\frac{\bar{u}}{\bar{U}}\,, \qquad
   {Q^\theta}_\Theta 
   ={\bar{F}^\theta}_\Theta
   =  1 \,, \qquad
   {Q^\phi}_\Phi 
   ={\bar{F}^\phi}_\Phi
   = 1 \,. \\
\end{aligned}
\end{equation}
%-----------------------------
}
%-----------------------------
\begin{equation}
   \mathbf{Q}(R)=\begin{bmatrix}
  \frac{\bar{u}(R)}{\bar{U}(R)} & 0  & 0  \\
  0 & 1  & 0  \\
  0 & 0  & 1
\end{bmatrix}\,.
\end{equation}
%-----------------------------

The accreting layer is not stress free due to the pressure exerted by the fluid. 
Let $\bm{\lambda}$ be the natural metric of the pre-stressed layers that are accreting to the solid. 
This metric $\bm{\lambda}$ is obtained by transforming the Euclidean metric $\mathbf{g}$ via a pre-deformation tensor $\bm{\Lambda}$ as $\lambda_{ij}={\Lambda^k}_i  ( g_{kl}\circ \bar{\varphi})\,{\Lambda^l}_j $. In this case, the material metric for the accreted layer is calculated by pulling back the natural metric $\bm{\lambda}$ via the accretion tensor $\mathbf{Q}$ \citep{Sozio2019}. Therefore, one has
%-----------------------------
\begin{equation}
   (\accentset{\mathcal{B}}{G}_0)_{IJ}
   ={Q^i}_I \,\lambda_{ij}\, {Q^j}_J 
   ={Q^i}_I \,{\Lambda^k}_i  \,( g_{kl}\circ \bar{\varphi}) \,{\Lambda^l}_j \,{Q^j}_J \,.
\end{equation}
%-----------------------------
In this example, it is assumed that
%-----------------------------
\begin{equation}
   \bm{\Lambda}(R)
   = \eta^2(R) \frac{\partial}{\partial r} \otimes \text{d}r 
   + \frac{\partial}{\partial \theta} \otimes \text{d}\theta 
   + \frac{\partial}{\partial \phi} \otimes \text{d}\phi\,,
\end{equation}
%-----------------------------
where the function $\eta^2(R)>0$ represents radial dilation if $\eta(R)>1$ and radial contraction if $\eta(R)<1$. Thus, the temperature-independent material metric at the time of accretion is written as
%-----------------------------
\begin{equation}
   \GS_0 
   =\begin{bmatrix}
  \varsigma^2(R) & 0  & 0  \\
  0 & \bar{r}^2(R)  & 0  \\
  0 & 0  & \bar{r}^2(R) \sin^2 \Theta
\end{bmatrix}\,,
\end{equation}
%-----------------------------
where $\varsigma(R)=\frac{\bar{u}(R) \eta^2(R)}{\bar{U}(R)} $.
Thus, the temperature-dependent material metric for the solid phase is written as
%-----------------------------
\begin{equation} \label{eq:SolidMatMetric}
  \GS
   =e^{2 \omega^\text{s} (T^\text{s}(R,t))}\begin{bmatrix}
 \varsigma^2(R) & 0  & 0  \\
  0 & \bar{r}^2(R)  & 0  \\
  0 & 0  & \bar{r}^2(R) \sin^2 \Theta
\end{bmatrix}\,,
\end{equation}
%-----------------------------
where the scalar function $\omega^\text{s} (T^\text{s})$ characterizes isotropic and homogeneous thermal expansion in the solid phase. The Jacobian of the deformation is written as
%-----------------------------
\begin{equation}
   J(R,t)
   =\begin{cases}
     \displaystyle \frac{\left[r^\text{f}(R,t)\right]^2 {r^\text{f}}_{,R}(R,t) }{e^{3 \omega^\text{f} (T^\text{f}(R,t))}\, R^2}    
     \,, \quad & 0\leq  R \leq  S(t) \,,\\[8 pt]
     \displaystyle \frac{\left[r^\text{s}(R,t)\right]^2{r^\text{s}}_{,R}(R,t)}{e^{3 \omega^\text{s} (T^\text{s}(R,t))} \bar{r}^2(R)\,\varsigma(R)}  
     \,, \quad & S(t)\leq  R \leq  R_0 \,.
    \end{cases}
\end{equation}
%-----------------------------
Further, $\displaystyle\alpha^\text{s}(T^\text{s})=\frac{\text{d}\omega^\text{s}(T^\text{s})}{\text{d}T^\text{s}}$ and $\displaystyle\alpha^\text{f}(T^\text{f})=\frac{\text{d}\omega^\text{f}(T^\text{f})}{\text{d}T^\text{f}}$ are the coefficients of thermal expansion in the solid and liquid phases, respectively.

%-----------------------------
%-----------------------------
\subsection{Balance laws}

\subsubsection{Conservation of mass}

The mass of the liquid and solid portions are calculated as\footnote{
Alternatively, one has $\displaystyle m^\text{f}(t)=\int_0^{s(t)} \varrho ^\text{f}(r,t) \, 4 \pi r^2 \,\text{d}r$, and $\displaystyle m^\text{s}(t)=\int_{s(t)}^{R_0} \varrho ^\text{s}(r,t) \, 4 \pi r^2 \,\text{d}r $.
} 
%-----------------------------
\begin{equation}
\begin{aligned}
	m^\text{f}(t)  
	&=\int_0^{S(t)} e^{3\omega^\text{f}\left(T^\text{f}(R,t)\right)} \rho ^\text{f}(R,t) \, 4 \pi R^2 \,\text{d}R 
	=\int_0^{S(t)}\rho_0^\text{f}(R) \, 4 \pi R^2 \,\text{d}R \,,  	 \\
	m^\text{s}(t)
	&=\int_{S(t)}^{R_0} e^{3\omega^\text{s}\left(T^\text{s}(R,t)\right)} \rho ^\text{s}(R,t) \, 4 \pi \bar{r}^2(R) \,\varsigma(R) \,\text{d}R 
	=\int_{S(t)}^{R_0}  \rho_0 ^\text{s}(R) \, 4 \pi \bar{r}^2(R) \,\varsigma(R) \,\text{d}R\,.
\end{aligned}
\end{equation}
%-----------------------------
Thus, the total mass of the system $m(t)=m^\text{f}(t) + m^\text{s}(t)$ is written as
%-----------------------------
\begin{equation}
	m(t)=\int_0^{S(t)}\rho_0^\text{f}(R) \, 4 \pi R^2 \,\text{d}R
	+\int_{S(t)}^{R_0}  \rho_0 ^\text{s}(R) \, 4 \pi \bar{r}^2(R) \,\varsigma(R) \,\text{d}R \,.
\end{equation}
%-----------------------------
Using the Leibniz rule, it can be shown that 
%-----------------------------
\begin{equation} \label{eq:mTotdot}
	\dot{m}(t)
	= 4\pi \left[\rho_0^\text{f}\left(S(t)\right) S^2(t) 
	-\rho_0 ^\text{s}\left(S(t)\right)  s^2(t) \,\varsigma\left(S(t)\right)\right]  \dot{S}(t) \,.
\end{equation}
%-----------------------------
As the mass of the entire body is conserved, $\dot{m}(t)=0$. Since $\dot{S}(t)$ is nonzero, it follows from \eqref{eq:mTotdot} that
%-----------------------------
\begin{equation}
	\rho_0^\text{f}\left(S(t)\right) S^2(t) 
	=\rho_0 ^\text{s}\left(S(t)\right)  s^2(t) \,\varsigma\left(S(t)\right) \,.
\end{equation}
%-----------------------------
The material continuity in the respective phases read\footnote{Note that
%-----------------------------
\begin{equation} \label{eq:materialmetric_timederivative}
	\operatorname{tr}_{\GS} \big({ {\GS}_{,t}}\big) 
	= 6 \dot{T}^\text{s}  \frac{\text{d}\omega^\text{s}}{\text{d}T^\text{s}} 
	= 6 \alpha^\text{s} \dot{T}^\text{s} 
	= 2 \beta^\text{s} \dot{T}^\text{s} \,,\quad\quad
		\operatorname{tr}_{\GL} \big({ {\GL}_{,t}}\big) 
	= 6 \dot{T}^\text{f}  \frac{\text{d}\omega^\text{f}}{\text{d}T^\text{f}} 
	= 6 \alpha^\text{f} \dot{T}^\text{f} 
	= 2 \beta^\text{f} \dot{T}^\text{f}\,,
\end{equation}
%-----------------------------
where the relations $\beta^\text{s}=3 \alpha^\text{s}$ and $\beta^\text{f}=3 \alpha^\text{f}$ have been used. Therefore, \eqref{eq:MaterialContinuityPhasewise_v1} follows from \eqref{eq:MatCont} and \eqref{eq:materialmetric_timederivative}.} 
%-----------------------------
\begin{equation} \label{eq:MaterialContinuityPhasewise_v1}
\begin{aligned}
	{\rho^\text{s}}_{,t}(R,t)
	+\beta^\text{s}\!\left(T^\text{s}(R,t)\right) {T^\text{s}}_{,t}(R,t) \rho^\text{s}(R,t)
	= 0 \,, \\
	{\rho^\text{f}}_{,t}(R,t)
	+\beta^\text{f}\!\left(T^\text{f}(R,t)\right) {T^\text{f}}_{,t}(R,t) \rho^\text{f}(R,t)
	= 0 \,.
\end{aligned}
\end{equation}
%-----------------------------
The density is assumed to be a function of temperature, i.e., $\rho^\text{s}(R,t)=\hat{\rho}^\text{s}(T^\text{s}(R,t))$ and $\rho^\text{f}(R,t)=\hat{\rho}^\text{f}(T^\text{f}(R,t))$. It follows that $\rho^\text{s}_0(R)=\hat{\rho}^\text{s}(T^\text{s}_0(R))$ and $\rho^\text{f}_0(R)=\hat{\rho}^\text{f}(T^\text{f}_0(R))$, where $T^\text{s}_0(R)$ and $T^\text{f}_0(R)$ are the reference temperatures for the solid and liquid phases, respectively. Thus, the respective continuity equations in \eqref{eq:MaterialContinuityPhasewise_v1} are rewritten as
%-----------------------------
\begin{equation}
	\frac{\text{d}\hat{\rho}^\text{s}}{\text{d} T^\text{s}}
	+\beta^\text{s} \hat{\rho}^\text{s}	= 0 \,,\qquad
	\frac{\text{d}\hat{\rho}^\text{f}}{\text{d} T^\text{f}}
	+\beta^\text{f}	\hat{\rho}^\text{f}= 0 \,.
\end{equation}
%-----------------------------
This is integrated to obtain
%-----------------------------
\begin{equation}
	\hat{\rho}^\text{s}(T^\text{s})
	=\hat{\rho}^\text{s}(T^\text{s}_0) -\int_{T^\text{s}_0}^{T^\text{s}} \beta^\text{s} (\tau) \, \hat{\rho}^\text{s}(\tau) \, \text{d}\tau	\,,\qquad
	\hat{\rho}^\text{f}(T^\text{f})
	=\hat{\rho}^\text{f}(T^\text{f}_0) -\int_{T^\text{f}_0}^{T^\text{f}} \beta^\text{f} (\tau) \, \hat{\rho}^\text{f}(\tau) \, \text{d}\tau \,,
\end{equation}
%-----------------------------
which are equivalent to
%-----------------------------
\begin{equation} \label{eq:dendityrelations}
	\rho^\text{s}(R,t)
	=\rho^\text{s}_0(R) -\int_{T^\text{s}_0(R)}^{T^\text{s}(R,t)} \beta^\text{s} (\tau) \, \hat{\rho}^\text{s}(\tau) \, \text{d}\tau	\,,\qquad
	\rho^\text{f}(R,t)
	=\rho^\text{f}_0(R) -\int_{T^\text{f}_0(R)}^{T^\text{f}(R,t)} \beta^\text{f} (\tau)  \, \hat{\rho}^\text{f}(\tau) \, \text{d}\tau \,.
\end{equation}
%-----------------------------
Note that $T^\text{f}_0(R)$ is the initial temperature of the liquid, while $T^\text{s}_0(R)$ represents the accretion temperature, i.e., $T^\text{s}_0(R)=T_m$.
\begin{remark}
To simplify the problem, it can be assumed that the liquid is initially at the solidification temperature, i.e., $T^\text{f}_0(R)=T_m$. Thus, there is no heat transfer in the liquid medium, i.e., $T^\text{f}(R,t)=T_m$. Thus, it follows from \eqref{eq:dendityrelations}$_2$ that $\rho ^\text{f}\left(R,t\right)=\rho ^\text{f}\left(R,0\right)= \rho ^\text{f}_0(R)=\rho ^\text{f}_0$, which is a constant for homogeneous fluids.\footnote{Alternatively, by substituting ${T^\text{f}}_{,t}=0$ into \eqref{eq:MaterialContinuityPhasewise_v1}$_2$, it is implied that ${\rho^\text{f}}_{,t}=0$.}  
Similarly, 
$\rho^\text{s}(R,t)	=\rho^\text{s}_0 -\int_{T_m}^{T^\text{s}(R,t)} \beta^\text{s} (\tau)  \, \hat{\rho}^\text{s}(\tau) \,  \text{d}\tau	$, where $\rho^\text{s}_0$ is a constant for homogeneous solids. As $T^\text{s}(S(t),t)=T_m$, it follows from \eqref{eq:MaterialContinuityPhasewise_v1}$_2$ that $\rho ^\text{s}\left(S(t),t\right)=\rho^\text{s}_0$. Therefore, one has
%-----------------------------
\begin{equation} \label{eq:acrretionfactor}
	\varsigma\left(S(t)\right) 
	=\frac{\rho_0^\text{f} \, S^2(t)}{\rho_0 ^\text{s} \, s^2(t)}  \,,
	\qquad\text{or equivalently,}\quad\quad
	\varsigma(R)
	=\frac{\rho_0^\text{f} \, R^2}{\rho_0 ^\text{s} \,\bar{r}^2(R)} \,.
\end{equation}
%-----------------------------
Thus
%-----------------------------
\begin{equation}
	\eta^2(S(t))
	=\frac{\rho_0^\text{f} \, S^2(t) \, U(t)}{\rho_0 ^\text{s} \, s^2(t) \, u(t) } \,,
	\qquad\text{or equivalently,}\quad\quad
	\eta^2(R)
	=\frac{\rho_0^\text{f} \, R^2  \bar{U}(R)}{\rho_0 ^\text{s} \,\bar{r}^2(R) \bar{u}(R)} \,.
\end{equation}
%-----------------------------
Further, the mass fraction solidified up to time $t$ is simplified as $\frac{m^\text{s}(t)}{m(t)}=1-\frac{S^3(t)}{R_0^3}$.
\end{remark}
%-----------------------------
\begin{remark}
%-----------------------------
The Jacobian is rewritten as
%-----------------------------
\begin{equation} \label{eq:jacobian_v2}
   J(R,t)
   =\begin{cases}
     \displaystyle \frac{\left[r^\text{f}(R,t)\right]^2 {r^\text{f}}_{,R}(R,t) }{e^{3 \omega^\text{f} (T^\text{f}(R,t))}\,
      R^2}    
     \,, \quad & 0\leq  R \leq  S(t) \,,\\[8 pt]
     \displaystyle \frac{\rho_0^\text{s} \left[r^\text{s}(R,t)\right]^2{r^\text{s}}_{,R}(R,t)}{\rho_0^\text{f} \, 
     e^{3 \omega^\text{s} (T^\text{s}(R,t))} R^2 } 
      \,, \quad & S(t)\leq  R \leq  R_0 \,.
    \end{cases}
\end{equation}
%-----------------------------
\end{remark}
%-----------------------------

\subsubsection{Heat equation}

Let $\mathbf{q}(R,t)$ denote the spatial heat flux in material coordinates, i.e., $\mathbf{q}(R,t)=\mathbf{h}(r(R,t),t)$, with $\mathbf{H}(R,t)$ being the material heat flux. In the model $\mathbf{H}= K \mathbf{G}^\sharp \text{d}T$, the radial components of $\mathbf{H}(R,t)$ and $\mathbf{q}(R,t)$ within the solid are as follows
%-----------------------------
\begin{equation}
	H^R (R,t)=- \frac{K (T (R,t)) \,T_{,R}(R,t)}{ e^{ 2 \omega^\text{s} (T (R,t))} \,
	\varsigma^2(R) }  \,,\quad \quad 
	q^r (R,t)=- \frac{1}{  \,\varsigma(R) } e^{ \omega^\text{s} (T (R,t))} K (T (R,t)) 
	\,T_{,R}(R,t)\,.
\end{equation}
%-----------------------------
Note that\footnote{Here, we have used the fact that
%-----------------------------
\begin{equation}
	 {H^A}_{|A}
	 =\left[ K(T)\,G^{AB}T_{,B}\right]_{|A}
	 =\frac{\text{d}K}{\text{d}T} G^{AB} T_{,A} T_{,B}
	 +K G^{AB} \left( T_{,AB}-{\Gamma^C}_{AB} T_{,C}  \right) \,.
\end{equation}
%----------------------------- 
In spherical coordinates, one has
%-----------------------------
\begin{equation}
	 \operatorname{Div}\mathbf{H}
	 =G^{RR} \left[ \frac{\text{d}K}{\text{d}T}  {T_{,R}}^2 + K T_{,RR} \right]
	 -K  T_{,R} \left[ G^{RR}\,{\Gamma^R}_{RR} 
	 + G^{\Theta\Theta}\,{\Gamma^R}_{\Theta \Theta} 
	 + G^{\Phi\Phi}\,{\Gamma^R}_{\Phi\Phi}   \right]  \,.	 
\end{equation}
%----------------------------- 
The Christoffel symbols ${\Gamma^C}_{AB}$ for the material metric $\GS$ are given in \eqref{eq:Christoffel_Gsolid_metric}.
} 
%-----------------------------
\begin{equation}
	 \operatorname{Div} \mathbf{H}
	 = \frac{1}{e^{2\omega} \,\varsigma^2}
	 \left[\left(\frac{\text{d}K}{\text{d}T} + \alpha^\text{s} K\right) {T_{,R}}^2 + K T_{,RR}
	 +\left(\frac{2\bar{r}'}{\bar{r}}-\frac{\varsigma'}{\varsigma}\right)K T_{,R}\right]  \,,
\end{equation}
%----------------------------- 
where the notation $(\cdot)':= \frac{\text{d}}{\text{d} R}(\cdot)$ has been used.
Therefore, the heat equation \eqref{eq:MatHeatEqRigidCond} inside the solid is written as
%-----------------------------
\begin{equation} \label{eq:MatHeatEqRigidCondRadial_v1}
	 K T_{,RR}
	 + \left(\frac{\text{d}K}{\text{d}T} + \alpha^\text{s} K\right) {T_{,R}}^2  
	 +\left(\frac{2\bar{r}'}{\bar{r}}-\frac{\varsigma'}{\varsigma}\right)K T_{,R}
	 = e^{2\omega^\text{s}} \,\varsigma^2 \rho\, c_E \,\dot{T} \,.
\end{equation}
%----------------------------- 
Let us assume that the heat conduction coefficient is independent of temperature, i.e., $K(T) = K^\text{s}$, a constant. Thus, using \eqref{eq:acrretionfactor}, the heat equation \eqref{eq:MatHeatEqRigidCondRadial_v1} is simplified as follows
%-----------------------------
\begin{equation} \label{eq:MatHeatEqRigidCondRadial_v2}
	 D^\text{s}_\text{f}  \Bigg \{T_{,RR}(R,t)
	 +\alpha^\text{s}(T(R,t))\, \big[T_{,R}(R,t)\big]^2  
	 +\left[\frac{4\bar{r}'(R)}{\bar{r}(R)}-\frac{2}{R}\right]  T_{,R}(R,t) \Bigg\}
	 = \frac{   R^4 \,\dot{T}(R,t)}{ \,\bar{r}^4(R) \,e^{\omega^\text{s}(T(R,t))}}   \,,
\end{equation}
%----------------------------- 
where the constant $D^\text{s}_\text{f} =\displaystyle\frac{K^\text{s} (\rho_0 ^\text{s})^3}{c_E (\rho_0^\text{f})^4   }$ is analogous to thermal diffusivity. Further, the temperature field $T^\text{s}(R,t)$ satisfies the following boundary conditions
%-----------------------------
\begin{equation} \label{eq:ThermalBCsolid}
\begin{aligned}
	-K^\text{s} T^\text{s}_{,R}(R_0,t)
	&= h_c \left[ T^\text{s}(R_0,t) -T_c \right]\,, \\
	T^\text{s}(S(t),t)
	&=T_m \,,
\end{aligned}
\end{equation}
%-----------------------------
where $h_c $ is the coefficient of heat transfer between the walls of the container and the solidified material. 
Thus, for the temperature field, we have a Neumann boundary condition near the fixed wall of the container and a Dirichlet boundary condition on the moving interface.

%----------------------------- 
\subsubsection{Stefan's condition}

The rate of mass transferred from liquid to solid phase is $\dot{m}^\text{s}(t)=-\dot{m}^\text{f}(t)$.
The rate of mass solidified is 
%-----------------------------
\begin{equation}
\begin{aligned}
	 \dot{m}^\text{s}(t)
	 =  -4\pi  \rho_0 ^\text{s}  s^2(t) \,\varsigma\left(S(t)\right) \dot{S}(t)
	 =  -4\pi  \rho_0 ^\text{s}  s^2(t) \,u(t)
	\,.
\end{aligned}
\end{equation}
%-----------------------------
Alternatively, $\dot{m}^\text{s}$ can be expressed as
%-----------------------------
\begin{equation}
	 \dot{m}^\text{s}(t)
	 = - \dot{m}^\text{f}(t)
	 = - 4\pi\rho^\text{f}_0  S^2(t) \,  U(t) \,.
\end{equation}
%-----------------------------
The time rate of heat released during solidification is $l\,\dot{m}^\text{s}(t)$. Further, the heat transferred into the solid medium is
%-----------------------------
\begin{equation}
	 -\int_{\Omega_t}\llangle \mathbf{H},\mathbf{N} \rrangle_{\GS} \text{d}A 
	 = -4 \pi S^2(t) \left[H^R N^R G_{RR}\right]\Big|_{R=S(t)}
	 =-4 \pi S^2(t)\, K(T_m) \,T_{,R} (S(t),t) \,.
\end{equation}
%-----------------------------
If the liquid is initially at the solidification temperature, there is no heat flux within it, and the heat entering the solid from the phase change interface is equal to the heat generated during solidification.
Thus, Stefan's condition is written as
%-----------------------------
\begin{equation}
	 S^2(t) \,K(T_m) \,T_{,R} (S(t),t)
	 =  l \, \rho^\text{s}_0  \,s^2(t) \, u(t) \,,
\end{equation}
%-----------------------------
or equivalently,
%-----------------------------
\begin{equation}
	K(T_m) \,T_{,R} (S(t),t)
	=  l \, \rho^\text{f}_0 \, \dot{S}(t) \,.
\end{equation}
%-----------------------------
Assuming a constant heat conduction coefficient $K(T) = K^\text{s}$, Stefan's condition is written as
%-----------------------------
\begin{equation} \label{eq:StefanConditionFinalVersion}
	T_{,R} (S(t),t)
	= L  \, \dot{S}(t) \,,
\end{equation}
%-----------------------------
where $L=\frac{\rho^\text{f}_0 l}{ K^\text{s} }$.

%-----------------------------
\subsubsection{Conservation of linear momentum in the solid portion}

The Cauchy stress tensor in the solid portion is related to the energy function $\check{W}(I_1,I_2,J,T)$ as follows\footnote{Note that the first Piola-Kirchhoff stress tensor $\mathbf{P}= J \bm{\sigma} \mathbf{F}^{-\star}$ is written as
%---------------------------------
\begin{equation}
	 \mathbf{P} =
	 \left[2 I_2\,\frac{\partial \check{W}}{\partial I_2}
	  + J\,\frac{\partial \check{W}}{\partial J}\right]  \mathbf{F}\mathbf{B}^{\sharp} 
	+2\left[ \frac{\partial \check{W}}{\partial I_1}\,\mathbf{b}^{\sharp} \mathbf{F}^{-\star}
	- J^2 \frac{\partial \check{W}}{\partial I_2}\, \mathbf{F}^{-\mathsf{T}}\mathbf{B}^{\sharp} \right]\,,
\end{equation}
%---------------------------------
where $\mathbf{b}^\sharp=\mathbf{F}\mathbf{G}^{\sharp} \mathbf{F}^\star$ and $\mathbf{B}^\sharp=\mathbf{F}^{-1}\mathbf{g}^{\sharp} \mathbf{F}^{-\star}$.
}
%---------------------------------
\begin{equation}
	\bm{\sigma} =
	  \left[2 J^{-1}I_2\,\frac{\partial \check{W}}{\partial I_2}
	  + \,\frac{\partial \check{W}}{\partial J}\right] \mathbf{g}^{\sharp} 
	+2\left[ J^{-1}\frac{\partial \check{W}}{\partial I_1}\,\mathbf{b}^{\sharp}
	- J \frac{\partial \check{W}}{\partial I_2}\,\mathbf{c}^{\sharp} \right]
	\,.
\end{equation}
%---------------------------------
Since $V^r(R,t)= r_{,t}(R,t)$, $V^\theta(R,t)= V^\phi(R,t)=0$, one has $A^r(R,t)= r_{,tt}(R,t)$, and $A^\theta(R,t)= A^\phi(R,t)=0$.\footnote{The Christoffel symbols for the Euclidean metric $\mathbf{g}$ are given in \eqref{eq:Christoffel_g_metric}.} Using \eqref{eq:Divsigma} and 
\eqref{eq:Christoffel_g_metric}, the radial equilibrium equation \eqref{eq:MatLinMomCons} is simplified to read
%-----------------------------
\begin{equation} \label{eq:radialequilibrium}
  \frac{\partial \sigma^{rr}}{\partial r}
  +\frac{2}{r} \sigma^{rr}
  -r \left[ \sigma^{\theta \theta}
  + \sin^2 \theta \,\sigma^{\phi \phi} \right]
  + \varrho\, b^r  
  =  \varrho \, r_{,tt}.
\end{equation}
%-----------------------------
The inertial effects can be ignored if the solidification process is slow, and hence  in the absence of body forces, it follows from \eqref{eq:radialequilibrium} that
%-----------------------------
\begin{equation} \label{eq:radialequilibriumSimple}
  \frac{\partial \sigma^{rr}}{\partial R}
  =\left[ \left( \sigma^{\theta \theta}
  + \sin^2 \theta \,\sigma^{\phi \phi} \right)r
  -\frac{2}{r} \sigma^{rr}\right]  \frac{\partial r}{\partial R}.
\end{equation}
%-----------------------------
In this example,\footnote{Recall that the components of $\mathbf{c}^\sharp$ and $\mathbf{c}^\flat$ are related as $c^{ab}=g^{am}c_{mn}g^{nb}$. Thus, the components $c^\sharp$ are $c^{ab}=g^{am}{F^{-A}}_m G_{AB} {F^{-B}}_n g^{nb}$.} 
%-----------------------------
\begin{equation} \label{eq:straintensorssolid}
   \mathbf{b}^\sharp
   =e^{-2 \omega^\text{s} }\begin{bmatrix} \displaystyle
\frac{ {r_{,R}}^2 }{ \varsigma^2} & 0  & 0  \\
  0 &  \displaystyle\frac{1}{ \bar{r}^2 } & 0  \\
  0 & 0  &  \displaystyle\frac{1}{ \bar{r}^2 \sin^2 \Theta }
\end{bmatrix}\,, \qquad
	\mathbf{c}^\sharp
   =e^{ 2 \omega^\text{s} }\begin{bmatrix} \displaystyle
\frac{ \varsigma^2 }{{r_{,R}}^2 } & 0  & 0  \\
  0 &  \displaystyle\frac{\bar{r}^2}{ r^4 } & 0  \\
  0 & 0  &  \displaystyle\frac{\bar{r}^2}{ r^4 \sin^2 \Theta }
\end{bmatrix}\,.
\end{equation}
%-----------------------------
Further, the principal invariants of $\mathbf{b}$ read\footnote{Here, we have used the fact that
%-----------------------------
\begin{equation}
\begin{aligned}
   I_1
   =g_{ab}{F^a}_A {F^b}_B G^{AB}  \,,\qquad
   I_2
   =\frac{1}{2}\left(I_1^2 - g_{mb}g_{na} {F^m}_M  {F^n}_N {F^a}_A {F^b}_B G^{AM} G^{BN} \right) \,.
\end{aligned}
\end{equation}
%----------------------------- 
} 
%-----------------------------
\begin{equation}
\begin{aligned}
   I_1
   &= e^{-2 \omega^\text{s}} \left[\frac{{r_{,R}}^2}{ \varsigma ^2} 
   +\frac{2r^2}{\bar{r}^2}\right] \,,\\
   I_2
   &=\frac{1}{2}\left[I_1^2 -  e^{-4 \omega^\text{s}}\left( \frac{{r_{,R}}^4}{\varsigma^4}   + \frac{2r^4}{\bar{r}^4}\right) \right]
   = \frac{r^2}{e^{4 \omega^\text{s}}\bar{r}^2}\left[  \frac{r^2}{ \bar{r}^2} + \frac{2{r_{,R}}^2}{ \varsigma ^2} \right]  \,.
\end{aligned}
\end{equation}
%-----------------------------
The Cauchy stress has the following nonzero components
%-----------------------------
\begin{equation} \label{eq:Cauchycomponents_v1}
\begin{aligned}
	\sigma^{rr}  
	&= 2 J^{-1}I_2\,\frac{\partial \check{W}}{\partial I_2}
	  + \,\frac{\partial \check{W}}{\partial J}
	+ 2\left[ \frac{   {r_{,R}}^2}{J e^{2\omega^\text{s}} \varsigma^2}\frac{\partial \check{W}}{\partial I_1} 
	- \frac{ J e^{2\omega^\text{s}} \varsigma^2 }{  {r_{,R}}^2 }\frac{\partial \check{W}}{\partial I_2} \right] \,,\\
	\sigma^{\theta\theta}  
	&=   \left[2 J^{-1}I_2\,\frac{\partial \check{W}}{\partial I_2}
	  + \,\frac{\partial \check{W}}{\partial J}\right]\frac{1}{r^2} 
	+ 2 \left[ \frac{1 }{J e^{2\omega^\text{s}} \bar{r}^2}\frac{\partial \check{W}}{\partial I_1} 
	- \frac{J e^{2\omega^\text{s}} \bar{r}^2  }{ r^4 }\frac{\partial \check{W}}{\partial I_2} \right] \,,\\
	\sigma^{\phi\phi}
	&=   \frac{\sigma^{\theta\theta} }{\sin^2 \Theta}\,.\\
\end{aligned}
\end{equation}
%-----------------------------
Substituting \eqref{eq:Cauchycomponents_v1} in \eqref{eq:radialequilibriumSimple}, one obtains\footnote{The relation $r^2 \left[ g^{\theta \theta}  + \sin^2 \theta \,g^{\phi \phi} \right]  =2 g^{rr}$ has been used here.
}
%-----------------------------
\begin{equation} \label{eq:SolidRadialEquilibrum_v1}
   \frac{\partial \sigma^{rr}}{\partial R}
  = \frac{4 r_{,R}}{r} \left[\frac{1}{J\, e^{2 \omega^\text{s}}}  
  \left(\frac{r^2}{\bar{r}^2} 
   - \frac{{r_{,R}}^2}{ \varsigma ^2}\right)\frac{\partial \check{W}}{\partial I_1}
   +  J\, e^{2 \omega^\text{s}}  \left(\frac{ \bar{r}^2 }{r^2} 
   - \frac{\varsigma ^2 }{ {r_{,R}}^2 }\right)\frac{\partial \check{W}}{\partial I_2} \right]\,.
\end{equation}
%-----------------------------
In the solid, one has $J=\displaystyle\frac{r^2 r_{,R}}{e^{3\omega^\text{s}}\bar{r}^2 \varsigma}$. 
Thus, \eqref{eq:SolidRadialEquilibrum_v1} is simplified as
%-----------------------------
\begin{equation} \label{eq:SolidRadialEquilibrum_v2}
   \frac{\partial \sigma^{rr}}{\partial R}
  =4\check{W}_1 e^{\omega^\text{s}  }
  \left[ \frac{\varsigma}{r} - \frac{\bar{r}^2 {r_{,R}}^2}{\varsigma r^3  }\right]
   +  \frac{4\check{W}_2}{e^{\omega^\text{s} }} \left[ \frac{{r_{,R}}^2}{\varsigma  r }-\frac{\varsigma r }{\bar{r}^2} \right] \,,
\end{equation}
%-----------------------------
where $\check{W}_i=\frac{\partial\check{W}}{\partial I_i} $ for $i =1,2$. 
Using \eqref{eq:acrretionfactor}, \eqref{eq:SolidRadialEquilibrum_v2} is rewritten as follows
%-----------------------------
\begin{equation} \label{eq:SolidRadialEquilibrum_v3}
    \sigma^{rr}_{,R}
  =4 \left[ \check{W}_1 e^{\omega^\text{s}  }
  - \frac{ \check{W}_2\, r^2}{e^{\omega^\text{s} } \bar{r}^2} \right]
  \left[ \frac{\rho_0^\text{f} \, R^2}{\rho_0 ^\text{s} \,\bar{r}^2 r } 
  - \frac{\rho_0^\text{s} \,\bar{r}^4\,{r_{,R}}^2}{\rho_0^\text{f} \, R^2 r^3 }\right] \,.
\end{equation}
%-----------------------------
Hence, \eqref{eq:SolidRadialEquilibrum_v3} can be integrated to obtain
%-----------------------------
\begin{equation} \label{eq:sigmarr_v2}
\begin{aligned}
  \sigma^{rr}(R,t)
  &= \sigma^{rr}(S(t),t) \\
  & \quad+ 4 \int_{S(t)}^R 
  \left[\check{W}_1(\xi,t) e^{\omega^\text{s}(T(\xi,t))}  
  - \frac{ \check{W}_2(\xi,t) r^2(\xi,t)}{e^{\omega^\text{s}(T(\xi,t))} \bar{r}^2(\xi)} \right]
    \bigg[ \frac{\rho_0^\text{f} \, \xi^2}{\rho_0 ^\text{s} \,\bar{r}^2(\xi) r(\xi,t) } 
  - \frac{\rho_0^\text{s} \,\bar{r}^4(\xi) \,{r_{,R}}^2(\xi,t) }{\rho_0^\text{f} \, \xi^2 r^3(\xi,t) }\bigg] \text{d}\xi\,.
\end{aligned}
\end{equation}
%-----------------------------

%-----------------------------
\begin{remark}
Note that $\sigma^{rr}(R,t)$ has to be continuous at $R=S(t)$ in order to satisfy the traction continuity across the phase change interface.
\end{remark}
%-----------------------------

%-----------------------------
\subsubsection{Conservation of linear momentum inside the liquid}

The Cauchy stress inside the liquid is related to the energy function $\breve{W}(J,T)$ as $\bm{\sigma}=\frac{\partial \breve{W}}{\partial J} \mathbf{g}^\sharp$, i.e.,
%-----------------------------
\begin{equation} 
    \bm{\sigma}
  =  \frac{\partial \breve{W}}{\partial J}
  \begin{bmatrix}
  1 & 0  & 0  \\
  0 & \frac{1}{r^2}  & 0  \\
  0 & 0  & \frac{1}{r^2 \sin^2 \theta}
  \end{bmatrix} \,. 
\end{equation}
%-----------------------------
Note that $r^2 \left[ \sigma^{\theta \theta}  + \sin^2 \theta \,\sigma^{\phi \phi} \right]  =2 \sigma^{rr}$. In the absence of inertial effects and body forces, the radial equilibrium equation is written as 
%-----------------------------
\begin{equation} \label{eq:RadialEqinLiquid_v0} 
   r  \frac{\partial \sigma^{rr}}{\partial R}
   = \left[ ( \sigma^{\theta \theta} 
   + \sin^2 \theta \,\sigma^{\phi \phi} )\, r^2
    - 2 \sigma^{rr}\right] \frac{\partial r}{\partial R} 
   =0 \,.
\end{equation}
%-----------------------------
Hence, it follows that $\frac{\partial \sigma^{rr}}{\partial R}$, i.e., $\sigma^{rr}(R,t)$ is independent of $R$. Moreover, one has
%-----------------------------
\begin{equation} \label{eq:RadialEqinLiquid} 
   \frac{\partial }{\partial R}\left( \frac{\partial \breve{W}}{\partial J} \right)
   =0  \,.
\end{equation}
%-----------------------------
If the liquid is initially at the melting temperature, then $T^\text{f}(R,t)=T_m$ and there is no heat transfer occurring inside the liquid during the entire process. Because there are no temperature changes, $\breve{W}$ and, consequently, $\frac{\partial \breve{W}}{\partial J}$ remain independent of temperature. Let us define the temperature-independent function $\breve{p}(J)$ as $\breve{p}=\frac{\partial \breve{W}}{\partial J}$, and denote $p(R,t)= \breve{p}(J(R,t))$. Since $\frac{\partial }{\partial R}\left( \frac{\partial \breve{W}}{\partial J} \right) = \frac{\text{d}\breve{p}}{\text{d}J} \frac{\partial J}{\partial R}$, it follows from \eqref{eq:RadialEqinLiquid} that $\frac{\partial J}{\partial R}=0$ \citep{podio1985cavitation}. Thus, $J$ is independent of $R$, which is indicated as $J(R,t)=J_0(t)$, for some function $J_0(t)>0$. Note that $\omega^\text{f}(T^\text{f})=0$ because $T^\text{f}=T_m$ throughout the process. Since \eqref{eq:jacobian_v2}$_1$ is simplified as  $J=\frac{r^2 r_{,R}}{ R^2}$, it is implied that inside the liquid one has
%-----------------------------
\begin{equation} \label{eq:radialSolutionLiq}
	r^3(R,t)
	=s^3(t)+J_0(t)\left[R^3- S^3(t)\right] \,.
\end{equation}
%----------------------------- 
Since $r(0,t)=0$, it follows from \eqref{eq:radialSolutionLiq} that $J_0(t)=\frac{s^3(t)}{S^3(t)}$, and hence $r(R,t)=\frac{R \,s(t)}{S(t)}$.\footnote{Furthermore, it is implied that ${r^\text{f}}_{,R}\left(R,t\right)=\frac{s(t)}{S(t)}$ and ${r^\text{f}}_{,t}\left(R,t\right)=\frac{R[ S(t)\dot{s}(t)- s(t)\dot{S}(t)]}{S^2(t)}$ inside the liquid. Thus, ${r^\text{f}}_{,R}\left(S(t),t\right)=\frac{s(t)}{S(t)}$ and ${r^\text{f}}_{,t}\left(S(t),t\right)=\dot{s}(t)-\frac{ s(t)\dot{S}(t) }{S(t)}$, which agrees with the fact that $\dot{s}(t)= {r^\text{f}}_{,R}(S(t),t)\, \dot{S}(t) + {r^\text{f}}_{,t}(S(t),t)$.
} Thus,
%-----------------------------
\begin{equation}
	\sigma^{rr}(R,t)
	=\sigma^{rr}(S(t),t)
	=\breve{p}\left(\frac{s^3(t)}{S^3(t)}\right) \,.
\end{equation}
%----------------------------- 

In our numerical examples, we consider the following temperature-independent energy function
%-----------------------------
\begin{equation}
	\breve{W}(J)
	= \pi^\text{f}_0  J
	+ \kappa^\text{f}_0 (J-1)^2  \,,
\end{equation}
%-----------------------------
where $\kappa^\text{f}_0$ denotes the the bulk modulus of the liquid at temperature $T_0$, while $\pi^\text{f}_0$ represents the initial pressure in the liquid \citep{ghosh2022elastomers}. Hence, $\breve{p}(J)=\pi^\text{f}_0+\kappa^\text{f}_0 (J-1)$, and $\bm{\sigma}=\left[\pi^\text{f}_0+\kappa^\text{f}_0 (J-1)\right] \mathbf{g}^\sharp$. If the liquid is initially stress-free, i.e., $\bm{\sigma}(R,0)=\mathbf{0}$, then, it can be deduced from $J_0(0)=1$ that $\pi^\text{f}_0=0$. Therefore, one has
%-----------------------------
\begin{equation} \label{eq:sigmarrLiquid}
	p(R,t)
	=\sigma^{rr}(R,t)
	=\kappa^\text{f}_0 \left[\frac{s^3(t)}{S^3(t)}-1\right] \,.
\end{equation}
%-----------------------------
This means that the Cauchy stress remains uniform in a compressible hyperelastic fluid in the absence of inertial effects, body forces, and heat flow.

%-----------------------------
%-----------------------------
\subsection{Stefan's problem for a neo-Hookean solid}

Consider the following energy function for a thermoelastic neo-Hookean solid \citep{Sozio2020} 
%-----------------------------
\begin{equation} \label{eq:energyfunc_coupled}
	\check{W}(I_1,J,T)
	= \left[ \frac{\mu^\text{s}_0}{2} (J^{-\frac{2}{3} }I_1 -3)
	+\frac{\kappa^\text{s}_0}{2} (J-1)^2
	\right]\frac{T}{T_0}
	-\kappa^\text{s}_0\,\beta^\text{s}_0\, (J-1)(T-T_0)
	\,,
\end{equation}
%-----------------------------
where $I_1= e^{-2 \omega^\text{s}} \left[ \frac{(\rho_0^\text{s})^2 \bar{r}^4 {r_{,R}}^2}{ (\rho_0^\text{f})^2 R^4} + \frac{2r^2}{\bar{r}^2}\right]$ and $J=\frac{\rho_0^\text{s}  r^2 \, r_{,R}}{\rho_0^\text{f} \, e^{3 \omega^\text{s}} R^2 }$. For more details, refer to Appendix \ref{App:ConstitutiveSolid}.
The nonzero components of Cauchy stress read 
%-----------------------------
\begin{equation} \label{eq:CauchyComp_v1}
\begin{aligned}
	\sigma^{rr}  
	=\check{W}_J
	+  \frac{ 2 \check{W}_1\,  {r_{,R}}^2}{J e^{2\omega^\text{s}} \varsigma^2}  \,,\qquad
	\sigma^{\theta\theta}  
	=   \frac{\check{W}_J}{r^2} 
	+  \frac{2 \check{W}_1}{J e^{2\omega^\text{s}} \bar{r}^2} \,,\qquad
	\sigma^{\phi\phi}
	=   \frac{\sigma^{\theta\theta} }{\sin^2 \Theta}\,,
\end{aligned}
\end{equation}
%----------------------------- 
where $\check{W}_J:=\frac{\partial \check{W}}{\partial J}$. 
These coefficients are calculated as follows
%-----------------------------
\begin{equation} \label{eq:StressCoeffNeoHookean_v1}
\begin{aligned}	
	\check{W}_1
	= \frac{\mu^\text{s}_0 T}{2 T_0} J^{-\frac{2}{3} }   \,, \qquad
    \check{W}_J
	=  \left[ \kappa^\text{s}_0(J-1)
	-\frac{\mu^\text{s}_0}{3} J^{-\frac{5}{3} }I_1 
	\right]\frac{T}{T_0}
	-\kappa^\text{s}_0\,\beta^\text{s}_0\,(T-T_0)	\,.
\end{aligned}
\end{equation}
%-----------------------------
Further, we assume that $\omega^\text{s}$ depends on the temperature as per the following relation (see \eqref{eq:TempDepofOmega}) 
%-----------------------------
\begin{equation} \label{eq:thermalexpansionrelation}
	e^{3 \omega^\text{s}(T) }
	= 1+ \beta^\text{s}_0 T^\text{s}_0\Big(1-\frac{T^\text{s}_0}{T}\Big)\,.
\end{equation}
%-----------------------------
Since $\sigma^{rr}$ is continuous across $R=S(t)$, it follows from \eqref{eq:sigmarr_v2} and \eqref{eq:sigmarrLiquid} that
%-----------------------------
\begin{equation} \label{eq:sigmarrSolidIntegralForm}
\begin{aligned}
  \sigma^{rr}(R,t)
  = \kappa^\text{f}_0 \left[\frac{s^3(t)}{S^3(t)}-1\right] 
  + 4 \int_{S(t)}^R \check{W}_1(\xi,t) e^{\omega^\text{s}(T(\xi,t))} 
   \left[ \frac{\rho_0^\text{f} \, \xi^2}{\rho_0 ^\text{s} \,\bar{r}^2(\xi) r(\xi,t) } 
  - \frac{\rho_0^\text{s} \,\bar{r}^4(\xi) \,{r_{,R}}^2(\xi,t) }{\rho_0^\text{f} \, \xi^2 r^3(\xi,t) }\right]\text{d}\xi \,.
\end{aligned}
\end{equation}
%----------------------------- 
Thus, using \eqref{eq:CauchyComp_v1}$_1$ and \eqref{eq:acrretionfactor}, \eqref{eq:sigmarrSolidIntegralForm} is rewritten as
%-----------------------------
\begin{equation} \label{eq:sigmarrSolidIntegralForm_v2}
\begin{aligned}
  \check{W}_J(R,t)
	+  \frac{ 2\rho_0^\text{s}\, \bar{r}^4(R) \,{r_{,R}}(R,t) }{\rho_0^\text{f}\, R^2 \, r^2(R,t)  }  \, \check{W}_1(R,t) \, e^{\omega^\text{s}(T(R,t))}
  = \kappa^\text{f}_0 \left[\frac{s^3(t)}{S^3(t)}-1\right] \qquad\qquad\qquad\\ 
  +\, 4 \int_{S(t)}^R \check{W}_1(\xi,t) e^{\omega^\text{s}(T(\xi,t))} 
   \left[ \frac{\rho_0^\text{f} \, \xi^2}{\rho_0 ^\text{s} \,\bar{r}^2(\xi) r(\xi,t) } 
  - \frac{\rho_0^\text{s} \,\bar{r}^4(\xi) \,{r_{,R}}^2(\xi,t) }{\rho_0^\text{f} \, \xi^2 r^3(\xi,t) }\right]\text{d}\xi\,,
\end{aligned}
\end{equation}
%----------------------------- 
where
%-----------------------------
\begin{equation} \label{eq:Solid_v2_coeff}
\begin{aligned}	
	\check{W}_1
	&= \frac{\mu^\text{s}_0 \, T \, e^{2 \omega^\text{s}}}{2 T_0} \left[\frac{ \rho_0^\text{f} \, R^2}{ \rho_0^\text{s} \, r^2 \, r_{,R} }\right]^{\frac{2}{3} }   \,, \\
    \check{W}_J
	&=  \frac{\kappa^\text{s}_0 T}{T_0} \left[ \frac{\rho_0^\text{s}  r^2 \, r_{,R}}{\rho_0^\text{f} \, e^{3 \omega^\text{s}} R^2 } - 1 \right]
	-\kappa^\text{s}_0\,\beta^\text{s}_0\,(T-T_0)	
	-\frac{\mu^\text{s}_0 T \,e^{3 \omega^\text{s}}}{3 T_0} \left[\frac{ \rho_0^\text{f} \,  R^2 }{ \rho_0^\text{s} \,  r^2 \, r_{,R} }\right]^{ \frac{5}{3} } 
	\left[ \frac{(\rho_0^\text{s})^2 \, \bar{r}^4 {r_{,R}}^2}{ (\rho_0^\text{f})^2 R^4} + \frac{2r^2}{\bar{r}^2}\right]  \,.
\end{aligned}
\end{equation}
%-----------------------------
Therefore, for the neo-Hookean solid, the moving boundary problem on the domain $R_0 \geq R \geq S(t)$ is written as\footnote{Recall that \eqref{eq:BVP_v1}$_1$ was obtained in \eqref{eq:sigmarrSolidIntegralForm_v2}, while \eqref{eq:BVP_v1}$_2$ restates the heat equation \eqref{eq:MatHeatEqRigidCondRadial_v2}, and \eqref{eq:BVP_v1}$_3$ is Stefan's condition \eqref{eq:StefanConditionFinalVersion}. The thermal boundary conditions are written in \eqref{eq:BVP_v1}$_4$ and \eqref{eq:BVP_v1}$_6$, while \eqref{eq:BVP_v1}$_5$ and \eqref{eq:BVP_v1}$_7$ are the kinematic boundary conditions. Finally, \eqref{eq:BVP_v1}$_8$ denotes the initial condition for the position of the moving interface.}
%-----------------------------
\begin{equation} \label{eq:BVP_v1}
\begin{dcases}
  \check{W}_J(R,t)
	+  \frac{ 2\rho_0^\text{s}\, \check{W}_1(R,t) \, e^{\omega^\text{s}(T(R,t))}\, \bar{r}^4(R) \,{r_{,R}}(R,t) }{\rho_0^\text{f}\, R^2 \, r^2(R,t)  }  
    = \kappa^\text{f}_0 \left[\frac{s^3(t)}{S^3(t)}-1\right] \\
   \hspace{128pt} 
   +4 \int_{S(t)}^R \check{W}_1(\xi,t) e^{\omega^\text{s}(T(\xi,t))} 
   \bigg[ \frac{\rho_0^\text{f} \, \xi^2}{\rho_0 ^\text{s} \,\bar{r}^2(\xi) r(\xi,t) } 
  - \frac{\rho_0^\text{s} \,\bar{r}^4(\xi) \,{r_{,R}}^2(\xi,t) }{\rho_0^\text{f} \, \xi^2 r^3(\xi,t) }\bigg]\text{d}\xi \,, \\
   \frac{K^\text{s} (\rho_0 ^\text{s})^3}{c_E (\rho_0^\text{f})^4   }\bigg [T_{,RR}(R,t)
	 +\alpha^\text{s}(T(R,t))\, \big[T_{,R}(R,t)\big]^2  
	 +\left[\frac{4\bar{r}'(R)}{\bar{r}(R)}
	 -\frac{2}{R}\right]  T_{,R}(R,t) \bigg]
	 = \frac{   R^4 \,\dot{T}(R,t)}{ \,\bar{r}^4(R) e^{\omega^\text{s}(T(R,t))}}  \,, \\
	T_{,R} (S(t),t)
	= \frac{\rho^\text{f}_0\,l}{ K^\text{s} } \, \dot{S}(t)  \,, \\
	T(S(t),t)=T_m \,, \\
	r(S(t),t)=\bar{r}(S(t))=s(t) \,, \\
	K^\text{s} T_{,R}(R_0,t)
	= h_c \left[ T_c - T(R_0,t) \right]\,, \\
	r(R_0,t)=R_0 \,,\\
	S(0)=R_0 \,,
\end{dcases}
\end{equation}
%----------------------------- 
where the temperature field $T(R,t)$, the radial placement map $r(R,t)$, and the location of the moving boundary $S(t)$ are unknown.
%-----------------------------
\begin{remark}
%-----------------------------------------------------------
On the moving boundary, \eqref{eq:sigmarrSolidIntegralForm_v2} is rewritten as
%-----------------------------
\begin{equation} \label{eq:tractionCont_v1}
\begin{aligned}
  \check{W}_J(S(t),t)
	+  \frac{ 2\rho_0^\text{s} \, s^2(t)  \,{r_{,R}}(S(t),t) }{\rho_0^\text{f}\, S^2(t)  }  \, \check{W}_1(S(t),t) 
  = \kappa^\text{f}_0 \left[\frac{s^3(t)}{S^3(t)}-1\right] \,,
\end{aligned}
\end{equation}
%----------------------------- 
where
%-----------------------------
\begin{equation}
\begin{aligned}	
	\check{W}_1(S(t),t)
	&= \frac{\mu^\text{s}_0 }{2} \left[\frac{ \rho_0^\text{f} \, S^2(t)}{ \rho_0^\text{s} \, s^2(t) \, r_{,R}(S(t),t) }\right]^{\frac{2}{3} }   \,, \\
    \check{W}_J(S(t),t)
	&=  \kappa^\text{s}_0 \left[ \frac{\rho_0^\text{s}  s^2(t) \, r_{,R}(S(t),t)}{\rho_0^\text{f} \, S^2(t) } - 1 \right]
	-\frac{\mu^\text{s}_0 }{3} \left[\frac{ \rho_0^\text{f} \,  S^2(t) }{ \rho_0^\text{s} \,  s^2(t) \, r_{,R}(S(t),t) }\right]^{ \frac{5}{3} } 
	\left[ \frac{(\rho_0^\text{s})^2 \, s^4(t)\, {r_{,R}}^2(S(t),t)}{ (\rho_0^\text{f})^2 S^4(t)} + 2\right]  \,.
\end{aligned}
\end{equation}
%-----------------------------
Therefore, traction continuity \eqref{eq:tractionCont_v1} across the moving interface $R=S(t)$ is written as $\mathcal{F}\left( \frac{s(t)}{S(t)}\,, r_{,R}(S(t),t)\right)=0$, where
%-----------------------------------------------------------
\begin{equation}
	\mathcal{F}(x,y):= \kappa^\text{s}_0 \left[ \frac{\rho_0^\text{s}  x^2 \, y}{\rho_0^\text{f}} - 1 \right]
	+\frac{2\mu^\text{s}_0 }{3} \left(\frac{ \rho_0^\text{s} \,  x^2 \,y  }{ \rho_0^\text{f} }\right)^{\frac{1}{3} } 
	\left[ 1-\frac{ (\rho_0^\text{f})^2 }{ (\rho_0^\text{s})^2 \, x^4 \, y^2 } \right] 
	+\kappa^\text{f}_0 \left[1-x^3\right]\,.
\end{equation}
%-----------------------------------------------------------
Since ${r^\text{f}}_{,R}\left(S(t),t\right)=\frac{s(t)}{S(t)}$, traction continuity implies an implicit relation between ${r^\text{s}}_{,R}(S(t),t)$ and ${r^\text{f}}_{,R}(S(t),t)$.
%-----------------------------------------------------------
\end{remark}
%-----------------------------

\paragraph*{Non-dimensionalization.}
%----------------------------- 
%----------------------------- 
\begin{table}   
\caption{Definitions of the scaled variables and the dimensionless parameters incorporated in \eqref{eq:LinMomEqnonDim}-\eqref{eq:ICBCnonDim}.}
\centering
\medskip
\begin{tabularx}{360 pt}{@{}lL @{}} 
\toprule
Category 
& Definitions    \\
\midrule
Independent variables 
& $\mathtt{R}=\frac{R}{R_0}$ \\[4pt]
& $\mathtt{t}= \frac{K^\text{s} (\rho_0 ^\text{s})^3 t}{c_E (\rho_0^\text{f})^4  R_0^2 }$    \\
\midrule 
\addlinespace
Dependent unknown variables \quad 
& 
$\mathtt{r}(\mathtt{R},\mathtt{t})=\frac{1}{R_0}r\left(R_0 \mathtt{R},\frac{c_E (\rho_0^\text{f})^4  R_0^2 \mathtt{t} }{ K^\text{s} (\rho_0 ^\text{s})^3 }\right)$  \\[4pt]
& $\mathtt{T}(\mathtt{R},\mathtt{t})=\frac{1}{T_c-T_m}\left[T\left(R_0 \mathtt{R},\frac{c_E (\rho_0^\text{f})^4  R_0^2 \mathtt{t} }{ K^\text{s} (\rho_0 ^\text{s})^3 }\right)-T_m \right]$  \\[4pt] 
& $\mathtt{S}(\mathtt{t})=\frac{1}{R_0}S\left( \frac{c_E (\rho_0^\text{f})^4  R_0^2 \mathtt{t} }{ K^\text{s} (\rho_0 ^\text{s})^3 }\right)$ \\
\midrule
\addlinespace
Dimensionless constant parameters 
& $\mathsf{p}= \frac{\mu_0^\text{s}}{\kappa_0^\text{f}}$ \\[4pt] 
& $\mathsf{q}= \frac{\kappa_0^\text{s}}{\kappa_0^\text{f}}$ \\[4pt]
& $\mathsf{f}= \frac{\rho_0^\text{s}}{\rho_0^\text{f}}$   \\[4pt]
& $\mathsf{a}=1-\frac{T_c}{T_m}$ \\[4pt]
&  $\mathsf{b}=\beta^\text{s}_0 [T_m - T_c] $\\[4pt]
& $\mathsf{h}=\frac{h_c R_0}{ K^\text{s}} $ \\[4pt]
& $\mathsf{L}=\frac{(\rho^\text{s}_0)^3 l }{(\rho^\text{f}_0)^3 c_E [T_m - T_c]} $ \\
\bottomrule
\end{tabularx}
\label{table:Nondimensionalization}
\end{table}
%-----------------------------
Let $0\leq\mathtt{R}\leq 1$ and $\mathtt{t}\geq 0$ be the dimensionless radial coordinate and time variable, respectively. The dimensionless radial placement map, temperature field and the location of phase-change interface are denoted by $\mathtt{r} (\mathtt{R},\mathtt{t})$, $\mathtt{T} (\mathtt{R},\mathtt{t})$ and $\mathtt{S} (\mathtt{t})$, respectively. These dimensionless quantities are defined in Table \ref{table:Nondimensionalization}. It follows from \eqref{eq:BVP_v1}$_1$ and \eqref{eq:Solid_v2_coeff} that for $1 > R\geq \mathtt{S} (\mathtt{t})$:\footnote{Since $\mathsf{a}=1-\frac{T_c}{T_m}$ and $\mathsf{b}=\beta^\text{s}_0 [T_m - T_c] $, it follows from \eqref{eq:thermalexpansionrelation} that $e^{ \omega^\text{s}(T) }=\left[\frac{1-(\mathsf{a}+\mathsf{b})\mathtt{T}}{1-\mathsf{a}\mathtt{T}}\right]^\frac{1}{3}$ and 
$[T_m-T_c]\,\alpha^\text{s}(T)=\frac{\mathsf{b}}{3[1-(\mathsf{a}+\mathsf{b})\mathtt{T}] [1-\mathsf{a}\mathtt{T}]}$.
}
%-----------------------------------------------------------
\begin{equation}  \label{eq:LinMomEqnonDim}
\begin{aligned}
	&\mathsf{q} [ 1-\mathsf{a} \mathtt{T} (\mathtt{R},\mathtt{t})] \left[\frac{\mathsf{f} \mathtt{r}^2(\mathtt{R},\mathtt{t}) \mathtt{r}_{,\mathtt{R}}(\mathtt{R},\mathtt{t}) [ 1-\mathsf{a} \mathtt{T} (\mathtt{R},\mathtt{t})]}{\mathtt{R}^2 [ 1 - (\mathsf{a}+\mathsf{b})\mathtt{T} (\mathtt{R},\mathtt{t})] }  - 1 \right] 
	+ \mathsf{p}\mathsf{b}\mathtt{T}(\mathtt{R},\mathtt{t})
	+1-\frac{ \mathtt{s}^3(\mathtt{t})}{\mathtt{S}^3(\mathtt{t})} \\
	& + \mathsf{p} [ 1-(\mathsf{a}+\mathsf{b})\mathtt{T} (\mathtt{R},\mathtt{t})]  \left(\frac{\mathsf{f} \mathtt{r}^2(\mathtt{R},\mathtt{t}) \mathtt{r}_{,\mathtt{R}}(\mathtt{R},\mathtt{t})}{\mathtt{R}^2}\right)^{\frac{1}{3}}
	\left[\frac{\bar{\mathtt{r}}^4(\mathtt{R})}{\mathtt{r}^4(\mathtt{R},\mathtt{t})}
	-\frac{1}{3}\left(\frac{2\,\mathtt{R}^4 }{ \mathsf{f}^2 \, \bar{\mathtt{r}}^2(\mathtt{R})\, \mathtt{r}^2(\mathtt{R},\mathtt{t})\, \mathtt{r}_{,\mathtt{R}}^2(\mathtt{R},\mathtt{t})}+1\right) \right] \\
	&- 2 \mathsf{p} \int_{\mathtt{S} (\mathtt{t})}^\mathtt{R} 
	[ 1-(\mathsf{a}+\mathsf{b}) \mathtt{T}(\zeta, \mathtt{t}) ] 	\left(\frac{\mathsf{f} \mathtt{r}^2(\zeta, \mathtt{t}) \mathtt{r}_{,\mathtt{R}}(\zeta, \mathtt{t})}{\zeta^2}\right)^{-\frac{2}{3}}
   \bigg[ \frac{  \zeta^2}{ \mathsf{f} \bar{\mathtt{r}}^2 (\zeta ) \mathtt{r} (\zeta, \mathtt{t}) } 
  - \frac{\mathsf{f} \,\bar{\mathtt{r}}^4(\zeta )  \,{\mathtt{r}_{,\mathtt{R}}^2(\zeta, \mathtt{t})} }{  \zeta^2 \mathtt{r}^3 (\zeta, \mathtt{t})  }\bigg]\text{d}\zeta
  =0 \,,
\end{aligned}
\end{equation}
%-----------------------------------------------------------
where $\bar{\mathtt{r}}(\mathtt{R})=\mathtt{r}(\mathtt{R},\mathtt{S}^{-1}(\mathtt{R}))$, $\mathtt{s}(\mathtt{t})=\mathtt{r}(\mathtt{S}(\mathtt{t}),\mathtt{t})$ and $\mathsf{a}$, $\mathsf{b}$, $\mathsf{f}$, $\mathsf{p}$, $\mathsf{q}$ are dimensionless constant parameters defined in Table \ref{table:Nondimensionalization}.\footnote{
The non-dimensionalized traction continuity condition across the moving interface reads
%-----------------------------------------------------------
\begin{equation}  
\begin{aligned}
	\mathsf{q} \left[\frac{\mathsf{f} \mathtt{s}^2(\mathtt{t}) \mathtt{r}_{,\mathtt{R}}(\mathtt{S}(\mathtt{t}),\mathtt{t})}{ \mathtt{S}^2(\mathtt{t})  }  - 1 \right] 
	+ \frac{2\mathsf{p}}{3} \left(\frac{\mathsf{f} \mathtt{s}^2(\mathtt{t}) \mathtt{r}_{,\mathtt{R}}(\mathtt{S}(\mathtt{t}),\mathtt{t})}{ \mathtt{S}^2(\mathtt{t})  }\right)^{\frac{1}{3}}
	\left[1	- \frac{ \mathtt{S}^4(\mathtt{t}) }{ \mathsf{f}^2 \, \mathtt{s}^4(\mathtt{t})\, \mathtt{r}_{,\mathtt{R}}^2(\mathtt{S}(\mathtt{t}),\mathtt{t})} \right] 
  =\frac{ \mathtt{s}^3(\mathtt{t})}{\mathtt{S}^3(\mathtt{t})} -1 \,.
\end{aligned}
\end{equation}
%-----------------------------------------------------------
}
%---------------------
\begin{figure}[t!]
\centering
\vskip 0.0in
\includegraphics[width=.27\textwidth]{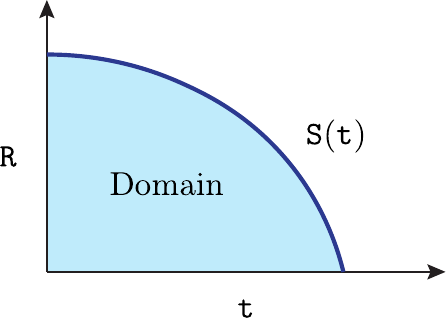}
\vskip 0.1in
\caption{A sketch of the domain $\{(\mathtt{R},\mathtt{t}): 0\leq \mathtt{S} (\mathtt{t}) \leq \mathtt{R} \leq 1=\mathtt{S} (0) \}$ for the non-dimensionalized moving boundary problem, where $\mathtt{S} (\mathtt{t})$ is an unknown. Additionally, the temperature field $\mathtt{T} (\mathtt{R},\mathtt{t})$ and the deformation field $\mathtt{r} (\mathtt{R},\mathtt{t})$ are unknown over this evolving domain.}
\label{fig:domain}
\end{figure}
%--------------------- 
Similarly, for $1 > R > \mathtt{S} (\mathtt{t})$, the heat equation \eqref{eq:BVP_v1}$_2$ is rewritten as 
%-----------------------------------------------------------
\begin{equation} \label{eq:HeatEqnonDim}
\begin{aligned}
	 \mathtt{T}_{,\mathtt{R}\mathtt{R}}(\mathtt{R},\mathtt{t}) 
	 -\frac{\mathsf{b} \mathtt{T}_{,\mathtt{R}} ^2 (\mathtt{R},\mathtt{t})}{3[1-(\mathsf{a}+\mathsf{b})\mathtt{T}(\mathtt{R},\mathtt{t})] [1-\mathsf{a}\mathtt{T}(\mathtt{R},\mathtt{t})]}   
	 +\left[\frac{4 \bar{\mathtt{r}}'(\mathtt{R})}{\bar{\mathtt{r}}(\mathtt{R})}
	 -\frac{2}{\mathtt{R}}\right] \mathtt{T}_{,\mathtt{R}}(\mathtt{R},\mathtt{t})
	 = \frac{   \mathtt{R}^4  [1-\mathsf{a}\mathtt{T}(\mathtt{R},\mathtt{t})]^\frac{1}{3}\mathtt{T}_{,\mathtt{t}}(\mathtt{R},\mathtt{t})}{ \bar{\mathtt{r}}^4(\mathtt{R} )  [1-(\mathsf{a}+\mathsf{b})\mathtt{T}(\mathtt{R},\mathtt{t})]^\frac{1}{3}} \,,
\end{aligned}
\end{equation}
%-----------------------------------------------------------
and, \eqref{eq:BVP_v1}$_{3-8}$ are rewritten as
%-----------------------------------------------------------
\begin{equation} \label{eq:ICBCnonDim}
\begin{aligned}
	\mathtt{T}_{,\mathtt{R}} (\mathtt{S}(\mathtt{t}),\mathtt{t})
	&= -\mathsf{L}  \frac{\text{d}\mathtt{S}(\mathtt{t})}{ \text{d}\mathtt{t} }  \,, \\
	\mathtt{T}(\mathtt{S}(\mathtt{t}),\mathtt{t}) 
	&= 0 \,, \\
	\mathtt{r}(\mathtt{S}(\mathtt{t}),\mathtt{t})
	&=\bar{\mathtt{r}}(\mathtt{S}(\mathtt{t}))
	 =\mathtt{s}(\mathtt{t})  \,, \\
	\mathtt{T}_{,\mathtt{R}}( 1,\mathtt{t})
	&=\mathsf{h} \left[ 1 - \mathtt{T}( 1,\mathtt{t}) \right]\,, \\
	\mathtt{r}(1,\mathtt{t})&=1 \,,\\
	\mathtt{S}(0)&=1 \,,
\end{aligned}
\end{equation}
%-----------------------------------------------------------
where $\mathsf{h}$, $\mathsf{L}$ are dimensionless constant parameters defined in Table \ref{table:Nondimensionalization}. Thus, \eqref{eq:LinMomEqnonDim}-\eqref{eq:ICBCnonDim} constitute the non-dimensionalized boundary-value problem on the evolving domain $\{(\mathtt{R},\mathtt{t}): 0\leq \mathtt{S} (\mathtt{t}) \leq \mathtt{R} \leq 1=\mathtt{S} (0) \}$ (Figure~\ref{fig:domain}).\footnote{Recall that $\mathbb{m}(\mathtt{t}):=1-\mathtt{S}^3(\mathtt{t})$ represents the mass fraction solidified.} 
%--------------------------------
Further, the physical components of the Cauchy stress in the solid are non-dimensionalized as $\mathring{ \sigma }^{ab}=\frac{\sigma^{ab} \sqrt{g_{aa}g_{bb}}  }{\kappa^\text{f}_0}$ (no summation).\footnote{Note that $\mathring{ \sigma }^{\theta\theta}=\mathring{ \sigma }^{\phi\phi}$.} Similarly, the pressure in the liquid, which is independent of $R$, is non-dimensionalized as $\mathring{p}(\mathtt{t})=\frac{p( t )}{\kappa^\text{f}_0}$.\footnote{It is implied from \eqref{eq:sigmarrLiquid} that $\mathring{p}(\mathtt{t})=1-\frac{\mathtt{s}^3(\mathtt{t})}{\mathtt{S}^3(\mathtt{t})}$. }

\begin{remark}
%--------------------------------
Note that \eqref{eq:HeatEqnonDim} is rewritten as 
%-----------------------------------------------------------
\begin{equation} \label{eq:HeatEqnonDim_v2forDG}
\begin{aligned}
	 \Bigg[ \frac{ \bar{\mathtt{r}}^4(\mathtt{R})}{ \mathtt{R}^2} \left[\frac{1-(\mathsf{a}+\mathsf{b})\mathtt{T}(\mathtt{R},\mathtt{t})}{1-\mathsf{a}\mathtt{T}(\mathtt{R},\mathtt{t})}\right]^\frac{1}{3}\mathtt{T}_{,\mathtt{R}}(\mathtt{R},\mathtt{t})\Bigg] _{,\mathtt{R}}
	 =   \mathtt{R}^2  \mathtt{T}_{,\mathtt{t}}(\mathtt{R},\mathtt{t})   \,,
\end{aligned}
\end{equation}
%-----------------------------------------------------------
which is integrated using \eqref{eq:ICBCnonDim}$_{1-6}$ to obtain
%-----------------------------------------------------------
\begin{equation} \label{eq:IntegratedHeat}
\begin{aligned}
	\mathsf{h}[1-\mathtt{T}(1,\mathtt{t})]\left[\frac{1-(\mathsf{a}+\mathsf{b})\mathtt{T}(1,\mathtt{t})}{1-\mathsf{a}\mathtt{T}(1,\mathtt{t})}\right]^\frac{1}{3}
	 +\frac{\mathsf{L} \mathtt{s}^4(\mathtt{t})\dot{\mathtt{S}}(\mathtt{t})}{ \mathtt{S}^2(\mathtt{t}) }  
	 = \int_{\mathtt{S}(\mathtt{t})}^1\xi^2 \, \mathtt{T}_{,\mathtt{t}}(\xi,\mathtt{t})\, \text{d}\xi  \,.
\end{aligned}
\end{equation}
%-----------------------------------------------------------
Since\footnote{This is implied from the fact that $\frac{\partial}{\partial\mathtt{t}}\int_{\mathtt{S}(\mathtt{t})}^1  \xi^2\,\mathtt{T}(\xi,\mathtt{t})\,\text{d}\xi=\int_{\mathtt{S}(\mathtt{t})}^1   \xi^2\,\mathtt{T}_{,\mathtt{t}}(\xi,\mathtt{t}) \,\text{d}\xi $.}
%------------------------------
\begin{equation}
	 \int_0^\mathtt{t}\left[\int_{\mathtt{S}(\tau)}^1 \xi^2 \,\mathtt{T}_{,\tau}(\xi,\tau)\text{d}\xi \right]\text{d}\tau
	 =\int_{\mathtt{S}(\mathtt{t})}^1  \xi^2\,\mathtt{T}(\xi,\mathtt{t}) \text{d}\xi\,,
\end{equation}
%------------------------------
it follows from \eqref{eq:IntegratedHeat} that\footnote{Note that the change of variable $\int_0^\mathtt{t}\frac{\mathtt{s}^4(\tau)\,\dot{\mathtt{S}}(\tau)}{ \mathtt{S}^2(\tau) } \text{d}\tau = \int_1^{\mathtt{S}(\mathtt{t})}\frac{\bar{\mathtt{r}}^4(\xi)\,\text{d}\xi}{ \xi^2} $ has been used here.}
%-----------------------------------------------------------
\begin{equation} \label{eq:StefanIntegralCoupled} 
\begin{aligned}
	\int_0^\mathtt{t} \mathsf{h}[1-\mathtt{T}(1,\tau)]\left[\frac{1-(\mathsf{a}+\mathsf{b})\mathtt{T}(1,\tau)}{1-\mathsf{a}\mathtt{T}(1,\tau)}\right]^\frac{1}{3} \text{d}\tau
	 = \int_{\mathtt{S}(\mathtt{t})}^1 \left[\xi^2\, \mathtt{T}(\xi,\mathtt{t})+\frac{ \mathsf{L}\,\bar{\mathtt{r}}^4(\xi)}{ \xi^2}\right] \text{d}\xi  \,.
\end{aligned}
\end{equation}
%-----------------------------------------------------------
Thus, Stefan's condition \eqref{eq:ICBCnonDim}$_{1}$ can be replaced with the integral constraint \eqref{eq:StefanIntegralCoupled}. 
\end{remark}
%--------------------------------

%--------------------------------
\begin{remark}
Let $\mathtt{z}\geq 0$ be the time when the layer with radial coordinate $\mathtt{R}$ solidifies and attaches to the shell, i.e., $\mathtt{R}=\mathtt{S}(\mathtt{z})$. Let $\rho$ and $\Upsilon$ denote the radial placement and temperature fields, respectively, expressed as functions of $\mathtt{z} $ and $\mathtt{t} $, i.e.,
%-----------------------------------------------------------
\begin{equation} \label{eq:TransformDef} 
	\rho(\mathtt{z},\mathtt{t}) 
	= \mathtt{r}(\mathtt{S}(\mathtt{z}),\mathtt{t}) \,, \qquad
	\Upsilon(\mathtt{z},\mathtt{t}) 
	= \mathtt{T}(\mathtt{S}(\mathtt{z}),\mathtt{t})  \,.
\end{equation}
%-----------------------------------------------------------
Thus, the heat equation \eqref{eq:HeatEqnonDim} is rewritten in terms of $\rho$ and $\Upsilon$ as\footnote{The following relations have been used here
%-----------------------------------------------------------
\begin{equation}  
	\mathtt{r}_{,\mathtt{R}}(\mathtt{R},\mathtt{t})
	=\frac{\rho_{,\mathtt{z}}(\mathtt{z},\mathtt{t})}{\dot{\mathtt{S}}(\mathtt{z})} \,, \qquad	
	\mathtt{T}_{,\mathtt{R}}(\mathtt{R},\mathtt{t})
	=\frac{\Upsilon_{,\mathtt{z}}(\mathtt{z},\mathtt{t})}{\dot{\mathtt{S}}(\mathtt{z})} \,, \qquad
	\mathtt{T}_{,\mathtt{R}\mathtt{R}}(\mathtt{R},\mathtt{t})
	=\frac{\Upsilon_{,\mathtt{z}\mathtt{z}}(\mathtt{z},\mathtt{t})}{\dot{\mathtt{S}}^2(\mathtt{z})}
	-\frac{\Upsilon_{,\mathtt{z}}(\mathtt{z},\mathtt{t})\,\ddot{\mathtt{S}}(\mathtt{z})}{\dot{\mathtt{S}}^3(\mathtt{z})} \,,
\end{equation}
%-----------------------------------------------------------
where $\mathtt{R}=\mathtt{S}(\mathtt{z})$. These are obtained by differentiating the definition \eqref{eq:TransformDef} with respect to $\mathtt{z}$. }
%-----------------------------------------------------------
\begin{equation} \label{eq:HeatEqnonDim_Transform_v1} 
\begin{aligned}
	 \Upsilon_{,\mathtt{z}\mathtt{z}} (\mathtt{z},\mathtt{t})
	& -\frac{\mathsf{b} \Upsilon_{,\mathtt{z}} ^2 
	 (\mathtt{z},\mathtt{t})}{3 [1-\mathsf{a}\Upsilon(\mathtt{z},\mathtt{t})]
	  [1-(\mathsf{a}+\mathsf{b})\Upsilon(\mathtt{z},\mathtt{t})]}   
	 +\left[\frac{4 \dot{\mathtt{s}}(\mathtt{z})}{\mathtt{s}(\mathtt{z})}
	 -\frac{2 \dot{\mathtt{S}}(\mathtt{z})}{\mathtt{S}(\mathtt{z})}
	 -\frac{ \ddot{\mathtt{S}}(\mathtt{z})}{\dot{\mathtt{S}}(\mathtt{z})}\right] 
	 \Upsilon_{,\mathtt{z}}(\mathtt{z},\mathtt{t}) \\
	 & = \frac{   \mathtt{S}^4(\mathtt{z}) \dot{\mathtt{S}}^2(\mathtt{z}) 
	 [1-\mathsf{a}\Upsilon(\mathtt{z},\mathtt{t})]^\frac{1}{3} 
	 \Upsilon_{,\mathtt{t}}(\mathtt{z},\mathtt{t})}{ \mathtt{s}^4(\mathtt{z})  
	 [1-(\mathsf{a}+\mathsf{b})\Upsilon(\mathtt{z},\mathtt{t})]^\frac{1}{3}} \,,
\end{aligned}
\end{equation}
%-----------------------------------------------------------
where $\mathtt{s}(\mathtt{t})=\bar{\mathtt{r}}(\mathtt{S}(\mathtt{t}))=\rho(\mathtt{t},\mathtt{t}) $, and thus $\dot{\mathtt{s}}(\mathtt{t})=\rho_{,\mathtt{z}}(\mathtt{t},\mathtt{t}) + \rho_{,\mathtt{t}}(\mathtt{t},\mathtt{t}) $. Note that \eqref{eq:HeatEqnonDim_Transform_v1} can be rearranged as follows
%-----------------------------------------------------------
\begin{equation}  \label{eq:HeatEqnonDim_Transform_v2} 
\begin{aligned}	   
	 \frac{\partial}{\partial \mathtt{z} }\left[\frac{ \mathtt{s}^4(\mathtt{z})  [1-(\mathsf{a}+\mathsf{b})\Upsilon(\mathtt{z},\mathtt{t})]^\frac{1}{3}\Upsilon_{,\mathtt{z}} (\mathtt{z},\mathtt{t})}{ \mathtt{S}^2(\mathtt{z}) \dot{\mathtt{S}}(\mathtt{z})[1-\mathsf{a} \Upsilon(\mathtt{z},\mathtt{t})]^\frac{1}{3} } \right] 
	 =   \mathtt{S}^2(\mathtt{z}) \dot{\mathtt{S}}(\mathtt{z}) \Upsilon_{,\mathtt{t}}(\mathtt{z},\mathtt{t}) \,.
\end{aligned}
\end{equation}
%----------------------------------------------------------- 
Similarly, \eqref{eq:LinMomEqnonDim} is rewritten as 
%-----------------------------------------------------------
\begin{equation} \label{eq:LinMomEqnonDim_Transform}
\begin{aligned}
	&\mathsf{q}  [ 1-\mathsf{a} \Upsilon(\mathtt{z},\mathtt{t}) ] \left[\frac{\mathsf{f} \rho^2(\mathtt{z},\mathtt{t}) \rho_{,\mathtt{z}}(\mathtt{z},\mathtt{t}) [1 - \mathsf{a} \Upsilon(\mathtt{z},\mathtt{t})  ]}{\mathtt{S}^2(\mathtt{z}) \dot{\mathtt{S}}(\mathtt{z})  [1 -(\mathsf{a}+\mathsf{b}) \Upsilon(\mathtt{z},\mathtt{t})  ]}  -1 \right] 
	+1-\frac{\mathtt{s}^3(\mathtt{z})}{\mathtt{S}^3(\mathtt{z})}  \\
	&+ \mathsf{p} [ 1- (\mathsf{a}+\mathsf{b}) \Upsilon(\zeta,\mathtt{t}) ] 
	\left(\frac{\mathsf{f} \rho^2(\mathtt{z},\mathtt{t}) \rho_{,\mathtt{z}}(\mathtt{z},\mathtt{t})}{\mathtt{S}^2(\mathtt{z}) \dot{\mathtt{S}}(\mathtt{z})}\right)^{\frac{1}{3}}
	\left[\frac{ \mathtt{s}^4(\mathtt{z})}{\rho^4(\mathtt{z},\mathtt{t})}
	-\frac{1}{3}\left(\frac{2\,\mathtt{S}^4(\mathtt{z}) \dot{\mathtt{S}}^2(\mathtt{z}) }{\mathsf{f}^2 \mathtt{s}^2(\mathtt{z})\rho^2(\mathtt{z},\mathtt{t}) \rho_{,\mathtt{z}}^2(\mathtt{z},\mathtt{t}) } +1\right) \right]  \\
	&+ 2 \mathsf{p} \int_{\mathtt{z} }^\mathtt{t} 
	[ 1- (\mathsf{a}+\mathsf{b}) \Upsilon(\zeta,\mathtt{t}) ] 
	\left(\frac{\mathsf{f} \rho^2(\zeta,\mathtt{t}) \rho_{,\mathtt{z}}(\zeta,\mathtt{t})}{\mathtt{S}^2 (\zeta) \dot{\mathtt{S}} (\zeta) }\right)^{-\frac{2}{3}} 
   \bigg[ \frac{ \mathtt{S}^2 (\zeta)}{ \mathsf{f} \mathtt{s}^2 (\zeta)  \rho (\zeta,\mathtt{t}) }
  - \frac{\mathsf{f} \,\mathtt{s}^4 (\zeta)  \rho_{,\mathtt{z}}^2(\zeta,\mathtt{t}) }{  \mathtt{S}^2 (\zeta) \dot{\mathtt{S}}^2(\zeta)\rho^3(\zeta,\mathtt{t})  }\bigg]\text{d}\zeta 
   =0\,,
\end{aligned}
\end{equation}
%-----------------------------------------------------------
and, \eqref{eq:ICBCnonDim} is rewritten as  
%-----------------------------------------------------------
\begin{equation}  \label{eq:ICBCnonDim_Transform}
\begin{aligned}
	\Upsilon_{,\mathtt{z}}(\mathtt{t},\mathtt{t})
	&= -\mathsf{L}\, \dot{\mathtt{S}}^2(\mathtt{t}) \,, \\
	\Upsilon(\mathtt{t},\mathtt{t}) 
	&= 0 \,, \\
	\rho( \mathtt{t} ,\mathtt{t} )
	&=  \mathtt{s}(\mathtt{t})   \,, \\
	\Upsilon_{,\mathtt{z}}(0,\mathtt{t})
	&=\mathsf{h}\, \dot{\mathtt{S}} (0)\left[ 1 - \Upsilon( 0 ,\mathtt{t}) \right]\,, \\
	\rho(0,\mathtt{t})&=1 \,,\\
	\mathtt{S}(0)&=1 \,.
\end{aligned}
\end{equation}
%-----------------------------------------------------------
Hence, the transformed system of equations \eqref{eq:HeatEqnonDim_Transform_v2}-\eqref{eq:ICBCnonDim_Transform} form a boundary-value problem over the fixed triangular domain $\{(\mathtt{z},\mathtt{t}): 0 \leq \mathtt{z} \leq \mathtt{t} \leq \mathtt{t}_{\text{end}} \}$, where $\mathtt{t}_{\text{end}}$ is the time taken for complete solidification, i.e., $\mathtt{S}(\mathtt{t}_{\text{end}})=0$.
%--------------------------------
\end{remark}
%--------------------------------

%--------------------------------
\begin{remark}
%--------------------------------
For $\mathtt{t}>0$, consider the ratio $\mathtt{u}=\frac{\mathtt{z}}{\mathtt{t}} \in [0,1]$. Let $\alpha$ and $\gamma$ denote the radial placement and temperature fields, respectively, expressed as functions of $\mathtt{u} $ and $\mathtt{t} $, i.e.,
%-----------------------------------------------------------
\begin{equation}  
	\gamma(\mathtt{u},\mathtt{t})=\Upsilon(\mathtt{u}\mathtt{t},\mathtt{t})\,, \quad\quad
	\alpha(\mathtt{u},\mathtt{t})=\rho(\mathtt{u}\mathtt{t},\mathtt{t})\,.
\end{equation}
%-----------------------------------------------------------
This change of variable transforms the triangular domain $\{(\mathtt{z},\mathtt{t}): 0 < \mathtt{z} \leq \mathtt{t} \leq \mathtt{t}_{\text{end}} \}$ into the rectangular domain $\{(\mathtt{u},\mathtt{t}): 0 \leq \mathtt{u} \leq 1, 0< \mathtt{t} \leq \mathtt{t}_{\text{end}} \}$. Further, the derivatives of $\rho$ and $\Upsilon$ are related to those of $\alpha$ and $\gamma$ as 
%-----------------------------------------------------------
\begin{equation}  
	\rho_{,\mathtt{z}}(\mathtt{z},\mathtt{t})
	=\frac{\alpha_{,\mathtt{u}}(\mathtt{u},\mathtt{t})}{  \mathtt{t} } \,, \qquad	
	\Upsilon_{,\mathtt{z}}(\mathtt{z},\mathtt{t})
	=\frac{\gamma_{,\mathtt{u}}(\mathtt{u},\mathtt{t})}{  \mathtt{t} } \,, \qquad
	\Upsilon_{,\mathtt{z}\mathtt{z}}(\mathtt{z},\mathtt{t})
	=\frac{\gamma_{,\mathtt{u}\mathtt{u}}(\mathtt{u},\mathtt{t})}{  \mathtt{t}^2 } \,, 
\end{equation}
%-----------------------------------------------------------
and,
%-----------------------------------------------------------
\begin{equation}  
	\Upsilon_{,\mathtt{t}}(\mathtt{z},\mathtt{t})
	=\gamma_{,\mathtt{t}}(\mathtt{u},\mathtt{t})-\frac{ \mathtt{u} }{  \mathtt{t} }\gamma_{,\mathtt{u}}(\mathtt{u},\mathtt{t}) \,,
\end{equation}
%-----------------------------------------------------------.
where $\mathtt{u}=\frac{\mathtt{z}}{\mathtt{t}}$. Note that the thermal and displacement boundary conditions in  \eqref{eq:ICBCnonDim_Transform} are expressed in terms $\alpha$ and $\gamma$ as follows
%-----------------------------------------------------------
\begin{equation}  \label{eq:ICBCnonDim_Transform_v2}
\begin{aligned}
	\gamma_{,\mathtt{u}}(1,\mathtt{t})
	&= -\mathsf{L}\mathtt{t}\,  \dot{\mathtt{S}}^2(\mathtt{t}) \,, \\
	\gamma(1,\mathtt{t})
	&= 0 \,, \\
	\alpha(1,\mathtt{t})
	&=  \mathtt{s}(\mathtt{t})   \,, \\
	\gamma_{,\mathtt{u}}(0,\mathtt{t})
	&=\mathsf{h}\, \dot{\mathtt{S}} (0)\,\mathtt{t}\left[ 1 - \gamma( 0 ,\mathtt{t}) \right]\,, \\
	\alpha(0,\mathtt{t})&=1 \,,\\
	\mathtt{S}(0)&=1 \,.
\end{aligned}
\end{equation}
%-----------------------------------------------------------
Thus, it follows from \eqref{eq:ICBCnonDim_Transform_v2}$_3$ that $\dot{\mathtt{s}}(\mathtt{t})=\alpha_{,\mathtt{t}}(1,\mathtt{t})$. Therefore, \eqref{eq:HeatEqnonDim_Transform_v1} is rewritten in terms of $\alpha$ and $\gamma$ as  %-----------------------------------------------------------
\begin{equation} \label{eq:HeatEqnonDim_Transform2} 
\begin{aligned}
	 \gamma_{,\mathtt{u}\mathtt{u}}(\mathtt{u},\mathtt{t})
	 &-\frac{\mathsf{b} \, \gamma_{,\mathtt{u}}^2(\mathtt{u},\mathtt{t})}{3 [1-\mathsf{a} \,\gamma(\mathtt{u},\mathtt{t})][1-(\mathsf{a}+\mathsf{b}) \,\gamma(\mathtt{u},\mathtt{t})]}   
	 +\left[\frac{4 \alpha_{,\mathtt{t}}(1,\mathtt{u}\mathtt{t})}{\alpha (1,\mathtt{u}\mathtt{t})}
	 -\frac{2 \dot{\mathtt{S}}(\mathtt{u}\mathtt{t})}{\mathtt{S}(\mathtt{u}\mathtt{t})}
	 -\frac{ \ddot{\mathtt{S}}(\mathtt{u}\mathtt{t})}{\dot{\mathtt{S}}(\mathtt{u}\mathtt{t})}\right]\mathtt{t} \gamma_{,\mathtt{u}}(\mathtt{u},\mathtt{t}) \\
	 &+ \frac{   \mathtt{S}^4(\mathtt{u}\mathtt{t})\, \dot{\mathtt{S} }^2(\mathtt{u}\mathtt{t}) [1- \mathsf{a} \,\gamma(\mathtt{u},\mathtt{t})]^\frac{1}{3} \left[ \mathtt{u}\gamma_{,\mathtt{u}}(\mathtt{u},\mathtt{t}) -\mathtt{t} \gamma_{,\mathtt{t}}(\mathtt{u},\mathtt{t}) \right] }{ \alpha^4 (1,\mathtt{u}\mathtt{t})  [1-(\mathsf{a}+\mathsf{b}) \,\gamma(\mathtt{u},\mathtt{t})]^\frac{1}{3}}   		=0\,.
\end{aligned}
\end{equation}
%-----------------------------------------------------------
Similarly, \eqref{eq:LinMomEqnonDim_Transform} is rewritten as 
%-----------------------------------------------------------
\begin{equation} \label{eq:LinMomEqnonDim_Transform2}
\begin{aligned}
	&\mathsf{q} [1-\mathsf{a} \gamma(\mathtt{u},\mathtt{t})] \left[\frac{\mathsf{f}  \alpha^2(\mathtt{u},\mathtt{t})\, \alpha_{,\mathtt{u}}(\mathtt{u},\mathtt{t}) [ 1-\mathsf{a} \gamma(\mathtt{u},\mathtt{t}) ] }{\mathtt{t} \,\mathtt{S}^2(\mathtt{u}\mathtt{t})\, \dot{\mathtt{S}}(\mathtt{u}\mathtt{t}) [ 1-(\mathsf{a}+\mathsf{b})\gamma(\mathtt{u},\mathtt{t}) ] } - 1 \right] 
	+1-\frac{\alpha^3(1,\mathtt{u}\mathtt{t})}{\mathtt{S}^3(\mathtt{u}\mathtt{t})}  \\
	& +\mathsf{p} [ 1-(\mathsf{a}+\mathsf{b})\gamma(\mathtt{u},\mathtt{t}) ] \left( \frac{\mathsf{f} \alpha^2(\mathtt{u},\mathtt{t})\, \alpha_{,\mathtt{u}}(\mathtt{u},\mathtt{t})}{\mathtt{t} \,\mathtt{S}^2(\mathtt{u}\mathtt{t})\, \dot{\mathtt{S}}(\mathtt{u}\mathtt{t})} \right)^{\frac{1}{3}} 
	\left[\frac{\alpha^4(1,\mathtt{u}\mathtt{t})}{\alpha^4(\mathtt{u},\mathtt{t})}
	-\frac{1}{3}\left( \frac{2\,\mathtt{t}^2 \,\mathtt{S}^4(\mathtt{u}\mathtt{t})\, \dot{\mathtt{S}}^2(\mathtt{u}\mathtt{t})}{\mathsf{f}^2 \alpha^2(1,\mathtt{u}\mathtt{t}) \alpha^2(\mathtt{u},\mathtt{t})\, \alpha_{,\mathtt{u}}^2(\mathtt{u},\mathtt{t})}+1\right)  \right]  \\
	&+ 2 \mathsf{p} \int_{\mathtt{u} }^1 
	[ 1-(\mathsf{a}+\mathsf{b}) \gamma(\nu,\mathtt{t}) ] 
	\left(\frac{\mathsf{f} \alpha^2(\nu,\mathtt{t})\, \alpha_{,\mathtt{u}}(\nu,\mathtt{t})}{\mathtt{t} \,\mathtt{S}^2(\nu\mathtt{t})\, \dot{\mathtt{S}}(\nu\mathtt{t})}\right)^{-\frac{2}{3}} 
   \bigg[ \frac{ \mathtt{S}^2 (\nu\mathtt{t})}{ \mathsf{f} \alpha^2(1,\nu\mathtt{t}) \, \alpha(\nu,\mathtt{t}) }
  - \frac{\mathsf{f} \alpha^4(1,\nu\mathtt{t}) \,\alpha_{,\mathtt{u}}^2(\nu,\mathtt{t}) }{ \mathtt{t}^2 \mathtt{S}^2 (\nu\mathtt{t})\, \dot{\mathtt{S}}^2(\nu\mathtt{t})\,\alpha^3(\nu,\mathtt{t})  }\bigg]\mathtt{t}\text{d}\nu
   =0\,.
\end{aligned}
\end{equation}
%-----------------------------------------------------------
Hence, \eqref{eq:HeatEqnonDim_Transform2}, \eqref{eq:LinMomEqnonDim_Transform2} and \eqref{eq:ICBCnonDim_Transform_v2} form a system of nonlinear PDEs coupled with an ODE,\footnote{Note that the integral equation \eqref{eq:LinMomEqnonDim_Transform2} can be differentiated with respect to $\mathtt{u}$ to get rid of the integral term. Thus, \eqref{eq:LinMomEqnonDim_Transform2} and \eqref{eq:HeatEqnonDim_Transform2} are second-order nonlinear PDEs in terms of the unknown fields $\alpha(\mathtt{u},\mathtt{t})$ and $\gamma(\mathtt{u},\mathtt{t})$. Similarly, \eqref{eq:ICBCnonDim_Transform_v2}$_1$ is an ODE in terms of the unknown function $\mathtt{S}(\mathtt{t})$. } with the unknown fields $\alpha(\mathtt{u},\mathtt{t})$, $\gamma(\mathtt{u},\mathtt{t})$, and $\mathtt{S}(\mathtt{t})$ over the rectangular domain $\{(\mathtt{u},\mathtt{t}): 0 \leq \mathtt{u} \leq 1, 0< \mathtt{t} \leq \mathtt{t}_{\text{end}} \}$. 
%--------------------------------
\end{remark}
%--------------------------------

%--------------------------------
\begin{remark}
%--------------------------------
Note that the standard heat equation is recovered by setting $\bar{r}(R) = R $ and $\alpha^\text{s}(T ) = 0 $ in  \eqref{eq:MatHeatEqRigidCondRadial_v2}, which is written as 
%-----------------------------
\begin{equation} 
	 D^\text{s}_\text{f}  \left [T_{,RR}(R,t)
	  +\frac{2}{R}  T_{,R}(R,t) \right]
	 = \dot{T}(R,t)  \,.
\end{equation}
%----------------------------- 
Therefore, in the absence of any elastic deformation or thermal expansion, the non-dimensionalized moving boundary problem reads
%-----------------------------------------------------------
\begin{equation}   \label{eq:NoDefSimpleHeat_nondim_v1}
\begin{aligned}
	 \mathtt{T}_{,\mathtt{R}\mathtt{R}}(\mathtt{R},\mathtt{t}) 
	 +\frac{2}{\mathtt{R}} \mathtt{T}_{,\mathtt{R}}(\mathtt{R},\mathtt{t})
	 &= \mathtt{T}_{,\mathtt{t}}(\mathtt{R},\mathtt{t}) \,, \\
	\mathtt{T}_{,\mathtt{R}} (\mathtt{S}(\mathtt{t}),\mathtt{t})
	&= -\mathsf{L}  \frac{\text{d}\mathtt{S}(\mathtt{t})}{ \text{d}\mathtt{t} }  \,, \\
	\mathtt{T}(\mathtt{S}(\mathtt{t}),\mathtt{t}) 
	&= 0 \,, \\
	\mathtt{T}_{,\mathtt{R}}( 1,\mathtt{t})
	&=\mathsf{h} \left[ 1 - \mathtt{T}( 1,\mathtt{t}) \right]\,, \\	
	\mathtt{S}(0)&=1 \,,
\end{aligned}
\end{equation}
%-----------------------------------------------------------
where $1 \geq \mathtt{R} \geq \mathtt{S}(\mathtt{t})$ and $\mathtt{t} \geq 0$. Here, $\mathsf{h}$ is the Biot number, and $\frac{1}{\mathsf{L}}$ is the Stefan number. The phase change problem \eqref{eq:NoDefSimpleHeat_nondim_v1} has been analyzed by \citet{london1943rate}, \citet{tao1967generalized}, \citet{shih1971analytical}, \citet{hill1983freezing}, and possibly others.
Furthermore, \eqref{eq:StefanIntegralCoupled} is simplified as
%-----------------------------------------------------------
\begin{equation} \label{eq:RigidStefanIntegral_v1}
\begin{aligned}
	 \mathsf{h}\,\mathtt{t}
	 + \frac{\mathsf{L}}{3} \left[\mathtt{S}^3(\mathtt{t})-1\right] 
	 -\int_0^\mathtt{t} \mathsf{h}\, \mathtt{T}(1,\tau)  \text{d}\tau
	 = \int_{\mathtt{S}(\mathtt{t})}^1 \xi^2\, \mathtt{T}(\xi,\mathtt{t})\,  \text{d}\xi    \,.
\end{aligned}
\end{equation}
%-----------------------------------------------------------
Alternatively, since \eqref{eq:NoDefSimpleHeat_nondim_v1}$_1$ is rewritten as $\left[ \mathtt{R}\mathtt{T}(\mathtt{R},\mathtt{t})\right]_{,\mathtt{R}\mathtt{R}}=\left[\mathtt{R}\mathtt{T}(\mathtt{R},\mathtt{t})\right]_{,\mathtt{t}}$, it can be shown using \eqref{eq:NoDefSimpleHeat_nondim_v1}$_{2-5}$ that Stefan's condition \eqref{eq:NoDefSimpleHeat_nondim_v1}$_{2}$ is equivalent to\footnote{Using \eqref{eq:NoDefSimpleHeat_nondim_v1}$_{2-4}$, it is implied that
%------------------------------
\begin{equation}
	\int_{\mathtt{S}(\mathtt{t})}^1  \left[ \mathtt{R}\mathtt{T}(\mathtt{R},\mathtt{t})\right]_{,\mathtt{R}\mathtt{R}} \text{d}\mathtt{R} 
	=\mathsf{h} + [1-\mathsf{h}] \mathtt{T}(1,\mathtt{t}) + \mathsf{L} \mathtt{S}(\mathtt{t})\dot{\mathtt{S}}(\mathtt{t})\,.
\end{equation}
%------------------------------
Therefore, since $\left[ \mathtt{R}\mathtt{T}(\mathtt{R},\mathtt{t})\right]_{,\mathtt{R}\mathtt{R}}=\left[\mathtt{R}\mathtt{T}(\mathtt{R},\mathtt{t})\right]_{,\mathtt{t}}$, \eqref{eq:RigidStefanIntegral_v2} follows from \eqref{eq:NoDefSimpleHeat_nondim_v1}$_{5}$ and the fact that 
%------------------------------
\begin{equation}
	\int_{\mathtt{S}(\mathtt{t})}^1   \mathtt{R}\mathtt{T}(\mathtt{R},\mathtt{t}) \text{d}\mathtt{R}
	= \int_0^\mathtt{t}\left[\int_{\mathtt{S}(\tau)}^1 \mathtt{R}\mathtt{T}_{,\tau}(\mathtt{R},\tau)\text{d}\mathtt{R} \right]\text{d}\tau\,.
\end{equation}
%------------------------------
}
%------------------------------
\begin{equation} \label{eq:RigidStefanIntegral_v2}
	\mathsf{h}\,\mathtt{t} 
	+ \frac{\mathsf{L}}{2}\left[\mathtt{S}^2(\mathtt{t})-1\right]
	+ (1-\mathsf{h}) \int_ 0 ^\mathtt{t}  \mathtt{T}(1,\tau) \text{d}\tau
	=\int_{\mathtt{S}(\mathtt{t})}^1\xi\,\mathtt{T}(\xi,\mathtt{t}) \,\text{d}\xi \,.
\end{equation}
%------------------------------
Thus, for a rigid conductor, Stefan's condition \eqref{eq:NoDefSimpleHeat_nondim_v1}$_{2}$ can be replaced with the integral constraint \eqref{eq:RigidStefanIntegral_v1}, or equivalently with  \eqref{eq:RigidStefanIntegral_v2}.
%--------------------------------
\end{remark}
%--------------------------------
%---------------------
\begin{figure}[t!]
\centering
\vskip 0.0in
\includegraphics[width=0.68\textwidth]{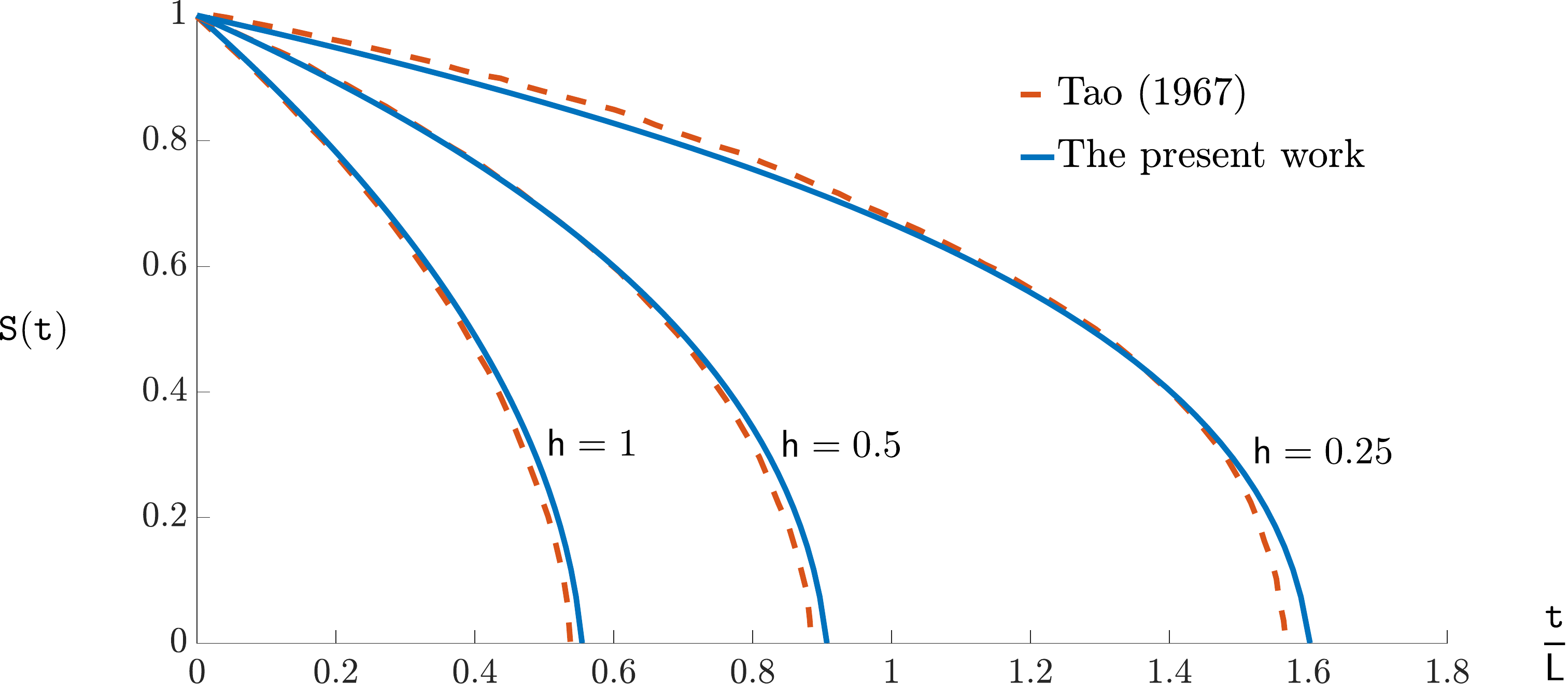}
\vskip 0.1in
\caption{The numerical solution for the evolution of the moving boundary obtained in the present work for the rigid conductor problem \eqref{eq:NoDefSimpleHeat_nondim_v1} with $\mathsf{L}=10$, is compared with that obtained by \citet{tao1967generalized}.}
\label{fig:TaoComparison}
\end{figure}
%---------------------

%---------------------
\begin{figure}[t!]
\centering
\vskip 0.0in
\includegraphics[width=0.96\textwidth]{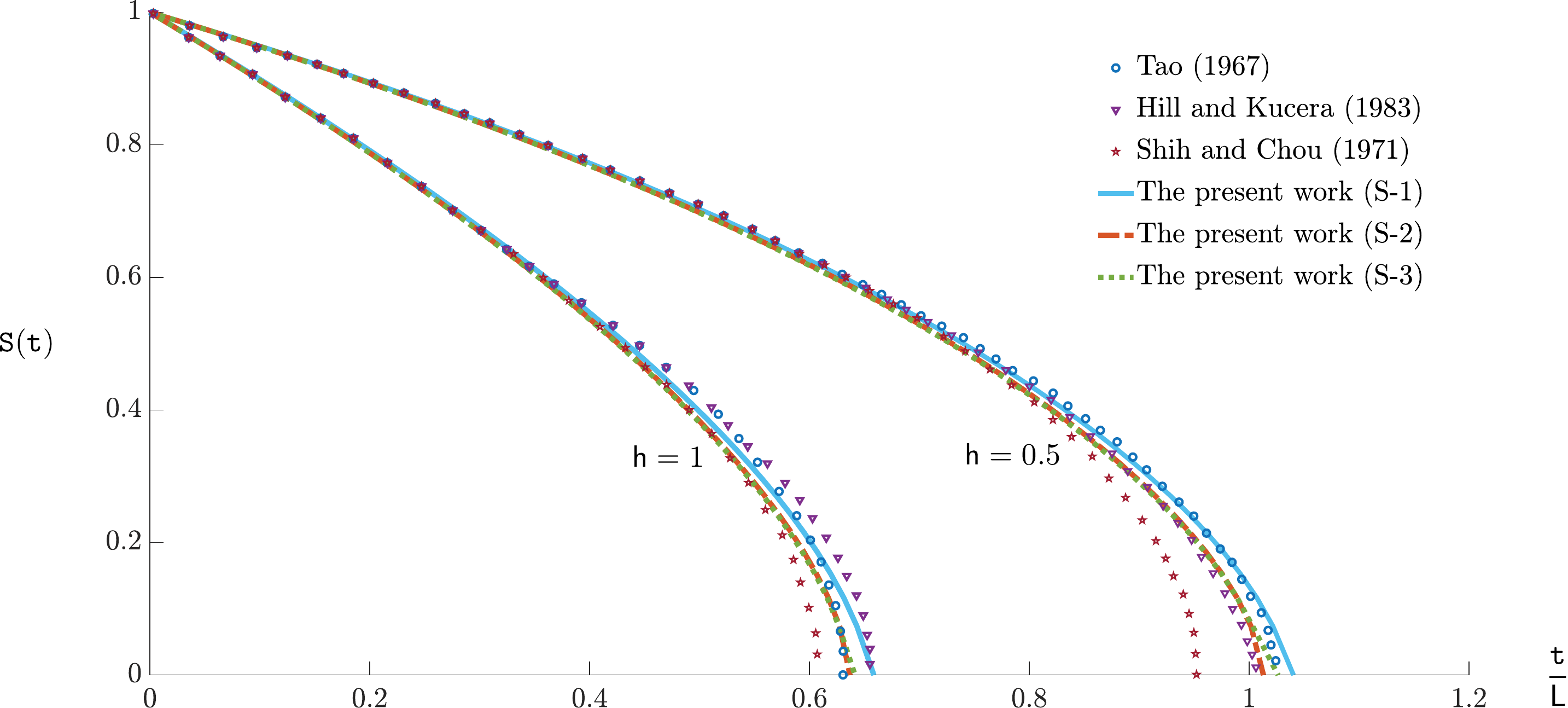}
\vskip 0.1in
\caption{The moving boundary evolution for the rigid conductor problem with $\mathsf{L}=2$ obtained in the present work is compared with numerical data from the literature \citep{tao1967generalized,shih1971analytical,hill1983freezing}. Here, S-1 refers to the solution of system \eqref{eq:NoDefSimpleHeat_nondim_v1}, while S-2 and S-3 correspond to the systems where \eqref{eq:NoDefSimpleHeat_nondim_v1}$_2$ is replaced by \eqref{eq:RigidStefanIntegral_v2} and  \eqref{eq:RigidStefanIntegral_v1}, respectively.}
\label{fig:AllComparision_rigid}
\end{figure}
%---------------------
%---------------------
\begin{figure}[t!]
\centering
\vskip 0.30in
\includegraphics[width=0.9\textwidth]{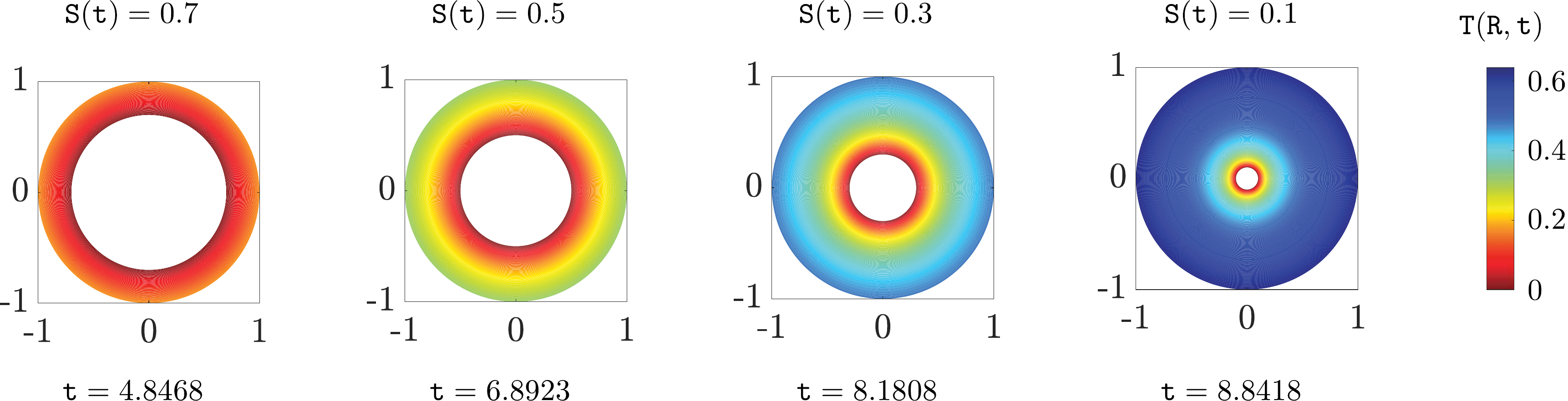}
\vskip 0.2in
\caption{The non-dimensionalized temperature field $\mathtt{T}(\mathtt{R},\mathtt{t})$ for the rigid conductor problem \eqref{eq:NoDefSimpleHeat_nondim_v1} with $\mathsf{h}=0.5$, $\mathsf{L}=10$ is depicted at various instances of time as the solidification interface moves inward.}
\label{fig:TemperaturesColor_rigid}
\end{figure}
%---------------------

%---------------------
%---------------------
\subsection{Residual stresses}

The solidification process is stopped at time $t_e$, when the the solid-liquid interface is at $R_e=S(t_e)$ in the reference configuration, or equivalently at $s(t_e)=\bar{r}(R_e)$ in the current configuration. Imagine that the accreted solid is drained of the remaining liquid and is allowed to reach a steady-state uniform temperature of $T_a<T_m$ in an ambient environment, while its inner and outer boundaries are traction-free. The resulting residually-stressed configuration is denoted by $\tilde{\mathcal{C}}\subset\mathcal{S}$. The material metric for the solid is written as
%-----------------------------
\begin{equation} \label{eq:SolidMatMetricResidual}
  \GS
   =e^{2 \omega^\text{s} (T_a)}\begin{bmatrix}
 \frac{(\rho_0^\text{f})^2 \, R^4}{(\rho_0 ^\text{s})^2 \,\bar{r}^4(R)} & 0  & 0  \\
  0 & \bar{r}^2(R)  & 0  \\
  0 & 0  & \bar{r}^2(R) \sin^2 \Theta
\end{bmatrix}\,.
\end{equation}
%-----------------------------
Recall that
%-----------------------------
\begin{equation} \label{eq:Residualexpansion}
	e^{3 \omega^\text{s}(T_a) }
	= 1+ \beta^\text{s}_0 T_m\Big(1-\frac{T_m}{T_a}\Big)\,.
\end{equation}
%-----------------------------
Note that since $\bar{r}(R)$ is now a known function defined on the interval $[R_e, R_0]$, determined from the solution of the IBVP during accretion, the material metric $\GS$ is considered to be given.

Let $\tilde{\varphi}:\mathcal{B}_{t_e}\rightarrow \tilde{\mathcal{C}}$ denote the deformation map corresponding to the residually-stressed configuration. In spherical coordinates $\tilde{\varphi}(R,\Theta,\Phi)=(\tilde{r}(R),\Theta,\Phi)$, where the placement map $\tilde{r}(R)$ represents the residual radial distortion. The deformation gradient reads
%-----------------------------
\begin{equation}
   \mathbf{F}(R)
	=\begin{bmatrix}
  \tilde{r}'(R) & 0  & 0  \\
  0 & 1  & 0  \\
  0 & 0  & 1
	\end{bmatrix}
\,.
\end{equation}
%-----------------------------
The Jacobian of the deformation is written as
%-----------------------------
\begin{equation}  
   J(R)
   = \frac{\rho_0^\text{s}\, \tilde{r}^2(R)\,\tilde{r}'(R)}{\rho_0^\text{f} \, e^{3 \omega^\text{s}(T_a) } R^2 } \,. 
\end{equation}
%-----------------------------
The strain tensors for this configuration are given as
%-----------------------------
\begin{equation} 
   \mathbf{b}^\sharp
   =e^{-2 \omega^\text{s}(T_a)  }\begin{bmatrix} \displaystyle
\left(\frac{ \rho_0^\text{s} \bar{r}^2 \tilde{r}' }{\rho_0^\text{f} R^2}\right)^2 & 0  & 0  \\
  0 &  \displaystyle\frac{1}{ \bar{r}^2 } & 0  \\
  0 & 0  &  \displaystyle\frac{1}{ \bar{r}^2 \sin^2 \Theta }
\end{bmatrix}\,, \quad\quad
	\mathbf{c}^\sharp
   =e^{ 2 \omega^\text{s}(T_a) }\begin{bmatrix} \displaystyle
\left(\frac{ \rho_0^\text{f} R^2}{\rho_0^\text{s} \bar{r}^2 \tilde{r}'  }\right)^2 & 0  & 0  \\
  0 &  \displaystyle\frac{\bar{r}^2}{ \tilde{r}^4 } & 0  \\
  0 & 0  &  \displaystyle\frac{\bar{r}^2}{ \tilde{r}^4 \sin^2 \Theta }
\end{bmatrix}\,.
\end{equation}
%-----------------------------
Further, the principal invariants of $\mathbf{b}$ read
%-----------------------------
\begin{equation} \label{eq:residual_principal_invariants}
   I_1
   = e^{-2 \omega^\text{s}(T_a) } \left[ \left(\frac{ \rho_0^\text{s}\, \bar{r}^2 \,\tilde{r}'}{ \rho_0^\text{f} R^2 } \right)^2 
   +2 \left(\frac{\tilde{r} }{\bar{r}}\right)^2\right] \,,\qquad
   I_2
   = e^{-4 \omega^\text{s}(T_a) }\left[  \left(\frac{\tilde{r}}{ \bar{r}}\right)^4 + 2\left(\frac{\rho_0^\text{s}\, \bar{r}\,\tilde{r}\,\tilde{r}'}{ \rho_0^\text{f} \,R^2}\right)^2 \right]  \,.
\end{equation}
%-----------------------------

%-----------------------------
\begin{example}[A neo-Hookean solid]
%-----------------------------
The thermoelastic neo-Hookean solid considered in \eqref{eq:energyfunc_coupled} is now at a constant temperature, and thus, is characterized by the temperature-independent energy function
%-----------------------------
\begin{equation}
	\tilde{W}(I_1,J)
	= \frac{T_a}{2T_m} \left[  \mu^\text{s}_0  [J^{-\frac{2}{3} }I_1 -3]
	+ \kappa^\text{s}_0  [J-1]^2
	\right]
	+\kappa^\text{s}_0\,\beta^\text{s}_0(T_m-T_a)\, [J-1]
	\,.
\end{equation}
%-----------------------------
The nonzero components of residual Cauchy stress $\tilde{\bm{\sigma}}(R)$ are written as 
%-----------------------------
\begin{equation}  \label{eq:residualstress}
\begin{aligned}
	\tilde{\sigma}^{rr}  
	=\tilde{W}_J
	+  \frac{ 2 (\rho_0^\text{s})^2 \tilde{W}_1\,  
	\bar{r}^4 (\tilde{r}')^2}{(\rho_0^\text{f})^2 J e^{2\omega^\text{s}(T_a)} R^4 }  \,,\qquad
	\tilde{\sigma}^{\theta\theta}  
	=   \frac{\tilde{W}_J}{\tilde{r}^2} 
	+  \frac{2 \tilde{W}_1}{J e^{2\omega^\text{s}(T_a)} \bar{r}^2} \,,\qquad
	\tilde{\sigma}^{\phi\phi}
	=   \frac{\tilde{\sigma}^{\theta\theta} }{\sin^2 \Theta}\,,
\end{aligned}
\end{equation}
%----------------------------- 
where the coefficients $\tilde{W}_1$ and $\tilde{W}_J$ are given as
%-----------------------------
\begin{equation} \label{eq:residualcoeff}
\begin{aligned}	
	\tilde{W}_1
	= \frac{\mu^\text{s}_0 T_a}{2 T_m} J^{-\frac{2}{3} }   \,, \qquad
    \tilde{W}_J
	=  \frac{T_a}{T_m} \left[ \kappa^\text{s}_0(J-1)
	-\frac{\mu^\text{s}_0}{3} J^{-\frac{5}{3} }I_1 
	\right]
	+\kappa^\text{s}_0\,\beta^\text{s}_0\,(T_m-T_a)	\,.
\end{aligned}
\end{equation}
%-----------------------------
The balance of linear momentum in the absence of body forces and inertial effects is simplified to yield the following radial equilibrium equation
%-----------------------------
\begin{equation} \label{eq:residualradialEquilibium}
  \frac{\text{d} \tilde{\sigma}^{rr}}{\text{d} R}
  =\left[ \left( \tilde{\sigma}^{\theta \theta}
  + \sin^2 \theta \,\tilde{\sigma}^{\phi \phi} \right)\tilde{r}^2
  - 2 \tilde{\sigma}^{rr}\right]  \frac{\tilde{r}'}{\tilde{r}}\,.
\end{equation}
%-----------------------------
Furthermore, the outer and inner boundaries are traction-free, i.e.
%-----------------------------
\begin{equation}   \label{eq:residualBC}
  \tilde{\sigma}^{rr} (R_0)
  =\tilde{\sigma}^{rr} (R_e)
  =0\,.
\end{equation}
%-----------------------------
It follows from \eqref{eq:residualstress}, \eqref{eq:residualcoeff} and \eqref{eq:residual_principal_invariants}$_1$ that \eqref{eq:residualradialEquilibium} is a nonlinear ODE in terms of $\tilde{r}(R)$, with the boundary conditions \eqref{eq:residualBC}. Thus, the problem of finding the residual stresses and distortions boils down to solving the boundary-value problem \eqref{eq:residualradialEquilibium}-\eqref{eq:residualBC} for the unknown function $\tilde{r}(R)$. This problem is then non-dimensionalized according to Table \ref{table:Nondimensionalization}.
The dimensionless radial displacement $\frac{\tilde{r}-R }{R_0}$ and the dimensionless physical components of the Cauchy stress $\mathring{ \tilde{\sigma} }^{ab}=\frac{\mathring{ \tilde{\sigma} }^{ab} \sqrt{g_{aa}g_{bb}}  }{\kappa^\text{f}_0}$ (no summation) in the residually-stressed configuration at a given dimensionless steady state temperature $\mathtt{T}_\mathtt{a}=\frac{T_a-T_m}{T_c-T_m}$ are illustrated in Figure \ref{fig:residualstress}.\footnote{Note that $\mathring{ \tilde{\sigma} }^{rr}=\frac{  \tilde{\sigma} ^{rr}  }{\kappa^\text{f}_0}$ and $\mathring{ \tilde{\sigma} }^{\phi\phi}=\mathring{ \tilde{\sigma} }^{\theta\theta}=\frac{\tilde{r}^2   \tilde{\sigma} ^{\theta\theta}   }{\kappa^\text{f}_0}$. }
%-----------------------------
\end{example}
%-----------------------------

%-----------------------------
%-----------------------------
\subsection{Numerical results and discussion}

Several numerical methods for the solution of moving boundary value problems have been proposed over the years \citep{rubinshteuin1971stefan,crank1984free}. 
In this work, we follow the approach of \citet{douglas1955numerical}, where for a specified space grid, the corresponding instances of time are calculated as the moving boundary assumes these discrete positions in progression.
It should be noted that the bijectivity of $S(t)$ is exploited here, allowing us to treat the time of accretion as the unknown.
For each unknown time step, the moving interface is first assigned a position. 
Treating the domain as fixed, we calculate the deformation and temperature fields, along with the instant of time for this interface location, by solving the conservation of linear momentum, transient heat equation, and Stefan's condition. 
This is implemented using a finite difference approximation (an implicit scheme) in \textsc{Matlab}.
The optimum time step that minimizes the residue from Stefan's condition to ensure a sufficiently small magnitude is calculated using \texttt{fminunc}, while the corresponding numerical solution for the radial equilibrium and the heat equation is simultaneously obtained using \texttt{fsolve}.
Extensive parametric studies are conducted by varying the numerical values of the dimensionless constants in Table \ref{table:Nondimensionalization}. The observations from the numerical results are qualitatively described in the following.

%---------------------
\begin{figure}[t!]
\centering
\includegraphics[width=0.7\textwidth]{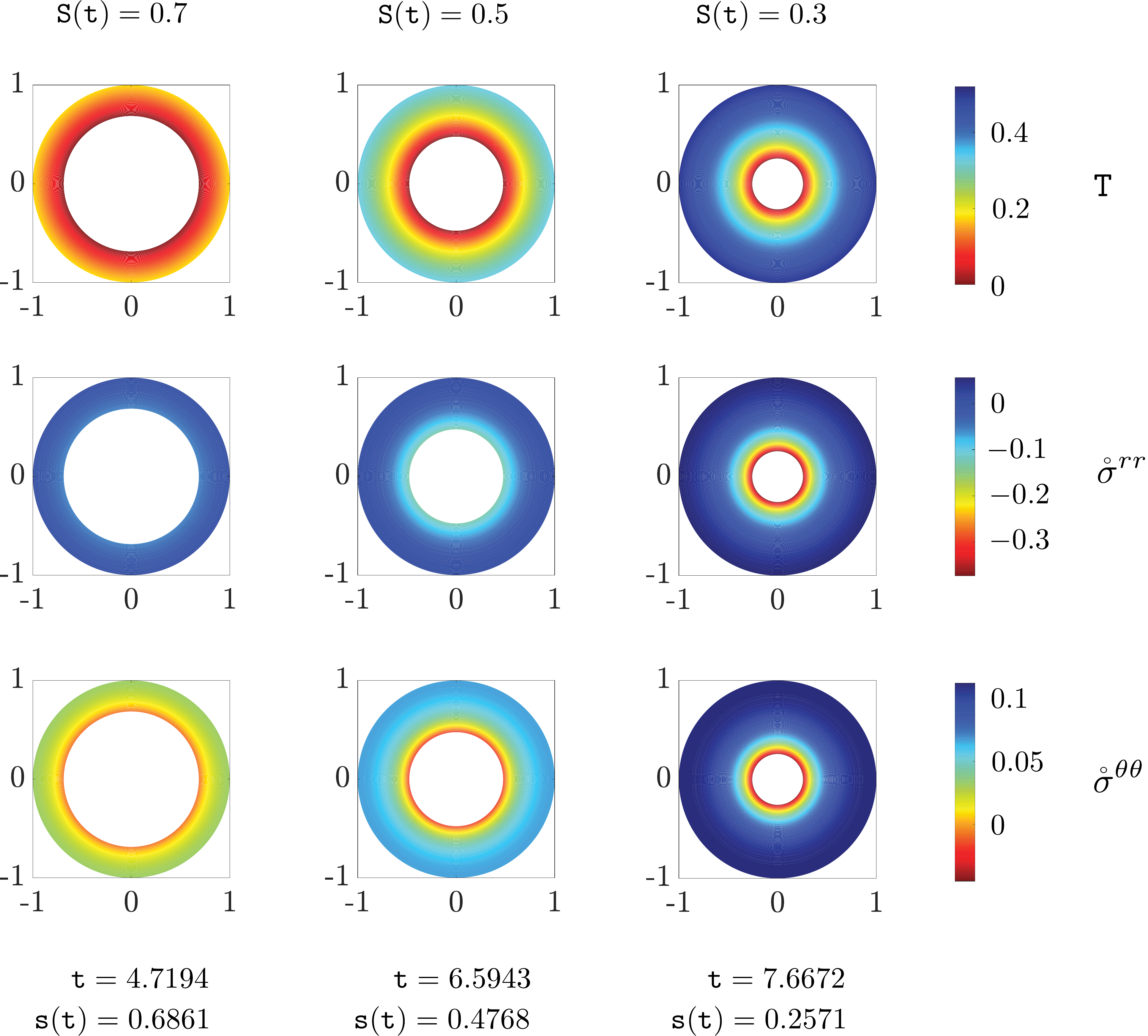}
\vskip 0.48cm
\caption{The non-dimensionalized temperature field $\mathtt{T}$, as well as the dimensionless physical components $\mathring{\sigma}^{rr}$ and $\mathring{\sigma}^{\theta\theta}$ of the Cauchy stress within the deformed solid are illustrated via color maps. These depictions are based on the solution of the general problem \eqref{eq:LinMomEqnonDim}-\eqref{eq:ICBCnonDim}, with  $\mathsf{f}=0.95$, $\mathsf{a}=0.8$, $\mathsf{b}=0.1$, $\mathsf{p}=1.1$, $\mathsf{q}=1.2$, $\mathsf{h}=0.5$, and $\mathsf{L}=10$, at various instances of time as the phase-change interface moves inward.}
\label{fig:coupledSolColorPlot}
\end{figure}
%---------------------

%\todo{Arash: Up to this point you have been citing figures as ``Figure" but in this section you are using ``Fig.". I let you decide which one to use but you need to be consistent throughout. I also think rearranging this discussion in an itemized format may make it easier for the reader. Let me know if you disagree.}

\begin{figure}[t!]
\centering
\vskip 0.64cm
\includegraphics[width=0.75\textwidth]{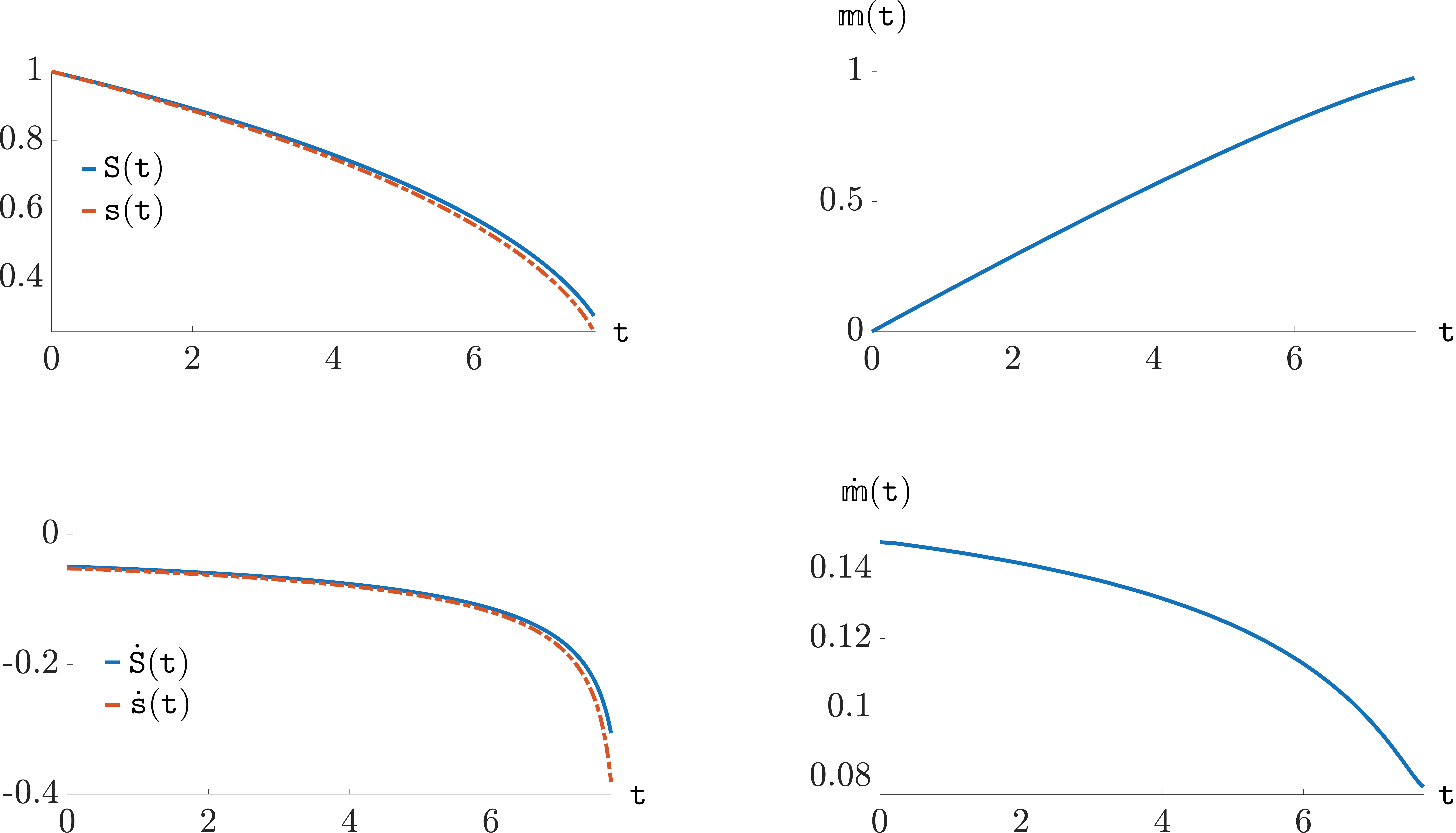}
\vskip 0.72cm
\caption{The motion of the solidification interface is illustrated. $\mathtt{S}(\mathtt{t})$ and $\mathtt{s}(\mathtt{t})$ denote its radial position in the reference and current configurations respectively, while $\dot{\mathtt{S}}(\mathtt{t})$ and $\dot{\mathtt{s}}(\mathtt{t})$ represent the respective velocities. $\mathbb{m}(\mathtt{t})$ denotes the fraction of the initial liquid mass solidified, and $\dot{\mathbb{m}}(\mathtt{t})$ represents the solidification rate. These figures depict the solution of the coupled problem \eqref{eq:LinMomEqnonDim}-\eqref{eq:ICBCnonDim} for  $\mathsf{f}=0.95$, $\mathsf{a}=0.8$, $\mathsf{b}=0.1$, $\mathsf{p}=1.1$, $\mathsf{q}=1.2$, $\mathsf{h}=0.5$, and $\mathsf{L}=10$.}
\label{fig:interface_coupledcase}
\end{figure}

\begin{figure}[t!]
%-----
\centering % <-- added
%-----    
\begin{subfigure}{0.44\textwidth}
  \includegraphics[width=\linewidth]{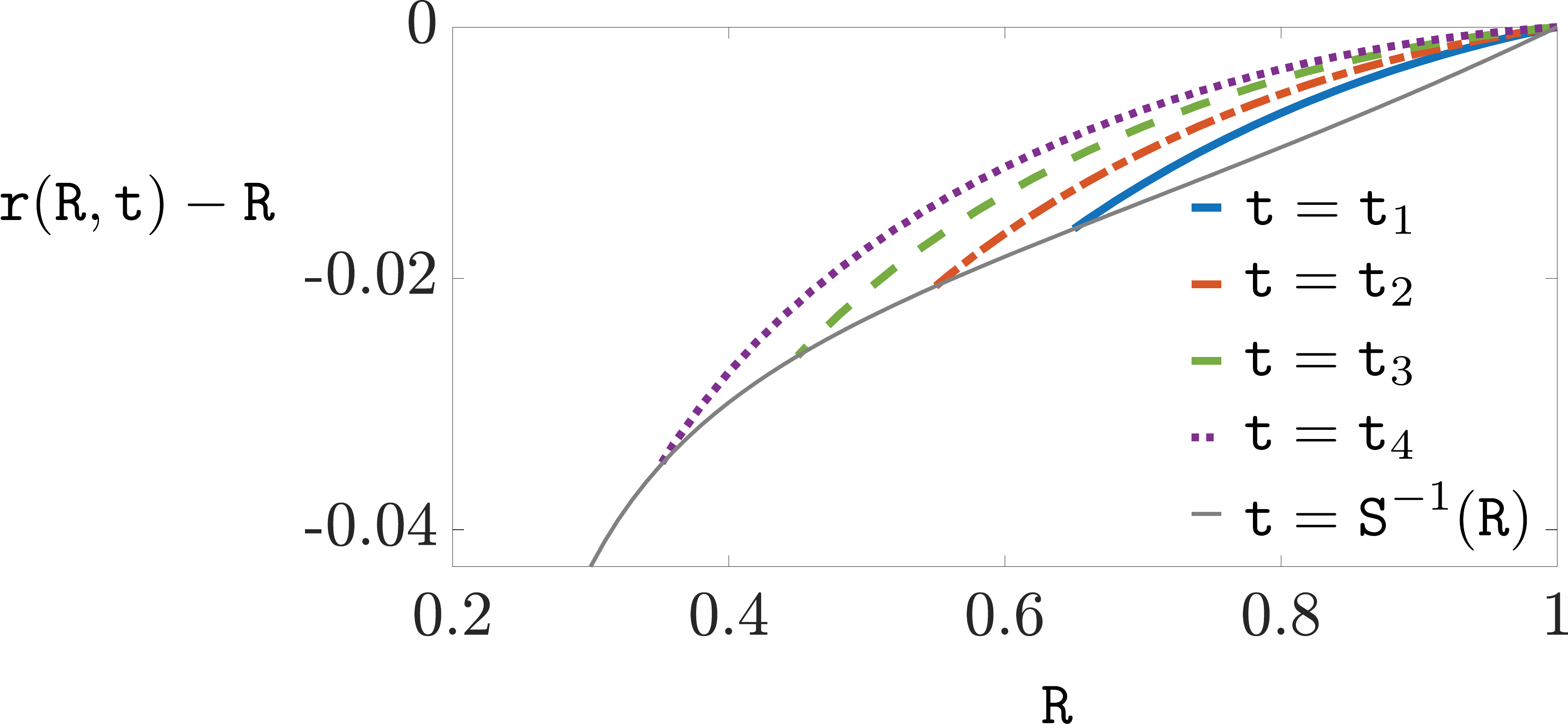}
  \caption{The variation of the displacement field with the material coordinate inside the solid.}
  \label{subfig:dispR}
\end{subfigure}
%-----
\hfil % <-- added
%-----
\begin{subfigure}{0.435\textwidth}
  \includegraphics[width=\linewidth]{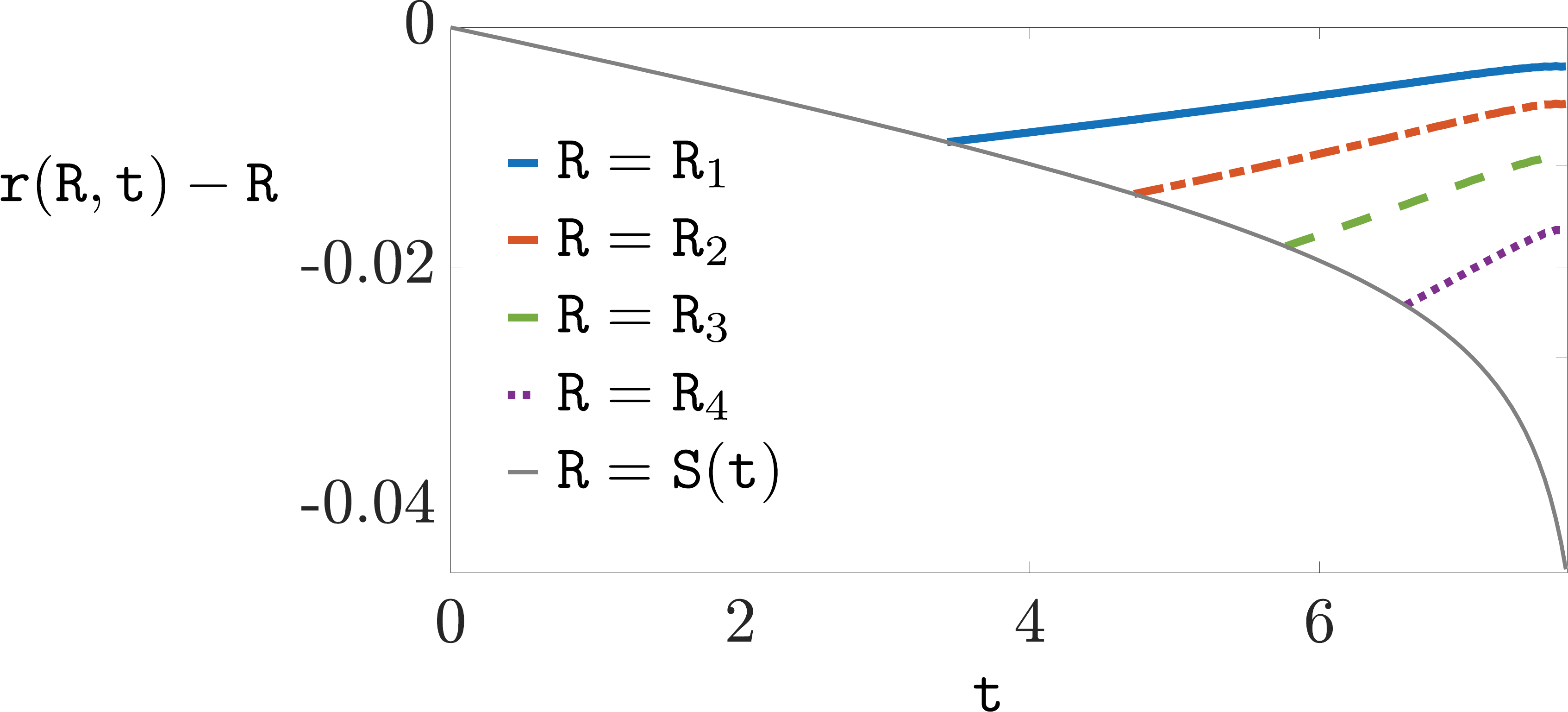}
  \caption{The time evolution of the radial displacement field within the solid.}
  \label{subfig:dispt}
\end{subfigure}
%-----
\medskip
\vspace{0.64cm}
%-----
\begin{subfigure}{0.41\textwidth}
  \includegraphics[width=\linewidth]{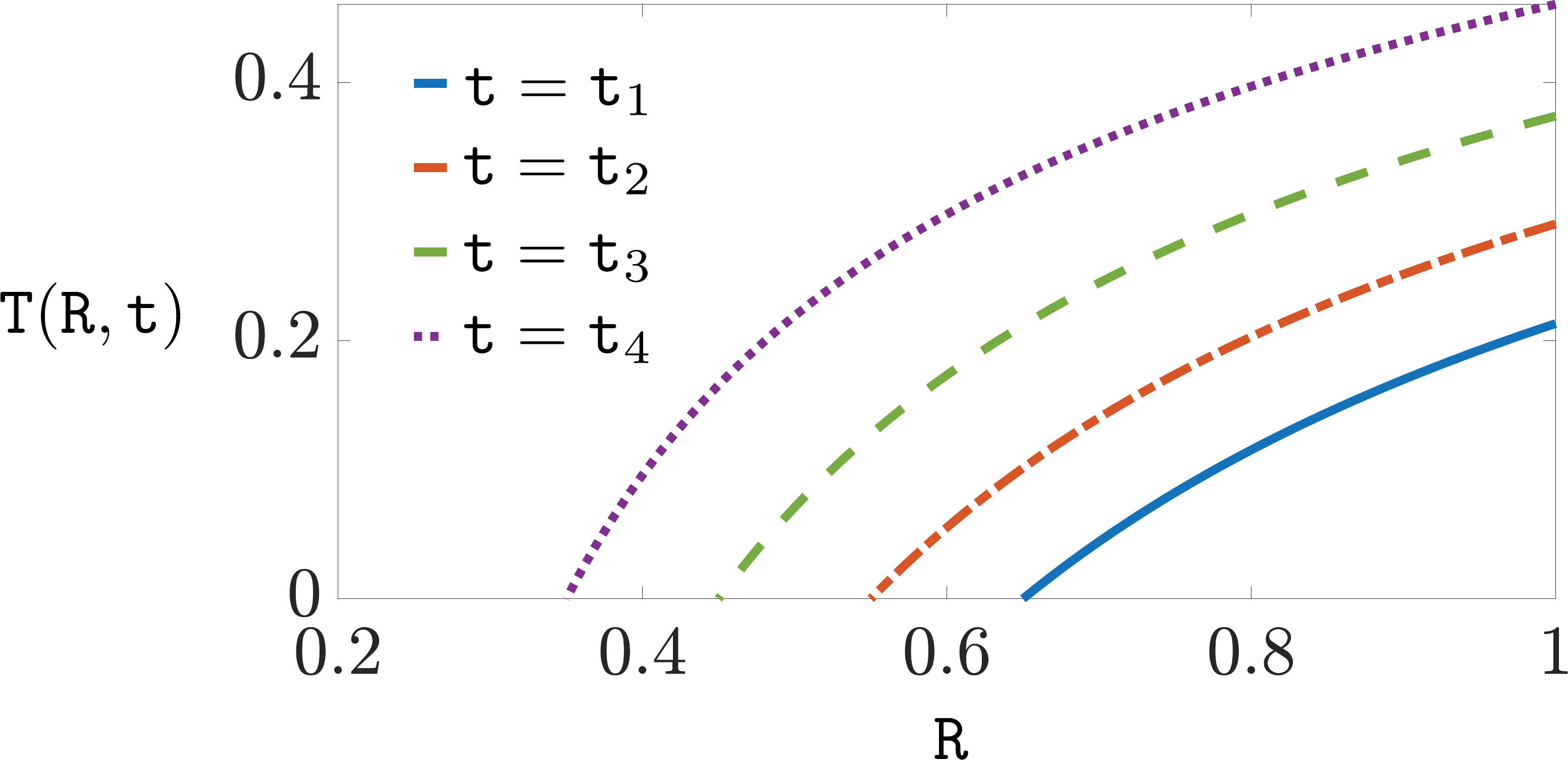}
  \caption{The variation of $\mathtt{T}(\mathtt{R},\mathtt{t})$ with $\mathtt{R}$ inside the solid.}
  \label{subfig:TempR}
\end{subfigure}
%-----
\hfil % <-- added
%-----
\begin{subfigure}{0.4\textwidth}
  \includegraphics[width=\linewidth]{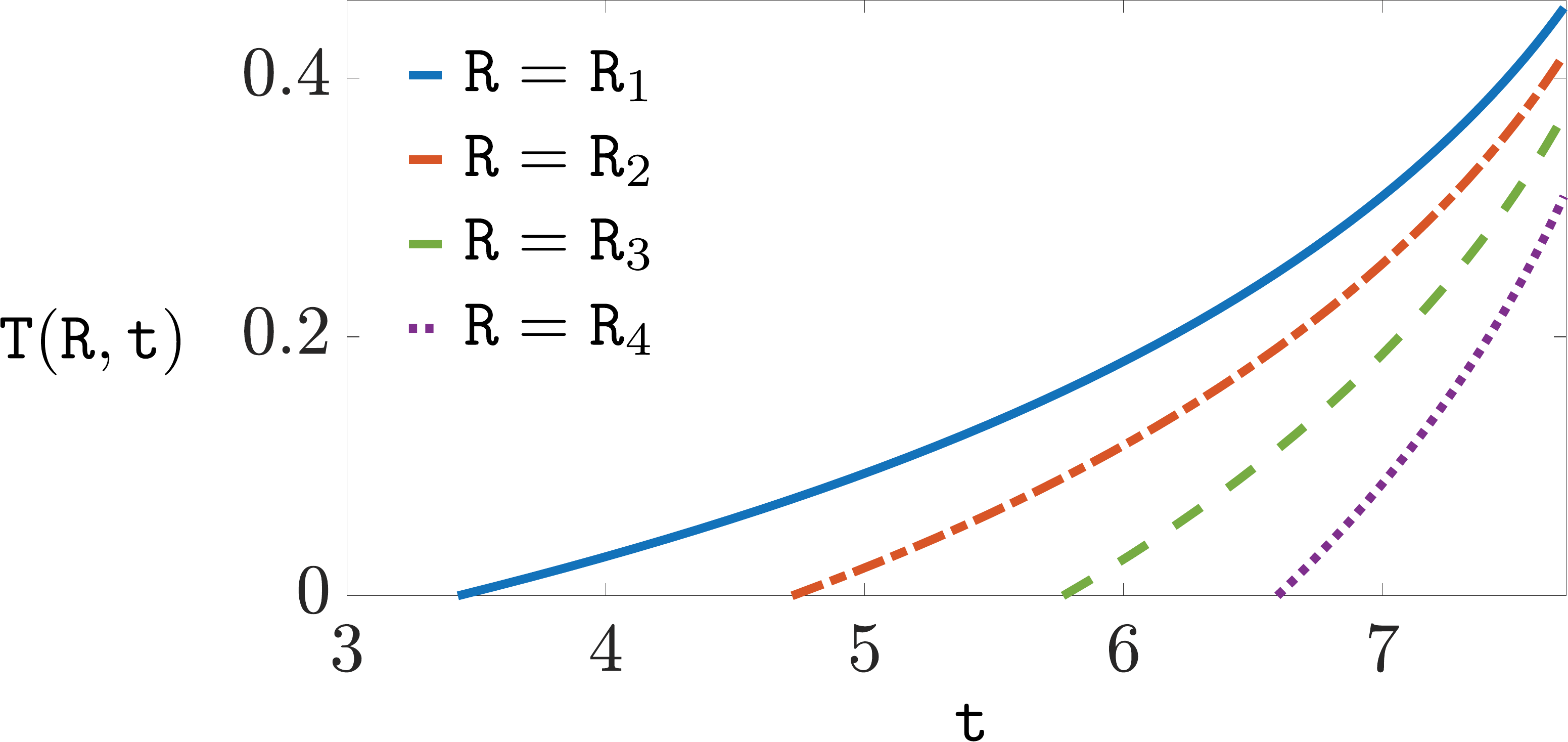}
  \caption{The time evolution of the temperature field within the solid.}
  \label{subfig:Tempt}
\end{subfigure}
%-----
\medskip
\vspace{0.64cm}
%-----
\begin{subfigure}{0.44\textwidth}
  \includegraphics[width=\linewidth]{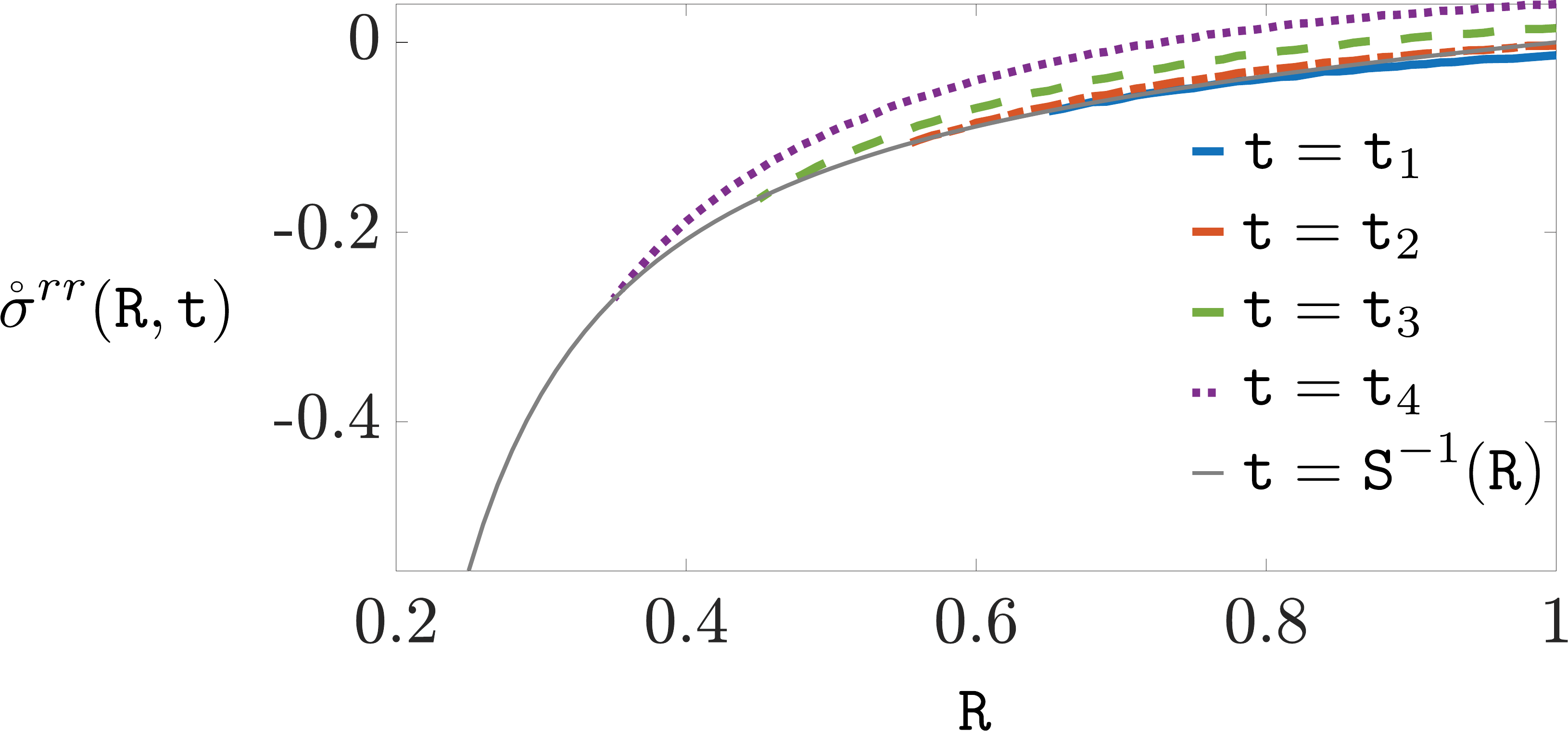}
  \caption{The dependence of $\mathring{\sigma}^{rr}$ on $\mathtt{R}$ within the solid.}
  \label{subfig:sigmarrR}
\end{subfigure}
%-----
\hfil % <-- added
%-----
\begin{subfigure}{0.435\textwidth}
  \includegraphics[width=\linewidth]{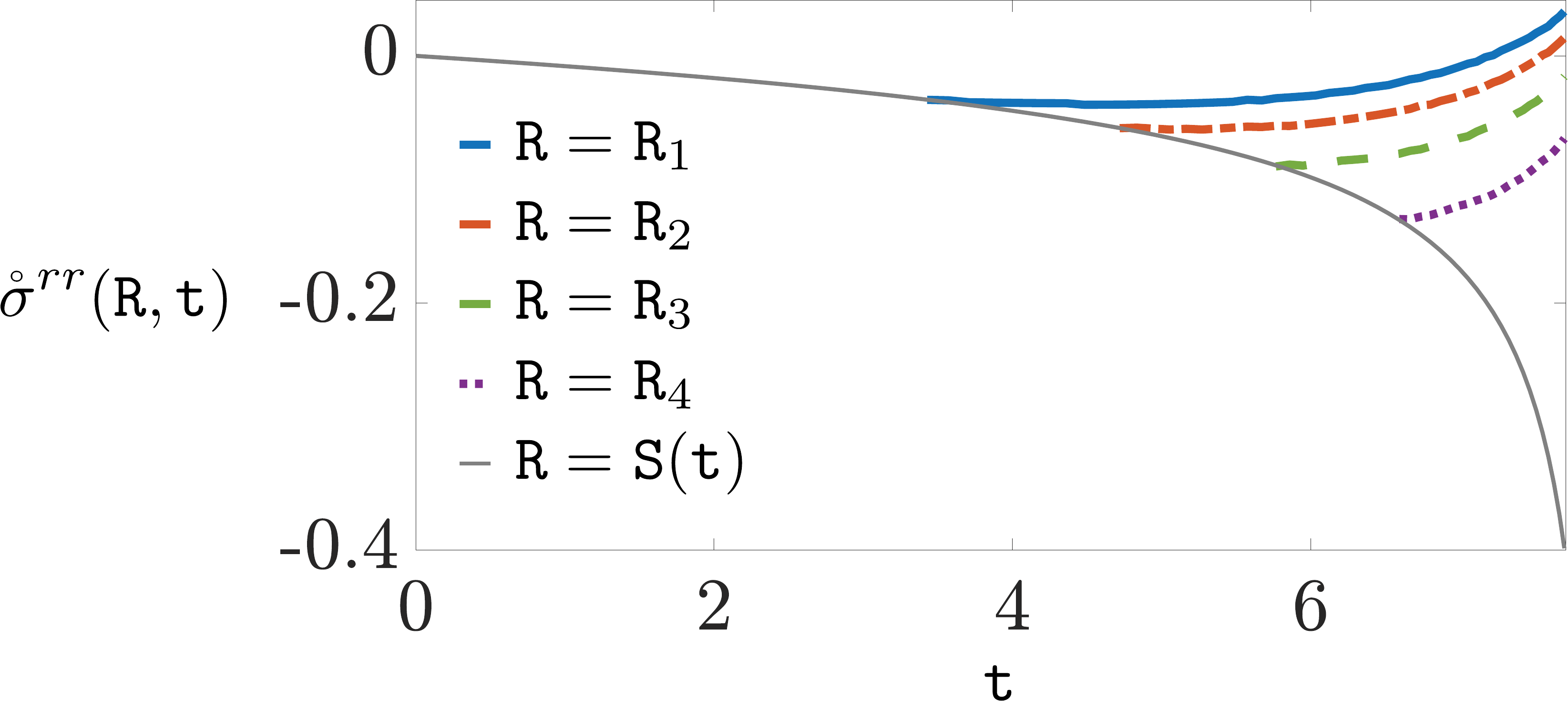}
  \caption{The time evolution of $\mathring{\sigma}^{rr}$ within the solid.}
  \label{subfig:sigmarrt}
\end{subfigure}
%-----
\medskip
\vspace{0.64cm}
%-----
\begin{subfigure}{0.44\textwidth}
  \includegraphics[width=\linewidth]{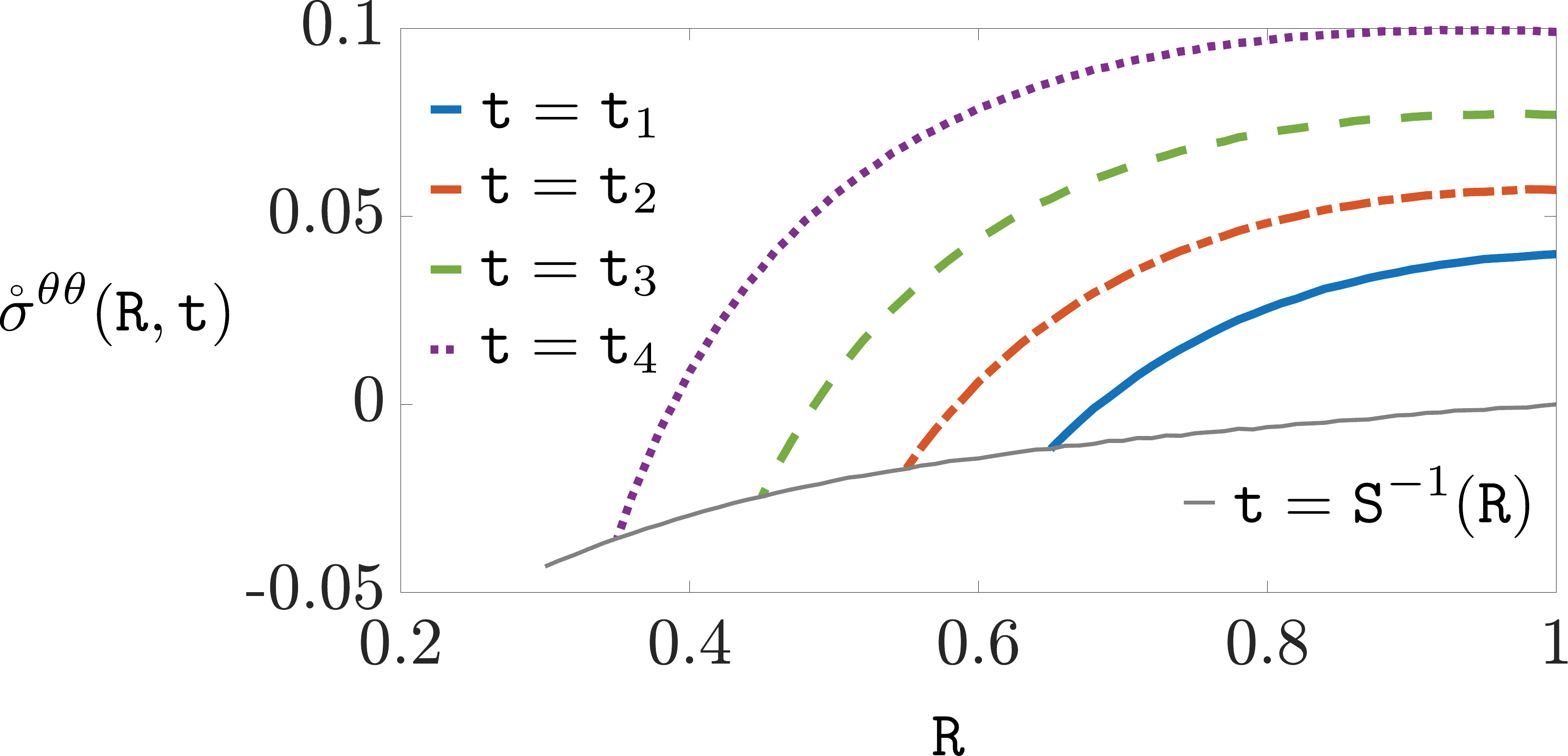}
  \caption{The dependence of $\mathring{\sigma}^{\theta\theta}$ on $\mathtt{R}$ within the solid.}
  \label{subfig:sigmattR}
\end{subfigure}
%-----
\hfil % <-- added
%-----
\begin{subfigure}{0.435\textwidth}
  \includegraphics[width=\linewidth]{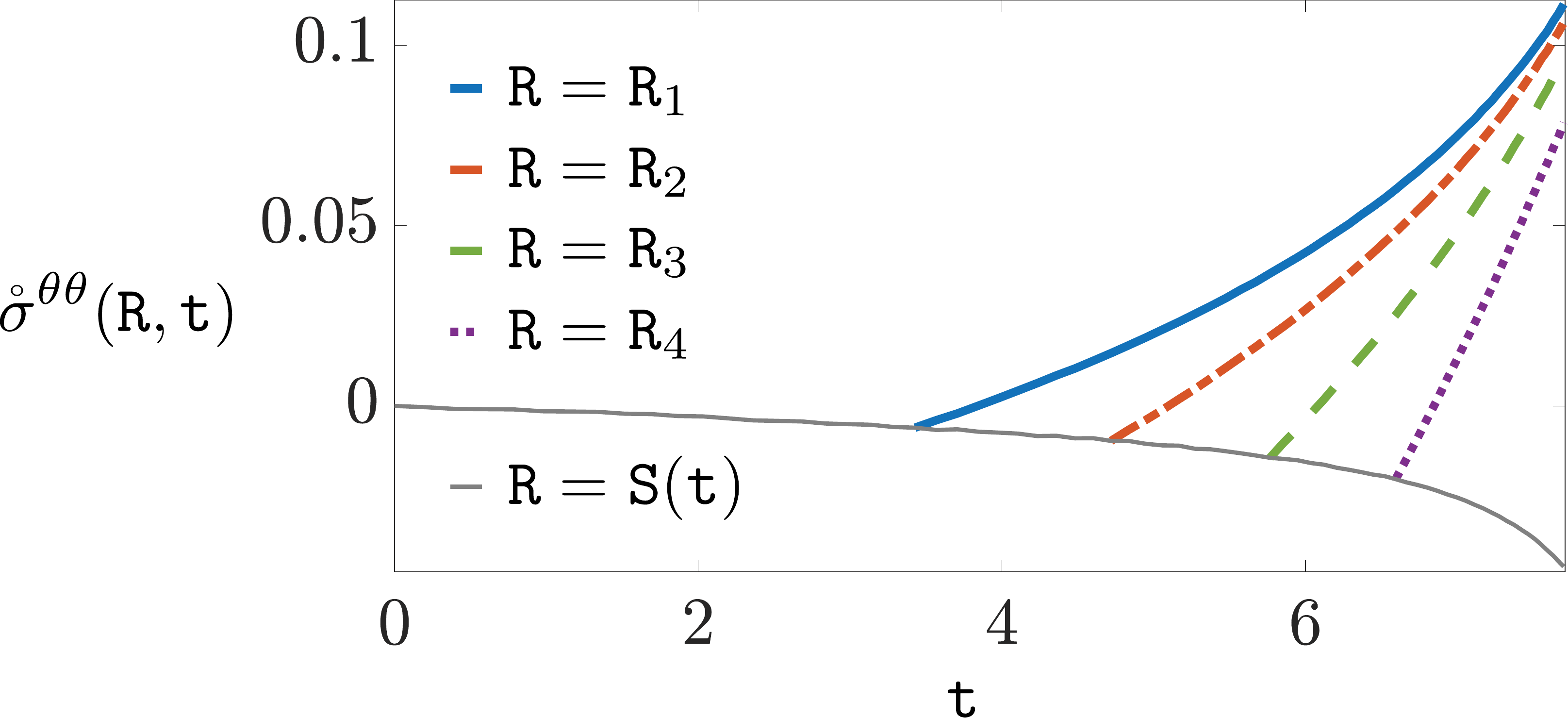}
   \caption{The time evolution of $\mathring{\sigma}^{\theta\theta}$ within the solid.}
  \label{subfig:sigmattt}
\end{subfigure}
%-----
\vspace{0.72cm}
\caption{The variation of the radial displacement field $\mathtt{r}(\mathtt{R},\mathtt{t})-\mathtt{R}$, temperature field $\mathtt{T}(\mathtt{R},\mathtt{t})$, and dimensionless physical components $\mathring{\sigma}^{rr}(\mathtt{R},\mathtt{t})$ and $\mathring{\sigma}^{\theta\theta}(\mathtt{R},\mathtt{t})$ of the Cauchy stress inside the accreting solid, with material radial coordinate $\mathtt{R}$ and time $\mathtt{t}$, is depicted. These illustrations are based on the solution of the general problem \eqref{eq:LinMomEqnonDim}-\eqref{eq:ICBCnonDim}, with  $\mathsf{f}=0.95$, $\mathsf{a}=0.8$, $\mathsf{b}=0.1$, $\mathsf{p}=1.1$, $\mathsf{q}=1.2$, $\mathsf{h}=0.5$, and $\mathsf{L}=10$. The spatial variation is shown for the instances $\mathtt{t}_1<\mathtt{t}_2<\mathtt{t}_3<\mathtt{t}_4$, corresponding to $\mathtt{S}(\mathtt{t}_1)=0.65$, $\mathtt{S}(\mathtt{t}_2)=0.55$, $\mathtt{S}(\mathtt{t}_3)=0.45$, $\mathtt{S}(\mathtt{t}_4)=0.35$, respectively. Similarly, the temporal evolution is depicted at the radii $\mathtt{R}_1=0.8$, $\mathtt{R}_2=0.7$, $\mathtt{R}_3=0.6$, and $\mathtt{R}_4=0.5$.}
%-----
\label{fig:Solutionwrttimeandrefpos}
%-----
\end{figure}

\begin{figure}[t!]
%-----
\centering % <-- added
%-----    
\begin{subfigure}{0.45\textwidth}
  \includegraphics[width=\linewidth]{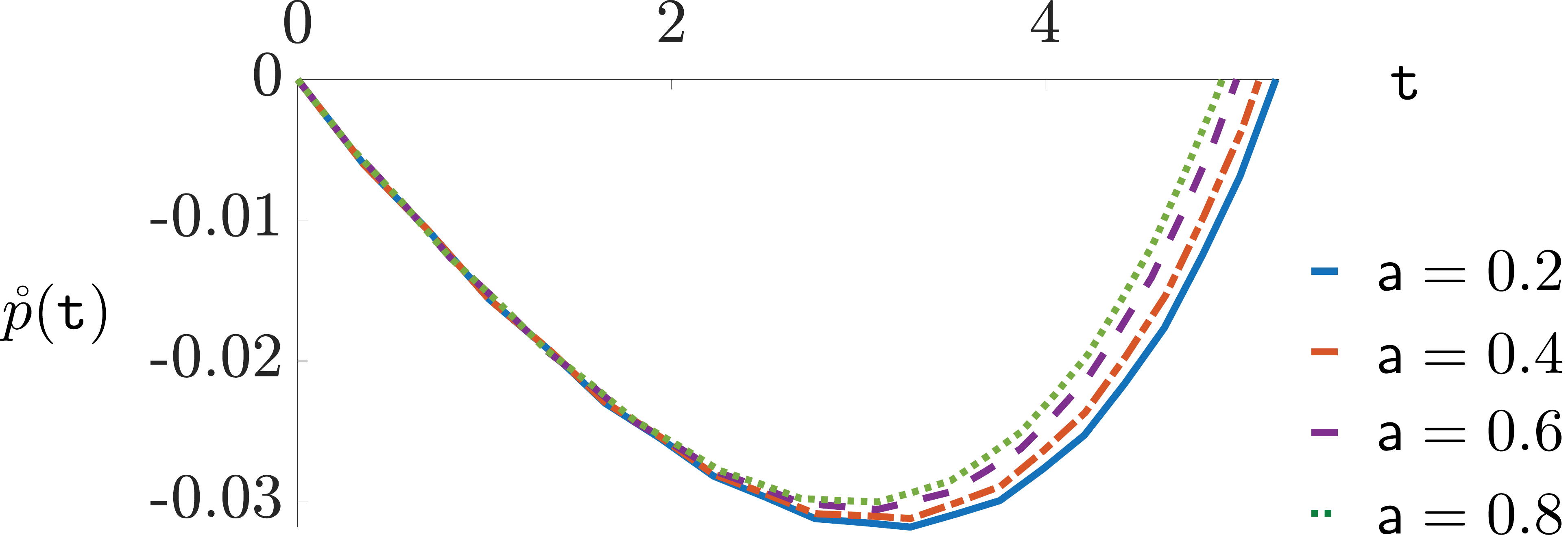}
  \vskip 0.16cm
  \caption{$\mathsf{a}$  is varied, while $\mathsf{f}=0.9$, $\mathsf{b}=1.5$, $\mathsf{p}=1.1$, $\mathsf{q}=1.2$, $\mathsf{h}=0.5$, and $\mathsf{L}=10$.}
  \label{subfig:varya_t1}
\end{subfigure}
%-----
\hfil % <-- added
%-----
\begin{subfigure}{0.42\textwidth}
  \includegraphics[width=\linewidth]{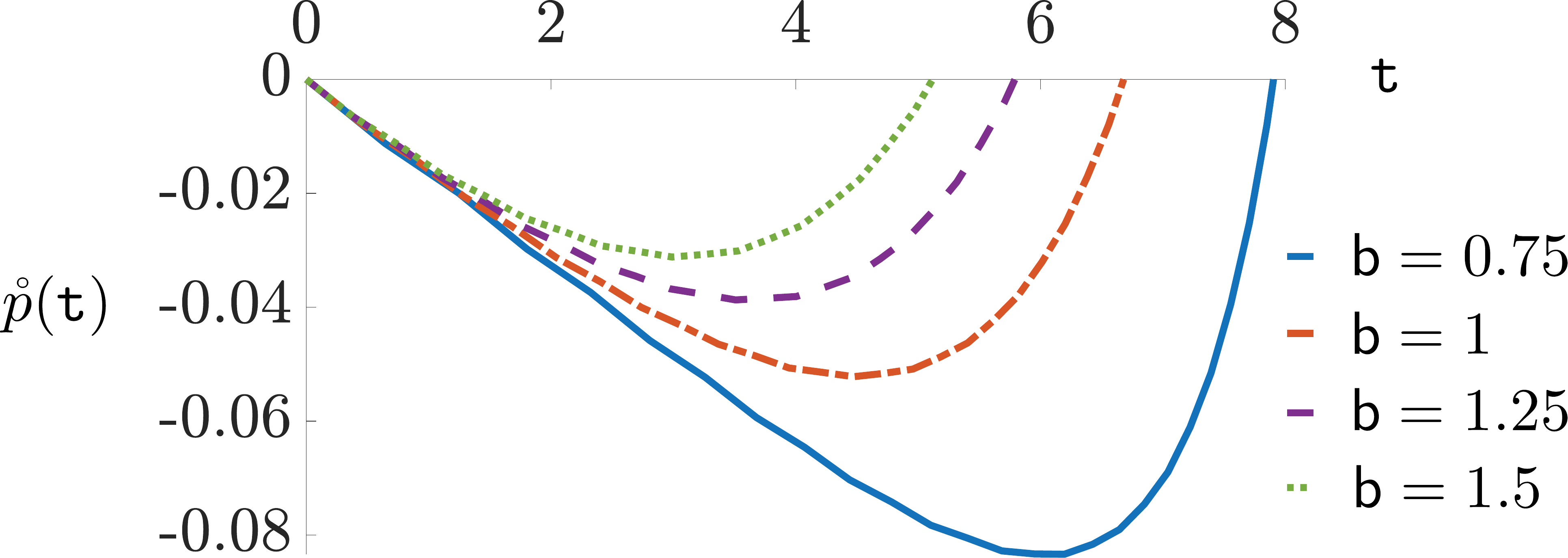}
  \vskip 0.16cm
  \caption{$\mathsf{b}$  is varied, while $\mathsf{f}=0.9$, $\mathsf{a}=0.5$, $\mathsf{p}=1.1$, $\mathsf{q}=1.2$, $\mathsf{h}=0.5$, and $\mathsf{L}=10$.}
  \label{subfig:varyb_t2}
\end{subfigure}
%-----
\medskip
\vspace{0.4cm}
%-----
\begin{subfigure}{0.45\textwidth}
  \includegraphics[width=\linewidth]{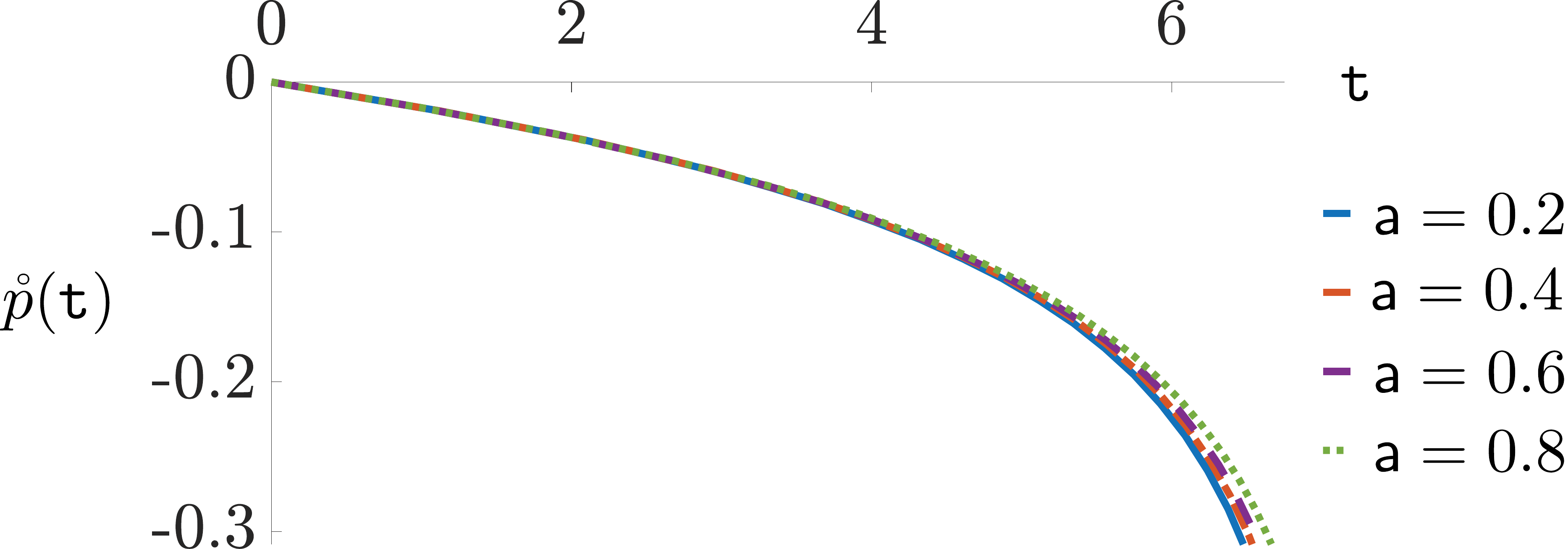}
  \vskip 0.16cm
  \caption{$\mathsf{a}$  is varied, while $\mathsf{f}=0.9$, $\mathsf{b}=0.25$, $\mathsf{p}=1.1$, $\mathsf{q}=1.2$, $\mathsf{h}=0.5$, and $\mathsf{L}=10$.}
  \label{subfig:varya_t2}
\end{subfigure}
%-----
\hfil % <-- added
%-----
\begin{subfigure}{0.415\textwidth}
  \includegraphics[width=\linewidth]{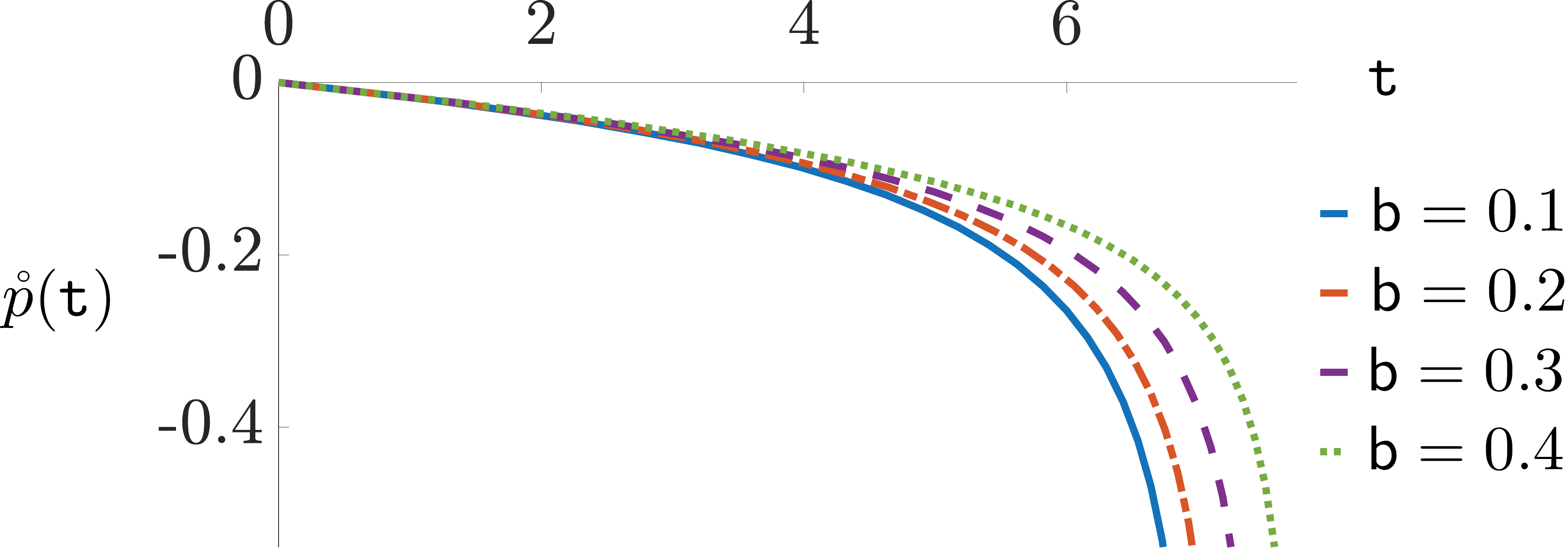}
  \vskip 0.16cm
  \caption{$\mathsf{b}$  is varied, while $\mathsf{f}=0.9$, $\mathsf{a}=0.5$, $\mathsf{p}=1.1$, $\mathsf{q}=1.2$, $\mathsf{h}=0.5$, and $\mathsf{L}=10$.}
  \label{subfig:varyb_t1}
\end{subfigure}
%-----
\medskip
\vspace{0.4cm}
%-----
\begin{subfigure}{0.44\textwidth}
  \includegraphics[width=\linewidth]{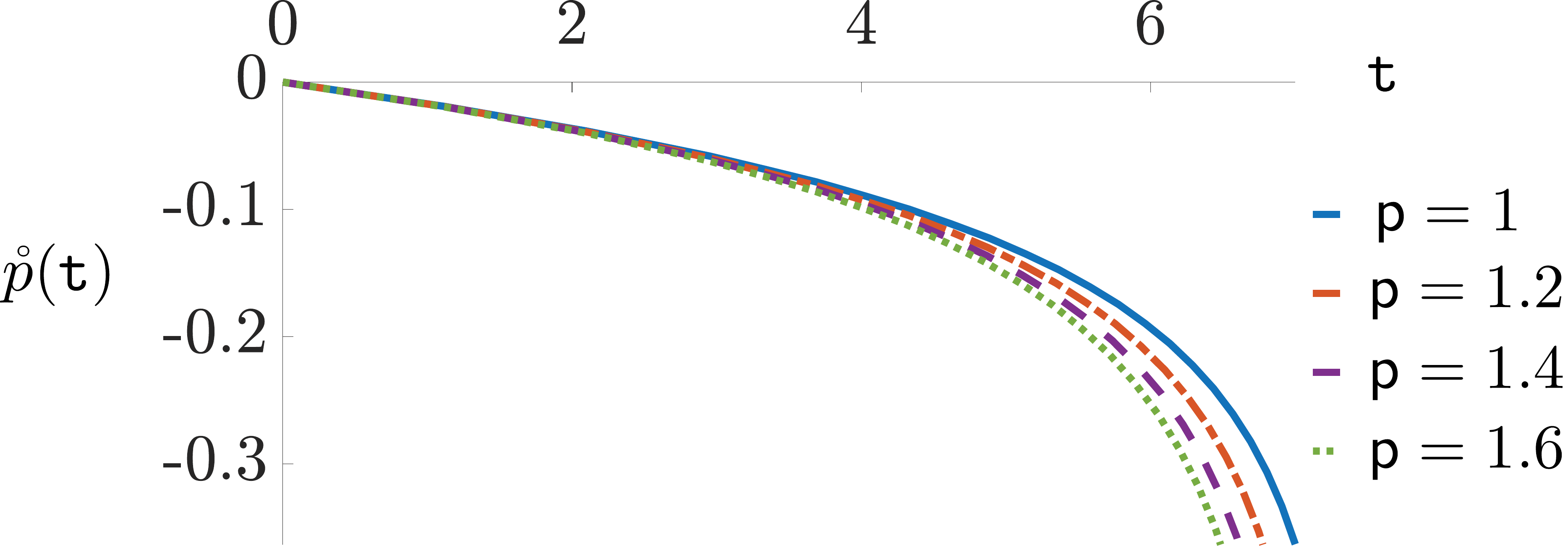}
  \vskip 0.16cm
  \caption{$\mathsf{p}$  is varied, while $\mathsf{f}=0.9$, $\mathsf{a}=0.8$, $\mathsf{b}=0.25$, $\mathsf{q}=1.2$, $\mathsf{h}=0.5$, and $\mathsf{L}=10$.}
  \label{subfig:varyp}
\end{subfigure}
%-----
\hfil % <-- added
%-----
\begin{subfigure}{0.44\textwidth}
  \includegraphics[width=\linewidth]{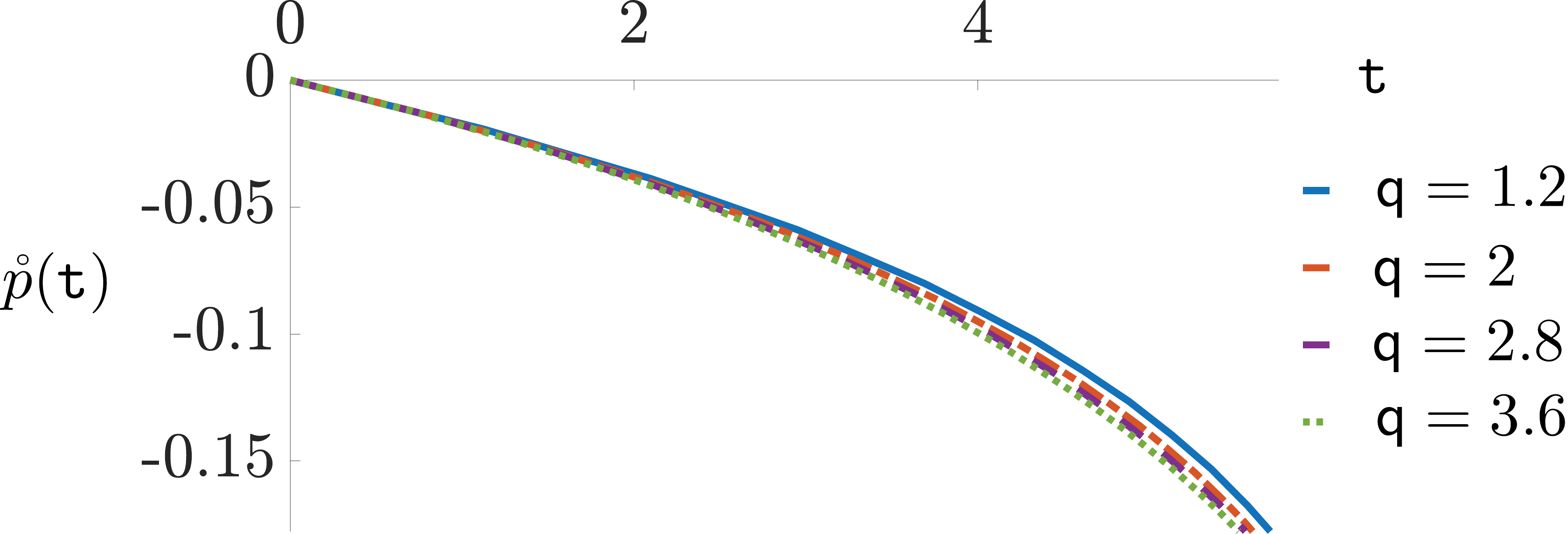}
  \vskip 0.16cm
  \caption{$\mathsf{q}$  is varied, while $\mathsf{f}=0.9$, $\mathsf{a}=0.8$, $\mathsf{b}=0.25$, $\mathsf{p}=1.1$, $\mathsf{h}=0.5$, and $\mathsf{L}=10$.}
  \label{subfig:varyq}
\end{subfigure}
%-----
\medskip
\vspace{0.4cm}
%-----
\begin{subfigure}{0.44\textwidth}
  \includegraphics[width=\linewidth]{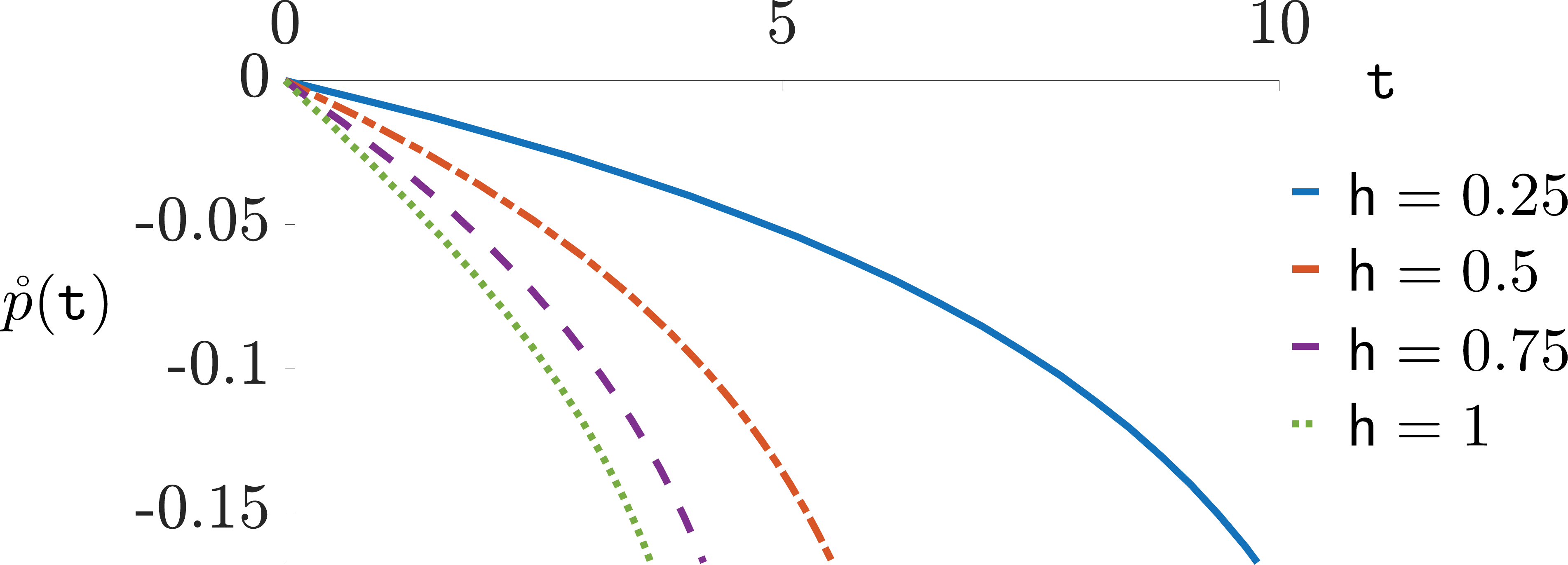}
   \vskip 0.16cm
  \caption{$\mathsf{h}$  is varied, while $\mathsf{f}=0.9$, $\mathsf{a}=0.8$, $\mathsf{b}=0.25$, $\mathsf{p}=1.1$, $\mathsf{q}=1.2$, and $\mathsf{L}=10$.}
  \label{subfig:varyh}
\end{subfigure}
%-----
\hfil % <-- added
%-----
\begin{subfigure}{0.44\textwidth}
  \includegraphics[width=\linewidth]{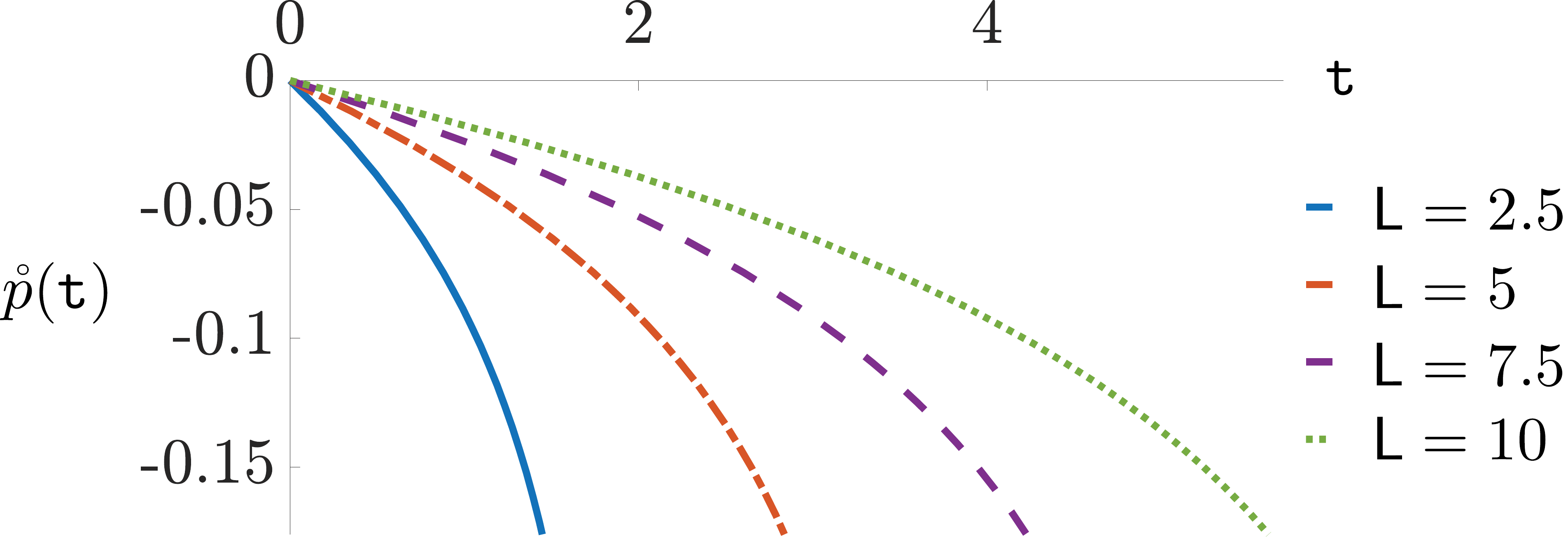}
   \vskip 0.16cm
  \caption{$\mathsf{L}$  is varied, while $\mathsf{f}=0.9$, $\mathsf{a}=0.8$, $\mathsf{b}=0.25$, $\mathsf{p}=1.1$, $\mathsf{q}=1.2$ and $\mathsf{h}=0.5$.}
  \label{subfig:varyL}
\end{subfigure}
%-----
\medskip
\vspace{0.4cm}
%-----
\begin{subfigure}{0.44\textwidth}
  \includegraphics[width=\linewidth]{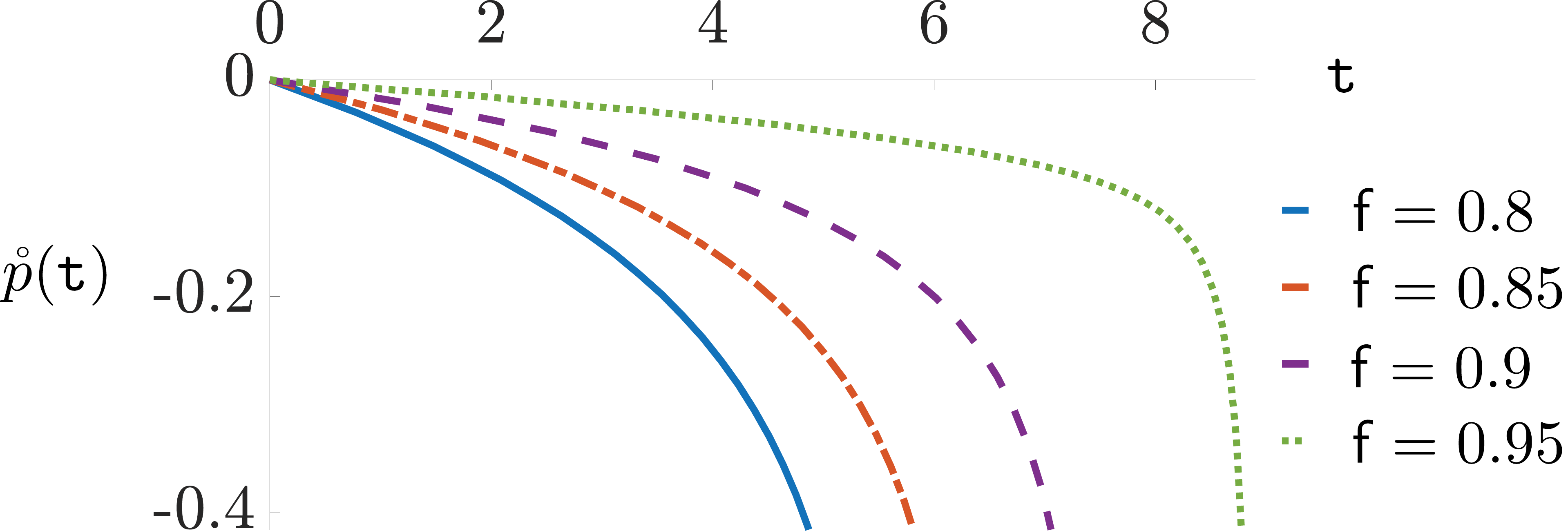}
  \vskip 0.16cm
  \caption{$\mathsf{f}$  is varied, while $\mathsf{a}=0.8$, $\mathsf{b}=0.25$, $\mathsf{p}=1.1$, $\mathsf{q}=1.2$, $\mathsf{h}=0.5$, and $\mathsf{L}=10$.}
  \label{subfig:varyf}
\end{subfigure}
%-----
\vspace{0.4cm}
\caption{The dimensionless parameters $\mathsf{f}$, $\mathsf{a}$, $\mathsf{b}$, $\mathsf{p}$, $\mathsf{q}$, $\mathsf{h}$, and $\mathsf{L}$ are varied to investigate their effects on the evolution of pressure inside the liquid.}
%-----
\label{fig:PressureEvolution}
%-----
\end{figure}

\begin{itemize}[topsep=2pt, leftmargin=10pt]
\item The radial speed of the interface, in both the reference and the current configurations, is observed to increase as the interface moves inward with time (Figure~\ref{fig:interface_coupledcase}). As expected, the  fraction of the initial liquid mass solidified increases over time. However, the rate of mass fraction solidified decreases with time (Figure~\ref{fig:interface_coupledcase}). These trends are similar to what has been observed in the rigid conductor case (see Figures~\ref{fig:TaoComparison} and \ref{fig:AllComparision_rigid}). The temperature field inside a rigid conductor is shown at different instances of time in Figure~\ref{fig:TemperaturesColor_rigid}. It should be noted that most numerical studies in the literature for the rigid conductor case only depict the motion of the interface \citep{tao1967generalized,shih1971analytical,hill1983freezing}. The experimental studies report the rate of solidification with the rate of change of mass fraction of the total initial liquid solidified with respect to time \citep{chan2006solidification}. This is possibly because the liquid inclusion tends to lose its spherical shape and concentricity with the previously accreted layers as the inclusion size decreases.  The trend we observe for the variation of mass fraction solidified qualitatively agrees with that of \citet{chan2006solidification}, although a direct comparison with the experimental data is not feasible due to the unavailability of a complete set of material properties of the substances used. The temperature field and the physical components of the Cauchy stress in the deformed solid for the coupled problem \eqref{eq:LinMomEqnonDim}-\eqref{eq:ICBCnonDim} are depicted in Figure~\ref{fig:coupledSolColorPlot}.

\item The symbol $\mathsf{f}$ denotes the ratio of the density of the undeformed solid to that of the liquid near the melting point. Solidification of a given mass of a liquid with $\mathsf{f}>1$ results in a reduction of the occupied volume. As the accretion surface moves inward, layers of liquid are replaced with denser solid layers, leading to a decrease in volume. Furthermore, since the container has fixed walls and, therefore, a fixed volume, the liquid inclusion naturally develops positive hydrostatic stress as soon as solidification begins, indicating possibility of cavitation. Although this is confirmed numerically, the observed data is excluded from figures as positive liquid pressure is not physically possible.
Moreover, it follows from $\mathring{p}(\mathtt{t})=1-\frac{\mathtt{s}^3(\mathtt{t})}{\mathtt{S}^3(\mathtt{t})}$ that a negative liquid pressure is equivalent to a negative radial displacement of the accreting layers (see Figures \ref{subfig:dispt} and \ref{subfig:dispR}).\footnote{Since $\mathtt{s}(\mathtt{t})=\mathtt{r}(\mathtt{S}(\mathtt{t}),\mathtt{t})$ is the position of the solidification interface in the deformed configuration, and $\mathtt{R}=\mathtt{S}(\mathtt{t})$ was its position in the initial liquid pool, $\mathtt{s}(\mathtt{t})-\mathtt{S}(\mathtt{t})$, or equivalently, $\bar{\mathtt{r}}(\mathtt{R})-\mathtt{R}$, denotes the radial displacement of an accreting layer.} 
As the solidification interface approaches the center, the magnitude of the displacement of the accreting layers increases rapidly, requiring it to decelerate and decrease swiftly to ultimately vanish at the center.\footnote{The time instant $\mathtt{t}_\text{c}$ marking the completion of solidification must satisfy $\mathtt{S}(\mathtt{t}_\text{c})=0$. Further, if this is achieved without cavitation, then $\mathtt{s}(\mathtt{t}_\text{c})=0$. Thus, $\mathtt{s}(\mathtt{t}_\text{c})-\mathtt{S}(\mathtt{t}_\text{c})=0$.}
Thus, the mesh near the center must be much finer; otherwise, numerical techniques that better accommodate such sudden fluctuations need to be employed.

\item The figures shown in this section are based on the assumption $\mathsf{f}<1$. With this assumption, an accreting layer with a given mass tends to occupy a greater volume upon solidification, compressing the liquid inclusion and resulting in negative hydrostatic stress. The magnitude of this negative hydrostatic stress increases with time as the solidification interface moves inward (Figure~\ref{fig:PressureEvolution}). The extra volume occupied by the solidifying layers piles up to create a significant gap, causing the pressure in the liquid to become highly compressive as the interface approaches the center.  

\item Surface stresses play a significant role for liquid inclusions smaller than a certain limit determined by the elastocapillarity length---the ratio of surface tension to the bulk modulus \citep{Bico2018}. In this paper we do not consider surface stress, and hence, do not report the numerical results for very small liquid inclusions. Although the process is halted a while before complete solidification, the rate of increase in the magnitude of liquid pressure is significantly high by the time this margin is reached.

\item Note that $\mathtt{T}=0$ at the melting point, and since $T_c<T_m$, $\mathtt{T}=\frac{T_m - T}{T_m - T_c}$ increases as the real temperature $T$ decreases (see Figures~\ref{fig:TemperaturesColor_rigid} and \ref{fig:coupledSolColorPlot}). The moving interface is always at the melting point, and the temperature decreases as one moves towards the fixed wall (Figure~\ref{subfig:TempR}). The temperature at a point decreases over time after it is accreted (Figure~\ref{subfig:Tempt}). Radial displacements are always negative, and the magnitude at any accreted point decreases over time (Figure~\ref{subfig:dispt}). At any instant, the magnitude of radial displacement is maximum at the moving boundary and decreases to zero at the fixed boundary (Figure~\ref{subfig:dispR}).  

\item Both $\mathring{\sigma}^{rr}$ and $\mathring{\sigma}^{\theta\theta}$ are negative near the moving boundary. $\mathring{\sigma}^{\theta\theta}$ increases as one moves away from the inclusion (i.e., decreases in magnitude), vanishes somewhere in between, and eventually becomes positive near the wall (Figure~\ref{subfig:sigmattR}). $\mathring{\sigma}^{rr}$ decreases in magnitude as one moves away from the inclusion but remains negative if the inclusion size is too large (Figure~\ref{subfig:sigmarrR}). However, when the interface has moved far enough from the wall, $\mathring{\sigma}^{rr}$ can be positive near the wall, decreasing to a negative value near the inclusion. At any accreted point, $\mathring{\sigma}^{rr}$ is initially negative and decreases in magnitude over time (Figure~\ref{subfig:sigmarrt}). For points closer to the fixed wall, $\mathring{\sigma}^{rr}$ eventually becomes positive as the inclusion size decreases. $\mathring{\sigma}^{\theta\theta}$ is initially negative for all accreted points, and quickly transitions to a positive value, except for the points accreted just before the process is halted (Figure~\ref{subfig:sigmattt}).

\item The dimensionless parameter $\mathsf{b}$ describes the thermal expansion properties of the solid relative to the temperature difference between its melting point and the cold wall temperature. A larger $\mathsf{b}$ implies a higher contraction of the solid for a given temperature drop.
It is observed that the rate of increase in liquid pressure magnitude is much faster for lower values of $\mathsf{b}$ (Figure~\ref{subfig:varyb_t1}). If $\mathsf{b}$ is too large, the liquid inclusion pressure decreases from zero until it reaches a minimum, and then increases until it becomes zero again (Figure~\ref{subfig:varyb_t2}).  Positive pressure solutions beyond this point are physically meaningless due to the possibility of cavitation and are therefore discarded. The reason behind this tendency of liquid cavitation, even with $\mathsf{f}<1$, is the extremely high thermal contraction in the colder layers closer to the container walls. $\mathsf{a}<1$ represents the ratio of the temperature difference between the cold container wall and the melting point of the liquid to the absolute melting temperature. Figures \ref{subfig:varya_t1} and \ref{subfig:varya_t2} describe the influence of $\mathsf{a}$ on the evolution of the liquid pressure within the inclusion for the two distinct categories of $\mathsf{b}$ discussed above.

\item The elastic material properties are captured by $\mathsf{p}$ and $\mathsf{q}$, which represent the shear and bulk modulus, respectively, of the solid near the melting point as compared to the liquid bulk modulus near the solidification temperature. The magnitude of pressure in the liquid inclusion rises faster with larger $\mathsf{p}$ and $\mathsf{q}$ values (see Figures~\ref{subfig:varyp} and \ref{subfig:varyq}). The specific latent heat of solidification appears only in the dimensionless constant $\mathsf{L}$, which is loosely interpreted as a measure of the latent heat released relative to the heat capacity of the solid. The heat transfer with the container walls is incorporated in the coefficient $\mathsf{h}$, loosely quantifying how much of the heat conducted towards the outer boundary of the accreted solid is transferred out into the cold wall.

\item The numerical variations in $\mathsf{a}$, $\mathsf{b}$, $\mathsf{p}$, and $\mathsf{q}$ used for the parametric studies do not significantly impact the solidification rate, indicating a lower sensitivity to these parameters (see Figures~\ref{subfig:varya_massfrac}, \ref{subfig:varyb_massfrac}, \ref{subfig:varyp_massfrac} and\ref{subfig:varyq_massfrac}). A value of $\mathsf{f}$ closer to $1$, with the liquid denser than the solid near melting, results in slower solidification (Figure~\ref{subfig:varyf_massfrac}); and the sensitivity to variations in $\mathsf{f}$ is moderate. The solidification rate is highly sensitive to $\mathsf{h}$ and $\mathsf{L}$. 
A larger $\mathsf{h}$ implies that the heat is able to flow more efficiently out of the solid into the container walls, facilitating in faster solidification (Figure~\ref{subfig:varyh_massfrac}). A smaller $\mathsf{L}$ implies less specific latent heat compared to the specific heat capacity, allowing the accreted solid to better absorb the heat released during solidification. This indirectly promotes outward heat conduction and results in a higher solidification rate (Figure~\ref{subfig:varyL_massfrac}). The higher rates of pressure drop in the liquid for larger $\mathsf{h}$ and smaller $\mathsf{L}$ values (see Figures~\ref{subfig:varyh} and \ref{subfig:varyL}) are attributed to the faster solidification rates.

\item The configuration obtained by detaching the accreted solid from the rigid walls of the cold container after a given time, removing any remaining unsolidified liquid, and subsequently cooling the solid to a uniform steady-state temperature is not stress-free (see Figures \ref{subfig:sigmarr_color} and \ref{subfig:sigmatt_color}). In this configuration, both the inner and outer boundaries are displaced inward relative to their positions in the initial liquid (Figure~\ref{subfig:disp_difftimes}). The inward displacement of the outer boundary is likely caused by thermal contraction. The inner layers experience highly negative $\mathring{\sigma}^{rr}$ during accretion, owing to the presence of a pressurized liquid inclusion. When the liquid is removed and the inner boundary becomes traction-free, the inner layers naturally tend to move apart to relieve the negative stress. Moreover, the closer the layer is to the inner boundary, the more pronounced this tendency becomes. In the residually-stressed configuration, $\mathring{\tilde{\sigma}}^{rr}$ is zero at the inner boundary, increases as one moves outward, reaches a maximum, and then decreases to vanish at the outer boundary (Figure~\ref{subfig:sigmarr_color}). The maximum value of $\mathring{\tilde{\sigma}}^{rr}$ is larger if the accretion process ends later (Figure~\ref{subfig:sigmarr_difftimes}). $\mathring{\tilde{\sigma}}^{\theta\theta}$ is negative at the inner boundary, increases as one moves outwards, eventually becoming positive at the outer boundary (Figure~\ref{subfig:sigmatt_color}). The variation in $\mathring{\tilde{\sigma}}^{\theta\theta}$ is larger if the solidification process is stopped later (Figure~\ref{subfig:sigmatt_difftimes}). Consequently, the outer boundary is prone to developing cracks, while the inner boundary is prone to buckling instabilities.

\end{itemize}

%\todo{By ``This peak $\mathring{\tilde{\sigma}}^{rr}$" do you mean ``The peak of $\mathring{\tilde{\sigma}}^{rr}$" or ``The maximum value of $\mathring{\tilde{\sigma}}^{rr}$"? \\
%Satya: Yes, rephrased it.}

%---------------------
%---------------------

%-----------------------------------------------------------
%-----------------------------------------------------------
%-----------------------------------------------------------
%-----------------------------------------------------------

%\todo{Arash: In the caption of Figure 11, you may want to replace $\mathtt{t}_1$, $\mathtt{t}_2$, $\mathtt{t}_3$, $\mathtt{t}_4$ by $\mathtt{t}_1<\mathtt{t}_2<\mathtt{t}_3<\mathtt{t}_4$?}

%-----------------------------------------------------------
%-----------------------------------------------------------
\begin{figure}[t!]
%-----
\centering % <-- added
%-----
\begin{subfigure}{0.45\textwidth}
  \includegraphics[width=\linewidth]{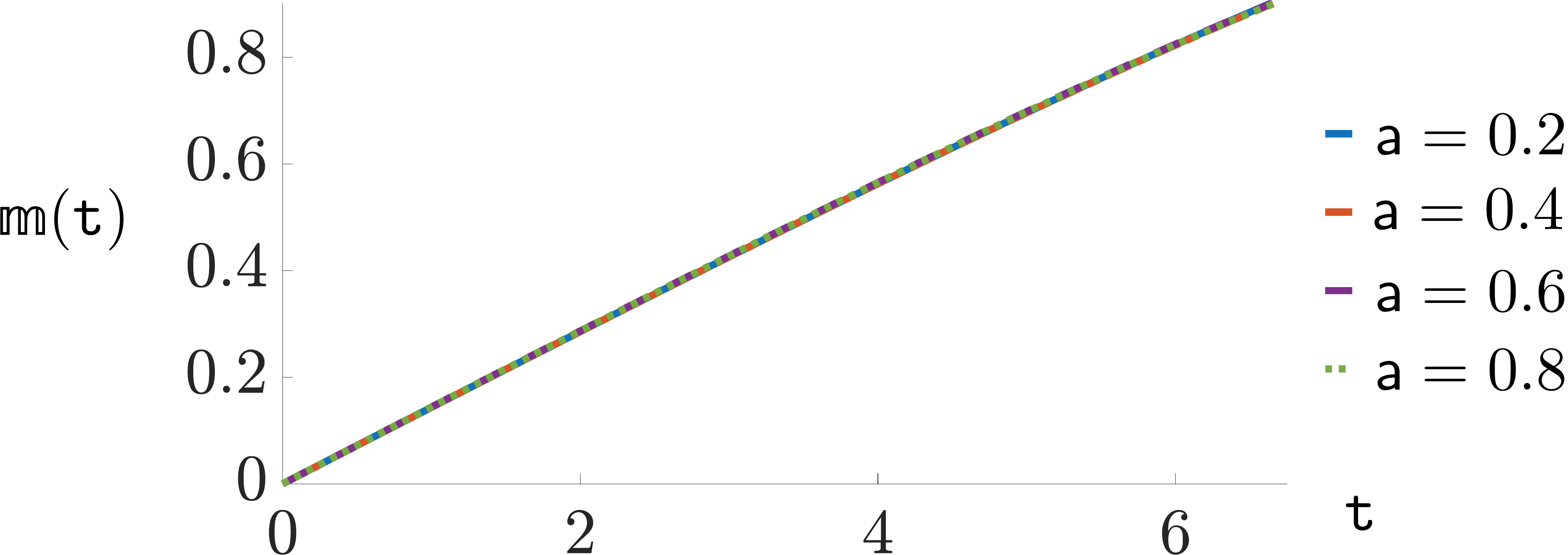}
  \vskip 0.16cm
  \caption{$\mathsf{a}$  is varied, while $\mathsf{f}=0.9$, $\mathsf{b}=0.25$, $\mathsf{p}=1.1$, $\mathsf{q}=1.2$, $\mathsf{h}=0.5$, and $\mathsf{L}=10$.}
  \label{subfig:varya_massfrac}
\end{subfigure}
%-----
\hfil % <-- added
%-----
\begin{subfigure}{0.415\textwidth}
  \includegraphics[width=\linewidth]{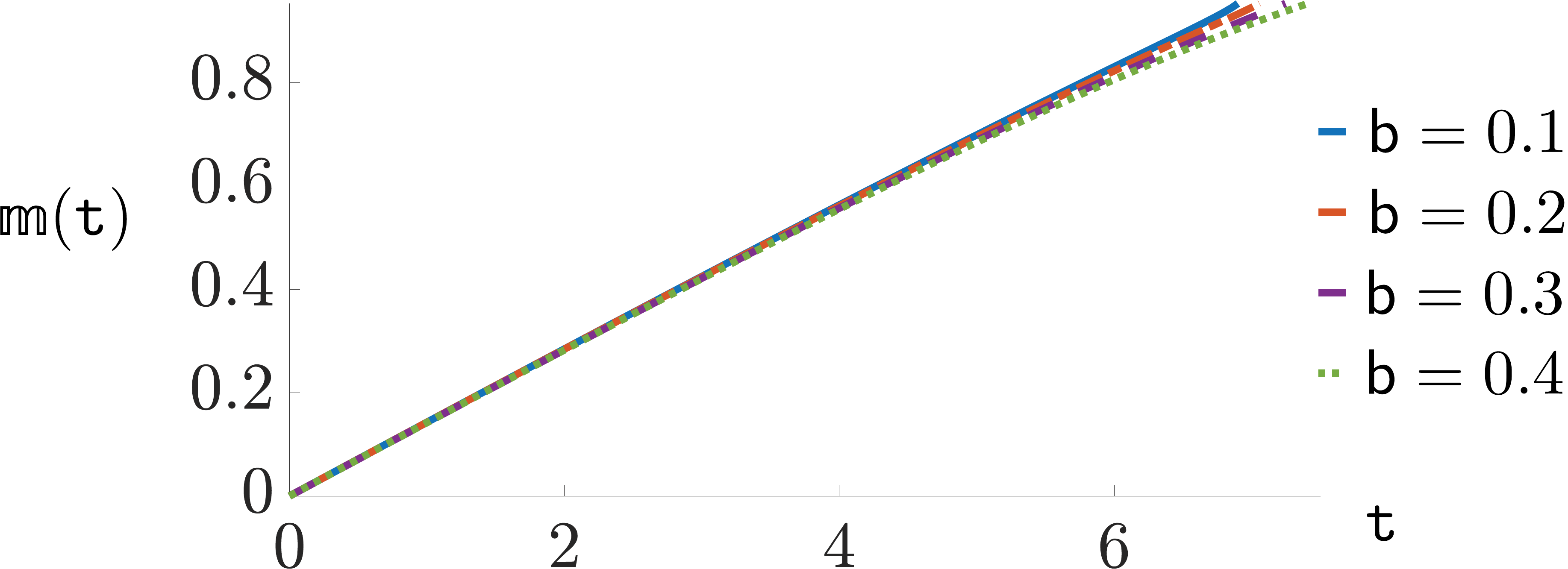}
  \vskip 0.16cm
  \caption{$\mathsf{b}$  is varied, while $\mathsf{f}=0.9$, $\mathsf{a}=0.5$, $\mathsf{p}=1.1$, $\mathsf{q}=1.2$, $\mathsf{h}=0.5$, and $\mathsf{L}=10$.}
  \label{subfig:varyb_massfrac}
\end{subfigure}
%-----
\medskip
\vspace{0.4cm}
%-----
\begin{subfigure}{0.44\textwidth}
  \includegraphics[width=\linewidth]{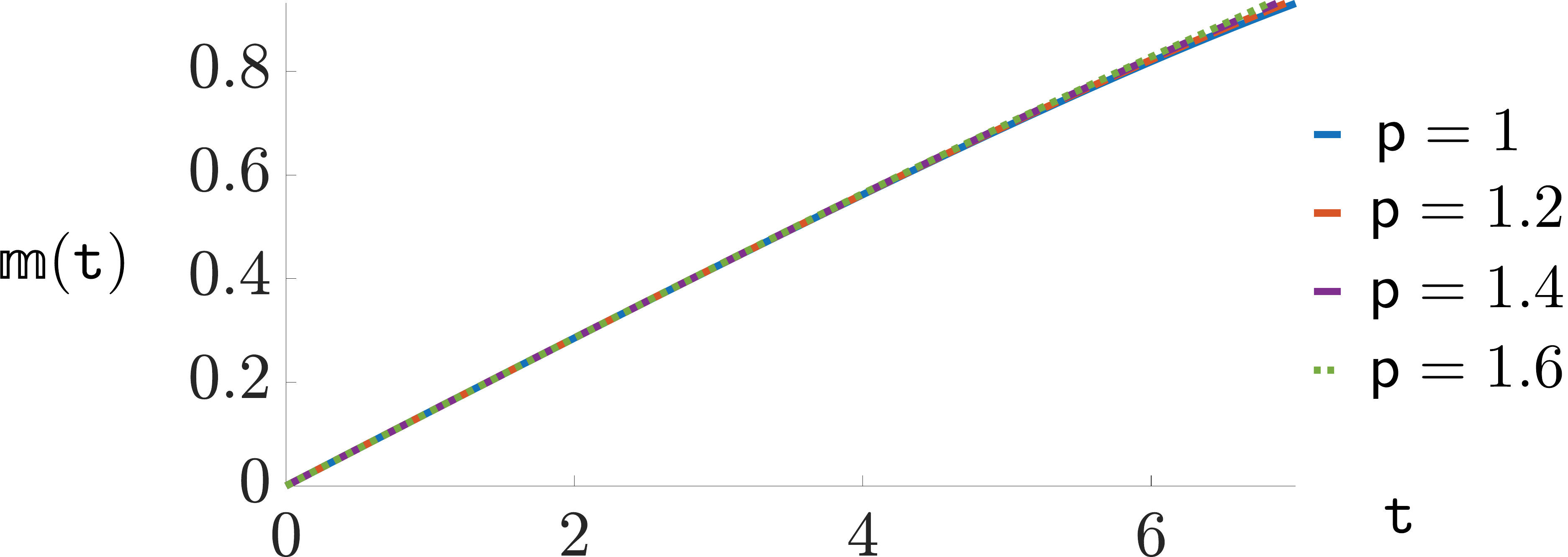}
  \vskip 0.16cm
  \caption{$\mathsf{p}$  is varied, while $\mathsf{f}=0.9$, $\mathsf{a}=0.8$, $\mathsf{b}=0.25$, $\mathsf{q}=1.2$, $\mathsf{h}=0.5$, and $\mathsf{L}=10$.}
  \label{subfig:varyp_massfrac}
\end{subfigure}
%-----
\hfil % <-- added
%-----
\begin{subfigure}{0.44\textwidth}
  \includegraphics[width=\linewidth]{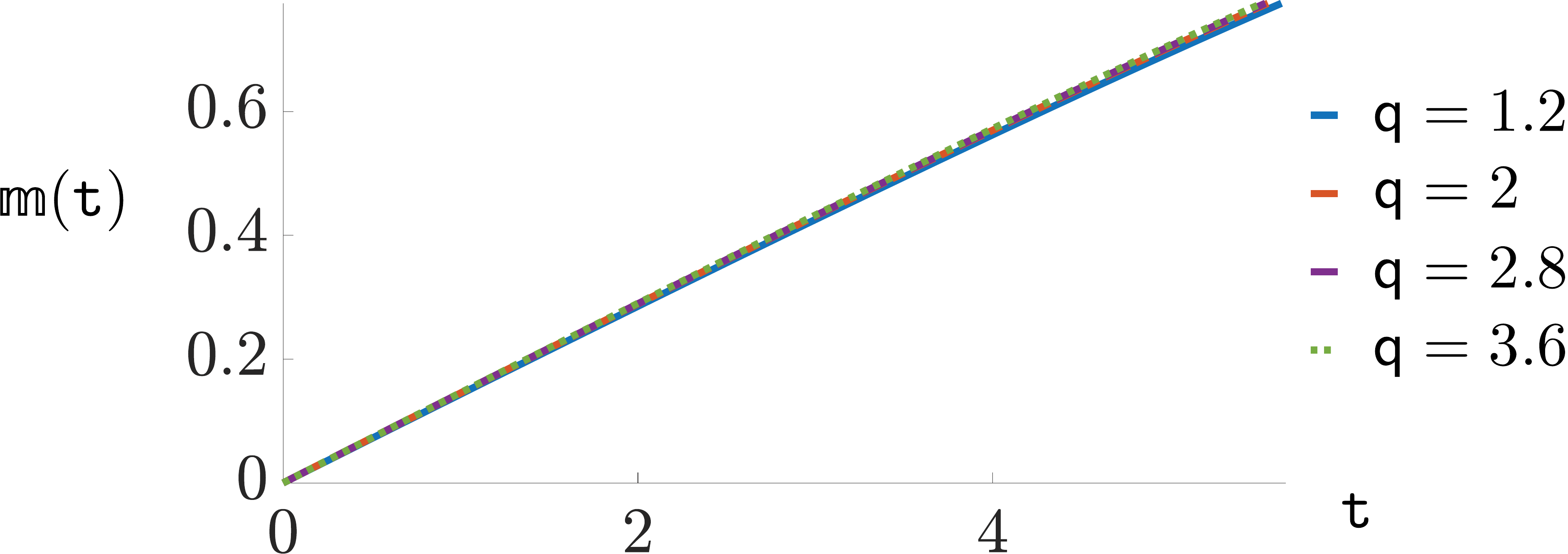}
  \vskip 0.16cm
  \caption{$\mathsf{q}$  is varied, while $\mathsf{f}=0.9$, $\mathsf{a}=0.8$, $\mathsf{b}=0.25$, $\mathsf{p}=1.1$, $\mathsf{h}=0.5$, and $\mathsf{L}=10$.}
  \label{subfig:varyq_massfrac}
\end{subfigure}
%-----
\medskip
\vspace{0.4cm}
%-----
\begin{subfigure}{0.44\textwidth}
  \includegraphics[width=\linewidth]{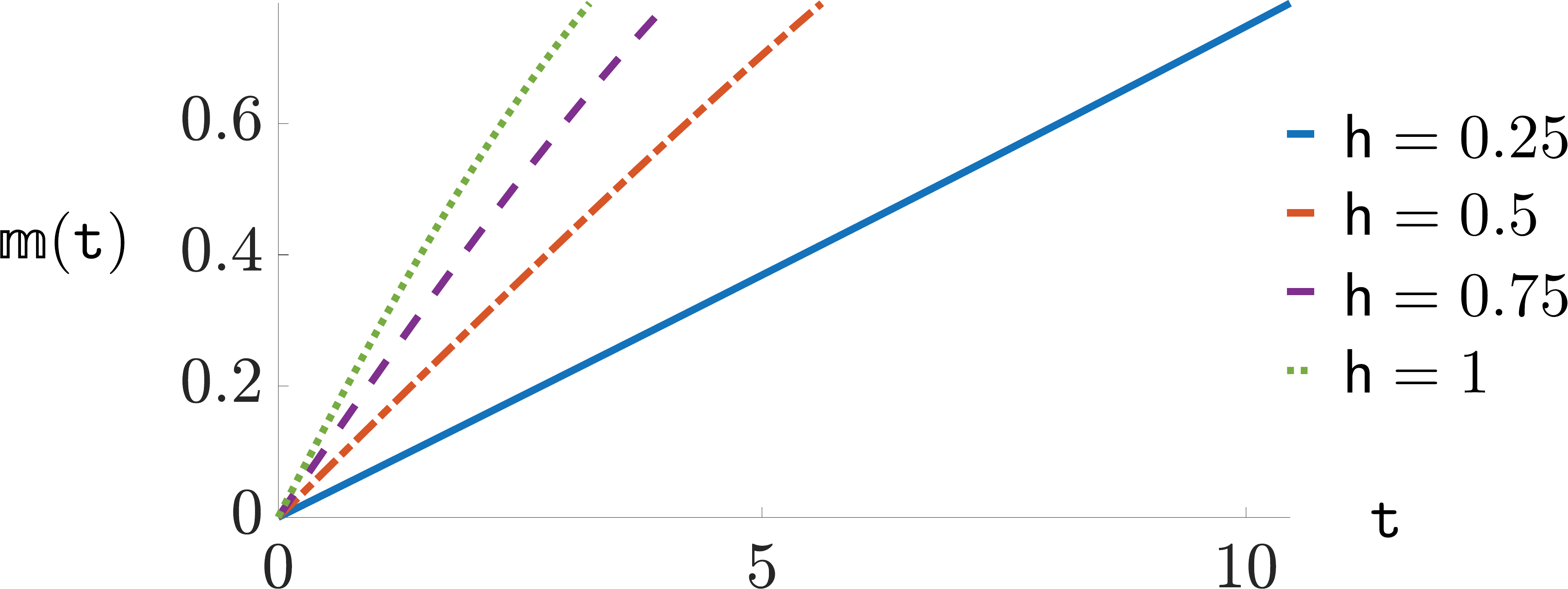}
   \vskip 0.16cm
  \caption{$\mathsf{h}$  is varied, while $\mathsf{f}=0.9$, $\mathsf{a}=0.8$, $\mathsf{b}=0.25$, $\mathsf{p}=1.1$, $\mathsf{q}=1.2$, and $\mathsf{L}=10$.}
  \label{subfig:varyh_massfrac}
\end{subfigure}
%-----
\hfil % <-- added
%-----
\begin{subfigure}{0.44\textwidth}
  \includegraphics[width=\linewidth]{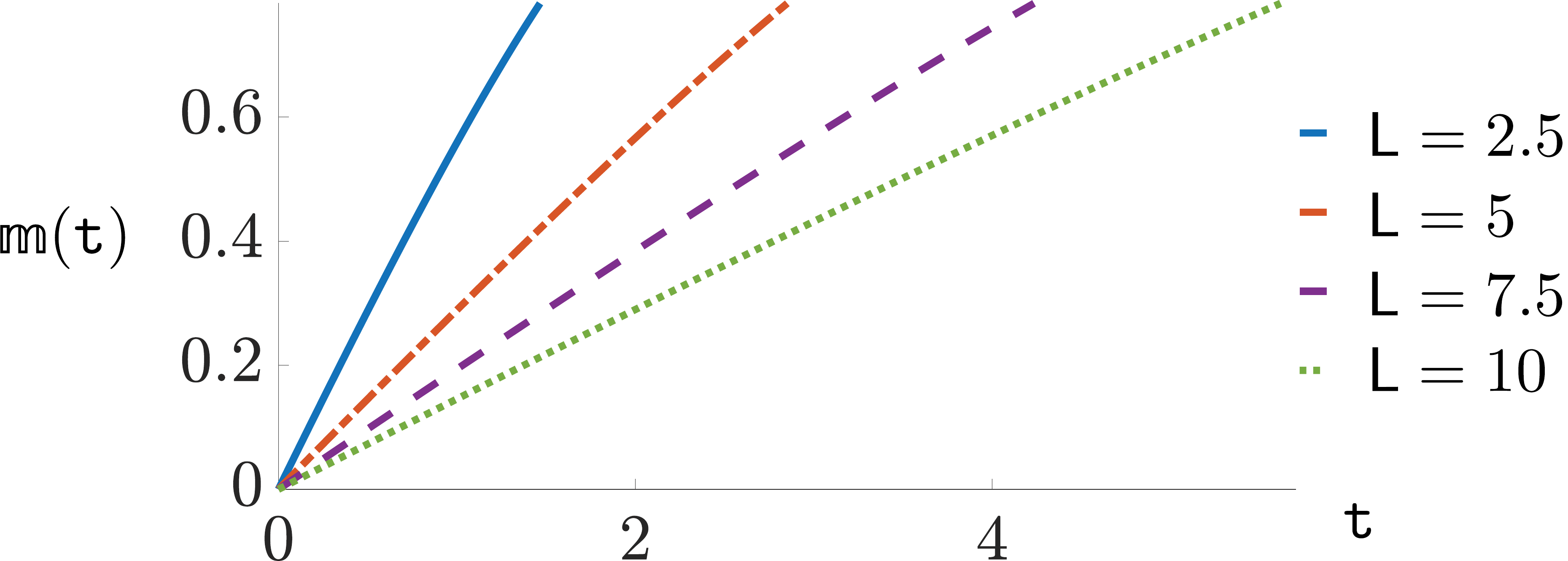}
   \vskip 0.16cm
  \caption{$\mathsf{L}$  is varied, while $\mathsf{f}=0.9$, $\mathsf{a}=0.8$, $\mathsf{b}=0.4$, $\mathsf{p}=1.1$, $\mathsf{q}=1.2$ and $\mathsf{h}=0.5$.}
  \label{subfig:varyL_massfrac}
\end{subfigure}
%-----
\medskip
\vspace{0.4cm}
%-----
\begin{subfigure}{0.44\textwidth}
  \includegraphics[width=\linewidth]{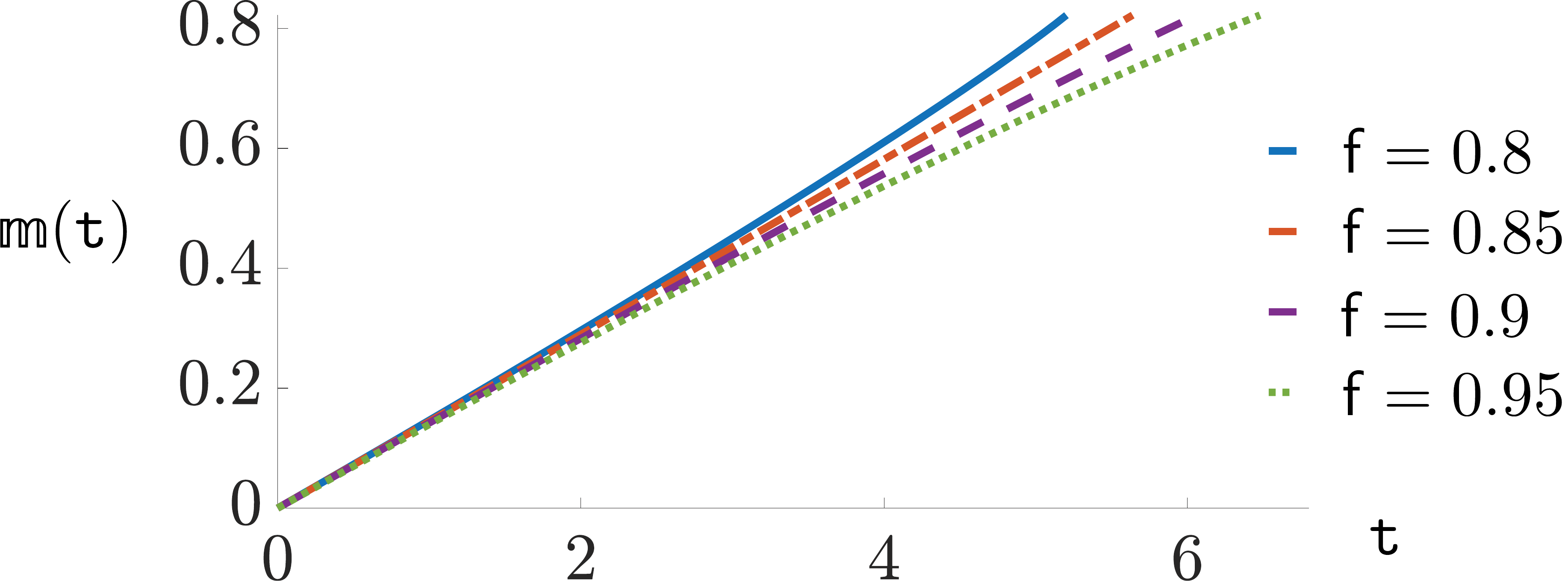}
  \vskip 0.16cm
  \caption{$\mathsf{f}$  is varied, while $\mathsf{a}=0.8$, $\mathsf{b}=0.25$, $\mathsf{p}=1.1$, $\mathsf{q}=1.2$, $\mathsf{h}=0.5$, and $\mathsf{L}=10$.}
  \label{subfig:varyf_massfrac}
\end{subfigure}
%-----
\vspace{0.4cm}
\caption{The dimensionless parameters $\mathsf{f}$, $\mathsf{a}$, $\mathsf{b}$, $\mathsf{p}$, $\mathsf{q}$, $\mathsf{h}$, and $\mathsf{L}$ are varied to investigate their effects on the accretion process. The assessment is based on $\mathbb{m}(\mathtt{t})$, which represents the fraction of the initial liquid mass solidified till time $\mathtt{t}$.}
%-----
\label{fig:massfrac}
%-----
\end{figure}
%-----------------------------------------------------------
%-----------------------------------------------------------

%\todo{Arash: In the captions of Figures 12 and 13, what do you mean by ``systematically varied"?}

%-----------------------------------------------------------
%-----------------------------------------------------------
%-----------------------------------------------------------
%-----------------------------------------------------------

%-----------------------------------------------------------
%-----------------------------------------------------------
\begin{figure}[t!]
%-----
\centering % <-- added
%-----
\begin{subfigure}{0.64\textwidth}
\centering
  \includegraphics[width=0.64\linewidth]{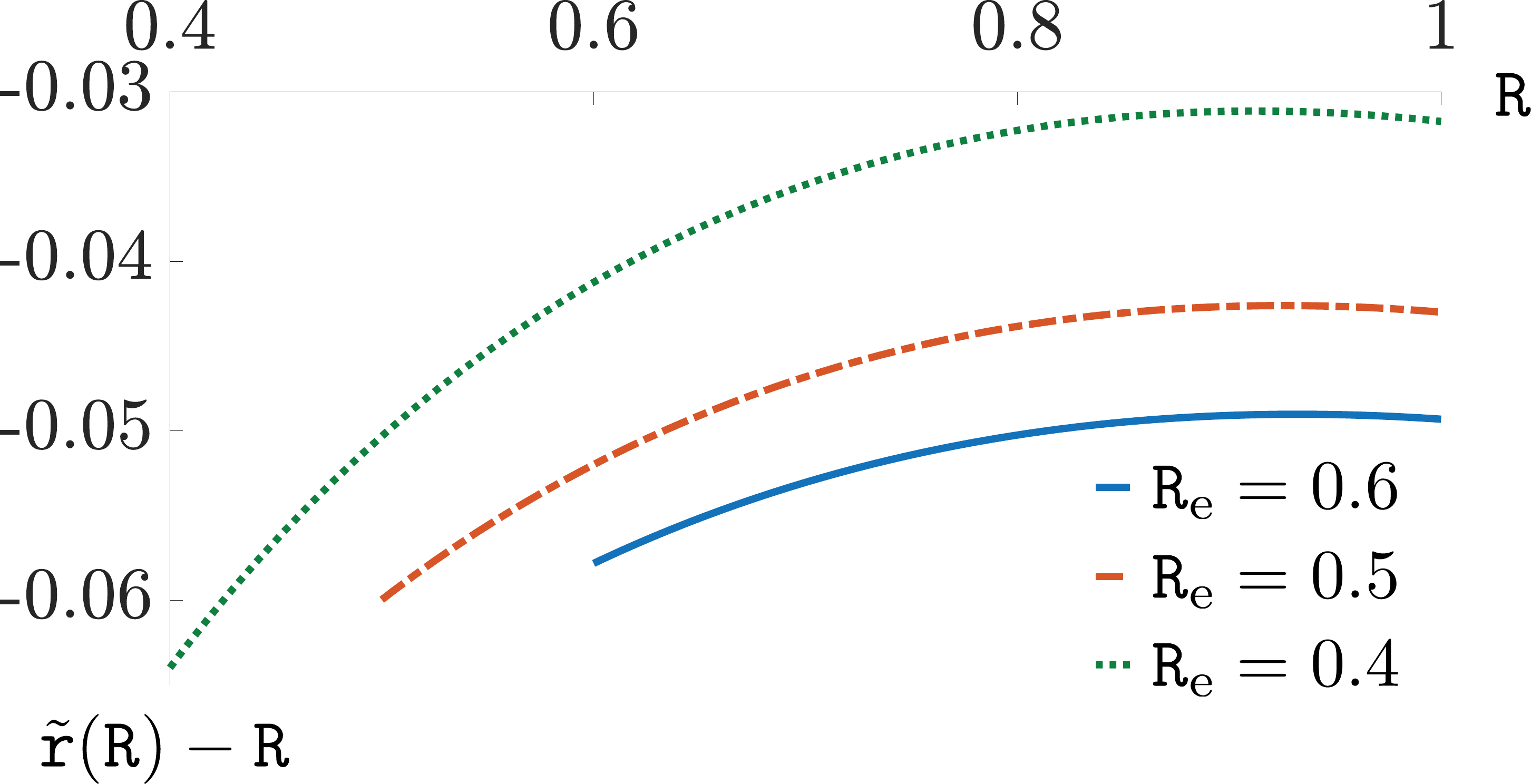}
  \vskip 0.16cm
  \caption{The displacement of a point in the residually-stressed configuration, relative to its position in the initial unsolidified liquid, is depicted as a function of the dimensionless material coordinate.}
  \label{subfig:disp_difftimes}
\end{subfigure}
%-----
\medskip
\vspace{0.4cm}
%-----
\begin{subfigure}{0.48\textwidth}
\centering
  \includegraphics[width=0.8\linewidth]{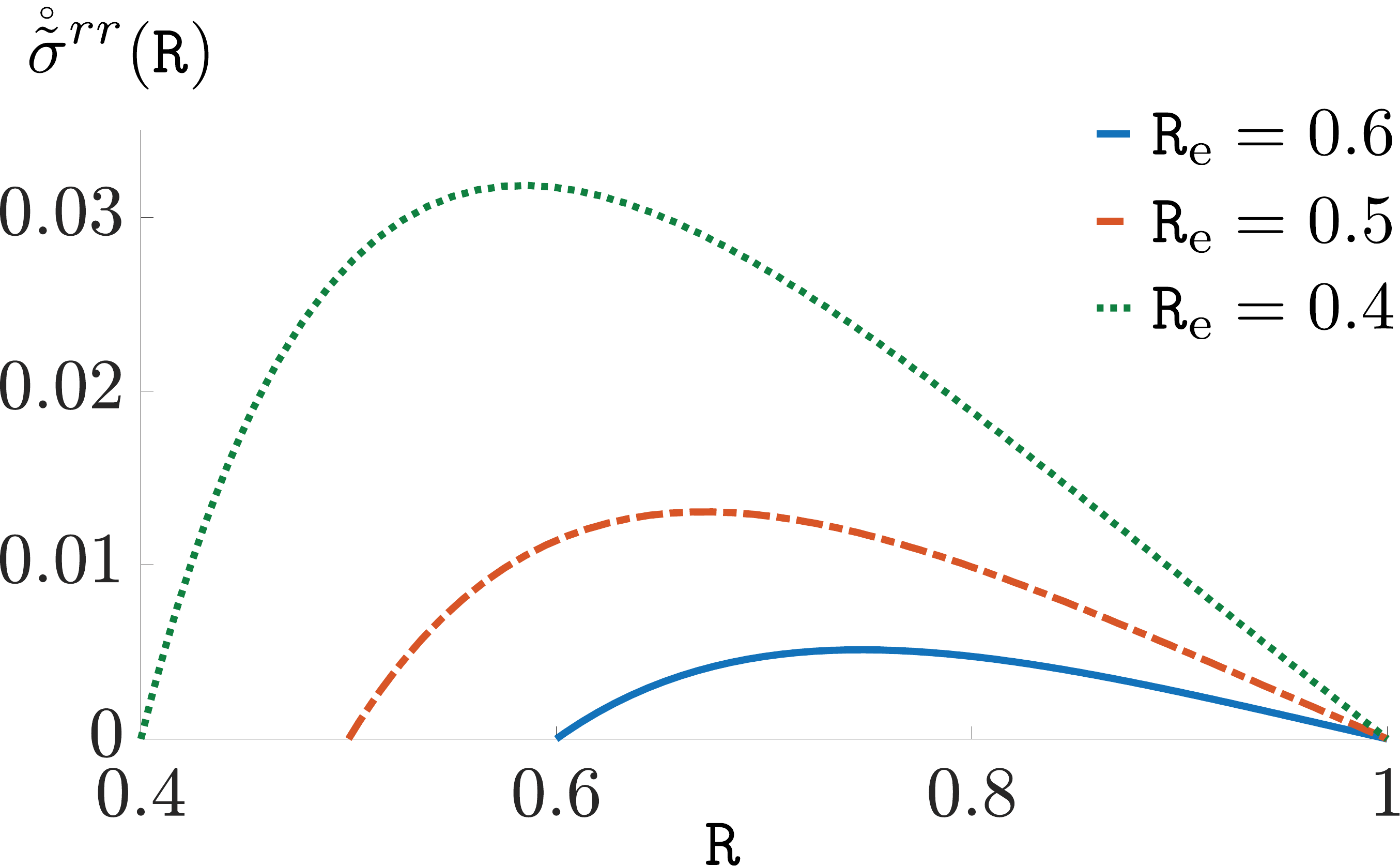}
  \vskip 0.16cm
  \caption{The dimensionless physical component $\mathring{\tilde{\sigma}}^{rr}$ of the Cauchy stress in the residually-stressed configuration is depicted as a function of the dimensionless material coordinate $\mathtt{R}$.}
  \label{subfig:sigmarr_difftimes}
\end{subfigure}
%-----
\hfill
%-----
\begin{subfigure}{0.48\textwidth}
\centering
  \includegraphics[width=0.8\linewidth]{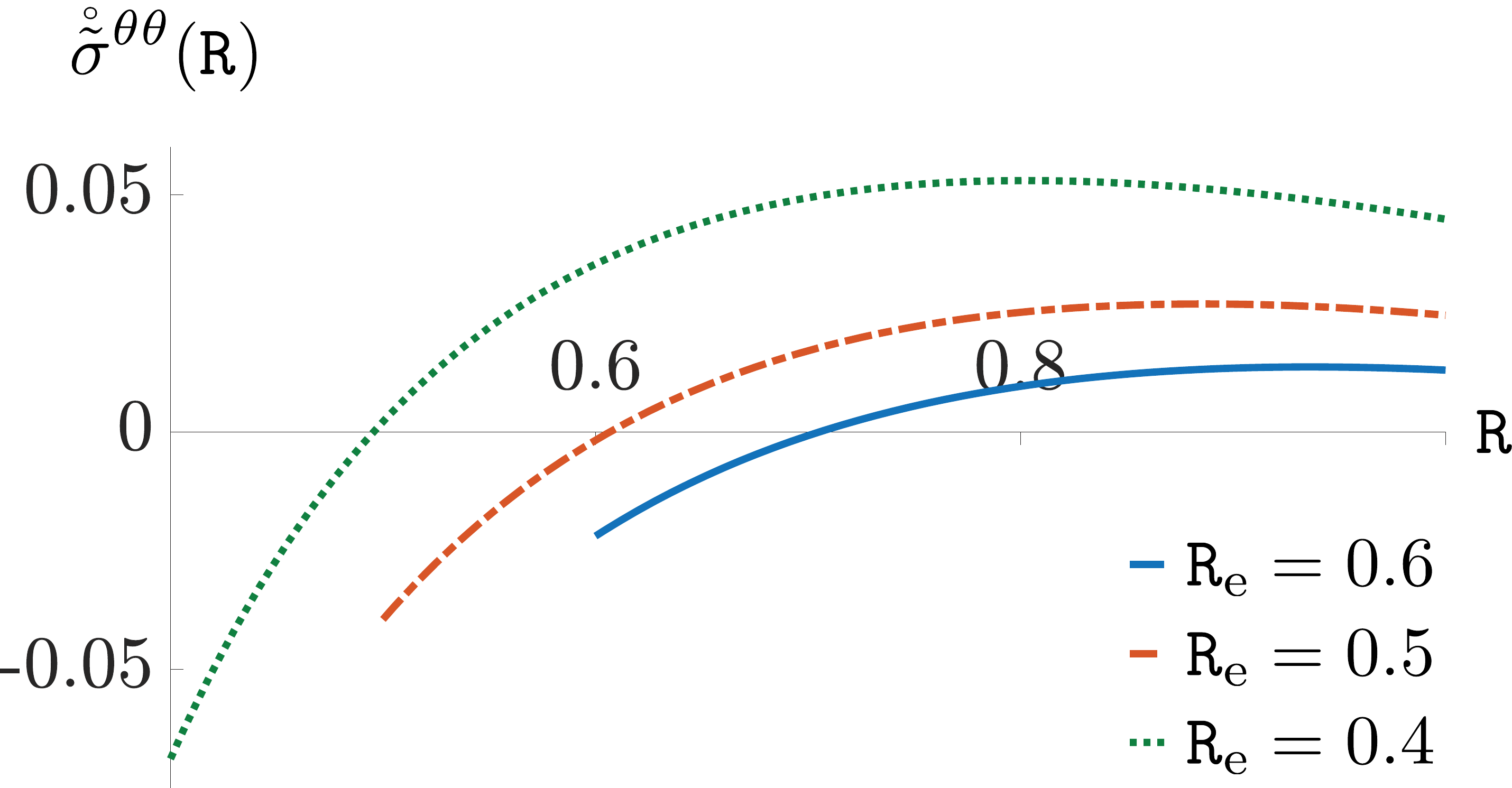}
  \vskip 0.48cm
  \caption{The dimensionless physical component $\mathring{\tilde{\sigma}}^{\theta\theta}$ of the Cauchy stress in the residually-stressed configuration is depicted as a function of the dimensionless material coordinate $\mathtt{R}$.}
  \label{subfig:sigmatt_difftimes}
\end{subfigure}
%-----
\medskip
\vspace{0.5cm}
%-----
\begin{subfigure}{0.48\textwidth}
\centering
\hspace{36pt}
  \includegraphics[width=0.62\linewidth]{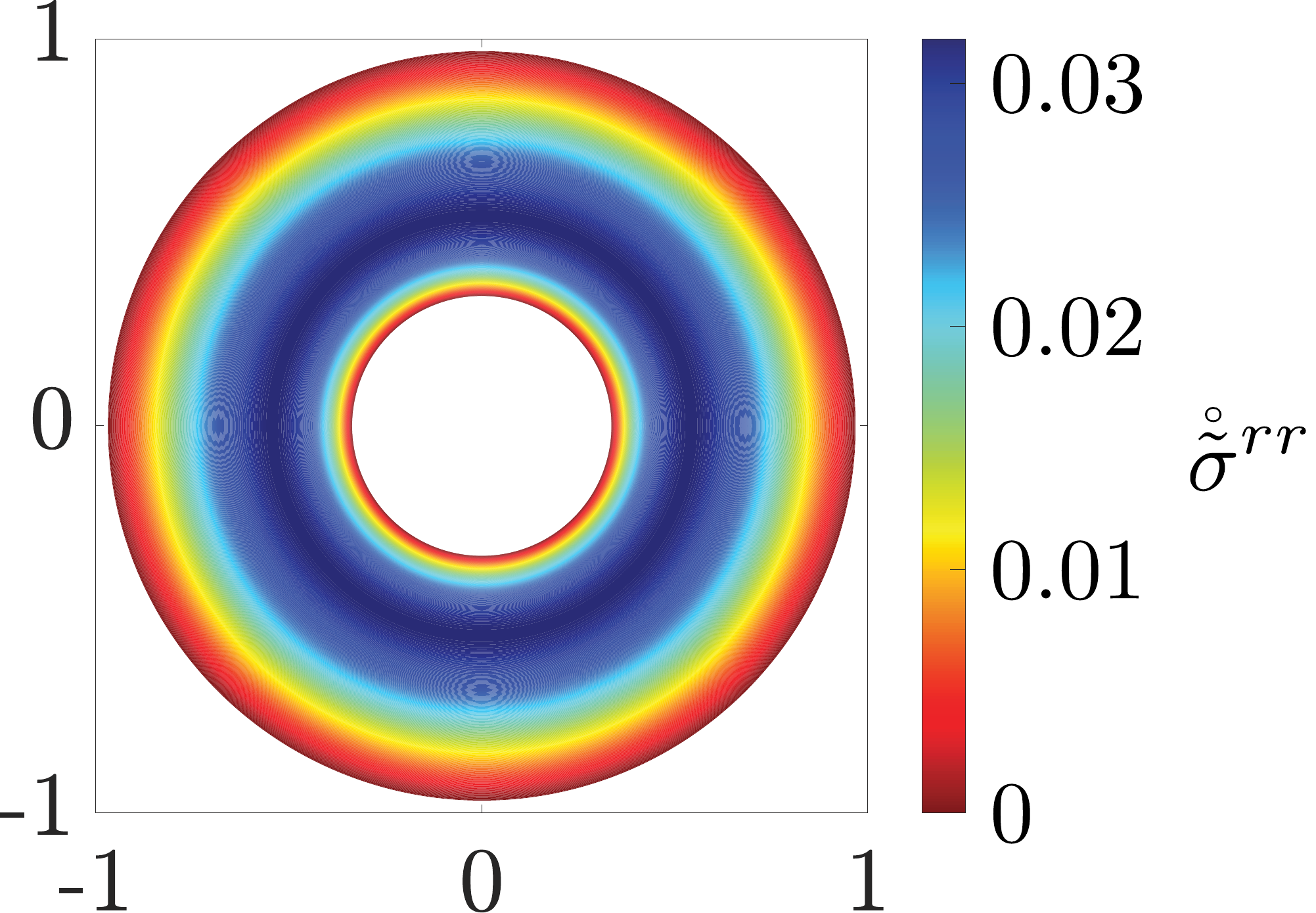}
   \vskip 0.16cm
  \caption{The residual stress $\mathring{\tilde{\sigma}}^{rr}$ in the body solidified till $\mathtt{R}_\mathtt{e}=0.4$ is illustrated as a color plot in the deformed configuration.}
  \label{subfig:sigmarr_color}
\end{subfigure}
%-----
\hfill
%-----
\begin{subfigure}{0.48\textwidth}
\centering
\hspace{36pt}
  \includegraphics[width=0.65\linewidth]{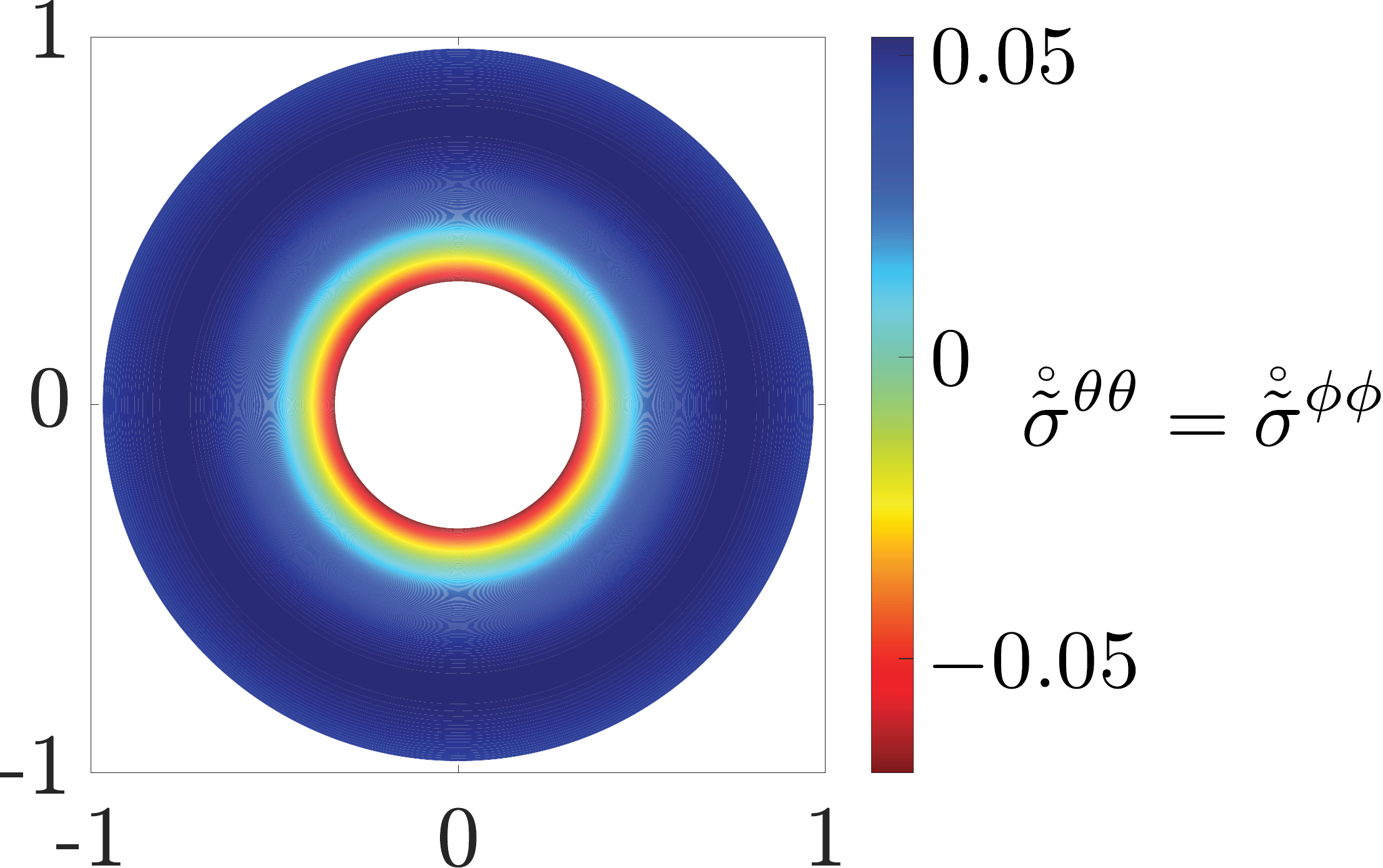}
   \vskip 0.16cm
  \caption{The residual stress $\mathring{\tilde{\sigma}}^{\theta\theta}=\mathring{\tilde{\sigma}}^{\phi\phi}$ in the body solidified till $\mathtt{R}_\mathtt{e}=0.4$ is illustrated as a color plot in the deformed configuration.}
  \label{subfig:sigmatt_color}
\end{subfigure}
%-----
\vspace{0.4cm}
\caption{The radial displacement $\tilde{\mathtt{r}}-\mathtt{R}$ and the physical components $\mathring{\tilde{\sigma}}^{rr}$, $\mathring{\tilde{\sigma}}^{\theta\theta}=\mathring{\tilde{\sigma}}^{\phi\phi}$ of the residual Cauchy stress in a body obtained by accretion till $\mathtt{R}_\mathtt{e}$ are shown here, after detaching it from the rigid cold walls, emptying the remaining liquid and cooling the accreted portion to a uniform steady-state temperature $\mathtt{T}_\mathtt{a}$. These solutions are based on the boundary-value problem \eqref{eq:residualradialEquilibium}-\eqref{eq:residualBC}, assuming $\mathsf{f}=0.95$, $\mathsf{a}=0.8$, $\mathsf{b}=0.1$, $\mathsf{p}=1.1$, $\mathsf{q}=1.2$, $\mathsf{h}=0.5$, $\mathsf{L}=10$ and $\mathtt{T}_\mathtt{a}=0.5$, post non-dimensionalization. The stopping times corresponding to material locations $\mathtt{R}_\mathtt{e}=0.6, 0.5$ and $0.4$ of the solidification interface at the end of accretion are $\mathtt{t}_\mathtt{e}=5.7663, 6.5943$ and $7.2242$, respectively.}
%-----
\label{fig:residualstress}
%-----
\end{figure}
%-----------------------------------------------------------
%-----------------------------------------------------------

%--------------------------------------------------------------
%--------------------------------------------------------------
\section{Conclusions} \label{Sec:Conclusions}

In this paper, the process of liquid-to-solid phase change was modeled as a thermoelastic accretion problem. Several simplifying assumptions were made, such as neglecting inertial effects in both phases, assuming the melting temperature to be independent of pressure (hydrostatic stress in the liquid), ignoring surface stresses, and assuming that the thermal conductivity and heat capacity of the solid are temperature-independent. Since the primary focus was on studying the solidification of a liquid inclusion, the liquid was assumed to be a compressible hyperelastic material, allowing deformation. The problem of determining the reference configuration as the solid portion of a deformable body grows by accretion has the following challenging aspects: first, determining the set of material points that are part of the solid (i.e., knowledge of the boundary location); second, determining the material metric at each point. The material metric depends on the state of deformation of the solidifying material during attachment and on the temperature evolution to account for the effects of thermal expansion. The boundary location, or the set of material points included in the solid at a given instant of time, is determined by the mass rate of solidification, which depends on the jump of the heat flux across the moving interface. Thus, this is a coupled nonlinear problem where the location of the boundary is an unknown, in addition to the deformation and temperature fields.

As a concrete example, the radially inward solidification of a liquid initially at the melting temperature was studied. The resulting moving boundary problem was numerically solved by treating the time of attachment map as an unknown, instead of the boundary location. In other words, for a given space grid, the time instances when the moving boundary crosses these grid points were calculated. This formulation enables one to study the deformation and stresses at any point inside the solid at any desired time, thus potentially highlighting critical zones prone to failures and instabilities. However, the solidification process is halted with a margin prior to completion due to multiple reasons. The numerical results become less accurate as one approaches the center, and they are also physically irrelevant as surface stresses, which become dominant for smaller inclusion sizes, are not considered in the formulation.  A detailed parametric study was performed by varying all the dimensionless constants. In all the numerical examples, the solid was assumed to be less dense than the liquid near the melting point, commonly observed in water and some polymers, though rare in metals. This assumption is essential to avoid cavitation inside the liquid in the context of a solidifying inclusion. However, even with this assumption of denser liquids, our numerical results show that cavitation might be possible in the case of extreme thermal contraction in the solid. The accreted body---once it is detached from the rigid container, drained of the remaining unsolidified liquid, and cooled to an ambient temperature---is residually-stressed, in general. The residually-stressed configuration and its residual stresses were computed numerically.
%, envisioning their crucial role in the identification of zones prone to cracks and instabilities.

The present study opens up several new avenues for future investigations.
The zero-displacement boundary condition imposes a volumetric constraint, potentially causing cavitation for materials denser in the solid phase near the melting point. This results from volume changes due to jump in density across phases. Replacing it with, for instance, an applied pressure on the outer boundary may enable one to study solidification across a broader range of materials.
Developing a variational theory to study instabilities in both solids and liquids during a solidification process is another extension. It is crucial to include surface stresses in the theory for a more realistic understanding of the physics and mechanics of solidification in simple, computationally feasible problems. Ultimately, a more general computational framework needs to be developed to fully utilize the benefits of this formulation for more complex geometries in real-world applications.

%---------------------------
%------------------------------
\section*{Acknowledgement}

We benefited from discussions with Aditya Kumar and Souhayl Sadik. This work was supported by NSF -- Grant No. CMMI~1939901.

%--------------------------------------
%-----------------------------------------
%\bibliographystyle{apa}
\bibliographystyle{abbrvnat}
\bibliography{ref}

\begin{thebibliography}{121}
\providecommand{\natexlab}[1]{#1}
\providecommand{\url}[1]{\texttt{#1}}
\expandafter\ifx\csname urlstyle\endcsname\relax
  \providecommand{\doi}[1]{doi: #1}\else
  \providecommand{\doi}{doi: \begingroup \urlstyle{rm}\Url}\fi

\bibitem[Abeyaratne and Knowles(1993)]{abeyaratne1993continuum}
R.~Abeyaratne and J.~K. Knowles.
\newblock A continuum model of a thermoelastic solid capable of undergoing
  phase transitions.
\newblock \emph{Journal of the Mechanics and Physics of Solids}, 41\penalty0
  (3):\penalty0 541--571, 1993.

\bibitem[Abeyaratne et~al.(2022{\natexlab{a}})Abeyaratne, Puntel, Recrosi, and
  Tomassetti]{abeyaratne2022surface}
R.~Abeyaratne, E.~Puntel, F.~Recrosi, and G.~Tomassetti.
\newblock Surface accretion of a pre-stretched half-space: Biot’s problem
  revisited.
\newblock \emph{Journal of the Mechanics and Physics of Solids}, 167:\penalty0
  104958, 2022{\natexlab{a}}.

\bibitem[Abeyaratne et~al.(2022{\natexlab{b}})Abeyaratne, Puntel, and
  Tomassetti]{abeyaratne2022stability}
R.~Abeyaratne, E.~Puntel, and G.~Tomassetti.
\newblock On the stability of surface growth: The effect of a compliant
  surrounding medium.
\newblock \emph{Journal of Elasticity}, pages 1--25, 2022{\natexlab{b}}.

\bibitem[Ateshian and Shim(2022)]{ateshian2022continuum}
G.~A. Ateshian and J.~J. Shim.
\newblock Continuum thermodynamics of the phase transformation of thermoelastic
  fluids.
\newblock \emph{arXiv preprint arXiv:2207.14158}, 2022.

\bibitem[Bachmann et~al.(2021)Bachmann, Obst, Knorr, Schm{\"o}lzer, Fruhmann,
  Witt, Osswald, Wudy, and Hinrichsen]{bachmann2021cavity}
J.~Bachmann, P.~Obst, L.~Knorr, S.~Schm{\"o}lzer, G.~Fruhmann, G.~Witt,
  T.~Osswald, K.~Wudy, and O.~Hinrichsen.
\newblock Cavity vat photopolymerisation for additive manufacturing of
  polymer-composite 3d objects.
\newblock \emph{Communications Materials}, 2\penalty0 (1):\penalty0 107, 2021.

\bibitem[Ba{\v{z}}ant et~al.(1997)Ba{\v{z}}ant, Hauggaard, Baweja, and
  Ulm]{bavzant1997microprestress}
Z.~P. Ba{\v{z}}ant, A.~B. Hauggaard, S.~Baweja, and F.-J. Ulm.
\newblock Microprestress-solidification theory for concrete creep. i: Aging and
  drying effects.
\newblock \emph{Journal of Engineering Mechanics}, 123\penalty0 (11):\penalty0
  1188--1194, 1997.

\bibitem[Bico et~al.(2018)Bico, Reyssat, and Roman]{Bico2018}
J.~Bico, {\'E}.~Reyssat, and B.~Roman.
\newblock Elastocapillarity: When surface tension deforms elastic solids.
\newblock \emph{Annual Review of Fluid Mechanics}, 50:\penalty0 629--659, 2018.

\bibitem[Binder(1987)]{binder1987theory}
K.~Binder.
\newblock Theory of first-order phase transitions.
\newblock \emph{Reports on Progress in Physics}, 50\penalty0 (7):\penalty0 783,
  1987.

\bibitem[Budak et~al.(1965)Budak, Vasil’ev, and
  Uspenskii]{budak1965difference}
B.~Budak, F.~Vasil’ev, and A.~Uspenskii.
\newblock Difference methods for solving certain boundary value problems of
  stefan type.
\newblock \emph{Chislennye metody v gazovoi dinamike}, pages 139--183, 1965.

\bibitem[Buffett et~al.(1992)Buffett, Huppert, Lister, and
  Woods]{buffett1992analytical}
B.~A. Buffett, H.~E. Huppert, J.~R. Lister, and A.~W. Woods.
\newblock Analytical model for solidification of the earth's core.
\newblock \emph{Nature}, 356\penalty0 (6367):\penalty0 329--331, 1992.

\bibitem[Buffett et~al.(1993)Buffett, Mathews, Herring, and
  Shapiro]{buffett1993forced}
B.~A. Buffett, P.~Mathews, T.~Herring, and I.~Shapiro.
\newblock Forced nutations of the earth: contributions prom the effects of
  ellipticity and rotation on the elastic deformations.
\newblock \emph{Journal of Geophysical Research: Solid Earth}, 98\penalty0
  (B12):\penalty0 21659--21676, 1993.

\bibitem[Buffett et~al.(1996)Buffett, Huppert, Lister, and
  Woods]{buffett1996thermal}
B.~A. Buffett, H.~E. Huppert, J.~R. Lister, and A.~W. Woods.
\newblock On the thermal evolution of the earth's core.
\newblock \emph{Journal of Geophysical Research: Solid Earth}, 101\penalty0
  (B4):\penalty0 7989--8006, 1996.

\bibitem[Caffarelli and Evans(1983)]{caffarelli1983continuity}
L.~A. Caffarelli and L.~C. Evans.
\newblock Continuity of the temperature in the two-phase stefan problem.
\newblock \emph{Archive for Rational Mechanics and Analysis}, 81\penalty0
  (3):\penalty0 199--220, 1983.

\bibitem[Carslaw and Jaeger(1959)]{Carslaw1959}
H.~Carslaw and J.~Jaeger.
\newblock \emph{Conduction of Heat in Solids}.
\newblock Clarendon Press, Oxford, 1959.

\bibitem[Chambre(1956)]{chambre1956dynamics}
P.~Chambre.
\newblock On the dynamics of phase growth.
\newblock \emph{The Quarterly Journal of Mechanics and Applied Mathematics},
  9\penalty0 (2):\penalty0 224--233, 1956.

\bibitem[Chan and Tan(2006)]{chan2006solidification}
C.~Chan and F.~Tan.
\newblock Solidification inside a sphere—{A}n experimental study.
\newblock \emph{International Communications in Heat and Mass Transfer},
  33\penalty0 (3):\penalty0 335--341, 2006.

\bibitem[Coussy(2005)]{coussy2005poromechanics}
O.~Coussy.
\newblock Poromechanics of freezing materials.
\newblock \emph{Journal of the Mechanics and Physics of Solids}, 53\penalty0
  (8):\penalty0 1689--1718, 2005.

\bibitem[Crank(1984)]{crank1984free}
J.~Crank.
\newblock \emph{Free and Moving Boundary Problems}.
\newblock Oxford University Press, 1984.

\bibitem[Crank and Gupta(1972)]{crank1972moving}
J.~Crank and R.~S. Gupta.
\newblock A moving boundary problem arising from the diffusion of oxygen in
  absorbing tissue.
\newblock \emph{IMA Journal of Applied Mathematics}, 10\penalty0 (1):\penalty0
  19--33, 1972.

\bibitem[Danilyuk(1985)]{danilyuk1985stefan}
I.~I. Danilyuk.
\newblock On the stefan problem.
\newblock \emph{Russian Mathematical Surveys}, 40\penalty0 (5):\penalty0 157,
  1985.

\bibitem[DebRoy et~al.(2018)DebRoy, Wei, Zuback, Mukherjee, Elmer, Milewski,
  Beese, Wilson-Heid, De, and Zhang]{debroy2018additive}
T.~DebRoy, H.~Wei, J.~Zuback, T.~Mukherjee, J.~Elmer, J.~Milewski, A.~M. Beese,
  A.~d. Wilson-Heid, A.~De, and W.~Zhang.
\newblock Additive manufacturing of metallic components--process, structure and
  properties.
\newblock \emph{Progress in Materials Science}, 92:\penalty0 112--224, 2018.

\bibitem[Deore et~al.(2021)Deore, Sampson, Lacelle, Kredentser, Lefebvre,
  Young, Hyland, Amaya, Tanha, Malenfant, et~al.]{deore2021direct}
B.~Deore, K.~L. Sampson, T.~Lacelle, N.~Kredentser, J.~Lefebvre, L.~S. Young,
  J.~Hyland, R.~E. Amaya, J.~Tanha, P.~R. Malenfant, et~al.
\newblock Direct printing of functional 3d objects using polymerization-induced
  phase separation.
\newblock \emph{Nature Communications}, 12\penalty0 (1):\penalty0 55, 2021.

\bibitem[Douglas(1957)]{douglas1957uniqueness}
J.~Douglas.
\newblock A uniqueness theorem for the solution of a stefan problem.
\newblock \emph{Proceedings of the American Mathematical Society}, 8\penalty0
  (2):\penalty0 402--408, 1957.

\bibitem[Douglas and Gallie(1955)]{douglas1955numerical}
J.~Douglas, Jr and T.~Gallie, Jr.
\newblock On the numerical integration of a parabolic differential equation
  subject to a moving boundary condition.
\newblock \emph{Duke Mathematical Journal}, 22:\penalty0 557--571, 1955.

\bibitem[Doyle and Ericksen(1956)]{DoyleEricksen1956}
T.~C. Doyle and J.~L. Ericksen.
\newblock Nonlinear elasticity.
\newblock \emph{Advances in Applied Mechanics}, 4:\penalty0 53--115, 1956.

\bibitem[Eckart(1948)]{Eckart1948}
C.~Eckart.
\newblock The thermodynamics of irreversible processes. {IV}. {T}he theory of
  elasticity and anelasticity.
\newblock \emph{Physical Review}, 73\penalty0 (4):\penalty0 373--382, 1948.

\bibitem[Elahinia et~al.(2016)Elahinia, Moghaddam, Andani, Amerinatanzi,
  Bimber, and Hamilton]{elahinia2016fabrication}
M.~Elahinia, N.~S. Moghaddam, M.~T. Andani, A.~Amerinatanzi, B.~A. Bimber, and
  R.~F. Hamilton.
\newblock Fabrication of niti through additive manufacturing: A review.
\newblock \emph{Progress in Materials Science}, 83:\penalty0 630--663, 2016.

\bibitem[Evans et~al.(1951)]{evans1951note}
G.~W. Evans et~al.
\newblock A note on the existence of a solution to a problem of stefan.
\newblock \emph{Quarterly of Applied Mathematics}, 9\penalty0 (2):\penalty0
  185--193, 1951.

\bibitem[Fasano and Primicerio(1979)]{fasano1979free}
A.~Fasano and M.~Primicerio.
\newblock Free boundary problems for nonlinear parabolic equations with
  nonlinear free boundary conditions.
\newblock \emph{Journal of Mathematical Analysis and Applications}, 72\penalty0
  (1):\penalty0 247--273, 1979.

\bibitem[Fedosejevs and Schneider(2022)]{fedosejevs2022sharp}
C.~S. Fedosejevs and M.~F. Schneider.
\newblock Sharp, localized phase transitions in single neuronal cells.
\newblock \emph{Proceedings of the National Academy of Sciences}, 119\penalty0
  (8):\penalty0 e2117521119, 2022.

\bibitem[Fekry(2023)]{fekry2023thermal}
M.~Fekry.
\newblock Thermal stresses in growing thermoviscoelastic cylinder and their
  evolution in the course of selective laser melting processing.
\newblock \emph{Zeitschrift f{\"u}r Angewandte Mathematik und Mechanik},
  103\penalty0 (2):\penalty0 e202100519, 2023.

\bibitem[Ghosh and Lopez-Pamies(2022)]{ghosh2022elastomers}
K.~Ghosh and O.~Lopez-Pamies.
\newblock Elastomers filled with liquid inclusions: Theory, numerical
  implementation, and some basic results.
\newblock \emph{Journal of the Mechanics and Physics of Solids}, 166:\penalty0
  104930, 2022.

\bibitem[Gough(1805)]{gough1805description}
J.~Gough.
\newblock A description of a property of caoutchouc, or indian rubber.
\newblock \emph{Memories of the Literacy and Philosophical Society of
  Manchester}, 1:\penalty0 288--295, 1805.

\bibitem[Gupta(2017)]{gupta2017classical}
S.~C. Gupta.
\newblock \emph{The Classical Stefan Problem: Basic Concepts, Modelling and
  Analysis with Quasi-Analytical Solutions and Methods}, volume~45.
\newblock Elsevier, 2017.

\bibitem[Gurtin et~al.(2010)Gurtin, Fried, and Anand]{gurtin2010mechanics}
M.~E. Gurtin, E.~Fried, and L.~Anand.
\newblock \emph{The Mechanics and Thermodynamics of Continua}.
\newblock Cambridge University Press, 2010.

\bibitem[Heinlein et~al.(1986)Heinlein, Mukherjee, and
  Richmond]{heinlein1986boundary}
M.~Heinlein, S.~Mukherjee, and O.~Richmond.
\newblock A boundary element method analysis of temperature fields and stresses
  during solidification.
\newblock \emph{Acta Mechanica}, 59\penalty0 (1-2):\penalty0 59--81, 1986.

\bibitem[Hill and Kucera(1983)]{hill1983freezing}
J.~M. Hill and A.~Kucera.
\newblock Freezing a saturated liquid inside a sphere.
\newblock \emph{International Journal of Heat and Mass Transfer}, 26\penalty0
  (11):\penalty0 1631--1637, 1983.

\bibitem[Holzapfel(2002)]{holzapfel2002nonlinear}
G.~A. Holzapfel.
\newblock \emph{Nonlinear Solid Mechanics: A Continuum Approach for Engineering
  Science}.
\newblock Kluwer Academic Publishers Dordrecht, 2002.

\bibitem[Holzapfel and Simo(1996)]{holzapfel1996entropy}
G.~A. Holzapfel and J.~Simo.
\newblock Entropy elasticity of isotropic rubber-like solids at finite strains.
\newblock \emph{Computer Methods in Applied Mechanics and Engineering},
  132\penalty0 (1-2):\penalty0 17--44, 1996.

\bibitem[Horvay(1962)]{horvay1962freezing}
G.~Horvay.
\newblock Freezing into an undercooled melt accompanied by density change.
\newblock In \emph{Proc. of the 4 US National Congres of Appl. Mech., Univ.
  California}, pages 1315--1325, 1962.

\bibitem[Isayev and Crouthamel(1984)]{isayev1984residual}
A.~Isayev and D.~Crouthamel.
\newblock Residual stress development in the injection molding of polymers.
\newblock \emph{Polymer-Plastics Technology and Engineering}, 22\penalty0
  (2):\penalty0 177--232, 1984.

\bibitem[Jaeger(1998)]{jaeger1998ehrenfest}
G.~Jaeger.
\newblock The {E}hrenfest classification of phase transitions: {I}ntroduction
  and evolution.
\newblock \emph{Archive for History of Exact Sciences}, 53:\penalty0 51--81,
  1998.

\bibitem[Jiang et~al.(2015)Jiang, Wang, Huang, He, Cui, Zhu, and
  Zheng]{jiang2015phase}
H.~Jiang, S.~Wang, Y.~Huang, X.~He, H.~Cui, X.~Zhu, and Y.~Zheng.
\newblock Phase transition of spindle-associated protein regulate spindle
  apparatus assembly.
\newblock \emph{Cell}, 163\penalty0 (1):\penalty0 108--122, 2015.

\bibitem[Joule(1859)]{joule1859v}
J.~P. Joule.
\newblock On some thermo-dynamic properties of solids.
\newblock \emph{Philosophical Transactions of the Royal Society of London},
  \penalty0 (149):\penalty0 91--131, 1859.

\bibitem[Klingbeil et~al.(2002)Klingbeil, Beuth, Chin, and
  Amon]{klingbeil2002residual}
N.~W. Klingbeil, J.~L. Beuth, R.~Chin, and C.~Amon.
\newblock Residual stress-induced warping in direct metal solid freeform
  fabrication.
\newblock \emph{International Journal of Mechanical Sciences}, 44\penalty0
  (1):\penalty0 57--77, 2002.

\bibitem[Kondo(1949)]{Kondo1949}
K.~Kondo.
\newblock A proposal of a new theory concerning the yielding of materials based
  on {R}iemannian geometry.
\newblock \emph{The Journal of the Japan Society of Aeronautical Engineering},
  2\penalty0 (8):\penalty0 29--31, 1949.

\bibitem[Kou(2015)]{kou2015criterion}
S.~Kou.
\newblock A criterion for cracking during solidification.
\newblock \emph{Acta Materialia}, 88:\penalty0 366--374, 2015.

\bibitem[Kumar et~al.(2021)Kumar, Gao, and Geubelle]{kumar2021analytical}
A.~Kumar, Y.~Gao, and P.~H. Geubelle.
\newblock Analytical estimates of front velocity in the frontal polymerization
  of thermoset polymers and composites.
\newblock \emph{Journal of Polymer Science}, 59\penalty0 (11):\penalty0
  1109--1118, 2021.

\bibitem[Kumar et~al.(2022)Kumar, Dean, Yourdkhani, Guo, BenVau, Sottos, and
  Geubelle]{kumar2022surface}
A.~Kumar, L.~M. Dean, M.~Yourdkhani, A.~Guo, C.~BenVau, N.~R. Sottos, and P.~H.
  Geubelle.
\newblock Surface pattern formation induced by oscillatory loading of frontally
  polymerized gels.
\newblock \emph{Journal of the Mechanics and Physics of Solids}, 168:\penalty0
  105055, 2022.

\bibitem[Labrosse et~al.(1997)Labrosse, Poirier, and
  Le~Mou{\"e}l]{labrosse1997cooling}
S.~Labrosse, J.-P. Poirier, and J.-L. Le~Mou{\"e}l.
\newblock On cooling of the earth's core.
\newblock \emph{Physics of the Earth and Planetary Interiors}, 99\penalty0
  (1-2):\penalty0 1--17, 1997.

\bibitem[Labrosse et~al.(2007)Labrosse, Hernlund, and
  Coltice]{labrosse2007crystallizing}
S.~Labrosse, J.~Hernlund, and N.~Coltice.
\newblock A crystallizing dense magma ocean at the base of the earth’s
  mantle.
\newblock \emph{Nature}, 450\penalty0 (7171):\penalty0 866--869, 2007.

\bibitem[Lam{\'e} and Clapeyron(1831)]{lame1831memoire}
G.~Lam{\'e} and B.~Clapeyron.
\newblock M{\'e}moire sur la solidification par refroidissement d’un globe
  liquide.
\newblock In \emph{Annales Chimie Physique}, volume~47, pages 250--256, 1831.

\bibitem[Landau(1936)]{landau1936theory}
L.~Landau.
\newblock The theory of phase transitions.
\newblock \emph{Nature}, 138\penalty0 (3498):\penalty0 840--841, 1936.

\bibitem[Li and Cohen(2024)]{li2024mechanical}
X.~Li and T.~Cohen.
\newblock Mechanical forces quench frontal polymerization: Experiments and
  theory.
\newblock \emph{Journal of the Mechanics and Physics of Solids}, 183:\penalty0
  105517, 2024.

\bibitem[London and Seban(1943)]{london1943rate}
A.~London and R.~Seban.
\newblock Rate of ice formation.
\newblock \emph{Transactions of the American Society of Mechanical Engineers},
  65\penalty0 (7):\penalty0 771--778, 1943.

\bibitem[Lotkin(1960)]{lotkin1960calculation}
M.~Lotkin.
\newblock The calculation of heat flow in melting solids.
\newblock \emph{Quarterly of Applied Mathematics}, 18\penalty0 (1):\penalty0
  79--85, 1960.

\bibitem[Lubarda(2004)]{Lubarda2004}
V.~A. Lubarda.
\newblock Constitutive theories based on the multiplicative decomposition of
  deformation gradient: Thermoelasticity, elastoplasticity, and biomechanics.
\newblock \emph{Applied Mechanics Reviews}, 57\penalty0 (2):\penalty0 95--108,
  2004.

\bibitem[Lychev and Fekry(2023{\natexlab{a}})]{lychev2023evaluation1}
S.~A. Lychev and M.~Fekry.
\newblock Evaluation of residual stresses in additively produced thermoelastic
  cylinder. part i. thermal fields.
\newblock \emph{Mechanics of Advanced Materials and Structures}, 30\penalty0
  (10):\penalty0 1975--1990, 2023{\natexlab{a}}.

\bibitem[Lychev and Fekry(2023{\natexlab{b}})]{lychev2023evaluation2}
S.~A. Lychev and M.~Fekry.
\newblock Evaluation of residual stresses in additively produced thermoelastic
  cylinder. part ii. residual stresses.
\newblock \emph{Mechanics of Advanced Materials and Structures}, 30\penalty0
  (10):\penalty0 1991--2000, 2023{\natexlab{b}}.

\bibitem[Marsden and Hughes(1983)]{MaHu1983}
J.~Marsden and T.~Hughes.
\newblock \emph{Mathematical Foundations of Elasticity}.
\newblock Dover, 1983.

\bibitem[Mazur(1970)]{mazur1970cryobiology}
P.~Mazur.
\newblock Cryobiology: The freezing of biological systems: The responses of
  living cells to ice formation are of theoretical interest and practical
  concern.
\newblock \emph{Science}, 168\penalty0 (3934):\penalty0 939--949, 1970.

\bibitem[McCue et~al.(2003)McCue, King, and Riley]{mccue2003extinction}
S.~W. McCue, J.~R. King, and D.~S. Riley.
\newblock Extinction behaviour for two--dimensional inward-solidification
  problems.
\newblock \emph{Proceedings of the Royal Society of London A}, 459\penalty0
  (2032):\penalty0 977--999, 2003.

\bibitem[Melamed(1958)]{melamed1958reduction}
V.~Melamed.
\newblock Reduction of the stefan problem to a system of ordinary differential
  equations.
\newblock \emph{Izv. Ac. Sci. USSR, geophys, set}, \penalty0 (7):\penalty0
  848--869, 1958.

\bibitem[Mercelis and Kruth(2006)]{mercelis2006residual}
P.~Mercelis and J.-P. Kruth.
\newblock Residual stresses in selective laser sintering and selective laser
  melting.
\newblock \emph{Rapid Prototyping Journal}, 12\penalty0 (5):\penalty0 254--265,
  2006.

\bibitem[Mukherjee et~al.(2017{\natexlab{a}})Mukherjee, Manvatkar, De, and
  DebRoy]{mukherjee2017mitigation}
T.~Mukherjee, V.~Manvatkar, A.~De, and T.~DebRoy.
\newblock Mitigation of thermal distortion during additive manufacturing.
\newblock \emph{Scripta Materialia}, 127:\penalty0 79--83, 2017{\natexlab{a}}.

\bibitem[Mukherjee et~al.(2017{\natexlab{b}})Mukherjee, Zhang, and
  DebRoy]{mukherjee2017improved}
T.~Mukherjee, W.~Zhang, and T.~DebRoy.
\newblock An improved prediction of residual stresses and distortion in
  additive manufacturing.
\newblock \emph{Computational Materials Science}, 126:\penalty0 360--372,
  2017{\natexlab{b}}.

\bibitem[Ogden(1992)]{ogden1992thermoelastic}
R.~Ogden.
\newblock On the thermoelastic modeling of rubberlike solids.
\newblock \emph{Journal of Thermal Stresses}, 15\penalty0 (4):\penalty0
  533--557, 1992.

\bibitem[Ole\u{\i}nik(1960)]{MR0125341}
O.~A. Ole\u{\i}nik.
\newblock A method of solution of the general stefan problem.
\newblock In \emph{Doklady Akademii Nauk}, volume 135, pages 1054--1057.
  Russian Academy of Sciences, 1960.

\bibitem[O'Neill(1983)]{o1983boundary}
K.~O'Neill.
\newblock Boundary integral equation solution of moving boundary phase change
  problems.
\newblock \emph{International Journal for Numerical Methods in Engineering},
  19\penalty0 (12):\penalty0 1825--1850, 1983.

\bibitem[Ozakin and Yavari(2010)]{ozakin2010geometric}
A.~Ozakin and A.~Yavari.
\newblock A geometric theory of thermal stresses.
\newblock \emph{Journal of Mathematical Physics}, 51:\penalty0 032902, 2010.

\bibitem[Pedroso and Domoto(1973{\natexlab{a}})]{pedroso1973inward}
R.~Pedroso and G.~Domoto.
\newblock Inward spherical solidification—solution by the method of strained
  coordinates.
\newblock \emph{International Journal of Heat and Mass Transfer}, 16\penalty0
  (5):\penalty0 1037--1043, 1973{\natexlab{a}}.

\bibitem[Pedroso and Domoto(1973{\natexlab{b}})]{pedroso1973state}
R.~Pedroso and G.~Domoto.
\newblock State of stress during solidification with varying freezing pressure
  and temperature.
\newblock \emph{Journal of Engineering Materials and Technology}, 95\penalty0
  (4):\penalty0 227--232, 1973{\natexlab{b}}.

\bibitem[Pielichowska and Pielichowski(2014)]{pielichowska2014phase}
K.~Pielichowska and K.~Pielichowski.
\newblock Phase change materials for thermal energy storage.
\newblock \emph{Progress in Materials Science}, 65:\penalty0 67--123, 2014.

\bibitem[Podio-Guidugli et~al.(1985)Podio-Guidugli, Vergara~Caffarelli, and
  Virga]{podio1985cavitation}
P.~Podio-Guidugli, G.~Vergara~Caffarelli, and E.~Virga.
\newblock Cavitation and phase transition of hyperelastic fluids.
\newblock In \emph{Analysis and Thermomechanics: A Collection of Papers
  Dedicated to W. Noll on His Sixtieth Birthday}, pages 401--416. Springer,
  1985.

\bibitem[Pradhan and Yavari(2023)]{pradhan2023accretion}
S.~P. Pradhan and A.~Yavari.
\newblock Accretion-ablation mechanics.
\newblock \emph{Philosophical Transactions of the Royal Society A}, 20220373,
  2023.

\bibitem[Rabin and Steif(1998)]{rabin1998thermal}
Y.~Rabin and P.~Steif.
\newblock Thermal stresses in a freezing sphere and its application to
  cryobiology.
\newblock \emph{Journal of Applied Mechanics}, 65\penalty0 (2):\penalty0
  328--333, 1998.

\bibitem[Rejovitzky et~al.(2015)Rejovitzky, Di~Leo, and
  Anand]{rejovitzky2015theory}
E.~Rejovitzky, C.~V. Di~Leo, and L.~Anand.
\newblock A theory and a simulation capability for the growth of a solid
  electrolyte interphase layer at an anode particle in a li-ion battery.
\newblock \emph{Journal of the Mechanics and Physics of Solids}, 78:\penalty0
  210--230, 2015.

\bibitem[Richmond and Tien(1971)]{richmond1971theory}
O.~Richmond and R.~Tien.
\newblock Theory of thermal stresses and air-gap formation during the early
  stages of solidification in a rectangular mold.
\newblock \emph{Journal of the Mechanics and Physics of Solids}, 19\penalty0
  (5):\penalty0 273--284, 1971.

\bibitem[Riley et~al.(1974)Riley, Smith, and Poots]{riley1974inward}
D.~Riley, F.~Smith, and G.~Poots.
\newblock The inward solidification of spheres and circular cylinders.
\newblock \emph{International Journal of Heat and Mass Transfer}, 17\penalty0
  (12):\penalty0 1507--1516, 1974.

\bibitem[Rongved(1954)]{rongved1954residual}
L.~Rongved.
\newblock \emph{Residual stress in glass spheres.}
\newblock PhD thesis, Columbia University., 1954.

\bibitem[Rubinsky et~al.(1980)Rubinsky, Cravalho, and
  Mikic]{rubinsky1980thermal}
B.~Rubinsky, E.~G. Cravalho, and B.~Mikic.
\newblock Thermal stresses in frozen organs.
\newblock \emph{Cryobiology}, 17\penalty0 (1):\penalty0 66--73, 1980.

\bibitem[Rubinstein(1947)]{rubinstein1947solution}
L.~Rubinstein.
\newblock On the solution of stefan's problem.
\newblock \emph{Bull. Acad. Sci. URSS. S{\'e}r. G{\'e}ograph.
  G{\'e}ophys.(Izvestia Akad. Nauk SSSR)}, 11:\penalty0 37--54, 1947.

\bibitem[Rubin{\v{s}}te{\u\i}n(1971)]{rubinshteuin1971stefan}
L.~Rubin{\v{s}}te{\u\i}n.
\newblock \emph{The Stefan Problem}.
\newblock American Mathematical Soc., 1971.

\bibitem[Rubinstein(1979)]{rubinstein1979stefan}
L.~Rubinstein.
\newblock The stefan problem: Comments on its present state.
\newblock \emph{IMA Journal of Applied Mathematics}, 24\penalty0 (3):\penalty0
  259--277, 1979.

\bibitem[Sadik and Yavari(2017{\natexlab{a}})]{Sadik2017}
S.~Sadik and A.~Yavari.
\newblock On the origins of the idea of the multiplicative decomposition of the
  deformation gradient.
\newblock \emph{Mathematics and Mechanics of Solids}, 22\penalty0 (4):\penalty0
  771--772, 2017{\natexlab{a}}.

\bibitem[Sadik and Yavari(2017{\natexlab{b}})]{Sadik2017Thermoelasticity}
S.~Sadik and A.~Yavari.
\newblock Geometric nonlinear thermoelasticity and the time evolution of
  thermal stresses.
\newblock \emph{Mathematics and Mechanics of Solids}, 22\penalty0 (7):\penalty0
  1546--1587, 2017{\natexlab{b}}.

\bibitem[Sadik and Yavari(2024)]{SadikYavari2024visco-anelasticity}
S.~Sadik and A.~Yavari.
\newblock Nonlinear visco-anelasticity.
\newblock 2024.

\bibitem[Shao et~al.(2020)Shao, Hanaor, Shen, and Gurlo]{shao2020freeze}
G.~Shao, D.~A. Hanaor, X.~Shen, and A.~Gurlo.
\newblock Freeze casting: {F}rom low-dimensional building blocks to aligned
  porous structures—a review of novel materials, methods, and applications.
\newblock \emph{Advanced Materials}, 32\penalty0 (17):\penalty0 1907176, 2020.

\bibitem[Shih and Chou(1971)]{shih1971analytical}
Y.-P. Shih and T.-C. Chou.
\newblock Analytical solutions for freezing a saturated liquid inside or
  outside spheres.
\newblock \emph{Chemical Engineering Science}, 26\penalty0 (11):\penalty0
  1787--1793, 1971.

\bibitem[Simo and Marsden(1984)]{SimoMarsden1983}
J.~Simo and J.~Marsden.
\newblock Stress tensors, {R}iemannian metrics and the alternative descriptions
  in elasticity.
\newblock In \emph{Trends and Applications of Pure Mathematics to Mechanics},
  pages 369--383. Springer, 1984.

\bibitem[Smith et~al.(2011)Smith, Burns, Xiong, and
  Dahn]{smith2011interpreting}
A.~Smith, J.~Burns, D.~Xiong, and J.~Dahn.
\newblock Interpreting high precision coulometry results on li-ion cells.
\newblock \emph{Journal of The Electrochemical Society}, 158\penalty0
  (10):\penalty0 A1136, 2011.

\bibitem[Soward(1980)]{soward1980unified}
A.~Soward.
\newblock A unified approach to {S}tefan’s problem for spheres and cylinders.
\newblock \emph{Proceedings of the Royal Society of London A}, 373\penalty0
  (1752):\penalty0 131--147, 1980.

\bibitem[Sozio and Yavari(2017)]{Sozio2017}
F.~Sozio and A.~Yavari.
\newblock Nonlinear mechanics of surface growth for cylindrical and spherical
  elastic bodies.
\newblock \emph{Journal of the Mechanics and Physics of Solids}, 98:\penalty0
  12--48, 2017.

\bibitem[Sozio and Yavari(2019)]{Sozio2019}
F.~Sozio and A.~Yavari.
\newblock Nonlinear mechanics of accretion.
\newblock \emph{Journal of Nonlinear Science}, 29\penalty0 (4):\penalty0
  1813--1863, 2019.

\bibitem[Sozio et~al.(2020)Sozio, Faghih~Shojaei, Sadik, and Yavari]{Sozio2020}
F.~Sozio, M.~Faghih~Shojaei, S.~Sadik, and A.~Yavari.
\newblock Nonlinear mechanics of thermoelastic accretion.
\newblock \emph{Zeitschrift f{\"u}r angewandte Mathematik und Physik},
  71:\penalty0 1--24, 2020.

\bibitem[Stanley(1971)]{stanley1971phase}
H.~E. Stanley.
\newblock \emph{Phase Transitions and Critical Phenomena}, volume~7.
\newblock Clarendon Press, Oxford, 1971.

\bibitem[Stefan(1891)]{stefan1891theorie}
J.~Stefan.
\newblock {\"U}ber die theorie der eisbildung, insbesondere {\"u}ber die
  eisbildung im polarmeere.
\newblock \emph{Annalen der Physik}, 278\penalty0 (2):\penalty0 269--286, 1891.

\bibitem[Stewartson and Waechter(1976)]{stewartson1976stefan}
K.~Stewartson and R.~Waechter.
\newblock On {S}tefan’s problem for spheres.
\newblock \emph{Proceedings of the Royal Society of London A}, 348\penalty0
  (1655):\penalty0 415--426, 1976.

\bibitem[Stojanovi{\'c}(1969)]{Stojanovic1969}
R.~Stojanovi{\'c}.
\newblock On the stress relation in non-linear thermoelasticity.
\newblock \emph{International Journal of Non-Linear Mechanics}, 4\penalty0
  (3):\penalty0 217--233, 1969.

\bibitem[Stojanovi{\'c} et~al.(1964)Stojanovi{\'c}, Djuri{\'c}, and
  Vujo{\v{s}}evi{\'c}]{Stojanovic1964}
R.~Stojanovi{\'c}, S.~Djuri{\'c}, and L.~Vujo{\v{s}}evi{\'c}.
\newblock On finite thermal deformations.
\newblock \emph{Archiwum Mechaniki Stosowanej}, 1\penalty0 (16):\penalty0
  103--108, 1964.

\bibitem[Tao(1967)]{tao1967generalized}
L.~C. Tao.
\newblock Generalized numerical solutions of freezing a saturated liquid in
  cylinders and spheres.
\newblock \emph{AIChE Journal}, 13\penalty0 (1):\penalty0 165--169, 1967.

\bibitem[Tien and Koump(1969)]{tien1969thermal}
R.~Tien and V.~Koump.
\newblock Thermal stresses during solidification on basis of elastic model.
\newblock \emph{Journal of Applied Mechanics}, 36\penalty0 (4):\penalty0
  763--767, 1969.

\bibitem[Tomassetti et~al.(2016)Tomassetti, Cohen, and
  Abeyaratne]{Tomassetti2016}
G.~Tomassetti, T.~Cohen, and R.~Abeyaratne.
\newblock Steady accretion of an elastic body on a hard spherical surface and
  the notion of a four-dimensional reference space.
\newblock \emph{Journal of the Mechanics and Physics of Solids}, 96:\penalty0
  333--352, 2016.

\bibitem[Truesdell and Noll(2004)]{truesdell2004non}
C.~Truesdell and W.~Noll.
\newblock \emph{The Non-Linear Field Theories of Mechanics}.
\newblock Springer, $3^{rd}$ edition, 2004.

\bibitem[Truesdell and Rajagopal(2000)]{truesdell2000introduction}
C.~Truesdell and K.~R. Rajagopal.
\newblock \emph{An Introduction to the Mechanics of Fluids}.
\newblock Springer Science \& Business Media, 2000.

\bibitem[Visintin(2008)]{visintin2008introduction}
A.~Visintin.
\newblock Introduction to {S}tefan-type problems.
\newblock \emph{Handbook of Differential Equations: Evolutionary Equations},
  4:\penalty0 377--484, 2008.

\bibitem[Vuik(1993)]{vuik1993some}
C.~Vuik.
\newblock Some historical notes about the {S}tefan problem.
\newblock Delft University of Technology, Faculty of Technical Mathematics and
  Informatics, 1993.

\bibitem[Wang and Truesdell(1973)]{wang1973introduction}
C.-c. Wang and C.~Truesdell.
\newblock \emph{Introduction to Rational Elasticity}, volume~1.
\newblock Springer Science \& Business Media, 1973.

\bibitem[Weiner and Boley(1963)]{weiner1963elasto}
J.~Weiner and B.~Boley.
\newblock Elasto-plastic thermal stresses in a solidifying body.
\newblock \emph{Journal of the Mechanics and Physics of Solids}, 11\penalty0
  (3):\penalty0 145--154, 1963.

\bibitem[Withers and Bhadeshia(2001{\natexlab{a}})]{withers2001residuala}
P.~J. Withers and H.~Bhadeshia.
\newblock Residual stress. {P}art 1--measurement techniques.
\newblock \emph{Materials Science and Technology}, 17\penalty0 (4):\penalty0
  355--365, 2001{\natexlab{a}}.

\bibitem[Withers and Bhadeshia(2001{\natexlab{b}})]{withers2001residualb}
P.~J. Withers and H.~Bhadeshia.
\newblock Residual stress. {P}art 2--nature and origins.
\newblock \emph{Materials Science and Technology}, 17\penalty0 (4):\penalty0
  366--375, 2001{\natexlab{b}}.

\bibitem[Wuttig et~al.(2017)Wuttig, Bhaskaran, and Taubner]{wuttig2017phase}
M.~Wuttig, H.~Bhaskaran, and T.~Taubner.
\newblock Phase-change materials for non-volatile photonic applications.
\newblock \emph{Nature Photonics}, 11\penalty0 (8):\penalty0 465--476, 2017.

\bibitem[Yang and Zhiwei(2009)]{yang2009injection}
W.~Yang and J.~Zhiwei.
\newblock Injection moulding of polymers.
\newblock \emph{Advances in Polymer Processing}, pages 175--203, 2009.

\bibitem[Yavari(2010)]{Yavari2010Growth}
A.~Yavari.
\newblock A geometric theory of growth mechanics.
\newblock \emph{Journal of Nonlinear Science}, 20\penalty0 (6):\penalty0
  781--830, 2010.

\bibitem[Yavari and Pradhan(2022)]{Yavari2022Torsion}
A.~Yavari and S.~P. Pradhan.
\newblock Accretion mechanics of nonlinear elastic circular cylindrical bars
  under finite torsion.
\newblock \emph{Journal of Elasticity}, pages 1--32, 2022.

\bibitem[Yavari and Sozio(2023)]{YavariSozio2023}
A.~Yavari and F.~Sozio.
\newblock On the direct and reverse multiplicative decompositions of
  deformation gradient in nonlinear anisotropic anelasticity.
\newblock \emph{Journal of the Mechanics and Physics of Solids}, 170:\penalty0
  105101, 2023.

\bibitem[Yavari et~al.(2023)Yavari, Safa, and
  Soleiman~Fallah]{Yavari2023Accretion}
A.~Yavari, Y.~Safa, and A.~Soleiman~Fallah.
\newblock Finite extension of accreting nonlinear elastic solid circular
  cylinders.
\newblock \emph{Continuum Mechanics and Thermodynamics}, pages 1--17, 2023.

\bibitem[Zabaras and Mukherjee(1987)]{zabaras1987analysis}
N.~Zabaras and S.~Mukherjee.
\newblock An analysis of solidification problems by the boundary element
  method.
\newblock \emph{International Journal for Numerical Methods in Engineering},
  24\penalty0 (10):\penalty0 1879--1900, 1987.

\bibitem[Zabaras et~al.(1990)Zabaras, Ruan, and Richmond]{zabaras1990front}
N.~Zabaras, Y.~Ruan, and O.~Richmond.
\newblock Front tracking thermomechanical model for hypoelastic-viscoplastic
  behavior in a solidifying body.
\newblock \emph{Computer Methods in Applied Mechanics and Engineering},
  81\penalty0 (3):\penalty0 333--364, 1990.

\bibitem[Zabaras et~al.(1991)Zabaras, Ruan, and
  Richmond]{zabaras1991calculation}
N.~Zabaras, Y.~Ruan, and O.~Richmond.
\newblock On the calculation of deformations and stresses during axially
  symmetric solidification.
\newblock \emph{Journal of Applied Mechanics}, 58\penalty0 (4):\penalty0
  865--871, 1991.

\bibitem[Zarek et~al.(2016)Zarek, Layani, Cooperstein, Sachyani, Cohn, and
  Magdassi]{zarek20163d}
M.~Zarek, M.~Layani, I.~Cooperstein, E.~Sachyani, D.~Cohn, and S.~Magdassi.
\newblock 3{D} printing of shape memory polymers for flexible electronic
  devices.
\newblock \emph{Advanced Materials}, 28\penalty0 (22):\penalty0 4449--4454,
  2016.

\end{thebibliography}
\appendix

%-----------------------------
%-----------------------------
\section{The first and second laws of thermodynamics and the heat equation} \label{AppendixA}

In this appendix, we derive the material and spatial heat equations from the fundamental laws of thermodynamics.

\paragraph{Material heat equation.} \label{AppendixA1}
Let $\mathcal{E}(X,t)$, $\Psi(X,t)$ and $\mathcal{N}(X,t)$ be, respectively, the specific internal energy, the specific free energy and the specific entropy in the material configuration. 
In thermoelasticity, deformation gradient is multiplicatively decomposed into elastic and thermal parts: $\mathbf{F}=\Fe\Ft$ \citep{Stojanovic1964,Stojanovic1969,Lubarda2004,Sadik2017}. Let us denote the induced Euclidean metric on $\mathcal{B}$ by $\Go=\mathbf{g}|_{\mathcal{B}}$. Internal energy and free energy explicitly depend on the elastic distortion $\Fe$:
%-----------------------------
\begin{equation} 	
	\mathcal{E}=\mathcal{E}(X,\mathcal{N},\Fe,\Go,\mathbf{g})\,,\qquad
	\Psi=\Psi(X,T,\Fe,\Go,\mathbf{g}) \,.
\end{equation}
%-----------------------------
Objectivity implies that
%-----------------------------
\begin{equation} \label{Free-Energy}
	\mathcal{E}=\hat{\mathcal{E}}(X,\mathcal{N},\Ce^\flat,\Go)\,,\qquad
	\Psi=\hat{\Psi}(X,T,\Ce^\flat,\Go) \,,
\end{equation}
%-----------------------------
where $\Ce^\flat=\Ft^*\mathbf{g}=\Ft^\star\mathbf{g}\Ft$. 

Conservation of energy for an arbitrary sub-body $\mathcal{U} \subset \mathcal{B}_t$ is written as\footnote{A term $\rho\frac{\partial \mathcal{E}}{\partial \mathbf{G}}\!:\!\frac{\partial \mathbf{G}}{\partial t}$ was included in the energy balance in \citep{Sadik2017Thermoelasticity}. It turns out that this term should not appear on the right-hand side of the energy balance. A detailed discussion is given in \citep{SadikYavari2024visco-anelasticity}.}
%-----------------------------
\begin{equation} \label{eq:MaterialEnergyConsVol}
	\frac{\text{d}}{\text{d}t} \int_{\mathcal{U} }\rho\left( \mathcal{E}+\frac{1}{2} \llangle \mathbf{V},\mathbf{V} 
	\rrangle_{\mathbf{g}}  \right)\text{d}V
	=\int_{\partial \mathcal{U}}\left( \llangle \mathbf{T},\mathbf{V} \rrangle_{\mathbf{g}} 
	- \llangle \mathbf{H},\mathbf{N} \rrangle_{\mathbf{G}}  \right)\text{d}A
	+\int_{ \mathcal{U}}\left(  \rho\,\llangle \mathbf{B},\mathbf{V} \rrangle_{\mathbf{g}} 
	+ R  \right)\text{d}V \,,
\end{equation}
%-----------------------------
where $\mathbf{T}(X,t)$ is the traction vector, $\mathbf{B}(X,t)$ is the body force (per unit mass) and $R(X,t)$ is a heat source/sink, i.e., $\int_{ \mathcal{U} }R  \text{d}V$ is the rate at which the heat is generated or destroyed in $\mathcal{U}$ \citep{Sadik2017Thermoelasticity,Yavari2010Growth}. Using \eqref{eq:MatCont}, \eqref{eq:MatLinMomCons} and the fact that $\mathbf{P}  \mathbf{N}^\flat =\mathbf{T}$ on $\partial \mathcal{U}$, \eqref{eq:MaterialEnergyConsVol} can be localized to read
%-----------------------------
\begin{equation} \label{eq:MatEnergyLocv1}
	\rho\, \dot{\mathcal{E}}
	= \mathbf{S}\!:\!\mathbf{D}
	-\operatorname{Div} \mathbf{H}    		
	+ R  \,,
\end{equation}
%-----------------------------
where $\mathbf{S}=\mathbf{P} \mathbf{F}^\star$ is the second Piola-Kirchhoff stress and $\mathbf{D}=\frac{1}{2}\dot{\mathbf{C}}^\flat$. The localized Clausius-Duhem inequality reads
%-----------------------------
\begin{equation} 
	  \rho \,\dot{\mathcal{N}} 
	+\operatorname{Div} \! \left( \frac{\mathbf{H}}{T} \right)
	-  \frac{R}{T} \geq 0  \,,
\end{equation}
%-----------------------------
which can be rewritten in terms of the rate of energy dissipation as 
%-----------------------------
\begin{equation} \label{eq:MatClausDuhemLocv1}
	 \dot\eta= \rho  T \dot{\mathcal{N}} 
	+\operatorname{Div} \mathbf{H} 
	- \frac{1}{T} \langle \text{d}T,\mathbf{H}\rangle
	-  R \geq 0  \,.
\end{equation}
%-----------------------------

Note that $\mathbf{C}^\flat=\varphi^*\mathbf{g}=(\Fe\Ft)^\star\mathbf{g}\,\Fe\Ft=\Ft^\star\Fe^\star\mathbf{g}\,\Fe\Ft=\Ft^\star\Ce^\flat\Ft=\Ft^*\Ce^\flat$. 
We assume an isotropic material, which is materially covariant \citep{MaHu1983}.\footnote{Extension of this analysis involves including structural tensors as arguments of the free energy \citep{YavariSozio2023}. In this paper, we restrict our analysis to isotropic materials.} Thus
%-----------------------------
\begin{equation} 
	\hat{\Psi}(X,T,\Ce^\flat,\Go)=\hat{\Psi}(X,T,\Ft^*\Ce^\flat,\Ft^*\Go)
	=\hat{\Psi}(X,T,\mathbf{C}^\flat,\mathbf{G}) \,,
\end{equation}
%-----------------------------
where $\mathbf{G}=\Ft^*\Go\Ft$ is the material metric. Therefore
%-----------------------------
\begin{equation} \label{Free-Energy-Rate}
	\dot\Psi=\frac{\partial\hat\Psi}{\partial T}\dot T
	+\frac{\partial\hat\Psi}{\partial \mathbf{C}^\flat}\!:\!\dot{\mathbf{C}}^\flat
	+\frac{\partial\hat\Psi}{\partial \mathbf{G}}\!:\!\dot{\mathbf{G}}
	=\frac{d \hat\Psi}{d T}\dot T
	+\frac{\partial\hat\Psi}{\partial \mathbf{C}^\flat}\!:\!\dot{\mathbf{C}}^\flat
	\,,
\end{equation}
%-----------------------------
where
%-----------------------------
\begin{equation} 
	\frac{d \hat\Psi}{d T} = \frac{\partial\hat\Psi}{\partial T}
	+\frac{\partial\hat\Psi}{\partial \mathbf{G}}\!:\!\frac{\partial \mathbf{G}}{\partial T} 
	\,.
\end{equation}
%-----------------------------
Recall that free energy, internal energy, entropy and temperature are related as
%-----------------------------
\begin{equation} \label{Energy-Free-Energy}
	\mathcal{E} = \Psi + T \mathcal{N}\,,
\end{equation}
%-----------------------------
which can be thought of as a change of variables from $(T,\mathbf{C}^\flat)$ to $(\mathcal{N},\mathbf{C}^\flat)$. 
Also, notice that $\mathcal{E}=\hat{\mathcal{E}}(X,\mathcal{N},\mathbf{C}^\flat,\mathbf{G})$, $\Psi=\hat{\Psi}(X,T,\mathbf{C}^\flat,\mathbf{G})$, and $\mathcal{N}=\hat{\mathcal{N}}(X,\mathbf{C}^\flat,\mathbf{G})$.
Taking partial derivatives of both sides with respect to $T$, one obtains
%-----------------------------
\begin{equation} 
	0 = \frac{\partial \Psi}{\partial T}  +  \mathcal{N}\,,	
\end{equation}
%-----------------------------
and hence
%-----------------------------
\begin{equation} \label{Entropy-Temperature}
	\mathcal{N} = - \frac{\partial  \Psi}{\partial T}  \,.	
\end{equation}
%-----------------------------
From \eqref{Energy-Free-Energy}, $\rho T \dot{\mathcal{N}}=\rho \dot{\mathcal{E}}-\rho \dot{T} \mathcal{N}-\rho \dot{\Psi}$. Substituting this relation and \eqref{eq:MatEnergyLocv1} into \eqref{eq:MatClausDuhemLocv1}, the rate of energy dissipation is simplified to read
%-----------------------------
\begin{equation} 
	 \dot\eta= \mathbf{S}\!:\!\mathbf{D} -\rho \dot{T} \mathcal{N} - \rho \dot{\Psi}
	- \frac{1}{T} \langle \text{d}T,\mathbf{H}\rangle \geq 0  \,.
\end{equation}
%-----------------------------
Using \eqref{Free-Energy-Rate} the energy dissipation rate is further simplified as
%-----------------------------
\begin{equation} 
\begin{aligned}
	 \dot\eta &= \frac{1}{2}\left[\mathbf{S}-\rho \frac{\partial\hat\Psi}{\partial \mathbf{C}^\flat}\right]
	 \!:\!\dot{\mathbf{C}}^\flat
	 -\rho \dot{T} \left[\mathcal{N} + \frac{d \hat{\Psi}}{d T}  \right]
	- \frac{1}{T} \langle \text{d}T,\mathbf{H}\rangle \\
	& = \left[\mathbf{S}-2\rho\frac{\partial\hat\Psi}{\partial \mathbf{C}^\flat}\right]\!:\!\mathbf{D} 
	-\rho \frac{\partial\hat\Psi}{\partial \mathbf{G}}\!:\!\frac{\partial \mathbf{G}}{\partial T}  \,\dot{T}
	- \frac{1}{T} \langle \text{d}T,\mathbf{H}\rangle \geq 0  \,,
\end{aligned}
\end{equation}
%-----------------------------
where \eqref{Entropy-Temperature} was used in the second equality.
Thus
%-----------------------------
\begin{equation} 
	 \mathbf{S}=2\rho\frac{\partial\hat\Psi}{\partial \mathbf{C}^\flat}\,,\qquad
	 \dot\eta=  -\rho \frac{\partial\hat\Psi}{\partial \mathbf{G}}\!:\!\frac{\partial \mathbf{G}}{\partial T}  \,\dot{T}
	 - \frac{1}{T} \langle \text{d}T,\mathbf{H}\rangle \geq 0  \,.
\end{equation}
%-----------------------------

From \eqref{Energy-Free-Energy}, $\rho\,\dot{\mathcal{E}} = \rho\dot{\Psi} + \rho \dot{T} \mathcal{N}+T \rho\,\dot{\mathcal{N}}=\mathbf{S}\!:\!\mathbf{D}+ \rho T \dot{\mathcal{N}}$. Substituting this back into the energy balance equation \eqref{eq:MatEnergyLocv1}, one obtains
%-----------------------------
\begin{equation} \label{N-dot}
	\rho T \dot{\mathcal{N}} 
	+\rho\, \dot{T}\, \frac{\partial\hat\Psi}{\partial \mathbf{G}}\!:\!\frac{\partial \mathbf{G}}{\partial T} 
	= \rho R -\operatorname{Div} \mathbf{H} \,.
\end{equation}
%-----------------------------
Using \eqref{Entropy-Temperature}, one writes
%-----------------------------
\begin{equation} 
	\dot{\mathcal{N}} = - \dot{T} \frac{d}{d T}\frac{\partial \hat{\Psi}}{\partial T}  
	- \frac{\partial^2 \hat{\Psi}}{\partial \mathbf{C}^\flat \partial T}:\dot{\mathbf{C}}^\flat
	\,.
\end{equation}
%-----------------------------
Thus, \eqref{N-dot} is written as
%-----------------------------
\begin{equation} \label{N-dot-2}
	\rho \left[ -T \frac{d}{d T}\frac{\partial \hat{\Psi}}{\partial T} 
	+\frac{\partial\hat\Psi}{\partial \mathbf{G}}\!:\!\frac{\partial \mathbf{G}}{\partial T} \right]  \dot{T}
	-\rho T \frac{\partial^2 \hat{\Psi}}{\partial \mathbf{C}^\flat \partial T}\!:\!\dot{\mathbf{C}}^\flat
	= \rho R -\operatorname{Div} \mathbf{H} \,.
\end{equation}
%-----------------------------
Note that 
%-----------------------------
\begin{equation} 
	-T \frac{d}{d T}\frac{\partial \hat{\Psi}}{\partial T} 
	+\frac{\partial\hat\Psi}{\partial \mathbf{G}}\!:\!\frac{\partial \mathbf{G}}{\partial T} 
	= -T \frac{d^2 \hat{\Psi}}{d T^2}
	+T\frac{d}{dT}\left[ \frac{\partial\hat\Psi}{\partial \mathbf{G}}\!:\!\frac{\partial \mathbf{G}}{\partial T} \right]
	+\frac{\partial\hat\Psi}{\partial \mathbf{G}}\!:\!\frac{\partial \mathbf{G}}{\partial T} 
	= -T \frac{d^2 \hat{\Psi}}{d T^2}
	+\frac{d}{dT}\left[T \frac{\partial\hat\Psi}{\partial \mathbf{G}}\!:\!\frac{\partial \mathbf{G}}{\partial T} \right]
	 \,.
\end{equation}
%-----------------------------
Hence, \eqref{N-dot-2} is simplified to read
%-----------------------------
\begin{equation} 
	\rho \left\{ -T \frac{d^2 \hat{\Psi}}{d T^2}
	+\frac{d}{dT}\left[T \frac{\partial\hat\Psi}{\partial \mathbf{G}}\!:\!\frac{\partial \mathbf{G}}{\partial T} \right]
	\right\}  \dot{T}
	+\operatorname{Div} \mathbf{H}
	= \rho R 
	+\rho T \frac{\partial^2 \hat{\Psi}}{\partial \mathbf{C}^\flat \partial T}:\dot{\mathbf{C}}^\flat \,.
\end{equation}
%-----------------------------
Therefore, the heat equation is written as
%-----------------------------
\begin{equation} \label{eq:MatHeatApp}
	\rho \,C_E \,\dot{T}   +\operatorname{Div} \mathbf{H} 
	=R + T\, \mathbf{M}\!:\! \mathbf{D} \,,
\end{equation}
%-----------------------------
where the specific heat capacity at constant strain $C_E$ and the referential thermal stress coefficient tensor $\mathbf{M}$ are given as 
%-----------------------------
\begin{equation}
	C_E	= -T \frac{d^2 \hat{\Psi}}{d T^2}
	+\frac{d}{dT}\left[T \frac{\partial\hat\Psi}{\partial \mathbf{G}}\!:\!\frac{\partial \mathbf{G}}{\partial T} \right]  \,,  
	\qquad 
	\mathbf{M} = 2 \rho \frac{\partial^2 \hat{\Psi}}{\partial \mathbf{C}^\flat \partial T}  \,.
\end{equation}
%-----------------------------
Here, the $2 \choose 0$-tensor $\mathbf{M}$ is also referred to as the stress-temperature moduli \citep{MaHu1983,truesdell2004non,holzapfel2002nonlinear,gurtin2010mechanics}.

\paragraph{Spatial heat equation.} \label{AppendixA2}
We next derive the spatial heat equation using the material heat equation \eqref{eq:MatHeatApp}.
Let $\psi(x,t)$ represent the specific free energy in the current configuration, with the corresponding constitutive response function denoted as $\psi(x,t)=\hat{\psi}(x,\mathcal{T}, \mathbf{g},\mathbf{c}^\flat)$. Since $\varphi_t(X)=x$, it follows that $\psi_t \circ \varphi_t= \Psi_t$. Moreover, the spatial and material response functions are related as 
%-----------------------------
\begin{equation}
	\hat{\psi}(x,\mathcal{T}, \mathbf{g},\mathbf{c}^\flat)
	=\hat{\Psi}\left(\varphi_t^{-1}(x),{\varphi_t}^* \mathcal{T},{\varphi_t}^* \mathbf{g}, {\varphi_t}^* \mathbf{c}^\flat\right)\,.
\end{equation}
%-----------------------------
Note that 
%-----------------------------
\begin{equation} 
	\frac{d \hat{\psi}}{d \mathcal{T}} = \frac{\partial\hat{\psi}}{\partial \mathcal{T}}
	+\frac{\partial\hat{\psi}}{\partial\mathbf{c}^\flat}\!:\!\frac{\partial \mathbf{c}^\flat}{\partial \mathcal{T}} 
	\,.
\end{equation}
%-----------------------------
The specific heat capacity at constant strain $c_E$ and the spatial thermal stress coefficient $\mathbf{m}$ in the current configuration are defined as 
%-----------------------------
\begin{equation}
	 	  c_E
	 	  = -T \frac{d^2 \hat{\psi}}{d \mathcal{T}^2}
	+\frac{d}{d\mathcal{T}}\left[\mathcal{T} \frac{\partial\hat\psi}{\partial \mathbf{c}^\flat}\!:\!\frac{\partial \mathbf{c}^\flat}{\partial \mathcal{T}} \right] \,,  
	 	  \quad \quad
	     \mathbf{m}
	     =2 \varrho \frac{\partial^2 \hat{\psi}}{\partial \mathbf{g}\,\partial \mathcal{T}}\,.
\end{equation}
%-----------------------------
Notice that $c_E= C_E$.\footnote{This is implied using $\frac{\partial\hat\psi}{\partial \mathcal{T}} = \frac{\partial \hat\Psi}{\partial  T }$, $\frac{d \hat\psi}{d \mathcal{T}} = \frac{d \hat\Psi}{d  T }$ and $\frac{\partial\hat\psi}{\partial\mathbf{c}^\flat}\!:\!\frac{\partial \mathbf{c}^\flat}{\partial \mathcal{T}}=\frac{\partial\hat\Psi}{\partial\mathbf{G} }\!:\!\frac{\partial \mathbf{G} }{\partial  T }$.  
} 
Further, since
%-----------------------------
\begin{equation} \label{eq:mMrelation}
	\frac{ \partial^2  \hat{\psi}}{ \partial g_{ab} \partial \mathcal{T}}
=  \frac{\partial C_{AB}}{\partial g_{ab}}\frac{\partial^2 \hat{\Psi}}{ \partial C_{AB} \partial T}
=  {F^a}_A {F^b}_B \frac{\partial^2 \hat{\Psi}}{ \partial C_{AB} \partial T}\,,
\end{equation}
%-----------------------------
the spatial and material thermal stress coefficients are related as
%-----------------------------
\begin{equation}
	J m^{ab}
	= 2 J \varrho\,\frac{ \partial^2  \hat{\psi}}{ \partial g_{ab} \partial \mathcal{T}}
	= 2 \rho \, {F^a}_A {F^b}_B \frac{\partial^2 \hat{\Psi}}{ \partial C_{AB} \partial T}
	= {F^a}_A {F^b}_B M^{AB} \,.
\end{equation}
%-----------------------------
Thus, $\mathbf{m}$ and $\mathbf{M}$ are related as $\mathbf{M}=J {\varphi_t} ^* \mathbf{m} $. Observe that\footnote{
The components of $\dot{\mathbf{C}}^\flat$ read 
%-----------------------------
\begin{equation}
	\dot{C}_{AB}
	= {F^a}_A {F^b}_B \frac{\partial g_{ab}}{\partial x^c} V^c
	+g_{ab} \left[ \frac{\partial V^b}{\partial X^B} {F^a}_A + \frac{\partial V^a}{\partial X^A} {F^b}_B  \right]
	=g_{ab} \left[ {V^b}_{|B} {F^a}_A + {V^a}_{|A} {F^b}_B  \right]\,,
\end{equation}
%-----------------------------
and the components of $\mathfrak{L}_\mathbf{v} \mathbf{g}$ are written as $(\mathfrak{L}_\mathbf{v} \mathbf{g})_{ab}= g_{ac} {v^c}_{|b} + g_{cb} {v^c}_{|a}$.
Since ${V^a}_{|B}= {v^a}_{|b} {F^b}_B $, the components of $\mathbf{D}= \frac{1}{2}\dot{\mathbf{C}}^\flat$ and $\mathbf{d}= \frac{1}{2}\mathfrak{L}_\mathbf{v} \mathbf{g}$ can be related as
%-----------------------------
\begin{equation} \label{eq:dDrelation}
	D_{AB}= d_{ab}\, {F^a}_A {F^b}_B \,,
\end{equation}
%-----------------------------
i.e., $\mathbf{D}=\varphi_t ^* \mathbf{d}$. Thus, \eqref{eq:mMrelation} and \eqref{eq:dDrelation} imply that $M^{AB}D_{AB}=J\, m^{ab}\,d_{ab}$.
}
%-----------------------------
\begin{equation} \label{eq:CouplingtermTransformation}
	   \mathbf{M} \!:\!  \mathbf{D} 
	  = J\, \mathbf{m} \!:\!  \mathbf{d}	 \,,
\end{equation}
%-----------------------------
where $\mathbf{d}= \frac{1}{2}\mathfrak{L}_\mathbf{v} \mathbf{g}$ is the Lie derivative of the spatial metric along the spatial velocity. 
Let $\mathcal{U}\subset \mathcal{B}_t$ ($\mathcal{U}\cap\partial\mathcal{B}_t=\emptyset$) and $\mathcal{P}_t=\varphi_t(\mathcal{U})$. Notice that $\partial\mathcal{P}_t=\varphi_t(\partial\mathcal{U})$. Using the divergence theorem in the deformed and the material manifolds, one has
%-----------------------------
\begin{equation} \label{eq:DivergenceTheorem}
	\int_{\mathcal{P}_t}  \operatorname{div}_\mathbf{g} \mathbf{h} \, \text{d}v 
	= \int_{\partial \mathcal{P}_t} \llangle \mathbf{h},\mathbf{n} \rrangle_{\mathbf{g}}
	\text{d}a \,, \quad \text{and} \qquad
	\int_{\partial \mathcal{U}} \llangle \mathbf{H},\mathbf{N} \rrangle_{\mathbf{G}}\text{d}A
	= \int_{\mathcal{U}} \operatorname{Div} \mathbf{H} \,\text{d}V\,.
\end{equation} 
%-----------------------------
Further, by the change of variables formula
%-----------------------------
\begin{equation} \label{eq:ChangeofVariable}
	 \int_{\partial \mathcal{P}_t} \llangle \mathbf{h},\mathbf{n} \rrangle_{\mathbf{g}}\text{d}a 
	=  \int_{\partial \mathcal{U}} \llangle \mathbf{H},\mathbf{N} \rrangle_{\mathbf{G}}\text{d}A.
\end{equation}
%-----------------------------
Hence, it is implied from \eqref {eq:MassCons}, \eqref{eq:CouplingtermTransformation},  \eqref{eq:DivergenceTheorem},  \eqref{eq:ChangeofVariable} and \eqref{eq:MatHeatApp} that\footnote{Here,
%-----------------------------
\begin{equation} 
\int_{\mathcal{P}_t}  {h^b}_{|b}  \text{d}v 
=  \int_{\partial \mathcal{P}_t} h^b n_b \text{d}a 
=  \int_{\partial \mathcal{U}} H^B N_B \text{d}A
= \int_{\mathcal{U}}  {H^B}_{|B}  \text{d}V\,,
\end{equation}
%----------------------------- 
and $\int_{\mathcal{U}}  R \text{d}V =\int_{\mathcal{P}_t}  r  \text{d}v$.
}
%-----------------------------
\begin{equation} \label{eq:ChangeofVariableHeatEquation}
	  \int_{\mathcal{P}_t} 
	  \left[\varrho \, c_E \dot{\mathcal{T}}
	  +\operatorname{div}_\mathbf{g} \mathbf{h} 
	  - \mathcal{T}  \mathbf{m} \!:\!  \mathbf{d} 
	  - r \right]
	  \text{d}v
	  =\int_{\mathcal{U}} 
	  \left[\rho \, C_E \dot{T}
	  +\operatorname{Div} \mathbf{H} 
	  - T \mathbf{M} \!:\!  \mathbf{D} 
	  - R \right]
	  \text{d}V
	  =0\,,
\end{equation}
%-----------------------------
which holds for an arbitrary sub-body $\mathcal{P}_t$. Therefore, the localized spatial heat equation reads
%-----------------------------
\begin{equation}
    \varrho \, c_E \dot{\mathcal{T}}
	  +\operatorname{div}_\mathbf{g} \mathbf{h} 
	  = \mathcal{T}  \mathbf{m} \!:\!  \mathbf{d} 
	  + r \,.
\end{equation}
%-----------------------------
Notice that $(\,)\dot{}=\frac{\partial}{\partial t}\big|_{X}(\,)$ represents the material time derivative.

%-----------------------------
%-----------------------------
\section{A constitutive model for thermoelastic solids} \label{App:ConstitutiveSolid}

We consider the following constitutive model for thermoelastic solids
%-----------------------------
\begin{equation}
	\check{\Psi}_{\text{s}}(I_1,J,T)
	= \left[ \frac{\mu^{\text{s}}_0}{2} (J^{-\frac{2}{3} }I_1 -3)
	+\frac{\kappa^\text{s}_0}{2} (J-1)^2
	\right]\frac{T}{T^{\text{s}}_0}
	-\kappa^{\text{s}}_0\,\beta^{\text{s}}_0\, (J-1)(T-T^{\text{s}}_0)
	-\rho \int_{T^{\text{s}}_0}^T \frac{T-\tau}{\tau}c_E(\tau)\, \text{d}\tau \,.
\end{equation}
%-----------------------------
Here, $\mu^{\text{s}}_0$, $\kappa^{\text{s}}_0$, and $\beta^{\text{s}}_0$ represent the shear modulus, bulk modulus, and volumetric thermal expansion coefficient of the solid at the reference temperature $T^{\text{s}}_0$ \citep{ogden1992thermoelastic, holzapfel1996entropy, Sadik2017Thermoelasticity}.\footnote{Recall that, in the thermally accreted part of the body, $T_0(X)$ represents the temperature during accretion, while in the initial body, it denotes the initial temperature.} 
The shear and the bulk moduli are assumed to evolve linearly with temperature, i.e.,
%-----------------------------
\begin{equation}
	\mu^{\text{s}}(T)=\frac{\mu_0^{\text{s}} T}{T^{\text{s}}_0} \,, \qquad
	\kappa^{\text{s}}(T)=\frac{\kappa_0^{\text{s}} T}{T^{\text{s}}_0} \,.
\end{equation}
%-----------------------------
Consider a homogeneous body initially at a uniform temperature $T_0$, which, when changed to another uniform temperature $T_1$, undergoes a stress-free volumetric deformation in the process. The state of stress for a purely volumetric deformation is quantified by $\sigma=\frac{1}{3} \operatorname{tr}\bm{\sigma}=\frac{\partial \check{\Psi}_{\text{s}}}{\partial J}$. Therefore, one has $\frac{\partial\check{\Psi}_{\text{s}}}{\partial J}=\kappa^\text{s}_0$, and hence it follows that 
%-----------------------------
\begin{equation}
	J=1+\beta^{\text{s}}_0 T^{\text{s}}_0 \left[1-\frac{T^{\text{s}}_0}{T}\right] \,.
\end{equation}
%-----------------------------
Further, since the Jacobian in such a process is $J=e^{\beta^\text{s}(T)}$, it is implied that 
%-----------------------------
\begin{equation} \label{eq:TempDepofOmega}
	e^{3 \omega^\text{s}(T) }
	= 1+ \beta^\text{s}_0 T^\text{s}_0\left[1-\frac{T^\text{s}_0}{T}\right]\,.
\end{equation}
%-----------------------------
Thus, the coefficient of thermal expansion at temperature $T$ is written as
%-----------------------------
\begin{equation} \label{eq:TempDepofBeta}
	\beta^\text{s}(T)
	= \frac{\beta^\text{s}_0 \left[\frac{T^\text{s}_0}{T}\right]^2}{1+ \beta^\text{s}_0 T^\text{s}_0\left[1-\frac{T^\text{s}_0}{T}\right]}\,,
\end{equation}
%-----------------------------
where the relation $\beta^\text{s} = 3\frac{\text{d}\omega^\text{s} }{\text{d}T}$ has been used. We shall use this model in our numerical examples. For more details, see \citep{Sadik2017Thermoelasticity}.

%-----------------------------
%-----------------------------
\section{Christoffel symbols of the spatial and material metrics}  \label{Sec:ChristoffelSymbols}

The nonzero Christoffel symbols for the spatial metric $\mathbf{g}$ \eqref{g-metric} read
%-----------------------------
\begin{equation} \label{eq:Christoffel_g_metric}
	\gamma^r_{\theta\theta}
	= -r \,,\qquad
	\gamma^r_{\phi\phi}
	= -r\sin^2 \theta  \,,\qquad
	\gamma^\theta_{r\theta}
	=\gamma^\phi_{r\phi}
	=\frac{1}{r}  \,,\qquad
	\gamma^\theta_{\phi \phi}
	=  -\sin\theta \cos\theta \,,\qquad
	\gamma^\phi_{\phi \theta}
	=\cot \theta\,.
\end{equation}
%-----------------------------
The nonzero Christoffel symbols for the material metric $\GL$ \eqref{eq:LiquidMatMetric} are listed as
%-----------------------------
\begin{equation}
\begin{aligned} 
     &\accentset{\circ}{\Gamma}^R_{RR} 
     =\alpha^\text{f} \, T_{,R}  \,, \qquad
	\accentset{\circ}{\Gamma}^R_{\Theta\Theta}
	= -\left[ \alpha^\text{f} \, T_{,R} +  \frac{1}{R}  \right] R^2   \,, \qquad
	\accentset{\circ}{\Gamma}^R_{\Phi\Phi}
	=  -\left[ \alpha^\text{f}\, T_{,R} +  \frac{1 }{R}  \right] R^2  \sin^2 \Theta \,, \\
	&\accentset{\circ}{\Gamma}^\Theta_{R\Theta}
	=\accentset{\circ}{\Gamma}^\Phi_{R\Phi}
	=  \alpha^\text{f}\, T_{,R} +  \frac{1}{R}    \,,\qquad
	\accentset{\circ}{\Gamma}^\Theta_{\Phi \Phi}
	= - \sin \Theta \cos \Theta   \,, \qquad
	\accentset{\circ}{\Gamma}^\Phi_{\Phi \Theta}
	= \cot \Theta\,,
\end{aligned}
\end{equation}
%-----------------------------
where $\alpha^\text{f}(T)=\frac{\text{d}\omega^\text{f}(T)}{\text{d}T}$. Here we have used the notation $(\cdot)':= \frac{\text{d}}{\text{d} R}(\cdot)$. 
Similarly, the nonzero Christoffel symbols for the material metric $\GS$ \eqref{eq:SolidMatMetric} are 
%-----------------------------
\begin{equation} \label{eq:Christoffel_Gsolid_metric}
\begin{aligned} 
     & \Gamma^R_{RR} 
     =\alpha^\text{s} \, T_{,R} +  \frac{\varsigma' }{\varsigma}  \,, \qquad
	\Gamma^R_{\Theta\Theta}
	= -\left[ \alpha^\text{s} \, T_{,R} +  \frac{\bar{r}\,' }{\bar{r}}  \right] \frac{\bar{r}^2 }{\varsigma^2}  \,, \qquad
	\Gamma^R_{\Phi\Phi}
	=  -\left[ \alpha^\text{s}\, T_{,R} +  \frac{\bar{r}\,' }{\bar{r}}  \right]\frac{\bar{r}^2 }{\varsigma^2} \sin^2 \Theta \,, \\
	& \Gamma^\Theta_{R\Theta}
	=\Gamma^\Phi_{R\Phi}
	=  \alpha^\text{s}\, T_{,R} +  \frac{\bar{r}\,'}{\bar{r}}    \,,\qquad
	\Gamma^\Theta_{\Phi \Phi}
	= - \sin \Theta \cos \Theta   \,, \qquad
	\Gamma^\Phi_{\Phi \Theta}
	= \cot \Theta\,,
\end{aligned}
\end{equation}
%-----------------------------
where $\alpha^\text{s}(T)=\frac{\text{d}\omega^\text{s}(T)}{\text{d}T}$ and $\varsigma(R)=\frac{\bar{u}(R) \eta^2(R)}{\bar{U}(R)}$.

%-----------------------------
%-----------------------------
\section{Constitutive equations for hyperelastic fluids} \label{App:Hyperelasticfluid}

The constitutive equation of an elastic fluid explicitly depends only on the mass density \citep{truesdell2004non,gurtin2010mechanics}.
The specific free energy function $\psi(X,T,\mathbf{F},\mathbf{G},\mathbf{g})$ for hyperelastic fluids can be expressed as a function of $J=\sqrt{\frac{\det\mathbf{g}}{\det\mathbf{G}}}\det \mathbf{F}$ \cite[p.~198]{wang1973introduction}, such that $\psi(X,T,\mathbf{F},\mathbf{G},\mathbf{g})=\hat{\psi}(X,T,J)$. Using this function, the Cauchy, the first and the second Piola-Kirchhoff stress tensors can be expressed as
%-----------------------------
\begin{equation}
	\bm{\sigma}
	= -\hat{p} \,\mathbf{g}^\sharp  
\,,\qquad 
	\mathbf{P}
	=-J \hat{p} \,\mathbf{g}^\sharp \mathbf{F}^{-\star}
\,,\qquad 
	\mathbf{S}
	=J \hat{p} \,\mathbf{F}^{-1}\mathbf{g}^\sharp \mathbf{F}^{-\star}\,,
\end{equation}
%-----------------------------
where $\hat{p}=-\frac{\partial\hat{\psi}}{\partial J}$. Since hydrostatic stresses are compressive in fluids, one must have $\frac{\partial\hat{\psi}}{\partial J}<0$. Note that
%-----------------------------
\begin{equation}
	(\operatorname{div}_\mathbf{g}\bm{\sigma})^a
	= -\left(\hat{p}\, g^{ab}\right)_{|b}
	=-g^{ab}\frac{\partial \hat{p}}{\partial x^b}
	= \left[\frac{\partial^2 \hat{\psi}}{\partial J^2} \frac{\partial J}{\partial x^b} 
	+\frac{\partial^2 \hat{\psi}}{\partial J \partial T} \frac{\partial T}{\partial x^b}
	+ \frac{\partial^2 \hat{\psi}}{\partial X^A \partial J} (F^{-1})^A{}_b \right]\,g^{ab} \,.
\end{equation}
%-----------------------------
For homogeneous fluids, $\hat{\psi}$ is independent of $X$, and hence, the term $\frac{\partial^2 \hat{\psi}}{\partial X^A \partial J}$ vanishes. In the absence of any heat flow, one can simply work with the smooth and strictly convex energy function $\breve{W}(J)$, with the property that $\lim_{J\rightarrow 0^+} \breve{W} = +\infty$. Since $(\operatorname{div}_\mathbf{g}\bm{\sigma})^a= \breve{W}''(J) \frac{\partial J}{\partial x^b} \,g^{ab} $, the balance of linear momentum reads
%-----------------------------
\begin{equation}
	\breve{W}''(J) \frac{\partial J}{\partial x^b} \,g^{bc}+ f^c=a^c \,,
\end{equation}
%-----------------------------
where $\breve{W}''= \frac{\text{d}^2 \breve{W}}{\text{d} J^2}$. Moreover, in the absence of inertial and body forces, one can assume that $\breve{W}''(J)> 0$. Thus, one concludes that $\frac{\partial J}{\partial x^b}=0$. Hence, in the absence of body and inertial forces, $J=J(t)$ in homogeneous fluids at constant temperature \citep{podio1985cavitation}.  For example, \citet{ghosh2022elastomers} considered the following energy function for hyperelastic liquid inclusions
%-----------------------------
\begin{equation}
	\breve{W}(X,J) =  J  \eta(X) +  \frac{\kappa}{2} [J-1]^2\,.
\end{equation}
%-----------------------------
It is implied that $\bm{\sigma}= \big[\eta  +  \kappa (J-1)\big] \mathbf{g}^\sharp$.\footnote{A similar constitutive relation for thermoelastic fluids has been proposed in \citep{ateshian2022continuum}.} Here, $\eta(X)$ represents the pressure in the undeformed liquid (which is not necessarily zero) while the bulk modulus $\kappa$ quantifies compressibility.
%-----------------------------

In the fluid mechanics literature, a compressible hyperelastic fluid is analogous to an unconstrained elastic fluid, which is characterized by the constitutive relation $\bm{\sigma}=-\mathcal{P}(\varrho)\, \mathbf{g}^\sharp$. It is known by several other names, such as, Euler fluid, ideal compressible fluid, perfect compressible fluid, inviscid compressible fluid, and barotropic fluid \citep[p.~44]{truesdell2000introduction}.

\end{document}